\documentclass[3p,singlecolumn,times,10pt]{elsarticle}

\usepackage{bm}
\usepackage{epsfig,subfigure}
\usepackage{latexsym,amssymb,amsmath,amsfonts} %
\usepackage[english]{babel} 
\usepackage[OT1]{fontenc}
\usepackage[colorlinks=true,linkcolor=black]{hyperref}
\usepackage{color,graphicx,graphics,wrapfig,epsf}
\usepackage{calligra}

\usepackage{ifpdf}
\usepackage{dcolumn}
\usepackage{graphicx}
\usepackage[numbers]{natbib}

\definecolor{red}{rgb}{1,0,0}

\def\p{\partial}

\def\+{^\dagger}

\def\<{\leftarrow}
\def\>{\rightarrow}

\def\({\left(}
\def\){\right)}

\def\a{\alpha} \def\b{\beta}   
\def\m{\mu} \def\n{\nu}   \def\l{\lambda} 
\def\D{\Delta}\def\G{\Gamma}

\newcommand{\benu}{\begin{enumerate}} 		\newcommand{\enu}{\end{enumerate}}
\newcommand{\bd}{\begin{dinglist}{0}}     \newcommand{\ed}{\end{dinglist}}
\newcommand{\bfig}{\begin{figure}[htbp]}  \newcommand{\efig}{\end{figure}}
        			
\newcommand{\bc}{\begin{center}} 				  \newcommand{\ec}{\end{center}}
\newcommand{\be}{\begin{equation}} 				\newcommand{\ee}{\end{equation}}
\newcommand{\bsub}{\begin{subequations}}  \newcommand{\esub}{\end{subequations}}
\newcommand{\ben}{\begin{eqnarray}} 			\newcommand{\een}{\end{eqnarray}}
\newcommand{\ba}[1]{\begin{array}{#1}} 		\newcommand{\ea}{\end{array}}
\newcommand{\bea}{\begin{equation}\begin{array}{rcl}}
\newcommand{\eea}{\end{array}\end{equation}}

\begin{document}
\title{Stellar structure models in modified theories of gravity: lessons and challenges}

\author{Gonzalo J. Olmo}
\address{Depto. de F\'isica Te\'orica and IFIC, Centro Mixto Universidad de Valencia-CSIC, \\ Burjassot-46100, Valencia, Spain.}
\address{Departamento de F\'isica, Universidade Federal da Para\'iba, 58051-900 Jo\~ao Pessoa, Para\'iba, Brazil.}
\ead{gonzalo.olmo@uv.es}

\author{Diego Rubiera-Garcia}
\address{Departamento de F\'isica Te\'orica and IPARCOS, Universidad Complutense de Madrid, E-28040
Madrid, Spain}
\ead{drubiera@ucm.es}

\author{Aneta Wojnar}
\address{N{\'u}cleo Cosmo-ufes \& PPGCosmo, Universidade Federal do Esp{\'i}rito Santo,
29075-910, Vit{\'o}ria, ES, Brasil}
\address{Laboratory of Theoretical Physics, Institute of Physics, University of Tartu, W. Ostwaldi 1, 50411 Tartu, Estonia}

\ead{aneta.wojnar@cosmo-ufes.org}

\date{\today}

\begin{abstract}
The understanding of stellar structure represents the crossroads of our theories of the nuclear force and the gravitational interaction under the most extreme conditions observably accessible. It provides a powerful probe of the strong field regime of General Relativity, and opens fruitful avenues for the exploration of new gravitational physics. The latter can be captured via modified theories of gravity, which modify the Einstein-Hilbert action of General Relativity and/or some of its principles. These theories typically change the Tolman-Oppenheimer-Volkoff equations of stellar's hydrostatic equilibrium, thus having a large impact on the astrophysical properties of the corresponding stars and opening a new window to constrain these theories with present and future observations of different types of stars. For relativistic stars, such as neutron stars, the uncertainty on the equation of state of matter at supranuclear densities intertwines with the new parameters coming from the modified gravity side, providing a whole new phenomenology for the typical predictions of stellar structure models, such as mass-radius relations, maximum masses, or moment of inertia. For non-relativistic stars, such as white, brown and red dwarfs, the weakening/strengthening of the gravitational force inside astrophysical bodies via the modified Newtonian (Poisson) equation may induce changes on the star's mass, radius, central density or luminosity, having an impact, for instance, in the Chandrasekhar's limit for white dwarfs, or in the minimum mass for stable hydrogen burning in high-mass brown dwarfs. This work aims to provide a broad overview of the main such results achieved in the recent literature for many such modified theories of gravity, by combining the results and constraints obtained from the analysis of relativistic and non-relativistic stars in different scenarios. Moreover, we will build a bridge between the efforts of the community working on different theories, formulations, types of stars, theoretical modellings, and observational aspects, highlighting some of the most promising opportunities in the field.

\end{abstract}
\maketitle

\tableofcontents


\section{Preamble}

\subsection{Introduction} \vspace{0.2cm}

\subsubsection{Astrophysical phenomenology beyond General Relativity} \vspace{0.2cm}

More than one hundred years after its formulation, Einstein's General Theory of Relativity (GR) still stands as an extremely successful theory of the gravitational interaction. It has passed tests ranging from submilimiter to solar system scales \cite{Will:2014kxa}, has been built in the concordance cosmological model \cite{Bull:2015stt}, and has received further confirmation from the recent detection of gravitational waves consistently interpreted as emitted out of binary mergers \cite{Barack:2018yly}, and by the imaging of the shadow of the supermassive object at the center of the M87 galaxy assumed to be a Kerr black hole \cite{Akiyama:2019eap}. Despite this, in the last decades there has been an enormous effort invested by the gravitational community at large to find alternatives to this framework following different directions. This effort is motivated by a plethora of reasons. On the theoretical side one finds the need for an ultraviolet completion of GR, seen as an effective theory \cite{Burgess:2003jk}, or the studies aimed at the resolution of space-time singularities deep inside black holes and in the early universe (see \cite{Senovilla:2014gza} for a pedagogical and historical discussion of this issue). On the phenomenological side, alternatives to the concordance model without dark sources has become a rich reservoir of proposals for extending GR, many of which have been severely constrained by the finding of the binary neutron star merger GW170817 \cite{TheLIGOScientific:2017qsa} and its electromagnetic counterpart GRB170817 \cite{Monitor:2017mdv}. Indeed, the reported constraint $\vert c_{T}/c -1 \vert \leq 1 \times 10^{-15}$  on the speed of propagation of gravitational waves as compared to the speed of light, has put into trouble many such proposals, see e.g. \cite{Ezquiaga:2017ekz} for a insightful analysis of this question.

Among the many different scenarios proposed in the literature to study gravitational phenomenology beyond GR, the one of stellar structure models represents both a powerful probe for testing gravity in its strong field regime \cite{Cottam:2007cd} and a promising avenue to find eventual deviations from GR predictions. The main target of this analysis are compact stars such as neutron stars and white dwarfs, which represent the end-state of stellar evolution of main-sequence stars, but also other types of stars such as brown and red dwarfs. Moreover, along the decades other stellar objects have appeared in the literature, such as hyperon stars \cite{Schaffner:1995th}, quark/hybrid stars \cite{Alford:2004pf,Baym:2017whm}, strange stars \cite{Alcock:1986hz}, and even more exotic objects such as boson stars \cite{Jetzer:1991jr}.

\subsubsection{Relativistic stars: neutron stars}  \vspace{0.2cm}

Neutron stars represent the most extreme packing of nuclear matter known in Nature, being the smallest and densest stars known so far. By a neutron star we roughly refer to an astrophysical object whose innermost region may reach from five to ten times the nuclear saturation density $\rho_s \approx 2.8 \times 10^{14}$gr/cm$^3$, its mass is of order $M \sim 1.4-2.0M_{\odot}$, and its radius lies in the range $R \sim 10-13$km. Though their structural properties can be roughly understood by assuming that they are composed of neutrons (hence their name) supported by their degeneracy pressure,  other baryonic components such as pions, kaons, meson condensates, and other hyperons may be present, forming a soup of such components in beta equilibrium and with charge neutrality \cite{STBook}. Moreover, neutron stars are expected to exhibit different phases of superfluidity/superconductivity \cite{Lombardo:2000ec}. All these ingredients together make neutron stars extremely hard objects to crack \cite{Lattimer:2004pg}. Indeed, as opposed to other kinds of stars where Newtonian gravity and well known nuclear laboratory physics suffice to study their structure and dynamics, neutron stars are truly relativistic objects. From a theoretical point of view one generally assumes the gravitational dynamics to be fully described by GR. Under this starting point, one of the main challenges of conventional astrophysics is the determination of the equation of state (EOS), that is, the relation between density and pressure within the different layers of a neutron star. Given the impossibility of direct access to such interiors, the physical reliability of any stellar structure model is determined via the comparison of suitable astrophysical predictions of such models with available observations. In implementing this program, one immediately faces the problem of the degeneracy of the EOS due to the inherent difficulties on the extrapolation of the behaviour of nuclear matter at the supranuclear densities reigning at the center of these stars, which are far from any conditions that can be achieved at terrestrial experiments. Among the many difficulties related to this extrapolation one finds that, at large enough densities, it is energetically favourable for some of the nucleons to combine to form hyperons which, yielding softer EOS, would reduce the maximum neutron mass thus producing a tension between nuclear physics and astronomical observations of neutrons in what is known as the hyperon puzzle (see \cite{Chatterjee:2015pua} for a recent discussion on this point). Further difficulties for the modelling include the formation of  ordered (crystalline) structures \cite{Glendenning:2001pe}, strong hadron-quark phase transitions \cite{Chatziioannou:2019yko,Bauswein:2018bma}, or deconfinement of quarks (quark stars). The combination and modelling of all these elements from the inner core outwards explain the broad range of models available in the market, each of them with their own set of macroscopic predictions. On the optimistic end, all these elements make neutron stars particularly suitable to test our theories about the interaction of gravity and matter at the highest densities achievable. A better knowledge of their populations, mass-radius relations, maximum masses, oscillations, moment of inertia, tidal deformability, etc, could help constrain numerous models of nuclear matter composition and also our theories of the gravitational interaction under the most extreme conditions.

Currently, the catalog of measured neutron masses upon observation of the timing of binary pulsar systems in datasets from X-ray  emissions at their surfaces covers dozens of members \cite{massNS}. The main conclusion is that, as measured for instance by Shapiro's delay method, neutron stars can be (at least) as massive as two solar masses, as in the pulsars PSR J0348+0432, with  $2.01 \pm 0.04 M_{\odot}$ \cite{J0348+0432} and PSR J1614-2230, with $1.97 \pm 0.04 M_{\odot}$ \cite{Crawford:2006xb}. Moreover, there are recent hints of even larger masses, such as in  PSR J2215+5135 with $2.27_{-0.15}^{+0.17}M_{\odot}$ \cite{LinShaCas18} and MSP J0740+6620 with  $2.17_{-0.09}^{+0.10}M_{\odot}$ \cite{Cromartie:2019kug}. The observational determination of neutron star radii has proved to be a more daunting task since it depends on many elements, such as the distance to the star, the uncertainties in the modelling of the crust \cite{Gamba:2019kwu}, the absorption of the interstellar medium, etc. \cite{Lattimer:2004nj}. Indeed, the measurement of neutron stars's radii is the target of new generations of X-ray missions \cite{Ozel:2016oaf}, which have constrained it to the range $R \approx 9.9-11.2$km for more than a dozen members. Unfortunately, simultaneous measurements of mass and radius for the same neutron star, which would allow to place constraints upon the EOS (and/or the gravity theory \cite{Eksi:2014wia}), are hard to achieve \cite{Miller:2014aaa,Psaltis:2013fha}. There are further observational weapons available to this challenge, such as the measurement of neutron stars' moment of inertia \cite{Raithel:2016vtt}, or as the analysis of gravitational wave and gamma ray burst data from binary neutron star mergers \cite{Abbott:2018exr,Bauswein:2017vtn,Ruiz:2017due,Shibata:2019ctb}. The latter allows to place constraints on tidal deformability \cite{Abbott:2018wiz} (and to infer the moment of inertia \cite{Landry:2018jyg}), on the EOS at supranuclear densities \cite{Wang:2018nye,Carney:2018sdv,Biswas:2019gkw,Vivanco:2019qnt,Essick:2019ldf} (for a recent review see \cite{Llanes-Estrada:2019wmz}), and on further properties \cite{Kumar:2019xgp}. Conversely, the combined measurements of radii and tidal deformability  can also place constraints on the EOS \cite{Fasano:2019zwm}. On the other hand, neutron star compactness can be determined via the measurement of gravitational redshift of the spectral lines emitted from the surface of the star \cite{Cottam:2002cu} (though it is subject to heavy observational limitations \cite{Cottam:2007cd}), or via quantities that depend on the star's tidal deformability, see e.g. \cite{Hinderer:2007mb,Paschalidis:2017qmb}. For some reviews of neutron stars' structure and their associated astrophysical and observational properties see \cite{Lattimer:2006xb} and the more recent book  \cite{Rezzollabook}.

The measurement of the mass of several neutron stars around and above $2M_{\odot}$ has introduced some tension between several realistic EOS and their corresponding predictions, ruling out many of the soft ones (mainly those including hyperons). The consideration of theories of gravity extending GR might help to alleviate/solve this problem via the additional gravitational corrections introduced in the corresponding hydrostatic equilibrium equations of these theories. Therefore, an immediate test of such gravitational proposals is to provide new mass-radius relations and maximum masses which are observationally accessible. Further suitable quantities in these models, which are harder to access to, are the moment of inertia and higher multipole moments, tidal deformability and Love numbers, etc. The consistency with observations of the predictions for each of these features can thus place constraints upon the space parameters of these theories and how well they fare as compared to GR predictions \cite{Berti:2015itd}. Unfortunately, as these gravitational theories typically introduce (from one to many) extra parameters and/or extra fields, the degeneracy problem inherent to the EOS of stellar structure in GR may actually worsen \cite{Shao:2019gjj}, which in turn makes neutron stars as an interesting as a dirty playground to test modifications of GR.

\subsubsection{Non-relativistic stars: white, brown, and red dwarfs} \vspace{0.2cm}

Beyond neutron stars, there exists a family of objects with better known chemical structure and thermodynamic properties which provide useful and complementary tests for modified theories of gravity. They are dwarf stars: low-mass stellar objects. Among them we find white dwarfs, which are the end-state of main-sequence stars whose masses are below the Chandrasekhar's  bound \cite{Chandra}, halting their collapse through the electron degeneracy pressure. On the other hand, brown dwarfs encompass a large variety of sub-stellar objects lying beyond the lower end of the main sequence in the Hertzsprung-Russell diagram with masses in the range $\sim 0.001-0.01M_{\odot}$ and central densities $\rho_c \sim 10-10^3$ gr/cm$^3$, mainly composed of molecular hydrogen and helium and which, not being massive enough to ignite stable thermonuclear reactions to fully compensate photospheric losses, eventually cool down into oblivion \cite{Burrows:1992fg,Burrows2}. Finally, red dwarfs are the most common star type in our galaxy, with a mass around $\sim 0.09-0.5M_{\odot}$  and a surface temperature $<4000$K. They are generated in a self-gravitational contraction process which is halted at the onset of thermonuclear fusion. These families of stars become valuable sources of information on modified theories of gravity since, being non-relativistic objects described by modifications of the Poisson equation, can be well modelled by polytropic EOS (i.e. simple polynomial relations between density and pressure), thus being more weakly depending on non-gravitational physics. In this sense, the weakening/strengthening of the gravitational force within matter sources predicted by many such theories yields significant deviations with respect to GR predictions on many aspects: masses, radius, luminosity, Chandrasekhar's limit (for white dwarfs) or the minimum hydrogen burning mass (for high-mass brown dwarfs), thus allowing to provide independent constraints on the additional parameters of modified gravity.

\subsubsection{This work: aim and limitations} \vspace{0.2cm}

The aim of this work is to review the current state-of-the-art within models of stellar structure of different modified theories of gravity, and discuss their corresponding predictions for individual stars, highlighting the lessons learned so far and the open challenges in the field. By modified gravity we mean any gravitational theory departing from GR either at the level of the action (for instance, via higher-order corrections in curvature scalars), at the level of the geometrical structure of the theory (for instance, adding non-metricity and/or torsion and/or other extra fields), or at the level of the fundamental principles (for instance, dropping Lorentz invariance).  We will try to build a bridge between the efforts of the community working in different modified theories of gravity, mathematical formalisms, types of stars, theoretical modelings, and observational aspects. We will not enter into the discussion of technical aspects of the numerical codes, focusing instead on the most relevant physical outcomes. In doing so, a strategic decision made in this work is to distinguish between the gravitational models formulated in a metric approach, where the affine connection is assumed to be given by the Christoffel symbols of the metric, the latter governing the system of modified gravitational field equations; and those formulated in the metric-affine (or Palatini) approach, where metric and connection are regarded as independent entities whose explicit forms are determined via a variational principle and resolution of its corresponding field equations. Each formulation has its own section devoted to the summary of the results achieved on it. Some of the gravitational models discussed here have been investigated in both formulations, and we review their phenomenology independently. For others, only one of the formulations is \emph{theoretically viable}, which we understand as being free of ghost-like or any other essential instabilities. Since the pool of theories studied in the literature is far larger in the metric approach than in the metric-affine one, in the former we shall introduce any suitable topic of research first before describing the different theories considered for it, while in the latter we shall take the opposite strategy by introducing first the few families of models considered in the literature and then go to describe the different results for each of them. The hybrid formulation will be included in the metric-affine one since the corresponding results are quite scarce. Our work will discuss both relativistic (neutron stars) and non-relativistic stars (white, brown and red dwarfs), and combine the results and constraints for the corresponding theory of gravity they are derived from.

A warning should be issued here: this is not a review on modified gravity \emph{per se}, meaning that we will not attempt to describe each gravitational theory in detail, but rather introduce their fundamental building blocks and assumptions, before heading right away to their predictions for the modifications to stellar structure description and their macroscopic signatures. For a broader and more detailed description of modified theories of gravity and their applications in different contexts the interested reader may profit from reading the following reviews: de Felice and Tsujikawa on $f(R)$ theories \cite{DeFelice:2010aj}, Sakstein on astrophysics of scalar-tensor theories \cite{Sakstein:2018fwz}, Capozziello and de Laurentis on many aspects of the theoretical foundations and applications of such theories \cite{CLreview}, Clifton et al. \cite{Clifton:2011jh}, Nojiri et al. \cite{Nojiri:2017ncd}, and Joyce et al. \cite{Joyce:2014kja} on cosmological aspects, Berti et al. on strong field modifications of GR and the dynamics of compact objects in these theories \cite{Berti:2015itd}, Beltran Jimenez et al. for Born-Infeld inspired modifications of gravity \cite{BeltranJimenez:2017doy}, and Heisenberg for theories based on additional scalar, vector, and tensor fields and their cosmological implications \cite{Heisenberg:2018vsk}. Nowadays, many such modified theories of gravity are heavily constrained from multimessenger  astronomy \cite{Ezquiaga:2018btd}.

As with any other work of this kind, though the selection of topics, the different gravitational models included herewith, and the depth of analysis carried out for each of them, has been done as fair as possible, it cannot but to reflect the bias of the authors regarding their favourite theories,  models, and topics. Any failure to properly include relevant literature or to give due author's credit can be only blamed on us, and we deeply apologize to anyone that feels his/her contribution to this subject has been unfairly overlooked and/or misrepresented. In this sense, we welcome any comment from readers to improve/enlarge/supplement the material presented here.

\subsection*{Notation} \vspace{0.2cm}

Unless explicitly stated we shall work in units $G=c=1$. In these units, $\kappa^2=8\pi$. Radial derivatives upon density or pressure will be denoted by primes or by a subindex $_r$, depending on the context. Quantities evaluated at the star center are denoted by a subindex $_c$, while the star surface carries the subindex $_S$ (for instance, $r_S$ denotes its radius). $M_{\star}$ denotes the absolute maximum value of the mass for a given EOS, usually expressed in units of solar's mass, $M_{\odot} \approx 1.9885 \times 10^{30}$kg, while $R_{\odot} \approx 6.957 \times 10^{5}$km is the solar radius. Curvature will be denoted by $R$ when in metric formalism, and by $\mathcal{R}$ in metric-affine formalism. We work in the mostly $+$ signature: $(-,+,+,+)$. Other suitable notations will be introduced when needed.

\subsection{Stellar structure models in General Relativity} \label{sec:stGR} \vspace{0.2cm}

General Relativity is based on a pseudo-Riemannian manifold equipped with a metric $g_{\mu\nu}$, which is a symmetric $4\times 4$ tensor, and an affine connection $\Gamma_{\mu\nu}^{\lambda}$, which does not transform as a tensor (though its antisymmetric part, called torsion, does). While the metric is related to local measurements in space (rods and clocks) thus defining notions of distances, angles, volumes, etc, the connection is linked to parallel transport and thus it is built on covariant derivatives. Despite being two mathematically and logically independent entities, in GR one traditionally assumes (in absence of torsion) that the affine connection is metric-compatible, namely\footnote{GR can be equivalently formulated in metric-affine spaces by regarding the affine connection to be independent of the metric (metric-affine or Palatini approach), and determined via a variational principle. In this case, for minimally coupled matter fields the corresponding equation fixes it to be given by the Levi-Civita connection plus a projective mode, $\Gamma_{\mu\nu}^{\lambda}= \{ _{\mu\nu}^{\lambda} \} + \xi_{\mu} \delta_{\nu}^{\lambda}$, the latter having no impact on field equations, solutions, or trajectories of test particles, and the predictions of the resulting theory is completely equivalent to metric GR. See however \cite{Bejarano:2019zco} for a critical discussion of the interpretation of some properties of specific solutions on both formalisms.},
\begin{equation}\label{eq:mccond}
\nabla_{\mu}^{\Gamma}(\sqrt{-g} g^{\alpha\beta})=0
\end{equation}
which means that the affine connection $\Gamma=\Gamma_{\mu\nu}^{\lambda}$ can be expressed as the Christoffel symbols of $g_{\mu\nu}$, that is, $\Gamma^\alpha_{\beta\gamma}\equiv \frac{g^{\alpha\rho}}{2}\left[\partial_\beta g_{\rho\gamma}+\partial_\gamma g_{\rho\beta}-\partial_\rho g_{\beta\gamma}\right]$. In this way, the dynamics of the theory is fully encapsulated by the metric. A variational principle can then be established from the  Einstein-Hilbert action (minimally) coupled to some matter fields
\begin{equation} \label{eq:EHaction}
\mathcal{S}_{GR}=\mathcal{S}_{EH}+\mathcal{S}_{m}=\frac{1}{2\kappa^2} \int d^4x \sqrt{-g} R(g) +\mathcal{S}_m[g_{\mu
\nu},\psi_m]
\end{equation}
where $g$ is the determinant of the metric, $R \equiv g^{\mu\nu}R_{\mu\nu}$ is the Ricci scalar, $R_{\mu\nu}$ is the (symmetric) Ricci tensor, and $\mathcal{S}_m$ denotes the matter action with $\psi_m$ labelling collectively the matter fields. Variation of the action (\ref{eq:EHaction}) with respect to the metric yields the  Einstein equations
\begin{equation}\label{eq:Einsteineqs}
G_{\mu\nu}(g_{\mu\nu})=\kappa^2 T_{\mu\nu}(\psi_m,g_{\mu\nu})
\end{equation}
where $T_{\mu\nu}=\frac{2}{\sqrt{-g}} \frac{\delta\mathcal{S}_M}{\delta g_{\mu\nu}}$ is the matter energy-momentum tensor. This represents a system of second-order field equations for the space-time metric $g_{\mu\nu}$ whose explicit form is obtained after some matter action $\mathcal{S}_m$ is specified, and which can be solved once some assumption(s) on the symmetries of the particular scenario under consideration are introduced, together with suitable boundary conditions.

For the discussion of the structural properties of spherically symmetric, static, stellar models, as the matter source one usually takes a perfect fluid described by the energy-momentum tensor
\begin{equation} \label{eq:PF}
T^{\mu\nu}=(\rho+P)u^{\mu}u^{\nu}+Pg^{\mu\nu}
\end{equation}
where $u^{\mu}$ is the timelike unit vector, $u^{\mu}u_{\mu}=-1$, while $\rho$ and $P$ are the energy density and pressure of the fluid, respectively. The most general (static) line element compatible with spherical symmetry can be written under the form
\begin{equation}\label{eq:ds2}
ds^2=-A(r)dt^2+\frac{1}{B(r)}dr^2+r^2d\Omega^2
\end{equation}
where $d\Omega^2=d\theta^2 + \sin^2 \theta d\varphi^2$ is the volume element of the two-spheres. Together with the standard matter  conservation equation, $\nabla_{\mu}T^{\mu\nu}=0$ (which is a consequence of Bianchi's identity, $\nabla_{\mu}G^{\mu\nu}=0$), the derivation of the stellar structure equations from the Einstein field equations (\ref{eq:Einsteineqs}) proceeds straightforwardly (see e.g. \cite{GlenBook} for details of this derivation) to yield a single equation\footnote{To see a similar derivation and subsequent properties of stellar structure in the presence of a cosmological constant, see e.g. \cite{Liu:2018jhd}.}:
\begin{equation} \label{eq:TOVGR}
P'=-(\rho+P) \frac{m(r)+4\pi P r^3}{r\left(r-2m(r)\right)}
\end{equation}
where the mass function is defined as
\begin{equation}
m(r)=4\pi \int_0^r d \tilde{r} \tilde{r}^2 \rho(\tilde{r})
\end{equation}
and interpreted as the mass enclosed within a sphere of radius $r$, from which the total mass of the star follows simply as $M=m(r_S)$, where $r_S$ is the star's radius. Note also that the metric functions in (\ref{eq:ds2}) follow immediately as
\begin{equation} \label{eq:B(r)normal}
\frac{A'(r)}{A(r)}=-\frac{2P'}{\rho+P} \hspace{0.1cm};\hspace{0.1cm} B(r)=1-\frac{2m(r)}{r}
\end{equation}
Eq.(\ref{eq:TOVGR}) is nothing but the well known Tolman-Oppenheimer-Volkoff (TOV) equation \cite{T,OV}  describing the interior of a relativistic, static, spherically symmetric star in hydrostatic equilibrium, where the gravitational pull is exactly counter-balanced by the interior pressure. This equation contains the Newtonian contribution, $m(r)/r^2$, acting on the shell of matter, and three additional relativistic corrections, and allows to determine the global properties of the star. It can be solved after assuming that pressure and density are uniquely related by a barotropic EOS, i.e., $P=P(\rho)$, which closes the system\footnote{This assumption implies that the contribution of temperature upon the EOS (as well as that of superfluid phases or magnetic fields) can be neglected, which is a valid approximation for the strongly degenerate matter inside neutron stars. However, in neutron stars short after birth temperature may play a major role, see e.g. \cite{ZhLuWa17}.}. Its (typically numerical) integration takes place via boundary conditions at the star's center $r=0$, namely, $\rho(0)=\rho_c$ (which gives $P(0)=P_c$ via the specific EOS chosen) and $m(0)=0$, and continues outwards until the star's surface is reached, namely, at the point $r_S$ at which $P(r_S)=0$, where the integrated solution is matched to an external (vacuum) Schwarzschild solution, and this provides the mass-radius relation $M-r_S$ for that EOS. Therefore, by varying the central density $\rho_c$ one obtains a sequence of mass-radius relations parameterized by $\rho_c$ describing the profile of static, spherically symmetric star's structure for each EOS.

It should be pointed out that the TOV equation (\ref{eq:TOVGR}) relies on the assumption of \emph{isotropy}, while certain phenomena, such as high-density relativistic interactions or superfluidity, are able to introduce a certain degree of anisotropy \cite{Carter:1998rn,KippWeiBook,Herrera:1997plx}. Anisotropic fluids are described by the energy-momentum tensor
\begin{equation} \label{eq:Tmunufluid0}
{T^\mu}_\nu=(\rho + P_{\perp}) u^{\mu}u_{\nu}+P_{\perp} {\delta^\mu}_{\nu} + (P_r-P_{\perp}) \chi^{\mu}\chi_{\nu} \ ,
\end{equation}
where $u^\mu$ and $\chi^{\mu}$ represent normalized timelike and spacelike vectors, respectively, and $P_r$ and $P_{\perp}$ denote the radial (in the direction of $\chi^\mu$) and tangential pressures (in the directions orthogonal to $\chi^\mu$). In that case, the TOV equation (\ref{eq:TOVGR}) picks up a new term $\Delta\equiv2(P_r-P_{\perp})/r$, and one needs a second EOS relating the tangential pressure with the density. While such stress anisotropies are generally negligible as compared to the pressure, it has been recently shown \cite{Raposo:2018rjn} with a simple EOS for $P_{\perp}$ that even very small anisotropies may induce significant changes on the mass and compactness of the star\footnote{Moreover, the consideration of anisotropies in the energy-momentum tensor may be also necessary in order to correctly address differentially rotating stars, see Sec. \ref{eq:sec:frm}.}.

The bottom line of the discussion above is that the predictions for the global properties of a neutron star  are largely dependent on the assumptions for the EOS at the supranuclear densities expected at the center of the star (which can be up to one order of magnitude higher), while low-energy nuclear physics is suitable only to model the crust. Unsurprisingly, a large number of such EOS for dense matter has been proposed in the literature along the decades, and the properties of the corresponding neutron stars have been thoroughly analyzed. Historically, the first  analysis of neutron star structure was carried out by Tolman \cite{T} and Oppenheimer and Volkoff \cite{OV}, finding that for the EOS of a free relativistic neutron gas the theoretical limit for a neutron star mass is $ M_{\star} \approx 0.72M_{\odot}$. This result is in open contradiction with the  threshold found first by Chandrasekhar \cite{Chandra}, assuming that the gravitational collapse is counterbalanced by the degeneracy pressure of a free electron gas, which yields a robust theoretical limit on the maximum mass that white dwarfs can have in order not to become a neutron star of $M_{\star} \approx 1.40M_{\odot}$. Over the years the maximum mass limit for a neutron star has significantly increased as high-density nuclear physics became better understood. It is now known that this limit critically depends on the assumptions made on the species of particles present in the plasma, in the way nuclear interactions are mediated, etc, but a trend has been observed in that, in general, for ``realistic" EOS, the mass that can be sustained against gravitational collapse is positively linked to its stiffness \cite{stifflimit}. However, many additional ingredients that come into play at supranuclear densities are known to ``soften" the EOS and, therefore, to lower the maximum mass\footnote{This is where the hyperonic puzzle appears, namely, the fact that for densities well above the nuclear saturation density hyperons play a non-negligible role, which effectively causes the EOS to be softer, see e.g. \cite{Bombaci:2016xzl} and references therein.}. Thus, typically these theoretical schemes struggle to reconcile their predictions with the observations of the heaviest neutron stars so far, with reported masses at or even above the two solar masses threshold \cite{J0348+0432,Crawford:2006xb,LinShaCas18,Cromartie:2019kug}. How massive can a neutron star made out of baryonic matter be?. Imposing causality (i.e., subluminical speed of sound), one finds the Rhoades-Ruffini limit  \cite{RhoRuf}, which amounts to $M_{\star} \approx 3.2M_{\odot}$ and $R\approx 13.4$km regardless of the details of the  EOS at high densities (though with some weak dependence on the central density), though further arguments based on stars with uniform density may slightly raise this limit. A recent review on the masses and radii of neutron stars can be found in \cite{Ozel:2016oaf}.

Another relevant parameter for neutron stars is their compactness, which is defined as $\mathcal{C} \equiv M/r_S$. In vacuum, a spherically symmetric black hole has its Schwarzschild radius at $r_S=2M$, thus obviously for a black hole, $\mathcal{C}=1/2$, which sets an absolute limit for any horizonless compact object. There is yet another upper theoretical limit on compactness for any  (independent of the EOS) spherical fluid model of a neutron star, called Buchdahl's limit \cite{Buchdahlpaper,WeinbergBook}, which is given by the maximum amount of mass that can be enclosed within a sphere without experiencing gravitational collapse, and which reads $\mathcal{C}<4/9$. This limit comes from imposing that the density must be decreasing from the star's center outward, namely, $\rho'(r)<0$, and such that the internal solution is matched to an external vacuum Schwarzschild solution. Therefore, compact objects violating the Buchdahl limit if $4/9<C<1/2$ presumably should collapse into a black hole\footnote{The case with cosmological constant was worked out in \cite{Mak:2001gg} and the charged case in \cite{Andreasson:2012dj}.}\footnote{The presence of anisotropies are known to make neutron stars more compact to the point of violating Buchdahl's limit, thus potentially yielding ultra-compact objects disguised as black holes, see \cite{Raposo:2018rjn} for a discussion.}\footnote{It is interesting to note that there is yet another bound on compactness which reads $1/3<\mathcal{C}<4/9$, corresponding to stars satisfying the Buchdahl limit and having a photon sphere, namely, an unstable circular null orbit. The existence of hypothetical ultracompact stars within GR and in extensions of it, might induce the period release of secondary gravitational waves (echoes) in the ringdown phase of black hole mimickers, which are potentially detectable by the LIGO/VIRGO collaboration, see however \cite{Urbano:2018nrs} for a critical discussion on this point.}. Typically, neutron stars lie on the range $\mathcal{C} \sim 0.1-0.2$, while any other star has $C \ll 0.1$. It should be stressed that the Buchdahl limit depends on the theory of gravity chosen, see e.g. \cite{Tsuchida:1998jw} for the case of scalar-tensor theories and \cite{Dadhich:2016fku} for Lovelock gravities, which may potentially become another test for the observational consistency of modified gravity.

Let us close this section by briefly addressing the stability problem, which is closely related to that of the maximum mass of neutron stars. The solutions of the TOV equations (\ref{eq:TOVGR}) describe stars in hydrostatic equilibrium; however one needs to further determine if a given equilibrium configuration is stable or not. It can be shown (see for example \cite{GlenBook}) that a turning point from stability to instability with respect to any radial oscillation is marked by the stationary equilibrium mass, namely,
\begin{equation}
 \frac{\partial M(\rho_c)}{\partial\rho_c}=0,
\end{equation}
that is, the mass must be an extremum at a given value of the central density. Now, stable equilibrium configurations are given by the condition
\begin{equation}
\frac{\partial M(\rho_c)}{\partial\rho_c}>0.
\end{equation}
In order to fully examine the stability problem, one has to solve the eigenvalue problem \cite{chandras} for amplitudes $u_n(r)$ of normal modes of vibration. The perturbations (in $r$) of adiabatic motion of the spherically symmetric star are given by $\delta r(r,t)=e^\nu u_n(r)e^{i\omega_nt}/r^2$, where $\omega_n$ is the star's oscillatory eigenfrequencies and $n$ is a mode index with $n=0$ being the fundamental mode. Thus, the equation characterizing the eigenvalue problem for $u_n(r)$ governing the $n$th mode is written in the Sturm-Liouville form as \cite{GlenBook}
\begin{equation}
 \frac{d}{dr}\left(\mathcal{P}\frac{du_n}{dr}\right)+(\mathcal{Q}+\omega^2_n\mathcal{W})u_n=0,
\end{equation}
where the functions denoted here by $\mathcal{P},\,\mathcal{Q},\,\mathcal{W}$ are expressed in terms of the equilibrium configuration of the star (\ref{eq:TOVGR}). The boundary conditions are $u_n(0)\sim r^3$ at the star's center while at the star's surface we have $\frac{du_n}{dr}\big\vert_{r_S}=0$. The solutions of the Sturm-Liouville equation are squared eigenfrequencies $\omega^2_n,\,(n=0,1,2,\ldots)$ which form an infinite sequence $\omega^2_0<\omega^2_1<\omega^2_2< \ldots$. If any of them is negative, then the frequency is purely imaginary, with the result that any perturbation of the star will grow exponentially $\sim e^{|\omega|t}$, leading to instability. Therefore, we can conclude that  the star will be stable for $\omega^2\geq0$ (all modes have to have real eigenfrequencies) and unstable for $\omega^2<0$. Because of the frequency sequence, the stability of the star will depend only on the sign of $\omega_0^2$ \cite{sagert}.

\section{Stellar structure models in metric formalism}

By \emph{metric} formalism/approach we refer to theories of gravity where the metric-compatibility condition (\ref{eq:mccond}) of GR is kept, but the action (\ref{eq:EHaction}) and/or other principles are modified. On the contrary, we shall refer to \emph{metric-affine} (or \emph{Palatini}) formulation when non-metricity, $Q_{\mu}^{\alpha\beta} \equiv \nabla_{\mu}g^{\alpha\beta} \neq 0$, is present (see Sec.\ref{sec:Pal} for the review of such theories). In this section we will start our considerations for modified theories of gravity in metric formalism with the case of $f(R)$ models, for which an extensive literature has been developed. Therefore we shall elaborate the corresponding formalism to some detail. More general gravitational theories and their consequences for stellar structure models will be subsequently described.

\subsection{Field equations and scalar field representation of $f(R)$ gravity} \label{sec:FSFR} \vspace{0.2cm}

The study of stellar objects in $f(R)$ theories of gravity requires a careful consideration of the modified dynamics that nonlinear Lagrangians involve. The purpose of this section is to present and discuss some basic properties that will be necessary to understand the various results found in the literature. The analysis presented here follows closely the discussion originally developed in \cite{Olmo:2006eh}. Let us start with the action
\begin{equation}\label{eq:def-f(R)}
\mathcal{S}=\mathcal{S}_{G}+\mathcal{S}_{m}=\frac{1}{2\kappa ^2}\int d^4 x\sqrt{-g}f(R)+\mathcal{S}_m[g_{\mu
\nu},\psi_m] \ .
\end{equation}
The field equations are obtained by variation of (\ref{eq:def-f(R)}) with respect to $g_{\mu\nu}$, and yield
\begin{equation}\label{eq:f-var}
f_RR_{\mu\nu}-\frac{1}{2}f g_{\mu\nu}-
\nabla_{\mu}\nabla_{\nu}f_R+g_{\mu\nu}\Box f_R=\kappa ^2T_{\mu
\nu }
\end{equation}
with the same definitions and conventions as in the previous section, and where $f_R\equiv df/dR$.  These equations contain two derivative operators acting on the scalar curvature $R$, which itself contains up to second-order derivatives of the metric. As a result, these theories are sometimes interpreted as fourth-order gravity theories. This interpretation, however, can be simplified by noticing that the higher-order derivatives always act on the object $f_R$. Moreover, the trace of
(\ref{eq:f-var}) takes the form
\begin{equation}\label{eq:trace-m}
3\Box f_R+Rf_R-2f=\kappa ^2T \ ,
\end{equation}
(where $T \equiv g^{\mu\nu}T_{\mu\nu}$ is the trace of the energy-momentum tensor) which represents a second-order equation for a scalar degree of freedom. This equation manifests the fact that for non-linear $f(R)$ Lagrangians the scalar curvature becomes a dynamical entity, forced to be continuous and differentiable, and with the terms $Rf_R-2f$ playing the role of self-interactions.  It is, however, more convenient to interpret this equation in terms of the canonical variable $\phi\equiv f_R$, such that $R$ becomes algebraically related to $\phi$ and the terms $R f_R-2f$ turn into $\phi V_\phi-2V$, with $V(\phi)=R f_R-f$. With this notation, the above equations take a standard scalar-tensor form\footnote{For an overall description of these theories, see e.g. \cite{stt}.}
\begin{eqnarray}
R_{\mu \nu }(g)-\frac{1}{2}g_{\mu \nu }R(g)&=&
\frac{\kappa^2}{\phi}T_{\mu\nu}-\frac{1}{2\phi}g_{\mu\nu}V(\phi)
+\frac{1}{\phi}\left[\nabla_\mu\nabla_\nu\phi-g_{\mu\nu}\Box \phi\right] \label{eq:Gab-phi} \\
3\Box \phi +2V(\phi)-\phi \frac{dV}{d\phi}&=&\kappa^2T \ , \label{eq:phi-ST}
\end{eqnarray}
corresponding to a Brans-Dicke theory
\begin{equation}
\mathcal{S}_G^J=\frac{1}{2\kappa^2}\int d^4 x \sqrt{-g}\left(\phi R-\frac{\omega(\phi)}{\phi}g^{\mu\nu}\nabla_{\mu} \phi \nabla_{\nu}\phi - 2V(\phi) \right) +S_m[g_{\mu
\nu},\psi_m] \ ,
\end{equation}
with parameter $\omega=0$ and a non-trivial potential. In this representation, it becomes apparent that in (\ref{eq:Gab-phi}) the matter and the scalar degree of freedom (either $\phi$ or $R$) act as sources for the metric. In Eq.(\ref{eq:phi-ST}) the matter sources the scalar field, which also feels self-interactions. Whenever a linearized approximation is possible, these self-interactions will be responsible for an effective mass for the scalar field, leading to Yukawa-type exponential corrections in the weak-field metric. If non-linearities become relevant in certain scenarios, a more detailed analysis is necessary to determine if screening (chameleon) effects may arise\footnote{See \cite{Brax:2017wcj,Zhang:2017srh,Sakstein:2018fwz} for an overall description on how to handle stellar structure models in screened modified gravities.}. One of such screening mechanisms is represented by the Vainshtein one \cite{Vainshtein:1972sx}, which is known to be valid within models with non-linear derivative interactions for the scalar field, and allows to suppress potential fifth-force effects within the solar system.

It should be stressed that $f(R)$ models are just a particular case of a more general family of theories where the additional gravitational interactions are mediated by a single scalar field, and dubbed as scalar-tensor theories. In the Jordan frame, these theories can be cast as
\begin{equation} \label{eq:stjordan}
\mathcal{S}_G^J=\frac{1}{2\kappa^2}\int d^4 x \sqrt{-g}\left(A^{-2}(\phi) R-g^{\mu\nu}\nabla_{\mu} \phi \nabla_{\nu}\phi - 2V(\phi) \right) +S_m[g_{\mu
\nu},\psi_m] \ ,
\end{equation}
which, via a conformal transformation $\tilde{g}_{\mu\nu} =A^{-2}(\phi) g_{\mu\nu}$, can be written in the (sometimes) more convenient Einstein frame form (for a broader discussion of the two frames of scalar-tensor theories see e.g. \cite{Flanagan:2004bz})
\begin{equation} \label{eq:steinstein}
\mathcal{S}_G^E=\frac{1}{2\kappa^2} \int d^4 x \sqrt{-\tilde{g}} \left(R(\tilde{g})-\tilde{g}^{\mu\nu}\tilde{\nabla}_{\mu} \psi \tilde{\nabla}_{\nu}\psi - 2\tilde{V}(\psi) \right) +S_m[A^2\tilde{g}_{\mu\nu},\psi_m] \ ,
\end{equation}
where the potential $\tilde{V}=A^{4}V$. As the matter fields are minimally coupled to the Jordan frame metric $g_{\mu\nu}$, this implies that clocks and rods measure time and distances in that metric. Different choices of $\phi$ and $V(\phi)$ in the Jordan frame lead to different implementations of the model; for instance, a well known choice is the Damour-Esp\'osito-Farese (DEF) model \cite{Damour:1992we}, which shall be used in the many of the applications below.

\subsubsection{Spherically symmetric stellar models} \label{sec:sssm} \vspace{0.2cm}

The finding of the stellar equilibrium equations in spherically symmetric $f(R)$ models parallels the GR analysis of Sec.\ref{sec:stGR} with the subtleties mentioned above. Therefore, for a perfect fluid source (\ref{eq:PF}), starting from the line element (\ref{eq:ds2}), the field equations that follow from (\ref{eq:Gab-phi}) are then
\begin{eqnarray}
\left[\frac{2}{r}+\frac{f_{R,r}}{f_R}\right]\frac{B_{r}}{2}-\frac{1-B}{r^2} &=& -\Bigg[\frac{\kappa^2}{f_R}\rho+\frac{R
f_R-f}{2f_R}+\frac{B}{f_R}\Bigg(f_{R,rr}+\frac{2}{r}f_{R,r}\Bigg)\Bigg]  \label{eq:Dr} \\
\frac{B}{2}\left[\frac{2}{r}+\frac{f_{R,r}}{f_R}\right]\frac{A_{r}}{A}-\frac{1-B}{r^2}&=&
\frac{\kappa^2}{f_R}P-\frac{R
f_R-f}{2f_R}- \frac{B}{r}\frac{f_{R,r}}{f_R} \label{eq:Br}
\end{eqnarray}
where $A_r \equiv dA/dr$ and so on.
The corresponding expression for the trace equation (\ref{eq:trace-m}) reads now
\begin{equation}\label{eq:trace-m2}
3B f_{R,rr}+\frac{\partial_r\left(r^2\sqrt{AB}\right)}{r^2\sqrt{A/B}}f_{R,r}+Rf_R-2f=-\kappa ^2(\rho-3P) \ .
\end{equation}
These equations must be supplemented with the conservation equation for the matter fields, $\nabla_{\mu}T^{\mu\nu}=0$, which yields the TOV equation
\begin{equation}\label{eq:TOV}
P'= -(\rho+P)\frac{A'}{2A}
\end{equation}
as in the GR case. If one is interested in considering anisotropies in the matter fields (\ref{eq:Tmunufluid0}) this last equation becomes
\begin{equation}\label{eq:TOVani}
P_r'= -(\rho+P_r)\frac{A'}{2A} + \frac{2}{r}(P_{\perp}-P_{r})
\end{equation}
which again is the same result as in the GR case. This is so because both in the isotropic and anisotropic cases, this equation follows directly from the matter conservation equation and the choice of the line element, thus being independent of the particular gravity theory considered.

The set of equations (\ref{eq:Dr}), (\ref{eq:Br}), (\ref{eq:trace-m2}) and (\ref{eq:TOV}) describe the internal structure of self-gravitating fluids in static equilibrium within metric $f(R)$ gravity theories. Out of these equations only three of them are independent, while we have four functions to be determined, $A(r),B(r),R(r),P(r)$; therefore, like in the GR case, the system is typically closed by choosing an EOS. In order to obtain solutions, in general one must apply numerical methods with consistent boundary conditions at the center and at some exterior radius. As already discussed, in GR the exterior radius is typically chosen as the surface of the star, $r=r_S$, a region where the pressure decays below a certain small reference value. However, in $f(R)$ theories one must first determine the right location to set this boundary value. This is so because the scalar degree of freedom encoded in the curvature has an associated energy density which could contribute both inside the region where the matter density is nonzero but also outside of the star. This energy may affect the asymptotic form of the metric, which is relevant for determining orbital motions. Therefore, the observable mass of the resulting star may be sensitive to the presence of the scalar degree of freedom and any attempt to define the observable mass in terms of an integral over the fluid density should be confronted with the numerical results in regions beyond the threshold where the fluid density can be regarded as negligible \cite{Resco:2016upv}. Therefore, the matching  to an asymptotically Schwarzschild solution, for further integration of the field equations for large radius in a vacuum scenario,  must be done carefully in order to obtain a physically meaningful mass that can be compared with observations\footnote{For instance, in scalar-tensor theories one can try to parameterize the deviation with respect to Schwarzschild metric using two constants related to the mass of the star and the strength of the scalar field, which effectively represents a Just spacetime  \cite{Damour:1992we} (see \cite{Silva:2018yxz} for further details), which can be directly constrained via observations of binary systems \cite{Horbatsch:2011nh}.}.

\subsubsection{Weak-field limit, PPN formalism, and matching to the exterior solution} \vspace{0.2cm}

Let us analyze the weak-field and slow-motion limit of $f(R)$ theories (Newtonian and post-Newtonian limits). In the far region outside of the sources, where $\rho$ and $P$ vanish, the above equations can be linearized and analytically solved to obtain an approximated asymptotic form for the metric and the scalar degree of freedom. Let us assume we are expanding around the asymptotically flat solution as $g_{\mu\nu} \simeq \eta_{\mu\nu} + h_{\mu\nu}$, so we can write the line element in this region using Schwarzschild-like coordinates as
\begin{equation} \label{eq:gpot}
ds^2=-(1-2\Phi)dt^2+(1+2\Psi)dr^2+r^2d\Omega^2
\end{equation}
where $\eta_{\mu\nu}$ is Minkowski spacetime and the two radial potentials are also assumed to fulfil the weak field approximation, i.e., $\Phi(r) \ll 1$, $\Psi(r)\ll 1$. To find the expressions of the potentials $\Phi,\Psi$, expanding (\ref{eq:Dr}), (\ref{eq:Br}),  and (\ref{eq:trace-m2}), assuming $f_R\approx f_R^B+\varphi$, with $f_R^B$ representing the background value of $f_R$ and $|\varphi|\ll f_R^B$, with $B(r)=1-\frac{2m(r)}{r}$ (such that $\Psi\equiv 2m(r)/r\ll 1$), the relevant equations boil down to
\begin{eqnarray}\label{eq:Dr-lin}
\frac{2}{r^2}m_{r} &=&V_B+\frac{1}{f_R^B}\left(\varphi_{rr}+\frac{2}{r}\varphi_{r}\right)\\
\frac{\Phi_{r}}{2}&=&-\frac{2m(r)}{r^2}+V_B r+\frac{2\varphi_{r}}{f_R^B}\label{eq:Br-lin}\\
\varphi_{rr}+\frac{2}{r}\varphi_{r}&=&  m^2_B\varphi \label{eq:frr-lin}
\end{eqnarray}
where we have defined the following background quantities (evaluated at $R=R_B$)
\begin{eqnarray}\label{eq:VB}
V_B&\equiv&\left. \frac{R f_R-f}{2f_R}\right|_{B}\\
m^2_B&=& \left.\frac{f_R-R f_{RR}}{3f_{RR}}\right|_{B} \label{eq:mass}
\end{eqnarray}
The solutions to the above equations can be classified depending on the sign of $m^2_B$. If $m^2_B>0$ then we have
\begin{eqnarray}
\varphi(r)&=&  \frac{{C_1}e^{-m_Br}}{r}\label{eq:growing}\\
m(r)&=&
{C_2}-\frac{{C_1}}{2f_R^B}[1+m_Br]e^{-m_Br}+\frac{V_B}{6}r^3 \label{eq:Mr-assol}\\
\Phi(r)&=&\frac{4{C_2}}{r}\left[1+\frac{{C_1}}{2{C_2}f_R^B}e^{-m_Br}\right]+\frac{2V_B}{3}r^2
\label{eq:psir-assol}
\end{eqnarray}
where an exponentially growing mode has been discarded on physical grounds to recover the Minkowskian asymptotics. For $m^2_B<0$, the exponential solutions turn into oscillating ones and the result is
\begin{eqnarray}
\varphi(r)&=&\frac{{{C}_1}\cos[\alpha_0+|m_B|r]}{r}\label{eq:oscillating} \\
m(r)&=&
{C_2}-\frac{{{C}_1}}{2f_R^B}\left(1+|m_B|r \tan[\alpha_0+|m_B| r]) \cos[\alpha_0+|m_B| r]\right)+\frac{V_B}{6}r^3 \label{eq:Mr-assol-osc}\\
\Phi(r)&=&\frac{4C_2}{r}\left(1+\frac{C_1(2f_R^B-1)}{2C_2f_R^B}\cos[\alpha_0+|m_B| r]\right)+\frac{2V_B}{3}r^2 \ .
\label{eq:psir-assol-osc}
\end{eqnarray}
Note that if the phase $\alpha_0$ is set to zero then one finds agreement with the results of the $m^2_B>0$ case in the limit $m_B\to 0$ (up to some redefinition of constants). From these results, it is apparent that in the discussion of isolated objects on scales sufficiently small to neglect the effective cosmological constant term $\frac{V_B}{3}r^2$, the mass function $m(r)$ will not be strictly a constant in any case. If $m_B^2>0$ one will observe an exponential decay towards an asymptotic constant value, while for $m_B^2<0$ an oscillatory behaviour around a constant is expected. The above discussion translates into a difficulty with the definition of the mass in $f(R)$ gravity, which is addressed in different works, as we shall see later.

The reader should note that the line element (\ref{eq:gpot}) is sometimes used in the literature to explore the weak field limit of gravitational theories. Though that choice is quite conventional to study perturbations in cosmology,  it is not the appropriate choice to determine the PPN parameters of a given theory because in weak-field scenarios, such as the solar system, a different (isotropic) gauge choice has been established (see \cite{WillsBook,Will:1972zz} for the foundations of PPN formalism). This aspect is not always accounted for in the literature, leading to some inaccuracies and confusions (see the discussion in \cite{Olmo:2006eh}). The choice of Schwarzschild gauge explains, in particular, the unconventional $m_B r e^{-m_B r}$ term in (\ref{eq:Mr-assol}), which is not present if the correct isotropic coordinates are taken. When the line element is expressed in the latter coordinates, the PPN $\gamma$ parameter can be directly read from the metric, leading to\footnote{Note that other authors \cite{Sotiriou:2008rp,Capozziello:2005bu} provide a different expression for this parameter $\gamma_{PPN} \equiv - \frac{\Psi(r)}{\Phi(r)}= 1-\frac{f_{RR}^2}{f_R+2f_{RR}^2}$ (working in Schwarzschild coordinates) . }
\begin{equation}\label{eq:gamma-PPN}
\gamma_{PPN}=\frac{3-F(r)}{3+F(r)} \ ,
\end{equation}
where
\begin{equation}
F(r)=\left\{\begin{array}{lr}
	e^{-m_B r} & \text{ if } m_B^2>0 \\
	\cos(m_B r) & \text{ if } m_B^2<0 \end{array}\right. \ .
\end{equation}
Historically, Chiba \cite{chiba} was the first to address the question of the solar system viability of $f(R)$ theories by considering the scalar-tensor version of the (CDTT) model proposed by Carroll et al. \cite{Carroll:2003wy}
\begin{equation} \label{eq:1/Rmodel}
f(R)=R-\frac{\mu^4}{R} \ ,
\end{equation}
where $\mu$ is the model parameter. Chiba concluded that the model was ruled out by Solar System data, contradicting the original claims of \cite{Carroll:2003wy} based on the assumption that the Schwarzschild solution should hold as a vacuum solution. Subsequently, many works  appeared which tried to either confirm this result or to find a way to overcome it. For instance, in Erickeck et al. \cite{Erickcek} it was reported that Schwarzschild-de Sitter (SdS) spacetime is not the unique vacuum solution in $1/R$ gravity when one matches the interior solution with an exterior one; thus Birkoff's theorem does not apply here\footnote{This is evident from our Eqs.(\ref{eq:Dr-lin}), (\ref{eq:Br-lin}), and (\ref{eq:frr-lin}), in which the scalar degree of freedom self-interacts with itself and generates a non-constant $m(r)$ and $\Phi(r)$ outside of the sources.}. Furthermore, they showed that, when taking the Newtonian limit of the theory, one cannot neglect the mass distribution $\rho(r)$ coming from the Sun at the region $r>R_{\odot}$, in agreement with our general statements for $f(R)$ theories. Thus the scalar curvature is not a constant outside the star. In fact, as shown in \cite{Olmo:2006eh}, in the linearized theory with $m_B^2>0$, the behavior of $R$ is given (using isotropic coordinates) by
\begin{equation}
R=  V_B+\frac{m_B^2 \kappa^2}{4\pi f_R^B}\int d^3\vec{x} \ '\frac{\rho(t,\vec{x}\ ')}{|\vec{x}-\vec{x}\ '|}e^{-m_B|\vec{x}-\vec{x}\ '|} \ .
\end{equation}
As expected, this expression recovers the GR behavior, $R=\kappa^2\rho$, in the limit to GR ($f_R\to 1$, $m_B\to \infty$, $V_B\to 0$). Chiba's conclusion regarding the $1/R$ model follows from (\ref{eq:gamma-PPN}) when the limit $m_B\to 0$ is considered (very light scalar field with cosmic-scale interaction range), which leads to\footnote{In the GR limit,  $m_B\to \infty$, one finds instead $ \gamma_{PPN}^{(R-\mu^4/R)}=1$. The transition from the GR behavior to the $1/R$ behavior is triggered by the evolving boundary conditions imposed by the asymptotic cosmology in which the local system is immersed.}
\begin{equation} \label{eq:gammametric}
\gamma_{PPN}^{(R-\mu^4/R)}=\frac{1}{2} \ .
\end{equation}
 This result is in open conflict with the Solar System experiments, which yield the constraint \cite{Bertotti:2003rm}\footnote{It has been recently argued \cite{Suvorov:2018frf} that in magnetized post-Newtonian stars with strong toroidal fields this constraint is further reduced down to $\gamma_{PPN}-1<8.0 \times 10^{7}$ for consistency with gravitational wave luminosity of the Vela pulsar \cite{Abbott:2017ylp}.}
\begin{equation}
\gamma_{PPN}-1 \lesssim 2.3 \times 10^{-5}
\end{equation}
 Similar conclusions were obtained by Kainulainen et. al. \cite{Kainulainen:2007bt}. In comparison to \cite{Erickcek}, they additionally considered a kind of dark matter halo surrounding stellar systems, which turns out not to alter the structure of the solutions, as opposed to the conclusion reached by Zhang on a previous work \cite{Zhang:2007ne}. Indeed, a wide range of boundary conditions at the center of the star, $r=0$, were discussed and the field equations were integrated outwards with the conclusion, up to the validity of the approximated equations (which assume pressureless scenarios), that the solution on the metric components in the asymptotic limit provides the general expression for the PPN parameter\footnote{An attempt to find a general parametrization for the departure from GR without selecting any particular extension of GR has been carried out by  Glampedakis et. al. \cite{Glampedakis:2015sua}, up to second order in the post-Newtonian theory for spherical stars.}
\begin{equation}
\gamma_{PPN}=\frac{4M-3(r\log[B])_{c}}{8M-3(r\log[B])_{c}}
\end{equation}
where $M$ is the total mass of the star. Thus, for any finite value of the metric function at the center of the star the equation above implies the result (\ref{eq:gammametric}), and thus the inconsistency with Solar System experiments, as discussed in \cite{chiba, Erickcek,Chiba:2006jp,Nojiri:2003ft}. This result is further reinforced by the fact that, should one desire to impose the result $\gamma_{PPN}=1$ at the exterior region, this would cause a blow up of the metric component at the center rendering the star an unstable system. Moreover, the numerical analysis of other $f(R)$ models, such as $f(R)=R-\mu^4/R +\alpha R^2$, $f(R)=R-\beta R^n$, or $f(R)=R+\alpha R^{1/2}$ \cite{Faulkner:2006ub}, seems to support the inconsistency of such models with Solar system experiments, as they produce similar results in that region as the $1/R$ one. This conclusion is consistent with (\ref{eq:gamma-PPN}) in the case of a very light scalar degree of freedom.

Let us consider now the matching to the exterior solution. In GR, Birkhoff's theorem guarantees that the external vacuum metric to a static, spherically symmetric star is given by the Schwarzschild (or SdS) solution. In the case of $f(R)$ theories, the fact that the scalar curvature satisfies the second-order differential equation (\ref{eq:trace-m}) implies that $R$ must  be a continuous and differentiable function everywhere, including the boundary layer between the interior of a star and its exterior. This explains why there are exponential/oscillatory tails associated to the linearized solutions presented above, as they allow for a smooth transition of the curvature. This contrasts with the situation found in GR, where the Ricci scalar can be a discontinuous function with a step-like behavior when going from the inside to the outside of massive bodies. Thus, even if the Schwarzschild solution is a valid exterior solution in $f(R)$ theories, in general, it will not be a unique solution because an infinite number of smooth deformations are possible (either exponential or oscillatory), which depend on the specific boundary conditions at the transition surface.

In spite of this, the approach of considering the Schwarzschild spacetime vacuum solution as an external boundary condition imposed from the onset was considered by Multamaki and Vilja \cite{mult} and by Henttunen et al. \cite{hen}. In \cite{mult} the authors discuss the form of a mass distribution which allows SdS metric to be a solution outside a star, namely
\begin{eqnarray}
A(r \rightarrow \infty)&=&B(r \rightarrow \infty) \to 1-\frac{2M}{r}-\frac{\Lambda}{3}r^2 \label{eq:SdS1}
\end{eqnarray}
where $\Lambda$ is the cosmological constant. This imposed form of the external solution provides a number of additional conditions for the higher derivatives of the metric components $A''_S$, $A'''_S$, $B'_S$, $B''_S$ on the surface $r=r_S$ of a star, in comparison to GR where one deals with $A_S$, $B_S$, and $A'_S$ only. Moreover, the curvature scalar is fixed there to some value $R(r_S)=R_0$ via the equation $ R_0f_R(R_0)=2f(R_0)$
and its first derivative is also fixed there. The analysis of this condition, in combination with the modified Einstein equations (\ref{eq:f-var}), provides more constraints on the surface for a general $f(R)$ theory such that $f_{RR}(R_0)\neq0$. Of special interest are those involving the pressure and energy density. For this purpose, let us bring at this point the polytropic EOS, which will be of great relevance for the discussion of Sec.\ref{sec:LEsection}. Such a polytropic EOS is given by the simple polynomial expression \cite{PolyBook}
\begin{equation} \label{eq:poly}
P=K \rho^{\Gamma}
\end{equation}
where $K$ is the polytropic constant which depends on the composition of the fluid, while it is also convenient to define the exponent $\Gamma=1+1/n$, with $n$ the polytropic index. There is a nice correspondence between different values of $n$ and their interpretation in terms of different states of the fluid; for instance, $\Gamma=5/3$ ($n=3/2$) and $\Gamma=4/3$ ($n=3$) correspond to the non-relativistic limit and the relativistic limit of a completely degenerate gas, respectively, which can be used to describe the low-density and high-density regions of neutron stars, respectively\footnote{Obviously, more elaborate models including, for instance, nuclear burning and metallicity effects, are needed to accurately describe the physics of relativistic stars.}. Moreover, different types of non-relativistic stars can be described by different polytropic indices; this is the case of  $n=3/2$, describing completely convective stars (such as red giants) and high-mass brown dwarfs \cite{Burrows:1992fg,KippenWeig}, and $n=3$, which is useful to describe high-mass white dwarfs \cite{STBook}. For this polytropic EOS, the above conditions on $f(R)$ theories lead to a lower limit on the polytropic index, namely, $n>1$, which is not found in the case of GR \cite{WeinbergBook}. Moreover, it is also found that the Schwarzschild fluid sphere (the radius where the interior and exterior solutions are matched at vanishing pressure) with constant density is not allowed in the case of $f(R)$ gravity.

Some simplifications can be achieved by considering $f(R)$ and $f_R$ as independent functions of the radial coordinate $r$ \cite{hen}. For each particular model this approach allows to write an additional constraint $f=f(f_R)$. However, since $f$ and $f_R$ are now not written as functions  of the scalar curvature, then the continuity equation (\ref{eq:TOV}) is not automatically satisfied, thus resulting into another independent differential equation. Together with the modified Einstein's equations (\ref{eq:Dr}) and (\ref{eq:Br}) this yields a set of independent, non-linear equations to be solved for the objects $\{f_R(r),A(r),B(r),\rho \}$ but, as opposed to the standard approach, one deals with second-order equations with six initial conditions (assuming a given polytropic EOS (\ref{eq:poly})). For the $1/R$ model (\ref{eq:1/Rmodel}) and its extension with an $R^2$ term, these equations can be solved numerically  starting from a given central density $\rho_c$ while preserving the correct asymptotic behavior. This requires to make use of the Lane-Emden equation (see Sec.\ref{sec:LEsection}) leaving just one free parameter, namely, the central curvature, which is obviously related to the form of $f_R$; varying it, one obtains different stellar masses and radii. Despite finding very different forms for the metric components as compared to the GR case, one goes back again to the result (\ref{eq:gammametric}), which further supports the problem of reconciling stellar structure models in  $f(R)$ gravity with Solar System experiments. On the other hand, demanding the SdS metric, Eq.(\ref{eq:SdS1}), to be an external solution yields divergences of density, central curvature, and metric components at the origin, although the consideration of different EOS at the core and the outer region could be able to improve this behaviour. A somewhat crude approach  to this problem  recently investigated by Cikintoglu \cite{Cikintoglu:2017jfh}, is to numerically seek for a boundary layer near the surface of the star where the higher-order terms are negligible. Outside the layer one imposes asymptotically the SdS solution, while the numerical method finds the interior solution which is consistent with the external one across the layer. Nonetheless, this method has the drawback of not being able to provide a single solution, since the resolution at the interior  depends on the value of the Ricci scalar at the star's surface. This is a manifestation of the lack of a theorem guaranteeing that all the the theories built this way will share the same solution at the matching surface.

The matching problem also appears when one considers spherical symmetric collapse of stellar solutions\footnote{For a broad discussion on the similarities and differences of models of collapse in GR and $R^2$ gravity see Astashenok et. al. \cite{Astashenok:2018bol}.} . Indeed, it was shown by Goswami et al. \cite{sunil} that the strong matching conditions on the stellar structure required in $f(R)$ gravity produce constraints which are unphysical. Indeed, known scenarios of collapsing matter in GR cannot be directly generalized to $f(R)$ gravity. More specifically, the smooth matching of the vacuum exterior solutions to the star's interior in any $f(R)$ model provides extra conditions demanding the collapsing matter to be inhomogeneous. An explicit analysis of this kind of inhomogeneous scenario was performed for the Starobinsky model $f(R)=R+\alpha R^2$ (which will appear later in many applications in this review). Since the additional matching conditions restrict the free functions of integration in the system, one concludes that these solutions can be unstable with respect to any matter perturbations in the star's interior, and one does not reach a consistent collapse description. Furthermore, doubts have been raised on the possibility of black hole formation via stellar collapse within $f(R)$ gravity since the standard Oppenheimer-Snyder-Datt model of this process \cite{OS39,Datt38} is not viable for these theories. This modelling seems to produce a naked singularity after the collapse as the apparent horizon does not appear fast enough,
violating the cosmic censorship hypothesis.

\subsection{Non-relativistic stars} \label{sec:LEsection} \vspace{0.2cm}

\subsubsection{Newtonian limit and the Lane-Emden equation} \vspace{0.2cm}

Let us now busy ourselves with non-relativistic stars. This includes white,  brown and red dwarfs.  We thus consider a static, spherically symmetric star described by the Newtonian regime of the TOV equation (\ref{eq:TOVGR}), where one has $P_r=(d\Phi/dr)\rho$, with $-\Phi$ representing the gravitational potential. Using suitable approximations for this regime, $P \ll \rho$, $4\pi r^3 P \ll m(r)$ and $2m(r)/r \ll 1$, the gravitational potential satisfies the Poisson equation $\nabla^2 \Phi=4\pi \rho$, which reads explicitly
\begin{equation} \label{eq:nlimit}
P'=-\frac{m(r)\rho(r)}{r^2} \rightarrow \frac{d}{dr}\left(\frac{r^2}{\rho}P'\right)=-4\pi r^2 \rho(r)
\end{equation}
This law simply expresses a link between the Newtonian potential with the mass function, neglecting contributions from relativistic particles. Likewise the full TOV equation, Eq.(\ref{eq:nlimit}) represents a closed system once an EOS is given, and yields a one-parametric family of solutions in terms of the central density $\rho_c \equiv \rho(0)$. Such a simple crude model turns out to represent a good approximation to some of the astrophysical properties of non-relativistic stars whose internal dynamics can be captured by the polytropic EOS of Eq.(\ref{eq:poly}), though further refinements require the use of more elaborate models \cite{HansenBook}. To work with dimensionless variables, let us introduce the following (canonical) redefinitions:
\begin{equation} \label{eq:redefpoly}
r=r_c\xi\hspace{0.1cm};\hspace{0.1cm}
\rho=\rho_c \theta^{n}\hspace{0.1cm};\hspace{0.1cm}
P=P_c \theta^{n+1} \hspace{0.1cm};\hspace{0.1cm}
r_c^2\equiv\frac{(n+1) P_c}{4\pi\rho_c^2}=\frac{K (n+1)}{4\pi }\rho_c^{\frac{1-n}{n}} \nonumber
\end{equation}
so that Eq.(\ref{eq:nlimit}) can be suitably cast as
\begin{equation}\label{le}
\frac{1}{\xi^2}\frac{d}{d\xi}\left(\xi^2 \frac{d\theta}{d\xi}\right)+\theta^{n}=0
\end{equation}
which is the well known Lane-Emden equation describing a spherically symmetric polytropic fluid with polytropic index $n$ in hydrostatic equilibrium. This equation has to be supplemented with the boundary conditions at the center of the star: $\rho_c \equiv \rho(0)$ and $\rho'(0)=0$, which translate into $\theta(0)=1$ and $\theta'(0)=0$. The Lane-Emden equation admits three exact solutions, namely,
\begin{eqnarray}
n&=&0 \rightarrow \theta(\xi)=1-\frac{\xi^2}{6} \rightarrow \xi_R=\sqrt{6} \label{eq:eta0} \\
n&=&1 \rightarrow \theta(\xi)=\frac{\sin(\xi)}{\xi} \rightarrow \xi_R=\pi \label{eq:eta1} \\
n&=&5 \rightarrow \theta(\xi)=\frac{1}{\sqrt{1+\frac{\xi}{3}}} \rightarrow \xi_R=\infty \label{eq:eta5}\end{eqnarray}
\begin{figure}[h!]
 \begin{center}
\includegraphics[height=5.0cm,width=10cm]{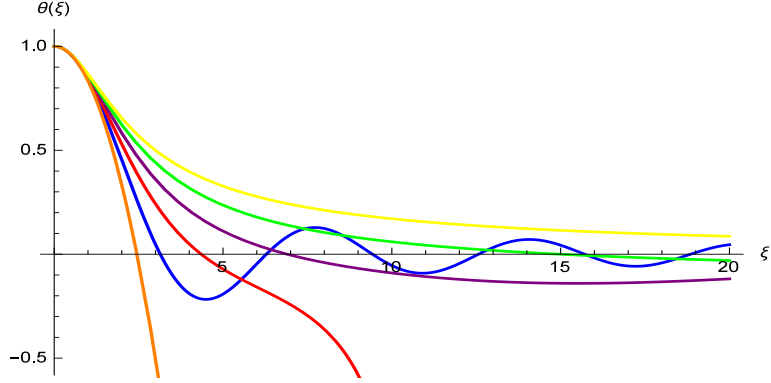}
\caption{The function $\theta(\xi)$ resulting from the integration of the Lane-Emden equation (\ref{le}) for $n=0$ (orange), $n=1$ (blue), $n=2$ (red), $n=3$ (purple), $n=4$ (green) and $n=5$ (yellow). The cut points with the $\xi$ axis, $\theta(\xi_R)=0$, allows to obtain the radius of the star as $r_S=r_c\xi_R$. For $n\geq 5$ no stable configurations can be found.}
\label{fig:LEGR}
 \end{center}
\end{figure}
In general, however, numerical methods will be needed in order to determine the shape of the  function $\theta(\xi)$ (see Fig.\ref{fig:LEGR}). For $n<5$ this function monotonically decreases crossing the zero $\xi=0$ several times. The first such zero defines the radius of the star, namely, $\theta(\xi_R)=0$, which is obtained as $r_S=r_c\xi_R$, and which increases monotonically with $n$ until $n=5$ is reached, for which the Lane-Emden equation provides an infinite radius. This way, given a polytropic index for the EOS (\ref{eq:poly}), the resolution of the Lane-Emden equation allows one to compute the star's mass, radius, and central density, for $n \geq 1$, as\footnote{Elaborating from here in the case of ultra-relativistic limit $\Gamma=4/3$ (matching to $\Gamma=5/3$ at low densities), one arrives to the well known Chandrasekhar's limit for the mass of a static carbon-oxygen white dwarf $M_{ch} \approx 1.40 M_{\odot}$.}
\begin{eqnarray}
M&=&4\pi\int_0^{r_S} r^2\rho(r)dr=4\pi r_c^3\rho_c w_{n} \label{eq:massle}\\
r_S&=&\gamma_{n}K^{\frac{n}{3-n}}M^{\frac{n-1}{n-3}} \label{eq:RSfor} \\
\rho_c&=&\delta_{n} \frac{3M}{4\pi r_S^3}
\end{eqnarray}
respectively, where the following constants have been introduced
\begin{eqnarray}
w_{n} &=& - \xi_R^2 \frac{d\theta}{d\xi}\Big\vert_{\xi=\xi_R } \label{eq:wn} \\
\gamma_n&=&(4 \pi)^{\frac{1}{n-3}}(n+1)^{\frac{n}{3-n}}w_n^{\frac{n-1}{3-n}} \xi_R \label{eq:gn}  \\
\delta_{n} &=&-\frac{\xi_R}{3d\theta/d\xi\vert_{\xi=\xi_R}} \label{eq:dn}
\end{eqnarray}

\subsubsection{Modifications to Lane-Emden equation. White dwarfs} \vspace{0.2cm}

The Lane-Emden equation has been generalized to the case of $f(R)$ gravity. One first such analysis was done by Capozziello et al. in \cite{capo}, and further refined by Farinelli et al. \cite{Farinelli:2013pza}. This analysis starts by writing down the modified Poisson equations in this case taking the Newtonian limit of Eq.(\ref{eq:f-var}), which yields
\begin{align}
\nabla^2\Phi+\frac{R^{(2)}}{2}+f_{RR}(0)\Delta R^{(2)}=&-8\pi\rho,\\
3f_{RR}(0)\nabla^2 R^{(2)}+R^{(2)}=&-8\pi \rho
\end{align}
where  $R^{(2)}$ is Ricci scalar expanded up to second order. Assuming a polytropic EOS (\ref{eq:poly}) one can combine the above system of equations into a  modified Lane-Emden equation
\begin{equation}\label{EM1}
 \frac{1}{\xi^2}\frac{d}{d\xi}\left(\xi^2 \frac{d\theta}{d\xi}\right)+\theta^n
 =\frac{m\xi_0}{8\xi}\int^{\xi/\xi_0}_0d\xi'\xi'
 \left( e^{-m\xi_0|\xi-\xi'|}-e^{-m\xi_0|\xi+\xi'|} \right)\theta(\xi')^n \nonumber
\end{equation}
where here $m^2=-\frac{1}{3f_{RR}(0)}$ (which is specified once a $f(R)$ model is given, so that the limit $m \rightarrow \infty$ corresponds to GR), and the constant $\xi_0= \sqrt{\frac{3}{16\pi A_n \Phi_c^{n-1}}}$ with $A_n=(1/K(n+1))^n$ . As compared to GR, the modified Lane-Emden equation (\ref{EM1}) possesses only one exact solution for $n=0$, which reads \cite{capo}
\begin{equation}\label{sol1}
 \theta^{(0)}_{f(R)}(\xi)=1-\frac{\xi^2}{8}+\frac{(1+m\xi)e^{-m\xi}}{4m^2\xi_0^2}\left( 1-\frac{\sinh{m\xi_0 \xi}}{m\xi_0 \xi} \right).
\end{equation}
Note that the GR limit for the above expression does not boil down to the known solution given by Eq.(\ref{eq:eta0}) when we set $m\rightarrow\infty$, that is, when $f(R)\rightarrow R$. This is caused by the different definition of the quantity $\xi_0$, which for GR is $\xi_0^{GR}=\sqrt{\frac{1}{4\pi A_n\Phi_c^{n-1}}}$ as compared to that used in the computation above. For the solution (\ref{sol1}) it is possible to find a constraint for the stellar radius $\xi=\xi_0 \xi_{f(R)}^{(0)}$, where $\xi_{f(R)}^{(0)}$ is the first zero of the solution (\ref{sol1}), and is obtained as
\begin{equation}
 \xi_{f(R)}^{(0)}=\sqrt{\frac{3\Phi_c}{2\pi}}\frac{1}{\sqrt{1+\frac{1+m\xi}{3}e^{-m\xi}}}
\end{equation}
which, for $m\rightarrow\infty$, recovers the Newtonian limit of GR, $\xi^{(0)}_{GR}=\sqrt{\frac{3\Phi_c}{2\pi}}$. In this case it is found that the radius is smaller than the one obtained in GR, which comes as a consequence of the fact that  the gravitational potential $-\Phi$ gives rise to a deeper potential well in comparison to the Newtonian one derived from GR. For $n=1$, numerical methods provide a solution with much the same features as the $n=0$ case. Farinelli et al. \cite{Farinelli:2013pza} also point out in their analysis to the potential existence of exotic objects in these models, that cannot be found as solutions of the GR case.

Using a different perturbative approach, Andr\'e and Kremer \cite{andre}, found that the modified Lane-Emden equation for the quadratic (Starobinsky) gravity\footnote{The current tightest observational constraint on $\alpha$ comes from the Gravity Probe B, which gives $\alpha \lesssim 2.3 \times 10^{5}$ ($ 5 \times 10^{11}$m$^2$ in physical units \cite{Naf:2010zy}).}
\begin{equation} \label{eq:fquadratic}
f(R)=R+\alpha R^2
\end{equation}
(where $\alpha$ is a constant with dimensions of length squared), can be written as a second-order differential equation
\begin{equation}\label{EM2}
\frac{1}{\xi^2}\frac{d}{d\xi}\left(\xi^2 \frac{d\theta}{d\xi}\right)+\theta^n
+\frac{1}{3m^2\xi_0^2}\frac{d^2\theta(\xi)^n}{d\xi^2}+\frac{2}{3m^2\xi_0^2}\frac{1}{\xi}\frac{d\theta(\xi)^n}{d\xi}=0 \nonumber
\end{equation}
where in this case $\xi_0=\sqrt{\frac{1}{4\pi A_n\Phi_c^{n-1}}}$ , $m^2=-1/(6\alpha)$ and, again, the Newtonian version is recovered in the limit $m\rightarrow\infty$, that is, $\alpha\rightarrow 0$. Solutions are found for the values of $n=\{1,3/2,3\}$, which can be used to model some types of neutron stars, red giants together with high-mass brown dwarfs, and fully radiative stars such as the Sun and those with degenerate nuclei, for example white dwarfs, respectively. This modification of the Lane-Emden equation in $f(R)$ gravity is treated here as a correction to the Newtonian solutions from known data on mass $M$ and radius $R_S$ which are regarded as observational data, that is
\begin{align}
r_S^{f(R)}=&(\xi)_{\theta=0}=\xi_{(n)},\\
 M^{f(R)}=&\left[ \left(-\xi^2\frac{d\theta}{d\xi}\right)+\frac{1}{3m^2\xi^2}\left(-\xi^2\frac{d\theta^n}{d\xi}\right) \right]_{\theta=0},
\end{align}
where we denote by $\xi_{(n)}$ as the first zero of the solution (\ref{EM2}) and the $f(R)$ corrections to the GR mass are apparent. Thus, the ratio between GR and $f(R)$ theories for radius and mass become \cite{andre}
\begin{equation}\label{corr}
 \frac{r_S}{r_S^{f(R)}}=\xi_0^{GR} \hspace{0.1cm};\hspace{0.1cm}
 \frac{M}{M^{f(R)}}=\xi_0^{GR} \Phi_0
\end{equation}
The expressions above allow to determine the ratio between the central density $\rho_c$ and its mean density $\bar\rho$ as
\begin{equation}
\frac{\rho_c}{\bar\rho}=\left( -\frac{3}{\xi}\frac{d\theta}{d\xi}-\frac{1}{\alpha^2 \xi}\frac{d\theta^n}{d\xi} \right)^{-1}_{\xi_{(n)}}
\end{equation}
On the other hand, using Eqs.(\ref{eq:poly}) and (\ref{corr}), the pressure is expressed as
\begin{equation}
P=\frac{\bar \rho}{(n+1)}\frac{M}{M^{f(R)}}\frac{R_S^{f(R)}}{R_S}\left( -\frac{3}{\xi}\frac{d\theta}{d\xi}-\frac{1}{\alpha^2 \xi}\frac{d\theta^n}{d\xi} \right)^{-1}_{\xi_{(n)}}\theta^{(n+1)}
\end{equation}
which provides the central pressure $P=P_c$ when
$\theta=1$. Assuming and ideal gas with central temperature defined by $T=\frac{P_c\mu m_{\mu}}{\rho_c k_B}\theta$, where $k_B$ is the Boltzmann constant, $\mu$ the atomic mass and $m_\mu$
the atomic mass unit, the above discussion allows to write the temperature as
\begin{equation}
 T=\frac{\mu m_\mu}{(n+1)k_B}\frac{M}{M^{f(R)}}\frac{r_S}{r_S^{f(R)}}\theta.
\end{equation}
Next, suitable stars are chosen\footnote{For $n=1$ the authors also make use of a neutron star given by the PSR J0348+0432 ($M=2.01M_{\odot}, r_S=1.87 \times 10^{-5} R_{\odot}$ \cite{J0348+0432}); however it should be pointed out that the assumption of the Newtonian limit for neutron stars raised serious doubts on the reliability of the corresponding results.}:  for $n=1.5$ a brown dwarf such as Teide 1 ($M=0.053M_{\odot},r_S=0.1R_{\odot}$ \cite{Teide1}) or a red giant star such as Aldeberan ($M=1.5M_{\odot}, r_S=44.2R_{\odot}$ \cite{RG1,RG2,RG3}), and for $n=3$ a white dwarf such as Sirius B ($M=1.5M_{\odot}, r_S=0.008R_{\odot}$ \cite{SiriusB}). A numerical resolution starts by approximating density, pressure, and temperature at the star's surface (where $\theta=0$) for several values of $\alpha$, and compare the obtained results in terms of the pressure with the estimations for these four candidates $P_c=\{5.01\times10^{34};\;10^{16};\;10^{8}; 4.95\times10^{24} \}$ (measured in Pa \cite{zhao}). The numerical integration shows that, for all these examples, the stronger the quadratic curvature corrections (i.e increasing $\alpha$), the lower the internal  density, pressure, and temperature become (save for the density in neutron stars, which remains unmodified).

Let us now consider the modifications to the Lane-Emden equation in Horndeski theories, which is the most general gravitational action with a single scalar field having second-order equations of motion. This family of theories are described by the action \cite{Horndeski:1974wa} (for a detailed account of the properties of Horndeski and beyond Horndeski models, see e.g. \cite{Langlois:2015cwa}):
\begin{equation} \label{eq:Horn}
\mathcal{S}=\sum_{i=2}^5 \int d^4x \sqrt{-g} \mathcal{L}_i(X,\phi)
\end{equation}
where the $\mathcal{L}_i$ terms represent the following contributions
\begin{eqnarray}
\mathcal{L}_2&=&G_2(\phi,X) \\
\mathcal{L}_3&=&G_3(\phi,X) \Box \phi \\
\mathcal{L}_4&=&G_4(\phi,X) R -2G_{4,X}(\phi,X)[(\Box \phi)^2-\phi_{\mu\nu}^2] \\
\mathcal{L}_5&=&G_5(\phi,X) G_{\mu\nu}\phi^{\mu\nu}+\frac{G_{5,X}}{3}[(\Box \phi)^3+2\phi_{\mu\nu}^3-3\phi_{\mu\nu}^2\Box \phi]
\end{eqnarray}
where the functions $G_i(X,\phi)$, which depend on both the scalar field $\phi$ and its kinetic term $X=\nabla_{\mu}\phi\nabla^{\mu}\phi$, characterize the particular member in the family (note that GR corresponds to $\{G_4=0,G_2=G_3=G_5\}$), and the notations $\phi_{\mu \ldots \nu} \equiv \nabla_{\mu}\ldots \nabla_{\nu} \phi $, and $\Box \phi \equiv g^{\mu\nu}\phi_{\mu\nu}$. This family can be further extended to a more general class of healthy theories having higher-order equations but such that the application of a number of hidden constraints prevent the propagation of the Ostrogradsky ghosts. It  includes as members the beyond Horndeski class \cite{Zumalacarregui:2013pma,Gleyzes:2014dya,Gleyzes:2014qga} and the even more general class called degenerate higher-order scalar-tensor (DHOST) theories \cite{Langlois:2015cwa}. The implementation of the Vainshtein mechanism\footnote{This mechanism allows to successfully \emph{hide} the extra scalar degree of freedom in solar system scales via non-linear effects, thus yielding $\gamma_{PPN}=1$. This guarantees that the effects on these theories will manifest only within astrophysical bodies through a modification of the Poisson equation.} for gravitational theories with a single extra scalar field has been studied in  \cite{Koyama:2013paa,DeFelice:2011th}.

For the subset of beyond Horndeski theories corresponding to the G$^3$ galileons \cite{Gleyzes:2014qga} (a Ostrogradski ghost-free generalization of the covariant quartic galileon G$^2$ \cite{Deffayet:2009wt}), it has been shown by Koyama and Sakstein \cite{Koyama:2015oma} that the generalized Lane-Emden equation in this case takes the form
\begin{equation} \label{eq:LEbH}
\frac{1}{\xi^2}\frac{d}{d\xi}\left[ \xi^2 \frac{d\theta}{d\xi} \left(\theta+\frac{\Upsilon}{4}\xi^2 \theta^{n} \right)  \right]+\theta^n=0
\end{equation}
where $\Upsilon$ is a parameter whose explicit form can be written in terms of functions defining the beyond Horndeski Lagrangian \cite{Kobayashi:2014ida}, though it can be always taken as a free arbitrary parameter. For the cubic galileon, this parameter combines the time derivative of the scalar field and the new mass scale appearing in the Lagrangian. The structure of the new contributions in Eq.(\ref{eq:LEbH}) suggests that $\Upsilon>0$ ($<0$) should induce a weakening (strengthening) of gravity, with relevant consequence for stellar structure models, as we shall see later. It has been shown by Saito et al. \cite{Saito:2015fza} that the formula (\ref{eq:LEbH}) holds indeed for any beyond Horndeski theory, and further argued that no physically sensible solutions for compact stars can be found for $\Upsilon <-2/3$ for any such theories. Wibisono and Sulaksono \cite{Wibisono:2017dkt} have shown that for main sequence stars described by a polytrope EOS (\ref{eq:poly}) with $\Gamma=4/3$ (and $K$ an analytic function of the gas pressure), positive  (negative) $\Upsilon$ tend to increase (decrease) the star's radius. As for the mass, it takes the same expression as the GR one, (\ref{eq:massle}), with the $\Upsilon$-corrections encoded through $w_n$ in Eq.(\ref{eq:wn}), which in GR it yields $w_n \simeq 2.018$ for that case. Moreover, it is further argued in \cite{Wibisono:2017dkt} that no Chandrasekhar's limit can be found in these theories for $\Upsilon<-0.1$ for a range of values of the polytropic parameter $\Gamma \in (1.25,1.75)$, and that stability of the corresponding stars via configuration entropy analysis constrains the admissible range of the beyond Horndeski parameter to $-0.6 \leq \Upsilon < 0.7$. On the other hand, Koyama and Sakstein \cite{Koyama:2015oma} integrate the generalized Lane-Emden equation (\ref{eq:LEbH}) for different values of $\Upsilon$ finding the luminosity function of main sequence stars, and compare it with GR results. They found a decrease of luminosity of $\sim 10\%$ for $\Upsilon \sim 0.2$ (thus yielding redder stars), an effect that affects low mass stars to a greater extent than high mass stars. Consistency with observations of solar brightness imposes $\Upsilon \lesssim \mathcal{O}(1)$, though this bound depends on further refinements of the modelling.

Let us now rewrite, for convenience, the  modified hydrostatic equilibrium equation for these G$^3$ type theories making explicit the gravitational constant \cite{Sakstein:2015zoa}:
\begin{equation} \label{heqst}
P'=-\frac{Gm(r)\rho(r)}{r^2}-\frac{\Upsilon}{4}G\rho\frac{d^2 m(r)}{dr^2}
\end{equation}
This equation can be rewritten as the standard hydrostatic equilibrium equation with an effective Newton constant
\begin{equation}
\frac{G_{eff}}{G}=1+\frac{\Upsilon}{4}\frac{r^2}{m(r)} \frac{d^2m(r)}{dr^2}
\end{equation}
Since $d\rho/dr<0$ inside the star, thus positive (negative) $\Upsilon$ will act to weaken (strengthen) gravity, as expected. Assuming a constant $\Upsilon$ and using $m(r)=4\pi r^3/3$, the hydrostatic equilibrium equation can be rewritten as
\begin{equation}
P'=-\frac{m(r)\rho}{r^2} \left[1+\frac{\Upsilon \pi r^3}{m(r)}\left(2\rho + r\frac{d\rho}{dr} \right)\right]
\end{equation}
Moreover, near the center of the star, approximating $m(r) \sim 4\pi \rho_c r^{3}/3$, Eq.(\ref{heqst}) becomes \cite{Saito:2015fza}
\begin{equation}
P'=-\frac{Gm(r)\rho_c}{r^2}\left(1+\frac{3\Upsilon}{2}\right)
\end{equation}
and thus the effective Newton's constant becomes $G_{eff}/G=1+3\Upsilon/2$, which again implies that $\Upsilon >-2/3$ in order to achieve equilibrium (nonetheless more stringent constraints for $\Upsilon$ have been achieved, as we shall see in Sec.\ref{sec:RS}). In solving the modified hydrostatic equilibrium equation  (\ref{heqst}), Koyama and Sakstein \cite{Koyama:2015oma} implement the stellar structure code MESA  \cite{Paxton:2010ji} to introduce several physical ingredients that are known to play a role in realistic modelling of main sequence stars. In this case, the decrease of luminosity with $\Upsilon$ persists and, in addition, it is found that a star with a certain metallicity $Z_1$ and a given $\Upsilon$ may mimic a GR-star of larger metallicity $Z_2$, which introduces an additional degeneracy in the predictions of beyond Horndeski theories as compared to GR. A potential probe of these theories would be thus to infer the metallicity of the star from measurements of the globular cluster on which a main sequence star is located, and comparison with its place within the Hertzsprung-Russell diagram in GR and in these theories.

White dwarfs in  beyond Horndeski theories of G$^3$-type have been considered recently \cite{Jain:2015edg,Saltas:2018mxc}.  In white dwarfs electrons become degenerate, and their Fermi pressure counteracts the inward pull of gravity. In such a case the energy density and pressure of degenerate electrons can be expressed as $\rho_e=m_e^4 \xi(x)$ and $P=m_e^4\psi(x)$, where $m_e$ is electron's mass, and $\xi(x)$ and $\psi(x)$ are some functions of the dimensionless Fermi momentum $x\equiv P_F/m_e$ (see \cite{STBook} for details on these functions). Considering a zero-temperature white dwarf  composed of fully ionized carbon 12C with energy density $\rho_C=m_Cm_e^3x^3/(18\pi^2)$, where $m_C$ is mass of ionized carbon, and using the fact that in these stars $P_c \ll P_e$, Kumar Jain et al. \cite{Jain:2015edg} use this setup to constrain the value of $\Upsilon$ as follows:
\begin{itemize}

\item Mass-radius relations in binary systems: $-0.18 \leq \Upsilon \leq 0.27$ at $1\sigma$, and $-0.48\leq \Upsilon \leq 0.54$ at $5\sigma$.

\item Chandrasekhar's limit: $\Upsilon  \geq 0.22$ at $1\sigma$,

\item Stability of rotating stars from maximum centrifugal force:  $-0.59 \leq \Upsilon \leq 0.50$ at $1\sigma$.

\end{itemize}
Therefore, the most stringent bound on this parameter comes from mass-radius relations, which must be compared with other bounds, for instance, coming from the minimum main sequence mass (see Section \ref{sec:RIM}).

A different approach to this issue is carried out by Saltas et al. \cite{Saltas:2018mxc}. Recasting the hydrostatic equilibrium equation (\ref{heqst}) in parametric form, one finds
\begin{equation} \label{eq:hydrox}
\frac{dx}{dr}= -\rho \frac{\frac{m(r)}{r^2} +2\pi \Upsilon r \rho}{\frac{dP}{dx}+\pi \Upsilon r^2 \rho\frac{d\rho}{dx}}
\end{equation}
where $x \equiv P_F/(m_e c)$. For a typical white dwarf its central pressure can be estimated to be $P_c \sim M^2/R^4$ so $x_c \sim \mathcal{O}(1)$. Using a modification of the EOS of a non-interacting electron gas \cite{KoesterChan} via a non-uniform distribution of electrons as well as taking into account electron-electron interactions (Hamada-Salpeter model \cite{Salpeter}), and using the catalog of temperature-dependent EOS of \cite{Parsons et al} plus a star's envelope modelling EOS (thus going well beyond a simple polytropic approximation) accounting for the atomic numbers $Z=4$ (helium), $Z=6$ (carbon), $Z=8$ (oxygen), the authors numerically solve the corresponding field equations for a range of values of $\Gamma \in [-0.66,3]$. As expected, for zero temperature stars, negative (positive) values of $\Upsilon$ yield smaller (larger) radius; for instance, while a carbon-core star of $M=1M_{\odot}$ yields in GR $r_S \approx 0.008 R_{\odot}$, a value of $\Upsilon=1.5$ yields instead $r_S \approx 0.017 R_{\odot}$, i.e. roughly a twofold size. When the temperature is switched on, massive carbon-core stars' radius get increased by a $\sim 10\%$, while for helium-core stars this increase can be as high as $\sim 40\%$. This clearly demonstrates that, for white dwarfs, simple EOS neglecting temperature effects may easily lead to wrong interpretations upon the role played by the modifications of gravity. Using this modelling for the envelope region of the star (which is well described by GR since the modified gravity effects are negligible enough there), the analysis of the hydrostatic equilibrium equation (\ref{eq:hydrox}) via numerical simulations and its comparison with the measurement of mass and radius of 26 white dwarfs reported in \cite{Parsons et al} (with uncertainties of $3\%$) yields an upper bound of $\Upsilon <0.14$ at $2\sigma$ and $\Upsilon <0.18$ at $5\sigma$.  On the negative branch, the combination of the analysis of Jain et al.\cite{Jain:2015edg} and Babichev et al. \cite{Babichev:2016jom} places the bound $\Upsilon >-0.48$ at $2\sigma$. Therefore, consistent analysis of white dwarfs structure via realistic modelling still leaves open space for Horndeski-type modifications of gravity, which would require reliable measurements of white dwarfs envelope thickness.  One could add anisotropic contributions to the hydrostatic equilibrium equation (\ref{heqst}) where, as in the GR and $f(R)$ cases, a new term $\Delta(r)\equiv2(P_r-P_{\perp})/r$ is picked up. As found by Chowdhury and Sarkar \cite{Chowdhury:2018qrf}, the presence of anisotropy modifies the bound $\Gamma>-2/3$ in a way which is dependent on the ansatz chosen for $\Delta$(r) \cite{HeinHil}, which might slightly increase the Chandrasekhar mass of the isotropic case to a factor $\sim 1.2-1.3$ within the current constraints for $\Upsilon$.

From the hydrostatic equilibrium equation of beyond Horndeski theories (\ref{heqst}), Cermeno et al. \cite{Cermeno:2018qed} propose a model-independent parametrization of the Newtonian limit for several families of modified theories of gravity (including, for instance, $f(R)$ and Yukawa-type theories) as
\begin{equation}
\nabla^2 \Phi=4\pi \mu \rho \hspace{0.2cm};\hspace{0.2cm} \Psi=\gamma \Phi
\end{equation}
where the physical meaning of the function $\mu(r)$ is to introduce an effective gravitational constant, $G_{eff}=G\mu$, while $\gamma(r)$  serves as a link between the two gravitational potentials. In this case, Eq.(\ref{eq:nlimit}) becomes
\begin{equation}
P'=-\frac{m(r)\rho(r)}{p_4 r^2} -\frac{\xi \rho(r)}{p_4}\left(\frac{m''(r)}{r^2}- 2\frac{m'(r)}{r^3}\right)
\end{equation}
where $p_4$ and $\xi$ are constants appearing in the parametrization of $\mu$ and $\gamma$ for each modified theory of gravity under consideration ($p_4=1$ corresponding to GR) . Using polytropic index $n=3$, mass-radius relations for white dwarfs and solar-type stars are built choosing the values $\vert \xi \vert = \{4,1,1/4\} \times 10^{18}$cm$^2$. For all these stars it is found that values of $\mu$ larger (smaller) than $1$ (GR case) yields more (less) massive stars. This effect is enhanced for larger values of (positive) $\xi$, for instance, $\vert \xi \vert = 4\times 10^{18}$cm$^2$ with $p_4=1.1$ yields $\sim 15 \%$ of increase as compared to GR, provided that $\xi \lesssim  1.225 \times 10^{18}$cm$^2$ for the validity of the perturbative approach to hold. For $70\%$-hydrogen/$30\%$-helium solar-type stars one also finds a positive correlation between larger values of $\xi$ and higher (but slight) values of the central temperature and density. Moreover, stellar luminosity is also affected due to a combination of the effects associated to either $p_4$ and $\xi$. However, for the values reported in this work this effect is just $\sim 1\%$, thus being difficult to observationally disentangle it from the star's own internal oscillations in luminosity. Using now white dwarfs, it is found that consistency with the Chandrasekhar mass of the white dwarf reported in \cite{Hachisu}, with an estimated mass $M\approx 1.37M_{\odot}$, determines an exclusion region in a $p_4-\xi$ diagram; moreover, the validity of the perturbative approach requires here $\xi \lesssim  4\times 10^{14}$cm$^2$. Within these constraints, the increases in mass for white dwarfs can only be significant for larger increases of $p_4$. This example highlights the opportunity present in model-independent parametrizations of modified gravity theories.

In Banerjee et al. \cite{Banerjee:2017uwz} several modified theories of gravity are constrained, including $f(R)$, the fourth-order gravity studied by Stelle \cite{Stelle:1977ry}, and scalar-vector-tensor gravity \cite{Moffat:2005si}. After computing the Newtonian limit of these three theories, the authors consider a simplified model of dwarf stars composed of completely ionized carbon-oxygen with the gas on its ground state. This yields an EOS
\begin{equation}
P=\frac{m_e^4}{8\pi^2} \left[x\sqrt{1+x^2}\left(\frac{2}{3}x^2-1\right)+\log \left(x+\sqrt{1+x}\right)\right]
\end{equation}
where $x$ satisfies $dx/dr=-\frac{\sqrt{1+x^2} 3\pi^2 \rho(r)\Phi(r)}{m_e^4x^4}$, with $\Phi(r)$ the effective Newtonian potential for each theory. In the ultrarelativistic ($x \gg 1$) and non-relativistic ($x \ll 1$) regimes, this EOS yield polytropic EOS (\ref{eq:poly}) with $n=3$ and $n=3/2$, respectively. Numerical integration of the corresponding equations on each case yields the following results:

\begin{itemize}

\item For scalar-vector-tensor theories, a repulsive term always dominates on its Newtonian limit for any value of the coupling constants of the theory, and therefore the Chandrasekhar mass will decrease in all cases.

\item For fourth-order gravity this mass is basically unconstrained with $x$, which in turns allows to put constraints on the model parameters via consistency with the suggestion on the existence of ``super-Chandrasekhar" white dwarfs\footnote{Super-Chandrasekhar white dwarfs have been also investigated within some extensions of massive gravity \cite{Sun:2019niy} and in non-commutative geometries \cite{Kalita:2019yaj}, finding a theoretical maximum limit of $2.9M_{\odot}$ in the former, and $2.6M_{\odot}$ in the latter.  } from Type Ia Supernovae with masses ranging from $2.1M_{\odot}$ to $2.8M_{\odot}$ \cite{Howell:2006vn}.

\item In the $f(R)$ case, assuming that the theory can be expanded around a certain value $R_0$ as $f(R)=c_0+c_1R+c_2 R^2+\ldots$, the Newtonian potential is parameterized in terms of $\xi \equiv \sqrt{c_1/(6c_2)}$ and $\delta \equiv c_1-1$ such that, assuming again a super-Chandrasekhar white dwarf yields the constraint $\delta <1.076$.

\end{itemize}

Furthermore, a statistical constraint analysis on the parameters of each model according to the catalogue of twelve dwarf stars of \cite{Holberg:2012pu} yields the results reported in Table I of \cite{Banerjee:2017uwz}.

White dwarfs have been also considered within $f(R,T)$ theories, where $T$ is the trace of the energy-momentum  tensor, by Carvalho et al. \cite{Carvalho:2017pgk} for the simple scenario $f(R,T)=R+2\lambda T$, where $\lambda<0$ is some parameter. To numerically solve the Newtonian equations in this scenario, the authors use the standard EOS describing completely ionized atoms within a sea of relativistic Fermi gas of electrons used by Chandrasekhar \cite{Chandra}. It is found a very slight increase $(\lesssim 5\%$) in the Chandrasekhar mass for values up to $\lambda=-4 \times 10^{-4}$, above which this mass tends to a plateau. The increases in the radius are much large, up to a factor two with almost a similar decrease in the central density.

Precise measurements of the heaviest white dwarfs ever observed, such as those of the Extreme Ultraviolet Explorer all-sky survey (EUVE) catalog, with several stars in the range $1.32-137M_{\odot}$ and, in particular, WD J1659+440, with an estimated mass $M \approx 1.41\pm 0.04 M_{\odot}$ \cite{Vennes} should be able to put constraints on the parameters of all the theories discussed above. For instance, Crisostomi et al. \cite{Crisostomi:2019yfo} considered white dwarfs in certain families of DHOST theories characterized by a single parameter $\varepsilon_G$, getting to the conclusion that in these theories the Chandrasekhar limit is modified as
\begin{equation}
M_{\mathrm{ch}} \approx 1.44 M_{\odot}\left(1+\varepsilon_{G}\right)^{-3 / 2}
\end{equation}
Therefore, compatibility with the heaviest stars of that catalog yields the constraint $\varepsilon_{G} \lesssim 0.034$.

To conclude this part of the section let us note that generalized uncertainty principles\footnote{This kind of modifications arise from the hypothesis of a minimal length in quantum gravity theories \cite{Garay:1994en}, which changes the invariant measure of the momentum phase space via a factor $(1+\beta p^2)$, with current experimental constraints $\beta_0 \equiv \beta M_P^2c^2<7.65 \times 10^{34}$.} have been incorporated in the  Chandrasekhar limit of (helium-composed) white dwarfs using different heuristic approaches with contradictory results. While Rashidi \cite{Rashidi:2015rro} computes the generalized Lane-Emden equation for an ultrarelativistic Fermi gas and finds that the Chandrasekhar limit is removed, this result is contested by Ong and Yao \cite{Ong:2018nzk}, while a more refined treatment by Mathew and Nandy using the full EOS of a completely degenerate electron gas finds that these corrections only increase slightly this limit to $M_{Ch} \approx 1.45M_{\odot}$ and $r_{S} \approx 600$km \cite{Mathew:2017drw} (see also the recent update on this topic in \cite{Mathew:2020wnx}).

\subsubsection{Minimum main sequence mass. Brown dwarfs} \label{sec:RIM} \vspace{0.2cm}

As we have seen above, beyond Horndeski theories  with $\Upsilon>0$ predict a weakening of the gravitational strength inside astrophysical bodies, due to a lowering of the core density and temperature within these theories as a consequence of the modification of the hydrostatic equilibrium equation (\ref{heqst}). In turn, this has a non-trivial impact on the minimum mass required for a star to burn hydrogen in a stable way. The latter occurs when a star forms from contraction of a sufficiently heavy gas cloud under its own self-gravity until high-enough temperatures and densities are reached at their core that allow sufficient thermonuclear fluid to be ignited to compensate the surface energy losses.  This scenario represents a perfect window to test the predictions of modified theories of gravity, since the degeneracies in the EOS that plague the stellar structure models of relativistic stars (see Sec.\ref{sec:RS} below) are not present for these ``stars" lying beyond the lower end of the main sequence (termed as brown dwarfs). Indeed, their main properties are weakly dependent on non-gravitational physics and can be well modelled by polytropic EOS (\ref{eq:poly}). Due to its relevance, let us detail this calculation a bit. First, one considers a crude analytical model which interpolates between the fully degenerate regime and the one where the main internal pressure comes from motion of the gas, and which is described by the following parameters \cite{Burrows:1992fg}
\begin{equation} \label{eq:polyMMHB}
\Gamma=\frac{5}{3} \rightarrow n=\frac{3}{2} \hspace{0.3cm}; \hspace{0.3cm} K=\frac{(3\pi^2)^{\frac{2}{3}}\hbar^2}{5m_e m_{H}^{\frac{5}{3}}\mu_e^{\frac{5}{3}}}\left(1+\frac{\alpha}{\eta}\right),
\end{equation}
where $m_e$ and $m_H$ are the electron and hydrogen masses, respectively,  $\mu_e$ is the number of baryons per electron (which for a $75\%$-hydrogen/$25\%$-helium fluid can be approximated as $\mu_e \approx 1.143$), the constant $\alpha \equiv \frac{5\mu_e}{2\mu} \approx 4.82$ (with $\mu$ the mean molecular mass), while $\eta$ is a sort of measure of the degeneracy pressure supporting the star, formally defined as the ratio
\begin{equation}
\eta \equiv \frac{\mu_F}{k_B T} =\frac{(3\pi^2)^{\frac{2}{3}}\hbar^2}{2m_{\rm e}m_{\rm H}^{\frac{2}{3}}k_{\rm B}}\frac{\rho^{\frac{2}{3}}}{\mu_{\rm e}^{\frac{2}{3}}T}.
\end{equation}
where $\mu_F$ is Fermi's energy and $T$ the star's temperature. For the case of cubic $G^3$ galileons, Koyama and Sakstein \cite{Koyama:2015oma} solve the Lane-Emden equation (\ref{eq:LEbH}) to obtain the stellar's mass and radius as well as the core density for the polytrope (\ref{eq:polyMMHB}). The recipe for calculating the minimum mass for a star to belong to the main sequence (MMSM) is provided by Burrows and Liebert \cite{Burrows:1992fg}, using the formula for the luminosity associated to a star fuelled by the self-limiting hydrogen burning process $\textrm{p}+\textrm{p}\rightarrow\textrm{d}+\textrm{e}^++\nu_e$, whose energy generation $\epsilon$ can be modelled by the analytical approximation\footnote{For a more general description of the complicated basis of energy generation in low mass stars see \cite{MacielBook}.}  $\epsilon_{\rm HB}=\epsilon_c\left(\frac{T}{T_c}\right)^s\left(\frac{\rho}{\rho_c}\right)^{u-1}$, with $s\approx 6.31$, $u\approx 2.28$, and $\epsilon_c=\epsilon_0
T_c^s\rho_c^{u-1}$, where the constant $\epsilon_0\approx 3.4\times10^{-9}$ ergs s$^{-1}$s$^{-1}$ \cite{Burrows:1992fg}. Integration of this energy generation rate over the star's radius yields the total luminosity:
\begin{equation} \label{eq:luminosity}
L_{HB}=4\pi r_c^3\rho_c\epsilon_c\int_0^{\xi_R}\xi^2\theta^{\frac{3}{2}u+s}d\xi.
\end{equation}
Using the modified Lane-Emden equation (\ref{eq:LEbH}) for these theories one finds that the solution near the center can be approximated as \cite{Sakstein:2015zoa,Sakstein:2015aac},
\begin{equation}
\theta(\xi) \approx 1-\left(1+\frac{3\Upsilon}{2}\right)\frac{\xi^2}{6} \approx \exp\left[-\left(1+\frac{3\Upsilon}{2}\right)\frac{\xi^2}{6}\right]
\end{equation}
which, introduced into Eq.(\ref{eq:luminosity}) and bearing in mind (\ref{eq:polyMMHB}) yields the explicit result\footnote{This approximation makes sense taking into account that most of the luminosity radiated by the star comes from the central region where  $\xi \ll 1$; indeed, using this approximation introduces errors of size of less than $1\%$ in the range $\Upsilon \sim 0-0.05$ \cite{Koyama:2015oma}.}:
\begin{equation}
L_{HB}=\frac{3\sqrt{3\pi}}{\sqrt{2} \omega_{3/2} \left[\left(1+\frac{3\Upsilon}{2} \right)\left(\frac{3}{2}u+s\right)\right]^{3/2}} \epsilon_c M
\end{equation}
This total luminosity must be matched to the luminosity at the photosphere, an equality which is achieved when the star is burning hydrogen in a stable way. The latter requires some extra information on the metallicity/gas transition at the photosphere, where the surface gravity can be taken to be constant. After some manipulations and using the generalized hydrostatic equilibrium equations (\ref{heqst}), Sakstein finds that the MMSM in these theories can be approximated as
\begin{equation}
0.376\frac{M_{MMHB}}{M_{\odot}} \approx  \left[\frac{1+\frac{\Upsilon}{2}}{100 \kappa_R}\right]^{0.11}\left(1+\frac{3\Upsilon}{2}\right)^{0.14}  \frac{\gamma_{3/2}^{1.32}\omega_{3/2}^{0.09}(\alpha+\eta)^{1.509}}{\delta_{3/2}^{0.51}\eta^{1.325}}  \label{eq:MMHBst}
\end{equation}
where  $\kappa_R$ is Rosseland's mean opacity, which for high-mass brown dwarfs can be approximated by a constant $\kappa_R \approx 10^{-2}$cm$^2$/g. The value $\Upsilon=0$ corresponds to GR, and in this $n=3/2$ case one has $\gamma_{3/2}=2.357,\delta_{3/2}=5.991,\omega_{3/2}=2.714$, and replacing these values into Eq.(\ref{eq:MMHBst}), one finds that the consistent solution of that equation with the lowest possible mass corresponds to $M_{MMSM} \approx 0.0865M_{\odot}$\footnote{Other analytic models show larger variations in this prediction: for instance, in \cite{Auddy}, by combining the modelling of the surface luminosity with with plasma phase transitions, this limit is placed in the range $M_{MMSM} \approx 0.064-0.087$.}, which is quite in agreement with numerical simulations which tend to slightly lower this value \cite{Kumarnum,Burrows:1992fg}. This bound is compatible with current observations; indeed, increasing values of $\Upsilon$ yields larger values of $M_{MMSM}$ in Eq.(\ref{eq:MMHBst}) such that for $\Upsilon \gtrsim 0.027$\footnote{A caveat comes here. In his work, Sakstein seems to compute $\omega_{3/2}$, $\delta_{3/2}$ and $\gamma_{3/2}$ using the values obtained in GR. However, as the resolution of the Lane-Emden equation (\ref{eq:LEbH}) in this case yields solutions different from those of GR (due to the presence of $\Upsilon$), these numbers should be different too, which should modify the value for $M_{MMSM}$ that the formula (\ref{eq:MMHBst}) yields, having a nontrivial impact on the  results. A second concern is that the $M-\Upsilon$ diagrams plotted in Refs.\cite{Sakstein:2015zoa,Sakstein:2015aac} do not match each other and, moreover, seem to be incompatible with formula (\ref{eq:MMHBst}) itself. A more careful analysis of this issue is needed in order to guarantee the consistency of these results. }
it overcomes the threshold $M_{MMSM} \approx 0.0930\pm 0.0008 M_{\odot}$, corresponding to the estimated mass of the lowest-mass M-dwarf star ever observed, Gl 866 C \cite{Segransan:2000jq}. Since the MMSM is weakly sensitive to potential degeneracies that may affect this computation (note in this sense, that the addition of rotation would slightly increase this mass), this places constraints upon some scalar-tensor theories. Indeed, the parameters of cosmological models of dark energy in beyond Horndeski theories of G$^3$ type are directly related, in an effective field theory approach, to $\Upsilon$ via the equation \cite{Saito:2015fza}
\begin{equation} \label{eq:Galpha}
\frac{\Upsilon}{4}=\frac{\alpha_H^2}{\alpha_H-\alpha_T-\alpha_B(1+\alpha_T)} \lesssim 0.0068
\end{equation}
where the coefficients $\{\alpha_H,\alpha_T,\alpha_B\}$ are found from the effective field theory description of dark energy on linear scales. In this sense, some cosmologically viable scalar-tensor theories can be excluded by the bound above, such as the beyond Horndeski covariant quartic galileon, which has $\Upsilon=1/3$. Note that, as discussed above, stable stellar models can only exist if $\Upsilon>-2/3$ \cite{Saito:2015fza}, but the consistency with mass-radius relations further narrows down this range to
\begin{equation}\label{eq:conhor}
-0.22\lesssim \Upsilon \lesssim 0.027
\end{equation}
for these theories to be observationally viable \cite{Babichev:2016jom}.

A similar computation of the MMSM mass for certain families of DHOST theories characterized by a single parameter $\varepsilon_{G}$  has been carried out by Crisostomi et al. \cite{Crisostomi:2019yfo} with the result
\begin{equation}
\left(1+\varepsilon_{G}\right)^{1.398} \frac{M}{0.1 M_{\odot}}\left(\frac{\kappa_{R}}{10^{-2}}\right)^{0.111}\gtrsim 0.9227
\end{equation}
which taking $\kappa_R=10^{-2}$ gives the bound for GR of $M_{MMSM} \approx 0.09227M_{\odot}$ which is a $\sim 10\%$ larger than the one obtained by Sakstein. In any case, via the same comparison with observational data of the lowest-mass stars, the computation above yields a constraint for the DHOST parameter $\varepsilon_{G}\gtrsim -0.0057$. 

There is a second limiting mass in brown dwarfs that offers complementary constraints and observational opportunities for brown dwarfs. This is the deuterium burning mass limit (MMDB), which divides giant planetary-mass objects and low-mass brown dwarfs. This value has a strong dependence on the assumed chemical composition (helium abundance, metallicity, etc) of the corresponding objects, namely, $M_{MMDB} \sim 0.010-0.015$ 
\cite{Spiegel:2010ju}. Recently Rosyadi et al. \cite{Rosyadi:2019hdb} have investigated this MMDM limit within beyond-Horndeski theories of $G^3$-type, using the compilation of brown dwarf masses and radii of the catalog of \cite{Bayliss2016}. Via $\chi^2$ analysis of these data they report the constraint on the $G^3$ parameter 
\begin{align}
 -0.565&<\Upsilon<0.234\;\;\;\text{up to}\;\;1\sigma\;\text{confidence level};\\
 -0.6&<\Upsilon<0.391\;\;\;\text{up to}\;\;5\sigma \;\text{confidence level}
\end{align}
Unfortunately, their analytical modelling of the MMDB limit within beyond-Horndeski gravity fails to yield the GR value, and would require a more refined calculation in order to provide reliable constraints on $\Upsilon$ from the combination of the MMSM and MMDB masses.

Further tests of these theories along these lines would be via systematic observations of the radius of low-mass brown dwarfs, which can be modelled via a polytropic EOS (\ref{eq:poly}) with $n=1$ \cite{Saumon:1995bn}, and then Eq.(\ref{eq:RSfor}) implies that the radius is independent of the mass, i.e., $r_S=\gamma_1 K^{1/2} \approx 0.1 R_{\odot}$. Thus, modifications of the Lane-Emden equation would change the value of $\gamma_1$ due to a different $\theta(\xi)$ coming from the resolution of that equation with respect to the GR one; in the case of beyond Horndeski theories this reads explicitly
\begin{equation}
r_S(\Upsilon)\approx 0.1 \frac{\gamma_1(\Upsilon)}{\gamma_1(\Upsilon=0)}R_{\odot}
\end{equation}
where $\gamma_1(\Upsilon=0) =\sqrt{\pi/2}$ corresponds to the GR case. At the maximum theoretically viable (though not observationally) value $\Gamma =-2/3$ one finds a  significant decrease of the brown dwarf radius to $r_S \approx 0.078R_{\odot}$ \cite{Sakstein:2015aac}, while for the branch $\Upsilon>0$, where the weakening of the gravitational force occurs, the radius is larger. This is quite a simple computation that offers a window for the parameters of many modifications of gravity to be constrained in the future via new missions able to carry out surveys of low mass stellar objects, such as GAIA \cite{Brown:2018dum}.   \\

\textit{Further comments} \vspace{0.2cm}

To conclude this section, let us mention that another suitable scenario to test these theories is that of stellar pulsations (i.e., oscillations around hydrostatic equilibrium configurations), which (for adiabatic oscillations) yields instabilities for $\Upsilon \gtrsim 49/6$ \cite{Sakstein:2016lyj} (already ruled out, as we just have seen). Considering polytropic EOS and keeping the mass unchanged, one finds that cubic galileons induce significant changes in the pulsation period of brown dwarfs. Cepheid stars are also suitable candidates to constrain the parameter $\Upsilon$ (and modified gravity at large) via distance and mass estimates in eclipsing binaries. This is the case for generic scalar-tensor theories, where the change in the oscillation period can reach a factor $\sim 30\%$ for matter couplings of order unity \cite{Sakstein:2013pda} as compared to GR. Cepheids are also useful to put stringent constraints upon chamaleon mechanism parameters \cite{Sakstein:2014nfa}, while the weakening of the gravitational force  may induce changes in the temperature of red giants of up to $\sim 100 K$ at the same luminosity, which could be measurable in certain dwarf galaxies \cite{Chang:2010xh}.

\subsection{Relativistic stars: mass-radius relations and maximum masses} \label{sec:RS} \vspace{0.2cm}

Neutron stars are truly relativistic objects, so for them one has to face the resolution of the TOV equations and their modifications within modified gravity (such as Eq.(\ref{eq:TOV}) for $f(R)$ theories)  in their full complexity. In GR, for each theoretical model of nuclear matter at supranuclear densities an EOS can be derived, which allows to numerically integrate the TOV equations to obtain a family of static, spherically symmetric solutions, corresponding to a certain range of values of the central density (and pressure).  The result of this computation can be typically extracted in terms of a characteristic mass-radius diagram on which, starting from some small value of the ratio $P(0)/\rho(0)$, the radius of the star decreases while the mass increases, until a saturation point is reached, corresponding to the configuration with maximum possible mass. Further increases of the mass will yield dynamically unstable configurations.  As it should be expected, the qualitative and quantitative details of this picture depend strongly on the EOS chosen and, in the context of modified gravity, also on the internal parameters of the specific model under consideration, thus allowing possibilities that run away from the simple description briefly depicted here.

Unsurprisingly, a widespread trend in relativistic stellar structure models within modified theories of gravity is to find stable configurations modifying the mass-radius relations as compared to those of GR and, in particular, yielding a maximum allowed mass that can be compared with observations. In this sense, the observations of several neutron stars with mass at the $2M_{\odot}$ threshold \cite{J0348+0432,Crawford:2006xb,LinShaCas18,Cromartie:2019kug} has put into trouble a number of EOS within GR, as we shall see below.
Regarding the radius, recent estimates from gravitational wave astronomy of neutron star mergers via hydrodynamic simulations with different EOS \cite{Bauswein:2013jpa} offer the bound $r_{S} \geq 9.60^{+0.14}_{-0.03}\text{km}$ for the minimum radius of a non-rotating star \cite{Bauswein:2017vtn}, which can be further refined by future simulations. The presence of gravitational corrections in these models  introduces a degeneracy with GR predictions since, in addition to the existence of numerous ``realistic'' EOS with different outcomes both for the mass-radius relations and for the maximum allowed masses, now we have extra parameters in the gravitational sector to play with. Due to this, several EOS ruled out in the context of GR resulting from their incompatibility with observations can be brought back to life in the context of modified gravity. At the same time, the analysis of numerical solutions and comparison with observational results may allow to constrain the extra parameter(s) of modified theories of gravity.

\subsubsection{An overview on equations of state} \label{sec:EOS} \vspace{0.2cm}

Neutron stars reach densities at their center well above the nuclear saturation density, $\rho_{s} \approx 2.8 \times 10^{14}$ g/cm$^3$ (typically from five to ten times). At such densities the reliability of many-body nuclear theory decreases quickly and there is little experimental information on the contributions of hyperons, quarks, phase transitions, etc. Thus, the EOS at these densities necessarily extrapolate calculations based on phenomenological models of the nuclear force as well as in effective field theory models of quantum chromodynamics, thus depending strongly on the assumptions upon which a microscopic theory of dense, strongly interacting matter is built. This modelling of the EOS is critical, as many of the macroscopic properties of neutron stars are strongly dependent on the ingredients included and the behaviour of the nuclear matter they are made of, and therefore they can be directly related to uncertainties on this extrapolation. One of the most important of such properties is its mass-radius relation, as it is (almost) uniquely defined by its EOS. The masses can be determined in binary systems via precise measurements of the effects on their companion's orbit, while Doppler shift of spectral transitions may provide the $M/r_S$ ratio. Recently, the observation of gravitational waves from the merger of a binary system of neutron stars also allowed to put stringent constraints upon the EOS of matter at supranuclear densities \cite{Abbott:2018exr,Annala:2017llu,Radice:2017lry}. This choice of EOS is doubly relevant here, as finer details in the modelling of the star's interior may significantly interfere with the predictions of the underlying gravitational theory extending GR.

Any EOS must satisfy a number of conditions to describe a physically viable star. These include:
\begin{itemize}
\item The fulfillment of the weak energy condition, which demands $\rho>0$ and $\rho +P>0$,
\item The Le Chatelier's principle for microscopic stability of matter (to avoid spontaneous local collapse of matter), which reads $P \geq 0$ and $dP/d\rho>0$,
\item A causality constraint upon the speed of perturbations to be lower than the speed of light, namely, $c_s\equiv (dP/d\rho)^{1/2} \leq 1$.
\item Obviously, consistency of the output of the corresponding numerical integration of the TOV equations with the maximum neutron star mass observed so far. In this work \emph{we shall take $M_{\star}=2M_{\odot}$ as the reference value for this threshold}.
\end{itemize}

There is a huge variety of such models in the literature. Generally speaking, EOS can be split into ``soft" and ``stiff", which refers to the behaviour of the pressure as more mass is captured, namely, the behaviour of the sound speed $c_S^2 \equiv dP/d\rho$. Thus, in soft EOS matter may be more effectively compressed, leading to smaller thresholds and to more compact stars while, the other way around, stiff EOS are less effective in this compression allowing a larger threshold with also larger radius. However, this broad description is influenced by other aspects on the modelling of the physical ingredients included (such as the appearance of hyperons/quarks at high energies which tend to soften the EOS \cite{GM1}), the symmetries assumed, or the properties of the theory of gravity under consideration. A generation of the most elaborate such realistic (but non-relativistic) EOS models are those based on experimental nuclear physics making use of many-body calculations for dense neutron matter at supranuclear densities.  Among them we underline several that will be extensively used in the text below. Then we find some well known models such as the FPS model \cite{FPS1,FPS2,Lorenz:1992zz,FPS3}, where the crust-core transition is realized through a sequence of phase transitions involving changes in nuclear shapes ($M_{\star} \approx 1.80 M_{\odot}$ \cite{STBook,Demorest10});  the SLy model of effective nuclear hamiltonian obtained from many-body calculations with a simple two-nucleon potential and a modelling of the crust-core transitions  as a weak first-order transition with a minor density jump \cite{SLy1,Sly2,Sly3,Sly4}, which typically admits analytical representations \cite{CamenzindBook} ($M_{\star} \approx 2.05M_{\odot}$ and $r_S \sim 10$ km); the APR4 EOS \cite{Lattimer12,Akmal:1998cf} uses instead a three-nucleon potential in addition to an Argonne 18 potential ($M_{\star} \approx 2.2M_{\odot}$), and its variant the WWF1 EOS \cite{Wiringa:1988tp}, which differs on the nucleon potential model used ($M_{\star} \approx 2.1M_{\odot}$ \cite{Akmal:1998cf}); or the BSk family of EOS (BSk19,BSk20,BS21) \cite{Goriely:2010bm,Chamel:2011aa} obtained from generalized Skyrme interactions supplemented with additional interacting and correcting terms ($M_{\star}\approx \{1.86,2.16,2.27\}M_{\odot}$ \cite{Potekhin:2013qqa}). These models only consider nucleons, while other constituents such as hyperons, pions/kaons, or condensates, which may play a role at very high densities, are incorporated in other EOS: the GM1 model \cite{GM1} based on relativistic mean field calculations ($M_{\star} \sim 2.0-2.4M_{\odot}$), and its extension GM1nph \cite{GM1nph} for cold neutron star matter in equilibrium including the effect of the baryon octet and electrons ($M_{\star} \sim 1.9-2.1M_{\odot}$); or the MPA1 \cite{MPA,Gungor:2011vq} obtained in the relativistic Dirac-Brueckner-Hartree-Fock formalism and relativistic mean field theory, accounting for the energetic contributions due to the exchange between pions and mesons ($M_{\star} \approx 2.5M_{\odot}$), as well as the related MS1 \cite{Mueller:1996pm} ($M_{\star} \approx 2.7M_{\odot}$).

A new state of matter that could be present inside the innermost regions of neutron stars is unconfined quark matter made of u, d, and s quarks (besides electrons) forming a colour superconductor, and yielding quark stars. One of the simplest such quark-core models is the MIT bag model \cite{JL79}, described by the simple EOS
\begin{equation}\label{quark}
P=a(\rho-4B)
\end{equation}
where the bag constant $B$ varies from $\sim 60-90$MeV/fm$^3$ depending on the specific modelling, while the parameter $a$ depends softly on the strange quark mass assumptions and the QCD coupling constant (typically $a \sim 0.28-0.33$, corresponding to $m_s=0$ and $m_s=250$MeV, respectively). This EOS is usually employed above a certain transition density $\rho \sim 1.0 \times 10^{15}$ g/cm$^3$, where the effects of quark matter cannot be neglected. These hybrid stars with a quark core and an hadronic outer layer can get to the $2M_{\odot}$ threshold provided that the model parameters are properly chosen \cite{Bonanno:2011ch}.

More realistic descriptions of nuclear matter inside neutron stars are built by joining together different polytropic phases on a sequence of different density intervals (piecewise EOS), namely (see e.g. \cite{Carney:2018sdv} for details)
\begin{equation}
P(\rho)=K_i \rho^{\Gamma_i}
\end{equation}
with particular $K_i$ and $\rho_{i-1}<\rho<\rho_i$ for each of them, and which are smoothly matched to each other at each transition density $\rho_i$ (so each density and pressure are everywhere continuous functions). Typically, it is also demanded that in these piecewise EOS there is a smooth transition from a high-density regime to the low-density one, namely,  that at low densities it recovers the well known EOS of neutron star crust \cite{Lattimer:2012nd}.  The analysis of these piecewise EOS at high densities can be systematized via parameterized phenomenological relations constrained by astrophysical observations on the most massive neutron stars observed so far \cite{Read:2008iy}. Let us also mention that Yagi and Nunes \cite{Yagi:2016bkt} have included a quite detailed literature on realistic EOS in their work on I-Love-Q relations, see figure 1 of that work for a quick glance on the mass-radius relations of many of the most popular EOS. In addition, Breu and Rezzolla \cite{Breu:2016ufb} report up to 28 different EOS motivated by different theoretical considerations that are compatible with the finding of neutron stars on the $2 M_{\odot}$ threshold. More recently, Oter et al. \cite{Oter:2019kig} provide tables of EOS constrained by laboratory physics and fundamental principles, without any reference to the underlying theory of gravity. We refer the reader to these works for a detailed account on the bibliography  associated to these realistic EOS.

\subsubsection{$f(R)$ gravity} \vspace{0.2cm}

After this brief overview on EOS, let us begin with the analysis of modified gravity models and their predictions for relativistic stars. Scalar-tensor theories, of which $f(R)$ is a particular case, are the most well studied models within stellar structure physics. Most of this analysis is restricted to matter sources satisfying $T=-\rho + 3P< 0$, which for a perfect fluid implies $P < \rho/3$, a condition typically satisfied by most EOS at the center of neutron stars. Nonetheless, this feature depends critically on some assumptions of the models, such as their softness or the appearance of hyperons/quark states. Indeed, some piecewise EOS allow for $T>0$ in some configurations, which would yield new phenomenology induced by nonperturbative effects, such as in spontaneous scalarization \cite{Damour:1993hw}, which will play a relevant role later. Moreover, it has been argued by Podkowka et al. \cite{Podkowka:2018gib} that the transition value $T=0$ at the star's center would yield a universal bound of these theories for the star's compactness $\mathcal{C}=0.262^{+0.011}_{-0.017}$ ($90\%$ confidence interval); therefore the observation of neutron stars at the $2M_{\odot}$ threshold and a radius $R \lesssim 11.2^{+1.0}_{-0.5}$km would imply that $T>0$ on the star's interior up to the confidence level above. Hereafter in this section we shall assume $T \leq 0$.

Let us thus consider in this section $f(R)$ gravity. The first studies in this context resorted to  a perturbative approach where density, pressure, mass function, and metric (and thus curvature) are expanded to linear order in some small parameter $\tau$ (which is related somehow to the extra gravitational corrections) as $\zeta \approx \zeta^{(0)}+\tau \zeta^{(1)} + \mathcal{O}(\tau^2)$, where the zero order terms $\zeta^{(0)}$ are assumed to satisfy the generalized Einstein equations, while the integration of the field equations must guarantee a mild growth of the perturbation terms to achieve stability of the solution.  A detailed analysis of the $f(R)$ equations in this case has been done by Aparicio Resco et al. \cite{Resco:2016upv}. There, by writing $f(R)=R+F(R)$ and assuming $F(0)=0$, the following theoretical stability requirement is found:
\begin{equation} \label{eq:stability}
\frac{1+F_R(0)}{3B(r)F_{RR}(0)}>0.
\end{equation}

Next we deal with the predictions of the quadratic $f(R)$ model (\ref{eq:fquadratic}).  For this model, Cooney et al. \cite{Cooney:2009rr} consider only the effect of higher-order corrections and take a polytrope (\ref{eq:poly}) with $\Gamma=9/5$ and numerically integrate (Runge-Kutta method) the generalized Einstein and TOV equations (\ref{eq:Dr}), (\ref{eq:Br}), (\ref{eq:trace-m2}) and (\ref{eq:TOV}) from the star's center with some initial free conditions upon the functions $A(0)$, $B(0)$ towards its surface, finding merely slight modifications to the mass-radius relation as compared to GR within the validity of the perturbative approximation, $-0.015<\alpha K^{5/4}<0.01$, with a minor increase of the maximum mass for $\alpha<0$. More interesting results are obtained by Orellana et al. \cite{Orellana:2013gn} using analytical representations of the Sly EOS with tabulated coefficients \cite{Sly4} and with a set of central densities $\rho \approx 10^{14.6}-10^{15.9}$ gr/cm$^3$ to Runge-Kutta-integrate the field equations.  Qualitatively similar density profiles and mass-radius relations are found as compared to GR ones, with larger (smaller) masses for $\alpha<0$($>0$), in agreement with the results of \cite{Cooney:2009rr}; for instance, for $\alpha=-0.2$km$^2$, the maximum mass can be raised up to $M_{\star} \sim 2.2M_{\odot}$, while the radius is only slightly reduced. The addition of  magnetic fields\footnote{For a general description of magnetized neutron stars in GR and how it influences the maximum mass, see e.g.  \cite{Chatterjee:2018prm}.} with values as high as $10^{16}-10^{17}$G for some suitable EOS \cite{Ryu:2010zzb} seems to have little  effect on these results \cite{Cheoun:2013tsa}. Besides these mild improvements on the maximum masses, these perturbative approaches have the limitation of being unable to take into account the large curvatures attained  at the star's center (which may overrule the validity of the perturbative approximation), as well as the impossibility to check the compatibility of the perturbative solution with the exact one. To obtain more reliable descriptions one needs to go to the fully non-linear regime and solve the exact generalized TOV equations.

A non-perturbative approach is implemented by Yazadjiev et al. \cite{Yazadjiev:2014cza} in the scalar-tensor representation in the Einstein frame, with standard boundary conditions at the star's center, $\rho(0)=\rho_c$ and $\phi'(0)=0$ (the latter to set regularity of the scalar field there), and via numerical integration of the field equations with a shooting method.  A Sly4, a piecewise APR4 \cite{FPS3}, and a FPS EOS are selected. Their main findings are that, by assuming $\alpha>0$, the maximum mass can be raised by a factor $\sim 10\%$ for suitable large values of $\alpha$. This result seems to be in contradiction with the perturbative analysis of  \cite{Cooney:2009rr,Orellana:2013gn, Arapoglu:2010rz}, where the positive $\alpha$ branch was reported to \emph{decrease} the maximum mass. This conflict lies in the exploration of different ranges of values of $\alpha$: while the perturbative approach relies on small enough values of $\alpha$ and the increase (decrease) of maximum masses for $\alpha<0$ ($\alpha>0$) is found, the authors of \cite{Yazadjiev:2014cza} report their results to be consistent with that range until a critical value of $\alpha$ is reached, above which such relation between mass and $\alpha$ is inverted. In another work on this model within the branch $\alpha<0$, Aparicio Resco et al. \cite{Resco:2016upv} use a shooting method demanding as boundary conditions at infinity the Schwarzschild solution up to a desired precision. The numerical integration of the field equations thus find the initial values $R(0)$ and $A(0)$ consistent with them. Using a blend of soft, middle and stiff samples of EOS provided in \cite{Hebeler:2013nza}, the authors find that the maximum mass of the star is basically unconstrained as $\vert \alpha \vert$ grows, while the radius of the star is significantly diminished (partially due to the energy contribution of the scalar field to the mass); due to this it is found that for large enough values of $\vert \alpha \vert$ the soft EOS yields larger masses than the middle and stiff ones. This problem is also attacked by Astashenok et al. \cite{Astashenok:2017dpo} with a different approach: regularity and finiteness for both density and pressure requires some conditions at the center of the star, $B_0=1, (A'/A)_{0}=0,R'(0)=0$, while imposing the matching with the asymptotic Schwarzschild solution (Eq.(\ref{eq:SdS1}) with $\Lambda=0$) requires $R(\infty)=0, B(\infty)=A(\infty)=1$\footnote{Should $A(\infty)$ go instead to some constant $A_{\infty}$, a simple re-scaling of the initial condition as $A(0)\rightarrow A(0)/A_{\infty}$ brings the numerical problem into the desired form.}. Therefore, assuming any EOS such that both density and pressure vanish beyond the star surface radius $r_S$, the numerical integration of the field equations takes place from $r_S$ outwards.
Due to the contribution of the matter fields outside the star surface to its total mass discussed in Sec.\ref{sec:sssm}, the interpretation of the total star's mass using Eq.(\ref{eq:B(r)normal}) must be treated with care. Here the authors parameterize the difference between the gravitational mass measured by an asymptotic observer, namely,  $M=m(r\rightarrow \infty)$, and the one enclosed within the star radius $M_S=m(r=r_S)$, by the \emph{gravitational sphere}, whose extra contribution to the mass is a crucial effect in order to obtain the right mass-radius relation in these models\footnote{For an extended discussion on the definition of the gravitational mass in models with $R^2$-corrections see \cite{Sbisa:2019mae}.}.  As EOS, analytical representations of Sly4 \cite{CamenzindBook} and of APR4, GM1, MPA1 \cite{Gungor:2011vq}, and also a GM1nph EOS are considered, besides quark stars (\ref{quark}) with $B=60$MeV/fm$^3$ and $a=0.31$. For $\alpha>0$ it is observed that the maximum allowed masses in the quadratic model (\ref{eq:fquadratic}) deviate from the GR ones by a correction of size $\propto \alpha^{1/2}$. Taking a value as high as $\alpha=20 \times 10^{10}$cm$^2$, this implies a slight increase of $2\%-4\%$ for the above EOS (in all cases one has $M_S<M$), peaking at $M_{\star}\approx 2.57M_{\odot}$ ($2.49M_{\odot}$ in the GR case) for the MPA1 EOS. Such slight mass increase fall well within the current observational errors of mass measurements. On the other hand, for the branch $\alpha<0$, taking MP1, AP4, SLy and GM1 EOS, the numerical integration fails to give a decreasing mass function of the radius, which implies that the Newtonian limit is not recovered far away from the star, and thus no stable configurations are found for the EOS above (in contradiction with the results of \cite{Yazadjiev:2014cza}). Astashenok et. al. \cite{Astashenok:2018iav} also analyze this model in the scalar-tensor representation using APR, SLy and GM1 and some realistic EOS with hyperons \cite{Miyatsu:2013hea}, solving the field equations with a Runge-Kutta-Merson fourth-order method. In the negative branch of $\alpha$, it is observed that the gravitational mass  grows unboundedly with the distance, deviating from the asymptotically Schwarzschild solution, which is argued by the authors to render this branch as physically unacceptable. For $\alpha>0$ a matching with the asymptotically Schwarzschild solution is possible, and the observed gravitational mass in terms of the mass of the star plus the mass of the gravitational sphere only increases slightly as compared to GR values for all these EOS, though the radius may be increased up to $\sim 10 \%$. Further refinements of these models such as the addition of electrically charged matter make use of unrealistic values of the electric charge inside compact stars \cite{Mansour:2017ovd} and thus are not described here\footnote{For some analysis of charged neutron matter within GR see e.g. \cite{Arbanil:2013pua}.}. 

Anisotropies of the matter fields incorporated in the TOV equations via Eq.(\ref{eq:TOVani}) have been also considered. If one parameterizes the anisotropy by  $\sigma \equiv P_{\perp}-P_r$, Folomeev \cite{Folomeev:2018ioy} uses the quasi-local EOS introduced by Horvat \cite{Horvat:2010xf}
\begin{equation}\label{eq:Horvat}
\sigma \equiv 2\lambda_{H}p_r \frac{M(r)}{r}
\end{equation}
and Bowers-Liang \cite{Bowers:1974tgi}
\begin{equation}\label{eq:BL}
\sigma \equiv \frac{1}{3}\lambda_{BL} (\rho+3p_r) (\rho+p_r)r^2\left(1-\frac{2(1-2M(r)/r)}{r}\right)^{-1}
\end{equation}
with the parameters $-2 \leq \{ \lambda_{H},\lambda_{BL} \} \leq 2$. Choosing EOS for the radial pressure to be FPS, SLy, BSk21 in terms of their analytical representations \cite{Sly4,Potekhin:2013qqa}, for values $\alpha= -5\times 10^{10}$cm$^2$ and $\alpha=-20 \times  10^{10}$cm$^2$, numerical resolution of the TOV equations yields a slight increase of $M_{\star}$ up to $ \sim 5\%$ for all these scenarios, with also slight decreases on the star's radius, though the central density may be increased up to a factor $2-3$ in some combinations of the selected parameters of these models. Let us also point out that a formulation of this theory including torsion and using APR4, MPA1, SLy, and WW1 EOS has been recently considered in Feola et. al. \cite{Feola:2019zqg}, finding that its effect is to decrease the corresponding total mass and compactness as compared to GR results.

The bottom line of the quadratic $f(R)$ model discussed above is that, while it is able to slightly increase the maximum mass in some cases (which may allow to re-consider FPS as a viable EOS in the context of these theories), these deviations are still comparable with the uncertainties present in the EOS for dense matter within GR and, therefore, do not help to solve the degeneracy problem, unless the EOS can be better constrained in the future by other means.

Stellar structure models in $f(R)$ gravity beyond the quadratic theory have been widely investigated in the literature. Astashenok et al. \cite{Astashenok:2013vza} consider three such models: the exponential one \cite{Cognola08}
\begin{equation} \label{eq:fRexponential}
f(R)=R+\beta R(\exp(-R/R_0)-1)
\end{equation}
(with $R_0$ and $\beta$ some constants); the logarithmic one
\begin{equation} \label{eq:fRlog}
f(R)=R+\alpha R^2(1+\beta \log[R/\mu^2])
\end{equation}
(with $\alpha, \beta$ and $\mu^2$ some constants); and the cubic one
\begin{equation} \label{eq:fRcubic}
f(R)=R+\alpha R^2(1+\gamma R)
\end{equation}
(with $\alpha$ and $\gamma$ some constants). This analysis makes use of analytical representations of SLy and FPS EOS (following the analysis of \cite{Camezind}), parameterized \{BSk19, BSk20, BSk21\} EOS as given by \cite{Potekhin:2013qqa}, and quark-core matter (\ref{quark}), solving numerically the generalized Einstein equations using a perturbative approach. For the exponential model (\ref{eq:fRexponential}), the quark-core and Sly EOS provide no significant deviations with respect to GR, but for the FPS one the maximum mass can be raised up to the $2M_{\odot}$ threshold. For the logarithmic model (\ref{eq:fRlog}), the Sly and FPS EOS can only support neutron stars with a central density below some critical value, yielding maximum masses $M_{\star} \approx 1.93 M_{\odot}$ (Sly) and $M_{\star} \approx 1.75  M_{\odot}$  (FPS), while a piecewise FPS+quark core EOS cannot get to the $2M_{\odot}$ threshold either. Thus, the logarithmic model with these EOS seem to be discarded by observations.

The logarithmic model (\ref{eq:fRlog}) has been considered by Alavirad and Weller \cite{Alavirad:2013paa}, where their focus is to verify that the chamaleon effect still allows $\gamma_{PPN} \simeq 1$ for this model, and to determine that the gravitational redshift at the star's surface, $z_s \equiv (1-2M/r)^{-1/2}-1$, may be significantly larger than in the GR case; for instance, taking $\beta=-0.05$ and $\alpha=10^{10}$cm$^2$ one finds an increase of $\sim 10 \%$ (note that Buchdahl's bound limits this redshift to $z_s \leq 2$). Regarding the structure of relativistic stars, they consider quark EOS (\ref{quark}) with $a=0.28$ and
$B=60$MeV/fm$^3$, finding an increase (decrease) of the maximum mass with negative (positive) $\alpha$ regardless of the value of $\beta$: for instance $\beta=-0.25$ and $\alpha= 10^{11}$cm$^2$ yields a $\sim 10 \%$ increase allowing this EOS to overcome the $2M_{\odot}$ threshold.

For the cubic model (\ref{eq:fRcubic}), the mass increases with negative $\gamma$ yielding stable stellar configurations with a maximum mass of $M_{\star} \approx 1.94 M_{\odot}$ and radius $r_S\approx 9.2\text{km}$ (significantly smaller than in GR) for the FPS EOS with $\gamma=-20$. Smaller stars are also found for BSk20 and AP4 EOS, with masses around $2M_{\odot}$ in some range of the parameters. A novel feature arises here, as a second branch of stable configurations at densities $\rho>10\rho_{ns}$ with no counterpart in GR, is found, representing stars with maximum masses $M_{\star}\approx1.90M_{\odot}$ (SLy), $M_{\star}\approx 2.0M_{\odot}$ (BSk20) and $M_{\star}\approx2.07M_{\odot}$ (AP4) and radius $r_S \approx 9\text{km}$. This perturbative analysis is extended to the full non-linear regime by shooting method-solving the generalized Einstein equations by Capozziello et al. \cite{Capozziello:2015yza} for the cubic model (\ref{eq:fRcubic}). The negativity of the trace of energy-momentum tensor, $T(r=0)=-\rho_c+3P_c<0$, constrains the maximum achievable central density to $\rho_c=\{2,1.25,1.35,1.7\} \times 10^{15}$ \text{gr/cm}$^{3}$ for $\{$BSk19,BSk20,BSk21,Sly$\}$ EOS. For the quadratic model (\ref{eq:fquadratic}), any non-vanishing value of $\alpha<0$ with these four EOS has the effect that at low central densities the maximum achievable mass is larger than in GR (with smaller radius), until a critical central density value, $\rho_c \sim 0.6-1.0 \times 10^{15}$, is reached (depending on the EOS), above with this effect is reversed. This is a sort of stretching-blending rotation of the $M-r_S$ shape with a transition at $M \sim (1-1.25) M_{\odot}$ and $r_S \sim (11-12.5)\text{km}$, with a similar effect for the $M-\rho_c$ curve. Therefore, in this analysis the quadratic model (\ref{eq:fquadratic}) also fails to raise the maximum mass $M_{\star}$ as compared to GR. However, when the cubic corrections (\ref{eq:fRcubic}) are switched on, it is found that negative (positive) values of $\gamma$ tend to increase (decrease) the maximum mass, but the improvement is meager.  The cubic model  has also been addressed by Astashenok et al. \cite{Astashenok:2014pua} in order  to alleviate the  hyperonic puzzle. Assuming $\alpha<0$ and using three parameterizations of the GM(nph) EOS \cite{GM1}, a perturbative procedure to solve the corresponding field equations shows that there is some room in the space of parameters $(\beta,\gamma)$ where the hyperonic puzzle can be solved; for instance, taking GM3nph and $\alpha=-0.22$ and $\alpha=-160$ one can rise the maximum mass to the $2M_{\odot}$ threshold.

The contribution of magnetic fields has been also considered within some of the models above, using a (slowly-varying field) parametrization \cite{Ryu:2010zzb}
\begin{equation}
B=B_S + B_0 [1-\exp(-\zeta(\rho/\rho_s)^{\eta})]
\end{equation}
with $B_S=10^{15}G$ the magnetic field at the star's surface. Using two regimes: $\{\zeta=0.02,\eta=2 \}$ (slowly-varying field) and $\{\zeta=0.005,\eta=3 \}$ (fast-varying field), Astashenok et al. \cite{Astashenok:2014gda} explore the effect of the different values of $B_0$ using the same EOS as in the previous work\footnote{In passing by we note that a similar parametrization for the magnetic fields within the context of $f(\mathcal{R}^2_{GB})=\beta \mathcal{R}^2_{GB}$, where $\mathcal{R}^2_{GB}$ is Gauss-Bonnet invariant, was considered by Astashenok et al. \cite{Astashenok:2014nua}, finding negligible increases in the maximum masses, but significantly higher central densities.}. In both the quadratic (\ref{eq:fquadratic}) and cubic (\ref{eq:fRcubic}) models, the maximum masses are very sensitive to the value of the reference magnetic field $B_0$ for both slowly-varying and fast-varying fields. Indeed from $B_0=0$ to $B_0=3 \times 10^{15}$G the maximum mass grows by a factor $2$, but the effect of the parameter $\alpha$ is meager; for instance, in the quadratic model one finds a mere $\sim 5\%$ for $\alpha=-5 \times 10^9$cm$^2$.  The main conclusion of this analysis is that magnetic fields within these theories cannot significantly alter the conclusions of the stellar structure models within GR.

Let us now consider the family of power-law models
\begin{equation} \label{eq:fRpower}
f(R)=R^{1+\epsilon}
\end{equation}
where the compatibility with observational abundances of light elements constrains its parameter $-0.017 \lesssim \epsilon \lesssim 0.0012$ \cite{Clifton05}. This model has been studied by Capozziello et al. \cite{Capozziello:2015yza}, and also by De Laurentis \cite{DeLaurentis:2018odx}. Let us assume, consistently with the constraints on $\epsilon$, that $\epsilon \ll 1$, so that the Lagrangian density can be expanded as $f(R)=R+\epsilon R \log[R] + \mathcal{O}(\epsilon^2)$\footnote{Notice that with the expansion the full non-perturbative behaviour of the theory is lost. Whether this has implications on losing some of the solutions of the corresponding equations or not is another issue.}. For negative values of $\epsilon$, the integration of the generalized Einstein equations with SLy, BSk20 and BSk21 EOS yield neutron stars which easily overcome the $2M_{\odot}$ threshold for $\vert \epsilon \vert  \sim  0.001$ \cite{Capozziello:2015yza}. Increasing further this value of $\vert \epsilon \vert$, De Laurentis shows \cite{DeLaurentis:2018odx}, using a different method based on Noether symmetries to implement the matching with the external solution, that $\{$BSk20,BSk21$\}$ EOS are able to reach maximum masses $M_{\star} \sim (2.8-3)M_{\odot}$ for $\vert \epsilon \vert \approx  0.008$ for both these EOS, with large fluctuations in the $M-r_S$ diagram depending on the exact value of  $\vert \epsilon \vert $, with significantly larger radius than in GR. To illustrate these results, for instance, BSk21 with $\vert \epsilon \vert \approx  0.008$ yields $r_S \approx 13.7$km, as compared to $r_S \approx 12.1$km of GR.

Another well known model in the literature is the Hu-Sawicki one \cite{HS07}
\begin{equation}
f(R)=R-m^2\frac{c_1(R/m^2)^n}{1+c_2(R/m^2)^n}
\end{equation}
which is well motivated in cosmological models of dark energy \cite{Tsujikawa:2007xu}, and where $m^2$, $c_1$, $c_2$ and $n$ are free parameters. Aparicio Resco et al. \cite{Resco:2016upv} consider this model with $n=1$, parameterized as $f(R)=-bR/(1+bR/d)$, with $b$ and $d$ some constants. For this model the stability condition (\ref{eq:stability}) yields two branches of solutions, $\{ \vert R \vert > d,b<1 \}$ and $\{ \vert R \vert < d,b>1 \}$. Exploration of the parameter space of the first branch reveals that it is either incompatible with the estimated value of the cosmological constant, or too close to the Schwarzschild solution to produce deviations with respect to GR. For the second branch, using the blend of soft, middle and stiff EOS of \cite{Hebeler:2013nza}, stable configurations are found whose masses, for $d<0$, generally grow with $b$ and are basically unconstrained from above, even to a greater extent than in the quadratic $f(R)$ case considered also in \cite{Resco:2016upv}, with a significant decrease of the radius of the star. Nonetheless,  whether such large values of the model parameters may be physical is uncertain, as no comparative analysis of the compatibility of such values with the observational constraints of the Hu-Sawicki model has been done yet.

Finally, for a recent and broad  analysis of general $f(R)$ theories in the equivalent Brans-Dicke representation using massless, constant mass, and a self-coupling potential for the scalar field and SLy and FPS EOS, see Kase and Tsujikawa \cite{Kase:2019dqc}.

\subsubsection{Scalar-tensor theories} \vspace{0.2cm}

A broad description of several aspects of stellar structure models within scalar-tensor theories with a single light scalar field and a negligible potential\footnote{As already mentioned, for these models, Cassini tracking yields a PPN parameter $\vert \gamma \vert < 2.3 \times 10^{-5}$ \cite{Bertotti:2003rm}, which translates into the constraint $a^2(\varphi) \equiv (A'(\varphi)/A(\varphi))^2<1.2 \times 10^{-5}$ for asymptotically values (i.e. outside of the solar system) of the field $\phi$.} in the Jordan frame was carried out by Horbatsch and Burgess \cite{Horbatsch:2010hj}, exploring perturbative scenarios and setting some main elements for the analysis of the Lane-Emden equation, spontaneous scalarization, and white dwarfs. A more specific analysis of scalar-tensor theories has been carried out by Cisterna et al. \cite{Cisterna:2015yla} by choosing the  George+John combination of models of the Fab Four family \cite{Charmousis:2011bf}. This model is equivalent  to GR with a non-minimal derivative coupling with the scalar field, which can be conveniently written as
\begin{equation} \label{eq:GJ}
\mathcal{S}=\int d^4x \sqrt{-g} \left[\frac{R-2\Lambda}{2\kappa^2}-\frac{1}{2}(\beta g^{\mu\nu} -\eta G^{\mu\nu})\nabla_{\mu}\phi\nabla^{\mu}\phi \right] +S_m[g_{\mu\nu},\psi_m]
\end{equation}
where $\eta$ and $\beta$ are some (real) parameters. Setting the ansatz for the scalar field
\begin{equation} \label{eq:parsca}
\phi(t,r)=Qt+F(r)
\end{equation}
where $Q \neq 0$ is a constant, and considering $\beta=\Lambda=0$, the corresponding TOV equations with boundary conditions $A(0)=A_0>0$, $A'(0)=0$, $B(0)=0$ and $P(0)=P_0$, imply that physically acceptable solutions can only be found if the following conditions hold:
\begin{equation}
 \frac{3\eta Q^2}{A_0} <\frac{1}{2\kappa^2} \hspace{0.2cm};\hspace{0.2cm}\frac{1}{4\pi \left(\frac{2\rho_c}{3P_c}-1\right)}<\frac{Q^2\vert \eta \vert}{A_0}  < \frac{1}{12\pi}
\end{equation}
Choosing a polytropic EOS (\ref{eq:poly}) with $K=123M_{\odot}^2$ and $n=2$ (motivated from \cite{Lattimer:2000nx}), the numerical integration proceeds from the center outwards until the star's surface, where the solution is matched to a Schwarzschild external solution with constant scalar field (which introduces the re-scaling $Q_{\infty}=Q/\sqrt{A_{\infty}}$)\footnote{As with similar models of this kind, this parametrization has the problem that neither $Q_{\infty} \rightarrow 0$ nor $\eta \rightarrow 0$ or $\infty$ yield the GR limit, so the sequence of $Q_{\infty}$ solutions is not smoothly connected with the GR one.}. Taking $\eta=\pm 1$ without loss of generality, a two-parametric $(P_c,Q_{\infty})$ family of solutions is obtained, finding that $\eta=+1$($-1$) yields more (less) massive stars with larger (smaller) radius. Indeed, in GR the above polytropic model peaks at $M_{\star} \approx 1.8 M_{\odot}$ and $r_S \approx 12.5$km, while for  $Q=0.08$ one finds instead $M_{\star} \approx 2.15M_{\odot}$ and $r_S \approx 13$km, though the range of allowed central pressures shortens with growing $Q_{\infty}$. This moderate improvement in the maximum mass with respect to GR results should be confirmed with more realistic EOS. Besides, this model has the drawback that the scalar field becomes imaginary in some regions of the solutions with $\eta=-1$, thus leading to a wrong-sign kinetic term which suggests the presence of ghost-like instabilities, demanding an investigation of the stability of the corresponding solutions.

Spontaneous scalarization, namely, the existence of stable stars with non-trivial values of the scalar field generated out of phase transitions, has been shown to have  a non-negligible impact on stellar oscillations at the neutron star's crust within scalar-tensor theories, since they might carry imprints of strong-field effects to be extracted via observations of gravitational waves resulting from the oscillations\footnote{Indeed, spontaneous scalarization in scalar-tensor models may left specific imprints in gravitational waves out of supernova core collapse, which may allow to constrain the parameter space of these theories \cite{Cheong:2018gzn}.}. Silva et al. \cite{Silva:2014ora} consider scalar-tensor theories with the parametrization\footnote{For this modelling it has been shown that spontaneous scalarization may take place provided that $\beta < -4.35$ in the static case \cite{Harada:1998ge} and $\beta < -3.9$ for rapidly rotating stars \cite{Doneva:2018ouu}.}\footnote{Palenzuela and Liebling \cite{Palenzuela:2015ima} have constructed a fully non-linear simulation of the stability of neutron stars with different compactness in scalar-tensor theories with $\beta>0$, finding that for very compact stars, $\mathcal{C} \geq 0.29$, the parameter $\beta$ is constrained to a maximum value $\beta \approx 90$, while for less compact stars, $\mathcal{C} < 0.27$, the parameter $\beta$ is essentially unconstrained. The existence of a potential instability in this $\beta>0$ branch is analyzed analytically and numerically by Mendes and Ortiz for some representative coupling functions in \cite{Mendes:2016fby}.}
\begin{equation} \label{eq:Aansatz}
A(\varphi)=e^{\frac{1}{2}\beta \varphi^2}
\end{equation}
The Einstein-frame representation and FPS and MS0 EOS \cite{Mueller:1996pm} for neutron's star core are used, while the crust is modelled with a KP \cite{Kobyakov:2013eta} and DH \cite{Sly3} EOS. The corresponding field equations are solved with a shooting-method with $\varphi(\infty)=0$. It is observed that, for $\beta=-6.0$, scalarized solutions with a non-vanishing value of the scalar field yield a $\sim 20\%$ increase in the maximum mass  for both FPS and MS0 EOS, with very little influence of the crust EOS, as expected. Next, when considering torsional (quasi-periodic) oscillations confined to the crust, they find the same master equation as in GR but with re-scaled wave frequencies and effective wave velocities determined by the scalar-tensor theory. These frequencies are found to depend on the stellar's mass and compactness as well as in the EOS chosen at the crust, and decrease with $\beta$. However, even for the closest values compatible with binary pulsar experiments, namely, $\beta=-4.5$ \cite{Freire:2012mg}, the corrections on the dominant frequencies are not large enough to overcome the uncertainties on the EOS; thus this modelling cannot be distinguished from GR regarding gravitational waves out of such oscillations. Sotani and Kokkotas \cite{Sotani:2017pfj} adopt instead a (stiff) parameterized EOS given by $P=P_t+v_s^2(\rho-\rho_t)$ where $\rho_t,P_t$ are the density and pressures at which the transition from the low-density regime (modelled according to the phenomenological approach of \cite{Oyamatsu:2006vd}) to the high-density one  (chosen as $2\rho_s$) takes place, and $v_s^2$ is the maximum sound speed in the latter regime (constrained as $1/3 \leq v_s^2 \leq 1$). It is observed that, for scalarized neutron stars with $\beta=-4.6$ and $\beta=-5.0$, small values of $v_s^2$ yield larger masses than in GR ($\sim 10\%$) until a critical value $v_s^2 \approx 0.62-0.64$ is reached, where this trend is reversed.

Quasi-normal modes of scalar-tensor theories with the parametrization (\ref{eq:Aansatz}), and also for
\begin{equation}
A(\phi)=[\cosh(\sqrt{3}\beta \phi)]^{\frac{1}{3\beta}}
\end{equation}
have been studied by Mendes and Ortiz \cite{Mendes:2018qwo}. Through a polytropic EOS modelling with a stiff core $\Gamma=3$ and a soft crust $\Gamma=2$, upon perturbations of the metric functions and the scalar field choosing $\beta=-5$ and $\beta=100$, a radically different spectrum of such quasi-normal modes is found. Indeed, the new scalar field modes have now much lower frequencies and shorter damping times than in GR, while their excitations may yield imprints potentially accessible via electromagnetic emission of magnetars \cite{Watts:2006mr}, or in gravitational waves out of binary neutron stars \cite{BaumShaShi}. A different approach is followed by Hendi et.al. \cite{Hendi:2015pua} by imposing an ansatz for the curvature scalar $R=e^{\alpha \phi(r)}$, and using an EOS derived from a AV18 potential, finding an increase as high as a factor two in the maximum masses for specific picks of $\alpha$ and the potential coefficients, though no analysis on the observational viability of such parameters is given.

A natural extension of the scalar-tensor theories arises when the scalar field is promoted from one to  $N$ fields, yielding tensor-multiscalar theories of gravity \cite{Damour:1992we}. Horbatsch et al
\cite{Horbatsch:2015bua} show that spontaneous scalarization also occurs on a simplified version of this theory with two scalar fields and a vanishing potential. The numerical integration of the corresponding field equations takes place with a similar parametrization as that of Eq.(\ref{eq:Aansatz}), suitably generalized to the present case, and further consider an EOS originally introduced by Novak in \cite{Novak:1997hw}. For some scalarized solutions with $\beta=-5.0$ they report an increase of the maximum mass as large as roughly $\sim 10-20 \%$ as compared to the GR case. The new freedom engendered by the $N$ scalar fields allows for new compact objects to exist \cite{Yazadjiev:2019oul,Doneva:2020afj}, in particular mixed phases of gravitational solitons and neutron stars as shown by Doneva and Yazadjiev \cite{Doneva:2019krb}, where several combination of different potentials, $A(\varphi)$ functions, and ranges of $\beta$ are considered. Moreover, if a non-trivial topology is introduced in the set of scalar fields, like $\mathbb{S}^3$, then a new kind of neutron star configuration - topological neutron star - arises, as shown by Doneva and Yazadjiev \cite{Doneva:2019ltb}. Using a piecewise APR4 EOS and characterizing such stars by their topological numbers, the authors make a quick analysis of some of their properties.

The stability of stellar models (see Sec.\ref{sec:stGR}) in quintessencial theories, which can be seen as the simplest scalar-tensor theories in the Jordan frame, i.e., Eq.(\ref{eq:stjordan}) with $A=1$ and an arbitrary potential, i.e., has been investigated in \cite{aneta1}
\begin{equation}\label{s-t}
\mathcal{S}=\frac{1}{2\kappa^2}\int d^4x\sqrt{-g}( R-2g^{\mu\nu}\nabla_\mu\phi\nabla_\nu\phi-2V(\phi))+\mathcal{S}_m[g_{\mu\nu},\psi_m].
\end{equation}
This action belongs to a class of theories for which a part of the field equations can be written as \cite{mim, mim2, mim3}
\begin{equation}\label{mod1}
 \sigma(\Psi^i)(G_{\mu\nu}-W_{\mu\nu})=\kappa^2 T_{\mu\nu},
\end{equation}
where the object $\sigma(\Psi^i)$ represents, in the general case, a coupling to the gravity sector, with $\Psi^i$ corresponding, for instance, to curvature invariants or other fields. In the specific case of (\ref{s-t}), we simply have $\sigma(\Psi^i)=1$.  The symmetric tensor $W_{\mu\nu}$ stands for additional geometrical terms which may appear in the specific theory under consideration. It turns out that for this class of gravitational theories one writes the generalized energy density and pressure in the case of a  static, spherically symmetric geometry (\ref{eq:ds2}) as
\begin{eqnarray}\label{gen_den}
Q(r) &\equiv&\rho(r)+\frac{\sigma(r)A(r)W_{tt}(r)}{\kappa^2},\\
\label{def2} \Pi(r)&\equiv& P(r)+\frac{\sigma(r)W_{rr}(r)}{\kappa^2 B(r)} \ ,
\end{eqnarray}
which allows to write the generalized TOV equations in this case as
\begin{equation}\label{tovgen}
  \left(\frac{\Pi}{\sigma}\right)'=-\frac{m(r)}{r^2}\left(\frac{Q}{\sigma}+\frac{\Pi}{\sigma}\right)
  \left(1+\frac{4\pi r^3\frac{\Pi}{\sigma}}{m(r)}\right)\left(1-\frac{2m(r)}{r}\right)^{-1}
	+\frac{2\sigma}{\kappa r}\left(\frac{W_{\theta\theta}}{r^2}-BW_{rr}\right)
\end{equation}
with the redefined mass function
\begin{equation}\label{mr}
m(r)= \int^r_0 4\pi \tilde{r}^2\frac{Q(\tilde{r})}{\sigma(\tilde{r})} d\tilde{r}.
\end{equation}
For the case of (\ref{s-t}), the corresponding metric field equations (\ref{mod1}) and those of the scalar field can be explicitly written as
\begin{eqnarray}
 G_{\mu\nu}+\frac{1}{2}g_{\mu\nu}\nabla_\alpha\phi\nabla^\alpha\phi-\nabla_\mu\phi\nabla_\nu\phi+g_{\mu\nu}V(\phi)&=&\kappa^2 T_{\mu\nu},\label{st1}\label{fieldEQ1}\\
 V'(\phi)-\Box\phi&=&0\label{fieldEQ2}.
\end{eqnarray}
The generalized TOV equations (\ref{tovgen}) in that case read
\begin{eqnarray}
 \Pi'&=&-\frac{m(r)}{Br^2}\Big(\Pi+Q\Big)\left(1+4\pi r\frac{\Pi}{m(r)}\right)-4\frac{C+V}{\kappa^2 r} \\
  m(r)&=&\int^r_0 4\pi \tilde r^2 Qd\tilde r \ ,
\end{eqnarray}
where we have introduced the definition $C=\frac{1}{2}B\phi'^2-V(\phi)$, while the generalized density (\ref{gen_den}) and pressure (\ref{def2}) take the form
\begin{eqnarray}\label{generF}
 Q&=&\rho(\tilde r)+ \kappa^{-2}(C+2V),\\
 \Pi&=& P(r)+\kappa^{-2}C.
\end{eqnarray}
It was shown in \cite{aneta1} that the stability theorem of Sec.\ref{sec:stGR} can be generalized for the above discussed minimally coupled scalar-tensor theory
with an arbitrary potential if one takes into account the extra terms appearing in (\ref{generF}) and the modified Klein-Gordon equation (\ref{fieldEQ2}). Thus, it turns
out that the stability will be achieved when the boundary term appearing during the derivations ($n(r)$ being the proper nucleon number density)
\begin{equation}\label{BT}
\int_0^\infty\partial^\mu \left( 4\pi r^2B^{-\frac{1}{2}}\frac{n(r)}{P(r)+\rho(r)} \delta\phi\partial_\mu\phi \right)dr
\end{equation}
vanishes. This will be so if both integrations at the upper and lower limits vanish. This sets conditions on the behaviour of $\partial\phi$, $\delta\partial_{\mu}\phi$, and other functions appearing in the integrand in such limits. Assuming that all functions, together with the derivatives of $\phi$ take finite values at zero, the lower limit of the integration will behave in the proper way due to the factor $r^2$. On the other hand, assuming the derivatives of the scalar field to vanish at infinity, the upper limit of the integration will also not contribute, providing the desired result. Therefore, in order to check if a stellar system is a stable one, one needs to carry out extra computations beyond performing  the well known procedure used in GR after choosing an EOS. Indeed, here one would additionally need to examine the extra terms entering in the mass  function $m(r)$: the scalar field $\phi$ and the form of its potential $V$ since the minimum of $m(r)$ may strongly depend on them.

\subsubsection{Horndeski family, beyond Horndeski, and extensions} \vspace{0.2cm}

The Horndeski family has faced some difficulties under the nearly simultaneous detection of the gravitational wave event GW170817 \cite{TheLIGOScientific:2017qsa} and its gamma-ray burst electromagnetic counterpart GRB170817A \cite{Monitor:2017mdv}, with two of its four free functions being strongly constrained. Nonetheless, it has been argued by Kobayashi and Hiramatsu \cite{Kobayashi:2018xvr} and by Langlois et al. \cite{Langlois:2017dyl}, that some families of models extending the  Horndeski  class, like degenerate higher-order scalar-tensor theories (DHOST) may overcome these constraints\footnote{Let us point out that DHOST have been recently constrained using helioseismology - via the acoustic oscillation imprints of fifth force effects within the solar interior \cite{Saltas:2019ius}.}. In this case, one needs to impose several restrictions on the five independent functions $\{A_1,A_2,A_3,A_4,A_5\}$ characterizing the theory to have $c_{GW}=c$ (see \cite{deRham:2016wji} for details). Now, assuming the standard parametrization for the scalar field (\ref{eq:parsca}), the TOV equations for a perfect fluid can be obtained after some algebra, though under a cumbersome form \cite{Kobayashi:2018xvr}. Next, standard boundary conditions at the center on the metric functions, scalar field, and density  are imposed, as well as a matching to an external Schwarzschild solution at the star's surface $P=0$. Choosing the quadratic Lagrangian \cite{Crisostomi:2017pjs}
\begin{equation}
\mathcal{L}_G=\left(\frac{1}{2\kappa^2} +\alpha (\nabla_{\mu} \phi \nabla^{\mu} \phi)^2\right)R - (8\alpha+\beta)\Box \phi \nabla^{\mu}\phi \nabla_{\mu}\nabla_{\nu}\phi \nabla^{\nu} \phi
\end{equation}
where $\alpha,\beta$ are some constants, and taking a polytropic EOS (\ref{eq:poly}) with $\Gamma=2$ and $K=123M_{\odot}^2$ (GR: $M_{\star} \approx 1.8M_{\odot} $ and $r_S \approx 12.5$km) the field equations are numerically integrated. For $\beta=0$, positive (negative) values of $\alpha$ lead to larger (smaller) maximum masses, until a maximum central density is found at a threshold $\alpha_c \approx 2 \times 10^{-4}/(\kappa^2 Q^4)$, above which no solution exists. A qualitatively similar effect exists for $\alpha=0$ and varying $\beta$. In both cases near the density/parameter threshold the maximum mass grows well above $3M_{\odot}$, while the radius increases significantly.

As already mentioned, Horndeski theories can be made compatible with solar system experiments by implementing the Vainshtein mechanism \cite{Vainshtein:1972sx} (see \cite{Babichev:2013usa} for a review), but this mechanism breaks down for beyond Horndeski theories inside astrophysical bodies, as the weak field limit functions (\ref{eq:gpot}) satisfy now \cite{Koyama:2015oma,Kobayashi:2014ida,Saito:2015fza}
\begin{eqnarray}
\frac{d\Phi}{dr}&=&-\frac{M}{r^2}-\frac{\Upsilon_1}{4}\frac{d^2M}{dr^2} \\
\frac{d\Psi}{dr}&=&-\frac{M}{r^2}+\frac{5\Upsilon_2}{4r}\frac{dM}{dr}
\end{eqnarray}
where the extra corrections by the factors $\Upsilon_1$ (governing deviations from Newton's law) and $\Upsilon_2$ (determining light bending) are non-vanishing when the Lagrangian density actually contains beyond Horndeski terms (for Horndeski theories one has $\Upsilon_1=\Upsilon_2=0$). These functions are constrained  as $\Upsilon_1=-0.11^{+0.93}_{-0.67}$  and $\Upsilon_2=-0.22^{+1.22}_{-0.19}$  by X-ray and lensing observations of galaxy clusters \cite{Sakstein:2016ggl}. Nonetheless we have already seen that Chandrasekhar's limit for white dwarfs in combination with the minimum mass for main sequence stars provides the constraint $-0.22<\Upsilon_1< 0.027$ (recall the discussion around Eq.(\ref{eq:conhor})). Relativistic stars in a quartic beyond Horndeski have been considered by Babichev et al. \cite{Babichev:2016jom}. In this case, $\Upsilon_1=\Upsilon_2=\Upsilon=-\frac{1}{3}(1-\sigma^2)$ (so $\gamma_{PPN}=1$), where $\sigma^2$ is a dimensionless quantity.  Using again the polytropic EOS (\ref{eq:poly}) with $\Gamma=2$ and $K=123M_{\odot}^2$, and taking a mild deviation with respect to GR given by $\Upsilon=-0.05$, from the hydrostatic equilibrium equation (\ref{heqst}) one finds a significant increase of the maximum mass: $M_{\star} \approx 2.5M_{\odot}$ and $R \approx 13$km. This feature is retained for more realistic EOS; for instance, Sly4 and BSk20 EOS with $\Upsilon=-0.05$ yield $M_{\star} \approx 2.9M_{\odot}$ and $r_S \approx 11$km (GR: $M_{\star} \approx 2M_{\odot}$, $r_S\approx 10$km), and $M_{\star} \approx 3.1M_{\odot}$ and $r_S \approx  11.5$km (GR: $M_{\star} \approx 2.2 M_{\odot}$, $r_S \approx 10.5$km), respectively. Within the current observational constraints for $\Upsilon$ the maximum mass can be further increased up to $5M_{\odot}$ (but with unacceptably large radii). In the positive branch of $\Upsilon$ lower masses and radii than in the GR case for all these EOS are found.

A family of models including a combination of quartic Horndeski and beyond Horndeski terms satisfying $c_{GW}=c$ (see \cite{Bettoni:2016mij} for an elaborate discussion on this point) was considered by Chagoya and Tasinato \cite{Chagoya:2018lmv}. The Lagrangian density is built from (\ref{eq:Horn}) and given by (where now we are using $X=-\frac{1}{2}\partial_{\mu}\phi\partial^{\nu}\phi$)
\begin{equation}
\mathcal{L}=X+G_4 R +\frac{G_{4,X}}{X} [(\partial^{\mu}\phi\partial_{\mu}\partial_{\nu}\phi\partial^{\nu}\phi)^2-(\partial^{\mu}\phi\partial_{\mu}\partial_{\nu}\phi\partial^{\nu}\phi) \Box \phi]
\end{equation}
and the field equations  are solved for different values of $Q$ in the ansatz (\ref{eq:parsca}). However, the EOS chosen is that of constant energy density, which is not realistic, though it allows for analytic solutions with fixed radius, finding that neutron stars can be up to twice as compact as their GR counterparts, though the compactness cannot exceed the $\mathcal{C}=4/9$ bound, as opposed to the case of vector-tensor theories (coupling of the metric to a vector field, \cite{Chagoya:2017fyl}). For the latter theories the TOV equations have been obtained by Momemi et al. \cite{Momeni:2016srq}, though they make use of unrealistic polytropic EOS and thus their results are highly doubtful, even in the context of GR.

\subsubsection{Einstein-Dilaton-Gauss-Bonnet  gravity}\vspace{0.2cm}

A unifying scenario to encompassing general scalar-tensor theories, $f(R)$ theories, Einstein-Dilaton-Gauss-Bonnet (EDGB) gravity (which belongs to the class of actions (\ref{eq:Horn})) and Chern-Simons gravity, was carried out by  Pani et al. \cite{Pani:2011xm}. Demanding second-order field equations, the corresponding Lagrangian density can be written as
\begin{eqnarray} \label{eq:EDGB}
\mathcal{L}&=& f_0(|\phi|)R+ f_1(|\phi|)\mathcal{R}^2_\text{GB}+ f_4(|\phi|)R_{\mu\nu\alpha\beta}{}^*R^{\mu\nu\alpha\beta} - \gamma(|\phi|)\partial_\mu\phi^*\partial^\mu\phi-V(|\phi|)+\mathcal{L}_\text{mat}\left[\psi_m,A^2(|\phi|)g_{\mu\nu}\right] \label{eq:EDGBaction}
\end{eqnarray}
where $^*R^{\mu\nu\alpha\beta}$ is the dual of the field strength tensor and $\mathcal{R}^2_\text{GB} \equiv R^2-4R_{\mu\nu}R^{\mu\nu}+R_{\mu\nu\alpha\beta}R^{\mu\nu\alpha\beta}$ denotes the Gauss-Bonnet invariant. Different choices for the functions $\{f_0,f_1,f_4,\gamma\}$ lead to the models mentioned above (see Table I of \cite{Pani:2011xm}).

The simplest member of this family is perhaps Gauss-Bonnet gravity, with $f_1=1$ and any other function vanishing, whose five-dimensional case was considered by Panotopoulos and Rinc\'on in \cite{Panotopoulos:2019zxv}. Casting the TOV equations in this case, they numerically integrate them for strange quark matter, where the EOS (\ref{quark}) is replaced by $P=\frac{1}{4}(\rho-5B)$ due to the presence of extra dimensions. The main reported result is that, for different values of the Gauss-Bonnet parameter, more compact stars are found than in the GR case. For more general members of the family (\ref{eq:EDGB}), static, spherically symmetric solutions, $\phi=\phi(r)$, with perfect fluids (\ref{eq:PF}) can be analytically worked out \cite{Pani:2011xm} and, imposing boundary conditions at the center, $\{m(0)=0,\rho(0)=\rho_c, A(0)=A_0, A'(0)=1 \}$, a shooting method from an asymptotically flat solution, $A(r \rightarrow \infty)=1$, fixes the constant $A(0)$, yielding a family of solutions parameterized by the central density $\rho_c$. Using DEF parameterization \cite{Damour:1996ke} in Eq.(\ref{eq:EF}) with $K=0.0195$ and $\Gamma=2.34$, besides FPS and APR EOS, the authors of \cite{Pani:2011xm} derive the $M$-$r_S$ diagrams of compact stars choosing $f_0=1/(2\kappa^2), V=0, f_1=(\alpha/16\pi)e^{\beta \phi},f_4=0$, with $\alpha>0$. In this combination no stable stellar models can be found above some maximum central density $\rho_c^{max}(\alpha,\beta)$, while the numerical integration shows that the maximum mass  slightly decreases with the product $\alpha \beta$; this effect  is more important for stiffer EOS than for softer ones.  For the FPS EOS with $\beta =\sqrt{2}$ \cite{Gross:1986mw} one finds $\alpha<23M_{\odot}^2$, which improves previous bounds on EDGB gravity from black holes \cite{Pani:2009wy}.

The EDGB family (\ref{eq:EDGBaction}) also allows for spontaneous scalarization. For instance, choosing the non-vanishing functions $f_0=1$,  $f_1=\eta \phi^2/8$, and $\gamma=1$, evidence of this spontaneous scalarization in the TOV equations using a neutron star with $M=1.4M_{\odot}$ and SLy4 EOS has been provided by Silva et al. \cite{Silva:2017uqg}. This result is further reinforced by Doneva and Yazadjiev \cite{Doneva:2017duq} working in the context of $f_0=1$, $f_1=\lambda^2 f(\phi)$ ($\lambda$ a constant) and $\gamma=1$, where spontaneous scalarization demands $\frac{df}{d\phi}(0)=0$ but $\phi \neq 0$. Considering a perturbation with respect to the GR solution (with $\phi=0$) and introducing the ansatz $f(\phi)=\frac{1}{2\beta} (e^{-\beta\phi^2}-1)$, where $\beta>0$ is some parameter ($\beta \rightarrow \infty$ recovering GR), the numerical integration takes place by imposing standard boundary conditions at the center of the star, $A'(0)=0$, $B'(0)=1$ and $\phi'(0)=0$, via a shooting method. For a piecewise polytropic approximation to a MPA1 EOS \cite{FPS3} it is found that, for fixed $\lambda$, $\beta$, and central density $\rho_c$ (bounded by some value depending on the model parameters), spontaneous scalarization tends to decrease the maximum mass, which in turn puts constraints on the parameters of this particular EDGB theory in combination with the EOS to be compatible with the $2M_{\odot}$ threshold. Similar results are obtained for the reversed sign model, $f(\phi)=-\frac{1}{2\beta} (e^{-\beta\phi^2}-1)$. Let us also mention that, for this theory, Bl\'azquez-Salcedo et al. \cite{Blazquez-Salcedo:2015ets} analyze its axial quasi-normal modes using eight realistic EOS, finding an increase in its frequency, which can raise a $\sim 10\%$ deviation as compared to GR for large enough values of the Gauss-Bonnet parameter, with a smaller damping time.

\subsubsection{Proca theories}  \vspace{0.2cm}

Neutron stars in generalized Proca theories have been considered by Kase et al. \cite{Kase:2017egk}. These theories implement a $U(1)$-symmetry breaking mechanism by adding vector fields with derivative couplings (entailing fifth-force effects, which need to be suppressed at solar system scales via the Vainhstein mechanism \cite{DeFelice:2016cri}), whose action can be generically written as
\begin{equation}
\mathcal{S}=\int d^4x \sqrt{-g} \left[ -\frac{1}{4} F_{\mu\nu}F^{\mu\nu}+\sum_{i=2}^6 \mathcal{L}_i + \mathcal{L}_m \right]
\end{equation}
where the field strength tensor shifts its partial derivatives to covariant ones, $F_{\mu\nu}=\nabla_{\mu}A_{\nu}-\nabla_{\nu}A_{\mu}$, picking up new (quadratic) terms in the vector potential $A_{\mu}$. The set of Lagrangians $\mathcal{L}_i$ is characterized by several independent functions $\{G_2,G_3,G_4,G_5,G_6,G_7\}$ which, in general, depend on both the vector potential and the field strength tensor (see \cite{Heisenberg:2014rta,Jimenez:2016isa} for details). For this action, spherically symmetric solutions can be analytically worked out, but their resolution needs specific ansatze for the functions above (besides the EOS itself). The case of cubic interactions, $G_{3}=\beta_3 X^p$ (supplemented by  $G_4=1/(2\kappa^2)$, which corresponds to GR), with $X=-\frac{1}{2}A_{\mu}A^{\mu}$, and $\{\beta_3,p>0\}$ some parameters, is considered by imposing the regularity conditions at the star's center, $A'(0)=B'(0)=\rho'(0)=P'(0)=A_0'(0)=0$ and an ansatz for the vector field $A_0=a_0+\sum_{i=2}^{\infty}a_ir^i$, such that inserting it into the corresponding field equations allows to fix the initial conditions of $A(0)$, $B(0)$, $P(0)$ and $A_0(0)$. Focusing on vector Galileons, $p=1$, and using the Damour-Esposito-Farese (DEF) parametrization \cite{Damour:1996ke}  of a polytropic model
\begin{equation} \label{eq:EF}
\rho=nm_b \left(\frac{n}{n_0}\right)\left(1 + \frac{K}{\Gamma-1}\left(\frac{n}{n_0}\right)^{\Gamma-1}\right) \hspace{0.1cm} ; \hspace{0.1cm} P=K(n_0m_b)^{\Gamma}\left(\frac{n}{n_0}\right)^{\Gamma}
\end{equation}
where $m_b=1.66\times 10^{-24}\,{\rm gr}$, $n_0=0.1\,{\rm fm}^{-3}$, and $n$ the baryonic number, the field equations are numerically integrated with the usual boundary condition $P(r_S)=0$,  and matched to the exterior solution. For the DEF parametrization (\ref{eq:EF}) with $K=0.013$ and $\Gamma=2.34$, in GR one has $M_{\star} \approx 1.67 M_{\odot}$, $r_S \approx 9.3$km and $\rho_c=3.5\times 10^{15}$gr/cm$^3$, while for cubic galileons \cite{Nicolis:2008in} the space of parameters with $\beta_3<0$ yields larger masses than in GR, allowing to overcome the $2M_{\odot}$ threshold  for some choices of the vector field constant $a_0$, while the radius enlarges to $r_S \gtrsim 12$km and the central density decreases; this effect is reversed for $\beta>0$. These enhancements of mass and radius are powered by both the strength  $\beta_3$ and the amplitude of $A_0$. A similar analysis can be carried out for the quartic coupling model $G_4=1/(2\kappa^2)+\beta_4 X^p$: for DEF EOS with $K=0.01$ and $\Gamma=2.34$, GR yields $M_{\star} \approx 1.51 M_{\odot}$, $r_S\approx 8.48$km and $\rho_c=4.1\times 10^{15}$gr/cm$^3$, while for the quartic model with $\beta_4<0$ the $2M_{\odot}$ threshold can also be easily overcome for some range of values of the vector parameter $a_0$ (even reaching to the range $\sim 3M_{\odot}$) and radius $r_S \gtrsim 11.8$km. Finally, for the case of intrinsic vector-scalar modes, with non vanishing $G_2,G_5,G_6$ terms, the vector field trivializes and no deviations with respect to GR results are obtained.

\subsubsection{Massive gravity} \vspace{0.2cm}

The de Rham-Gabadadze-Tolley massive gravity (for an overview of massive gravity see e.g. \cite{Hinterbichler:2011tt}) is an extension of GR where a second metric is introduced to generate a set of non-trivial terms able to remove any ghost mode.  The resulting ghost-free theory is mediated by a massive spin-2 field, able to avoid any conflict with solar system experiments thanks to an implementation of the Vainshtein mechanism (for a review of this theory see e.g. \cite{deRham:2014zqa}). The TOV equations of one of its minimal versions were found after a lengthy calculation and analyzed for a particular set of the free parameters  by Katsuragawa et al. \cite{Katsuragawa:2015lbl}, and further extended to include a cosmological constant by Kareeso et al. \cite{Kareeso:2018xum}, and read
\begin{equation}
\frac{dP}{dr} = -\frac{(\rho +P)\Big[\Big(8\pi P-\frac{2}{3}
\Lambda\Big)r^3 + \gamma r^2 +2m(r)\Big]}{2r^2\Big[1 - 2\frac{m(r)}{r} - \frac{\Lambda}{3}r^2 + \gamma r\Big]}.
\end{equation}
where $\gamma$ is a parameter combining the different constants of massive gravity, in particular, the mass squared of the graviton\footnote{This constant also induces changes in the Buchdahl limit depending on its sign, see \cite{Kareeso:2018xum}  for details.}. Choosing quark matter with the standard parametrization (\ref{quark}) and $a=0.28, B=60$MeV/fm$^3$, the maximum mass is reduced from $M_{\star} \approx 1.8 M_{\odot}$ to $M_{\star}  \approx 1.2 M_{\odot}$, while for parameterized SLy EOS it is also reduced from $M_{\star}  \approx 2.1 M_{\odot}$ to $M_{\star}  \approx 1.6 M_{\odot}$ \cite{Katsuragawa:2015lbl}. This does not rule out this class of models, since the pick of coefficients chosen in this work is far from being unique, needless to say the EOS itself. For instance, using nucleon-nucleon interacting potential with a charge-dependence of Argonne AV18 \cite{Wiringa:1994wb} and refined Reid-93 \cite{Stoks:1994wp} type EOS\footnote{The same EOS have been considered within a generalization of doubly special relativity incorporating curvature corrections, known as gravity's rainbow \cite{Magueijo:2002xx}. In this case, Hendi et al. \cite{Hendi:2015vta} find also an increase of the maximum mass with a single function parameterizing gravity's rainbow, up to $M_{\star} \approx 2.8 M_{\odot}$.}, Hendi et al. \cite{Hendi:2017ibm} show that the maximum mass is an increasing function of massive gravity's parameter $m^2c_2$, being able to reach masses above $3M_{\odot}$ (GR: $M_{\star} \approx 1.68 M_{\odot}$, $r_S \approx 8.42$km), accompanied by larger radius ($r_S \approx 11$km) and higher central densities. These results are further complemented by the finding that, for white dwarfs, the Chandrasekhar mass (GR: $M \approx 1.4 M_{\odot}$) can be raised above $3M_{\odot}$ for viable values of the massive gravity parameters \cite{EslamPanah:2018evk}.

An extension of massive gravity where the requirement for the auxiliary metric to be non-dynamical is removed, results into massive bigravity \cite{Hassan:2011vm}. This theory was briefly discussed by Enander and Mortsell \cite{Enander:2015kda} with the main result that the Vainshtein mechanism is also applicable for these theories. Aoki et al. \cite{Aoki:2016eov} find several classes of solutions in this setting. Only some of those described by polytropic EOS (\ref{eq:poly}) with $\Gamma=2$ are able to yield $M_{\star} \approx 2M_{\odot}$, which may allow to provide strong constraints on the spaces of parameters of these theories.

Yet another generalization of massive gravity is the  mass-varying massive gravity, where the mass of the graviton varies depending on the value of the scalar field \cite{Huang:2012pe}. Neutron stars in this model were studied by Sun and Zhou \cite{Sun:2019niy} using analytic fits of APR and SLy EOS. It is found that, depending on the value of the mass, $m$, appearing on a simple quartic potential, the maximum mass in both EOS can significantly increase by a factor $\sim 10 \%-20\%$ (with negligible modifications in the corresponding radius), though whether the parameters chosen are observationally viable is not discussed.

\subsubsection{Teleparallel gravity} \vspace{0.2cm}

Another family of extensions of GR is based upon the teleparallel equivalent of GR,  where the main geometrical ingredient is torsion instead of curvature \cite{Maluf:2013gaa}. Stellar structure models within $f(\mathcal{T})$ gravity, where $\mathcal{T}$ is the torsion scalar, has been recently considered by Ilijic and Sossich \cite{Ilijic:2018ulf}, where the TOV equations for the quadratic teleparallel model
\begin{equation}
f(\mathcal{T})=\mathcal{T}+\frac{\alpha}{2}\mathcal{T}^2
\end{equation}
and a polytropic EOS (\ref{eq:poly}) with $\Gamma=2$ are obtained and subsequently solved with suitable boundary conditions. $f(\mathcal{T})$ theories also suffer from the ambiguous definition of the mass of the star \cite{DeBenedictis:2016aze} (since the vacuum solution is presently unknown in closed form) already discussed in the $f(R)$ case. To overcome it, the authors of \cite{Ilijic:2018ulf} compute the total particle number $N=-4\pi \int_0^{r_S} n(r) r^2/B(r)$ (where $n=dN/dV$ is the particle number density) as a rough estimation of the star's mass, and find, via numerical integration, that positive (negative) $\alpha$ yields less (more) particles that can be supported against the gravitational pull as compared to GR, producing an increase (decrease until a critical value of $a$, then increase) of the star's radius; for $a=10(-10)$, this effect increases (decreases) the radius up to a factor $\sim 40\%$ ($\sim 70 \%$). However, this analysis does not allow one to make any claims on the behaviour of the mass in this model. Further simplified modelling with anisotropic fluids within this theory can be found in Deb et al. \cite{Deb:2018voz}.

\subsubsection{Mimetic gravity} \vspace{0.2cm}

In mimetic gravity, the space-time metric is re-parameterized using an auxiliary metric $g_{\mu\nu}=-\hat{g}^{\rho\sigma}\partial_{\rho}\phi\partial_{\sigma}\phi \hat{g}_{\mu\nu}$ with the auxiliary scalar field $\phi$ satisfying the condition $g^{\rho\sigma}(\tilde{g}_{\mu\nu},\phi)\partial_{\rho}\phi \partial_{\sigma}\phi=-1$ (see \cite{Sebastiani:2016ras} for a review of this theory). Astashenok and Odintsov \cite{Astashenok:2015qzw} consider mimetic $f(R)$ models given by the action
\begin{equation}
\mathcal{S}=\frac{1}{2\kappa^2} \int d^4 x \sqrt{-g} \left[ f(R(g_{\mu\nu})) -V(\phi) +\beta g^{\mu\nu}\partial_{\mu}\phi\partial_{\nu} \phi \right] + \mathcal{S}_m(g_{\mu\nu},\psi_m)
\end{equation}
where  $V(\phi)$ is the scalar potential, while the last term in the gravity part is a Lagrange multiplier enforcing the constraint on the scalar field. Assuming $\phi=\phi(r)$ and a perfect fluid (\ref{eq:PF}) as the matter source, the TOV equations for this theory are found, and after going to the scalar-tensor representation the analytic fitting of the SLy4 EOS \cite{CamenzindBook} allows to solve them. For the case of mimetic GR with a potential $V(\phi)=A/\phi^2$ one finds slightly higher masses ($\sim 10 \%$) than in the GR case for $A>0$ depending on the value of the mimetic scalar field $\phi(0)$. For a potential $V(\phi)=A e^{C\phi^2}$ negligible increases are reported. For quark EOS (\ref{quark}) with i) $a=0.28, B=60$MeV/fm$^3$ and ii) $a=1/3, B=60$MeV/fm$^3$ and $V(\phi)=A/\phi^2$, one also finds very slight increases of the maximum mass. Finally, for mimetic $f(R)=R+\alpha R^2$ gravity with the choices above the increase of the maximum mass does not reach the $10\%$ level.

\subsubsection{Lorentz invariance-breaking theories} \vspace{0.2cm}

A model-independent characterization of tiny violations of Lorentz invariance and the associated set of TOV equations within a standard model extension framework has been recently introduced by Xu et. al. \cite{Xu:2019gua}. Predictions on the mass-radius relations from such violations have been incorporated into several explicit theories. For instance, Horava-Lifshitz gravity is a proposal for an ultraviolet completion of GR at the price of dropping Lorentz invariance at those scales \cite{Horava:2009uw}. The theory may be formulated to depend on a single parameter $\omega$, heavily constrained by light deflection \cite{Liu:2010hzb}. For this theory, using APR4, MPA1, MS1 and WFF1 \cite{Wiringa:1988tp} EOS (all of which yield $ M\gtrsim 2M_{\odot}$ in GR), Kim et al. \cite{Kim:2018dbs} build  a set of solutions for constrained values of this parameter, finding very significant enhancements of neutron star's maximum mass (though with only slight modifications of the maximum radius), which may raise even above $3M_{\odot}$ for $\omega=2 \times 10^{-12}$cm$^{-2}$.  Another theory violating Lorentz invariance is Einstein-Aether, which is the most general covariant action of the spacetime metric and the extra field $u^{\mu}$ \cite{Eling:2004dk}. PPN consistency imposes strong constraints upon the parameters of this theory \cite{Eling:2007xh}, and using three samples of soft A18 (GR: $M_{\star} \approx 1.67 M_{\odot}$), medium A18$\delta$vUIX (GR: $M_{\star} \approx 2.2 M_{\odot})$, and hard A18UIX EOS \cite{Farhi}, an overall decrease ($6\%-15\%$) of the maximum mass for all of them is found. Therefore, these EOS within these models are more severely constrained than in GR due to the $2M_{\odot}$ bound. For the low-energy limit of this theory Barausse \cite{Barausse:2019yuk} computes neutron star's ``sensitivities", namely, a parameterized violation of the strong equivalence principle, finding no such violations  at post-Newtonian order in compact binaries.

\subsubsection{Braneworld models} \vspace{0.2cm}

Relativistic stars within braneworld models of Randall-Sundrum (RS) type have been also analyzed in the literature. This model consists on a five-dimensional bulk such that elementary particles are constrained to move in the embedded four dimensional space-time (the brane), while the graviton is free to explore the (warped) extra dimension \cite{Randall:1999ee,Randall:1999vf}. A first take on this subject is the one by Wiseman \cite{Wiseman:2001xt}, where the full Einstein equations with cosmological constant of RS are solved for a static, spherically symmetric metric on the brane but with axial symmetry in the bulk. Creek et al. \cite{Creek:2006je} consider known bulk solutions of Schwarzschild-Anti deSitter type and look for trajectories in the brane corresponding to solutions of (embedded) TOV equations, where the corresponding energy-momentum tensor carries new terms due to the bulk contributions. This entails the existence of new matching conditions at the stellar surface departing from the GR one, and which do not have a unique solution on the brane  \cite{Ovalle:2007bn}. For instance, in \cite{Ovalle:2013xla} an exact analytical solution is found, corresponding to non-uniform interior structure of Tolman IV-type, finding hints that compactness is reduced due to the bulk effects, in agreement with the results of \cite{Garcia-Aspeitia:2014pna}, where the Buchdahl limit becomes $5/18<M_{Buch}<4/9$ depending on the brane tension.

\subsubsection{Non-minimal gravity-matter couplings} \vspace{0.2cm}

Models with non-minimal curvature-matter couplings have been the subject of recent investigations in extensions of GR \cite{Bertolami:2008ab}. This is for instance the case of $f(R,T)$ theories, originally introduced by Harko et al. \cite{Harko:2011kv}. In this case, the field equations (\ref{eq:f-var}) pick up $f_T$-dependent terms as compared to (\ref{eq:f-var}) as:
\begin{equation}\label{frt2}
f_R(R,T)R_{\mu\nu}-\frac{1}{2}f(R,T)g_{\mu\nu}+(g_{\mu\nu}\Box-\nabla_\mu\nabla_\nu)f_R(R,T)=8\pi T_{\mu\nu}-f_T(R,T)T_{\mu\nu}-f_T(R,T)\Theta_{\mu\nu},
\end{equation}
where $\Theta_{\mu\nu}\equiv g^{\alpha\beta}\frac{\delta T_{\alpha\beta}}{\delta g^{\mu\nu}}$. These theories have the distinctive feature of breaking the conservation of the stress-energy tensor, i.e.,
\begin{equation}
\nabla^{\mu}T_{\mu\nu}=\frac{f_T(R,T)}{8\pi -f_T(R,T)}[(T_{\mu\nu}+\Theta_{\mu\nu})\nabla^{\mu}\ln f_T(R,T)
+\nabla^{\mu}\Theta_{\mu\nu}-(1/2)g_{\mu\nu}\nabla^{\mu}T].
\end{equation}
To fully cast the dynamics of these theories one needs to specify not just the $T_{\mu\nu}$ but also the form of the matter Lagrangian $\mathcal{L}_m$. For a perfect fluid there is an ambiguity in that choice \cite{Faraoni:2009rk} (see \cite{Avelino:2018rsb} for a critical discussion on this issue). Let us take here $\mathcal{L}_m=-P$, which is the one used by Moraes et al. \cite{Moraes:2015uxq} to derive the modified TOV equations for a  simple family of models $f(R,T)=R+2\lambda T$, with $\lambda$ some constant, which read
\begin{equation} \label{eq:TOVem}
P'=-(\rho+P) \frac{4\pi P r +\frac{m(r)}{r^2}-\frac{\lambda(\rho-3P)}{2}}{\left(1-\frac{2m(r)}{r}\right)\left[1-\frac{\lambda}{8\pi+2\lambda}\left(1-\frac{d\rho}{dP}\right)\right]}
\end{equation}
where now the mass function reads as $m'(r)=4\pi r^2 \rho+\frac{\lambda(3\rho-P)}{2}r^2$. The $\lambda$-corrections to the standard TOV equation (\ref{eq:TOVGR}) are now apparent where, to obtain stable configurations, the term within brackets in the denominator has to be positive. Thus, proceeding as usual with boundary conditions at the star's center, $m(0)=0,\rho(0)=\rho_c,P(0)=P_c$, while matching to an external Schwarzschild metric at the star's surface, and choosing as EOS a polytrope (\ref{eq:poly}) with $\Gamma=5/3$, and strange quark matter (\ref{quark}) with $a=0.28$ and $B=60$MeV/fm$^3$, the TOV equations are Runge-Kutta solved, resulting in a two-parametric $\{\rho_c,\lambda\}$ family of solutions\footnote{The unphysical case of polytropic EOS with $n\rightarrow \infty$ (isotropic EOS) has been considered by Das et al. \cite{Das:2016mxq} within the same family of models.}. It is found that the maximum mass of the star increases with $\lambda$, while the corresponding radius decreases; for instance, for $\lambda=0.2$, polytropic stars increase from $M_{\star}\approx 1.419M_{\odot}$ at $\rho_c \approx 10.97\rho_s$ to $M_{\star}\approx 1.538M_{\odot}$ at $\rho_c \approx 6.969\rho_s$, while for quark matter one needs to go to higher values of $\lambda$ for similar increases; for instance, choosing $\lambda=2.0$ one goes from $M_{\star}\approx 1.760M_{\odot}$ at $\rho_c \approx 8.917\rho_s$ to $M_{\star} \approx 2.017M_{\odot}$ at $\rho_c \approx 6.836\rho_s$. Thus, this model seems to be able to improve GR results offering the possibility to reach the $2M_{\odot}$ threshold for the EOS discussed here. Strange stars in this same model have been examined by Deb et. al. \cite{Deb:2018gzt} where the emphasis is put on the physical effect of the presence of a high electric field at the strange star's surface \cite{Usov:2004iz}. Using $a=1/3$ and $B=83$MeV/fm$^3$, and writing a modification of the TOV equations (\ref{eq:TOVem}) with an spherically symmetric electromagnetic field and a choice of a simple monotonically decreasing function of the density inside the star, the field equations are numerically integrated. Taking astrophysically reasonable values for the electric charge  \cite{Usov:2004iz}, the authors report that negative (positive) values of $\lambda$ raise (lower) the maximum mass with an increase (decrease) of the corresponding radius. For instance, $\lambda=-0.8$ yields $M_{\star} \approx 3.166 M_{\odot}$ and $r_S \approx 11.38$km at $\rho_c \approx 7.648 \rho_s$ (GR: $M_{\star} \sim 2.8M_{\odot}$ and $ r_S \sim 10$km). Further aspects of stellar structure solutions in these models have been considered by Zubair et al.  \cite{Zubair:2015gsb}. The (charged) anisotropic case was analyzed in detail by Maurya and collaborators in \cite{Maurya:2019hds,Maurya:2019sfm}, where radius, redshift, mass-radius relation and maximum compactness for several strange star modelings are discussed, though the modifications as compared to GR results are found to be tiny unless the coupling parameter $\vert\lambda \vert$ is pushed to $\sim \mathcal{O}(1)$. Let us also note that the choice $\mathcal{L}_m=\rho$ for the fluid's Lagrangian density was considered by Carvalho et al. \cite{Carvalho:2019gzs} using a simple EOS $P=0.28\rho$, finding significant increases of $M_{\star}$ for $\lambda<0$. Finally, quark stars in an extension of the kind $f(R,T)=R+\alpha R^2 + \omega RT$ with bag constant $B=60$MeV/fm$^3$ and $a=0.28$ were analyzed by Mathew et al. \cite{Mathew:2020zic}.

Let us mention another two theories featuring the non-conservation of the energy-momentum tensor. The first one  corresponds to Rastall's theory \cite{Rastall1,Rastall2}, where $\nabla_{\mu}T^{\mu\nu}=\alpha \nabla^{\nu}R$  and with $\alpha$ known as Rastall's parameter. This case was considered by Oliveira et al. \cite{Oliveira:2015lka} using a polytropic EOS with $\Gamma=2$, and \{BSk19, BSk20, BSk21\} EOS, finding only a slight increase of
the maximum mass with negative $\alpha$ provided that only small deviances from the conservation law of the energy-momentum tensor are allowed. Hansraj and
Banerjee \cite{Hansraj:2018reh} also consider Rastall gravity with polytropic EOS (\ref{eq:poly}) and $\Gamma=1$, with inconclusive results. The second theory is unimodular gravity \cite{UG}, whose generalized TOV equations were obtained in \cite{Astorga-Moreno:2019uin}, finding constant density stars as well as some polytropic stars, with larger compactness than their GR counterparts.

\subsubsection{Action-dependent Lagrangian theories} \vspace{0.2cm}

In an action-dependent Lagrangian theory \cite{lazo} one deals with an additional dissipative term which is constructed with the Christoffel symbols of the metric and a coupling four-vector $\lambda_\mu$ being related to cosmological bulk viscosity in an expanding background. The existence of the static spherically symmetric configuration is strictly related to the form of the four-vector $\lambda_\mu$; if the only non-vanishing component is $\lambda_\theta$, then  modifications to the gravitational component of pressure in the TOV equations are found to be  \cite{julio}
\begin{align}\label{m2TOV}
 P'(r)=
 &-\frac{m(r)}{r^2}(\rho+P)\left(1+\frac{4\pi r^3\left(P-\frac{\lambda_0}{\kappa r^2}\right)}{m(r)}\right)\left(1-\frac{2m(r)}{r}\right)^{-1},\\
 m(r)=&\int_0^r\left(4\pi r^2\rho -\frac{\lambda_0}{2}\right)d\tilde{r},
 \end{align}
where $\lambda_0=\lambda_\theta \cot{\theta}$ is a dimensionless parameter. Fabris et al. \cite{julio} show that  for the simple polytropic EOS (\ref{eq:poly}) with $\Gamma=2$ and $K=\kappa_0$  a fitting constant, positive (negative) $\lambda_0$ yield more (less) massive stars. Analogously as in GR, unstable configurations appear beyond the central density of the maximum value of the mass. The dust matter collapse was also briefly studied in \cite{julio}: the use of Darmois junction conditions for the matching provides a set of equations to be satisfied. One also will deal here with a physical singularity at $r=0$ and an event horizon at
$
r=\frac{2M_0+\lambda_0r_S}{(\lambda_0+1)},
$
where $M_0=m(r_s)$, which for a particular values of $\lambda_0$ may disappear leaving a naked singularity at $r=0$.

\subsubsection{Further considerations} \vspace{0.2cm}

Let us quickly mention now the analysis of the TOV equations of other modified theories of gravity where the research is more scattered. Such is the case of a natural extension of the quadratic $f(R)$ model in Eq.(\ref{eq:fquadratic})  via additional corrections in the (symmetrized) Ricci-squared invariant, leading to the  quadratic Lagrangian\footnote{A Kretchsman scalar, $R_{\mu\nu\alpha\beta}R^{\mu\nu\alpha\beta}$, can also be added but, using the Gauss-Bonnet invariant in four spacetime dimensions, $\mathcal{R}^2_\text{GB} \equiv R^2-4R_{\mu\nu}R^{\mu\nu}+R_{\mu\nu\alpha\beta}R^{\mu\nu\alpha\beta}$, it can always be removed from this action via a redefinition of the constants.}:
\begin{equation} \label{eq:quadraticRR}
f(R,R_{\mu\nu}R^{\mu\nu})=R+\alpha (R^2 + b R_{\mu\nu}R^{\mu\nu})
\end{equation}
where $b$ is some dimensionless number. To overcome the well known trouble with ghosts in this theory, Santos \cite{Santos:2011ye} assumes the quadratic terms in (\ref{eq:quadraticRR}) to be supplemented by another (logarithmic) term which is extremely small in the weak field limit and whose mission is to ``eat up" the Ricci-square correction, therefore removing its effects in solar system experiments. Assuming a polytropic EOS (\ref{eq:poly}) with $\Gamma=5/3$, the generalized TOV equations are numerically solved for values $\alpha^{1/2}=0.96$km and $b=-2$, finding that the mass peaks at $M_{\star}\approx 0.8M_{\odot}$ for $\rho_c \approx 0.72\times 10^{18}$kg/m$^3$. More realistic EOS are employed in the perturbative approach of Deliduman et al. \cite{Deliduman:2011nw}. Expanding the metric as $g_{\mu\nu}=g_{\mu\nu}^{(0)}+\beta g_{\mu\nu}^{(1)} + \mathcal{O}(\beta^2)$, and similarly for density, pressure and mass function, the corresponding TOV equations are Runge-Kutta solved using $\{$FPS, AP4, SLy, MS1\cite{Mueller:1996pm}, MPA1, GS1\cite{Glendenning:1997ak}$\}$ EOS. In units of $\beta_{1}=\beta/10^{7}$m$^2$, the compatibility with the $2M_{\odot}$ threshold as well as with measured mass-radius relations of neutron stars demands the following intervals of $\beta_{1}$ for the six EOS above: $\{ -4 \lesssim \beta_1 \lesssim -2; -2 \lesssim \beta_1 \lesssim -4;-1 \lesssim  \beta_1 \lesssim  2;$ excluded$; 4 \lesssim \beta_1 \lesssim 10; -5 \lesssim \beta_1 \lesssim -3 \}$. As the validity of this approach breaks down for values $\vert \beta \vert \sim 10^{8}$m$^2$ for masses around the $2M_{\odot}$ threshold, this casts doubts on the validity of the predictions of this modelling.

Other models considered in the literature are: $f(R,\mathcal{R}^2_\text{GB} )$  gravity, which has been studied with anisotropic fluids but isotropic EOS \cite{Shamir:2017yza,Shamir:2017rjz,Shamir:2018mjs}, (anisotropic) stars in non-minimal gravity-electromagnetic coupling models of type $f(R)F_{\mu\nu}F^{\mu\nu}$ \cite{Sert:2018rjv} and even with more involved couplings \cite{Sharif:2019txi}, other models with non-minimally coupled scalar fields with several realistic EOS \cite{Arapoglu:2019mun}, corrections in the Lagrangian density in scalars of the energy-momentum tensor such as $T_{\mu\nu}T^{\mu\nu}$ \cite{Nari:2018aqs,Akarsu:2018zxl} with polytropic and strange quark matter (all these yielding slight to moderate  increases on the maximum allowed mass depending on the EOS chosen); and scalar-tensor-vector theories \cite{Armengol:2016mzu}, where FPS, SLy and BSK21 EOS yield small to moderate increases of $M_{\star}$ for astrophysically viable model parameters.

Let us conclude this section by mentioning that a semiclassical implementation of hypothetical quantum gravitational corrections via the approach of Parker \cite{Parker:1993dk} in the s-wave Polyakov approximation in the Boulware state \cite{BirrelDavies}  was used by Carballo-Rubio \cite{Carballo-Rubio:2017tlh}  to compute the   impact  upon the TOV equations as
\begin{equation}
P'\left(1-\frac{\ell_{\rm P}^2}{2r}\frac{P'}{\rho+P}\right)= -(\rho+P)\frac{4\pi r P+\frac{m(r)}{r^2}}{1-\frac{2m(r)}{r}}\label{eq:semitov}
\end{equation}
where a new term in Planck's length, $l_P^2 \equiv \hbar G/c^3$, has been picked up as compared to  (\ref{eq:TOVGR}). Such corrections might significantly raise the central density of the star, potentially allowing new classes of relativistic stars not present in GR, such as black stars \cite{Barcelo:2007yk}, though explicit models are yet to be explored. \\

This concludes our analysis of static stellar structure models of modified theories of gravity in metric formalism.

\subsection{Rotating stellar models} \vspace{0.2cm}

Let us now review  rotating stars. It is known that in GR the addition of rotation may increase the maximum mass for a typical realistic EOS by a factor $\sim 15\%-20 \%$ and the radius by $ \sim 30\%-40 \%$, decreasing the central density by $\sim 20\%$ (for a detailed account on this issue, see \cite{Stergioulas:2003yp}), an effect further enhanced for strange matter EOS. In this sense, Breu and Rezzolla \cite{Breu:2016ufb}  have provided compelling arguments and numerical evidence by which, assuming uniform rotation, the maximum mass of a rotating star, $M_{\star}^R$, can be expressed in terms of the maximum mass of the non-rotating one, $M_{\star}$, as $M_{\star}^R \simeq (1.203 \pm 0.022) M_{\star}$, for more than 40 (analytic and tabulated) EOS and for the largest angular momentum that can be attained. This is a quite striking result, as EOS are built on different assumptions leading to different maximum mass in the non-rotating case yet the relative increase of the mass turns out to be approximately universal for all these EOS. The next important comment is that while in spherically symmetric solutions the focus is typically on mass-radius relations, in rotating solutions one is more interested on the predictions for the moment of inertia, since the deviations from GR are typically much larger than those of the mass. Moreover, the moment of inertia can be extracted from observations of relativistic binary pulsars, for instance, those of PSR J0737-3039 may allow to constrain the moment of inertia of the primary pulsar by a $10\%$ of error \cite{Lattimer:2004nj}. Such measurements place constraints on the combination of the EOS and the modified gravity parameters which, in combination with other tests (for instance, via the I-Love-Q relations, see below), may help alleviate the degeneracy issue of these models.

\subsubsection{Slowly rotating models} \vspace{0.2cm}

Theoretical studies of rotating stars within modified gravity  start with the simplified  slowly rotating case.  The idea of this approximation is to generate the effects of rotation in terms of a small perturbation on a static, spherically symmetric background extending the line element (\ref{eq:ds2}) via a first-order perturbation in the star's angular velocity $\Omega$. Let us consider the slowly rotating axially symmetric line element \cite{Hartle:1967he} (check Yagi and Yunes \cite{Yagi:2016bkt} for further technical details)
\begin{equation}\label{eq:ds2rot}
ds^2=-A(r)dt^2+\frac{1}{B(r)}dr^2
+ r^2(d\theta^2+\sin^2 \theta (d\varphi - (\omega(r,\theta) + \mathcal{O}(\Omega^3))dt)^2)
\end{equation}
where $\omega(r)\equiv d\varphi/dt$ is the angular velocity of the local inertial frames related to the dragging of the rotation. As the slowly rotating approximation considers only linear order terms in $\Omega$, this is a reliable approximation for rotational frequencies of a few hundreds of Hz, which turns out to be the case for most of the observed neutron stars so far since they typically quickly spin down after their formation. Within this approximation, the spherically symmetric TOV equations (\ref{eq:TOVGR}) must be supplemented by the $(_{\theta}{^t})$ component of the Einstein's equations, which reads
\begin{equation}
\frac{1}{r^4} \frac{d}{dr} \left(r^4 j \frac{d\bar{\omega}}{dr}\right)+\frac{4}{r} \frac{dj}{dr} \bar{\omega}=0
\end{equation}
where $j(r) \equiv \sqrt{B/A}$. This equation implies that the centrifugal force acting upon the star depends on the difference $\bar{\omega}(r) \equiv \Omega -\omega(r)$. The corresponding TOV equations in this case are integrated from the center (where now one also has $\omega(0)=1$), outwards until the star's surface is reached and matched to an external solution. The equation for the angular velocity requires that for consistency there:
\begin{equation} \label{eq:omegaasymp}
\bar{\omega}(r_S) \rightarrow \Omega- \frac{2J}{r_S^3} \hspace{0.2cm} ;\hspace{0.2cm}   \omega'(r_S)  \rightarrow  \frac{6J}{r_S^4}
\end{equation}
where $J$ is the star's total angular momentum. This also allows to compute the star's moment of inertia as
\begin{equation} \label{eq:inertia}
I=\frac{8\pi}{3} \int_0^{r_S}  (\rho+P) r^4 \sqrt{\frac{1}{A(r) B(r)}}\left(\frac{\bar{\omega}}{\Omega} \right)dr
\end{equation}
Typically, the moment of inertia increases for low masses until a certain point near the maximum mass is reached, after which it begins to decrease. Therefore,  many investigations have focused on the modifications to the moment of inertia in extensions of GR. Measuring the moment of inertia represents a daunting task as one needs to compute the spin-orbit contribution to the orbit of neutron star binaries \cite{Lattimer:2004nj}, and could be available in the near future via direct observations of periastron in double binary pulsars \cite{Lattimer:2004nj,Kehl:2016mgp} as well as within the context of gravitational wave emission via tidal polarizability of neutron star-black hole mergers  \cite{Friedman:1997uh}.

In the context of general scalar-tensor theories, a complete toolkit for the modelling of X-ray pulse profiles of rotating neutron stars has been recently released by Silva and Yunes \cite{Silva:2018yxz}. As the resulting waveform can significantly deviate from the GR counterpart, this opens the possibility to constraining a large number of such models via observations of binary pulsars. Let us start considering particular cases. The quadratic $f(R)$ model (\ref{eq:fquadratic}) was discussed by Staykov et al. \cite{Staykov:2014mwa} using the scalar-tensor representation defining the new scalar field  $\varphi=\frac{\sqrt{3}}{2}\log \phi$ with $\phi \equiv f_R$.
The equation for the relative angular velocity $\bar{\omega}$ decouples from the other equations, and satisfies boundary conditions at the center $d\bar{\omega}/dr(0)=0$, and asymptotically as given by (\ref{eq:omegaasymp}); for the remaining field equations the standard conditions upon density, pressure, and scalar field are imposed. The moment of inertia in these theories is obtained as
\begin{equation} \label{eq:stinertia}
I=\frac{8\pi}{3} \int_0^{r_S} A^4(\varphi) (\rho+P) r^4 \sqrt{\frac{1}{A(r) B(r)}}\left(\frac{\bar{\omega}}{\Omega} \right)dr
\end{equation}
Using SLy4 and FPS EOS, as well as quark  matter (\ref{quark}) with $ a=1/3$ and  $B=60$MeV/fm$^3$, corresponding to SQSB60 EOS \cite{GondekRosinska:2008nf}, in the static case numerical resolution of the field equations by shooting-method yields slight increases with $\alpha$ of the maximum mass ($ \sim 10-20\%$) and the radius for all these EOS; for the quark matter EOS the mass-radius relation below $1M_{\odot}$ does not significantly differ from GR, and one has to push near the maximum mass to observe any such deviations. Within the slowly rotating approximation, pushing the constant $\alpha$ to scales $ \vert \alpha \vert \lesssim 10^{4}$ (satisfying Gravity Probe B constraints \cite{Naf:2010zy}) introduces deviations in the moment of inertia whose significance grows with larger masses, peaking at an increase of $ \sim 30-40\%$ (similar as those in $\bar{\omega}$ for $M \approx 2M_{\odot}$ once any of the above EOS is fixed). This is a simple example showing that precise measurements of the moment of inertia to come in the next few years (reducing errors down to
 $\lesssim 10\%$) may help breaking the degeneracy between the EOS of compact stars and the parameters of modified gravity. Further reanalysis of the quadratic model and general scalar-tensor theories is employed by Staykov et al. \cite{Staykov:2015mma} to address neutron star glitches. Using FPS, SLy, APR4 and MS1 EOS the authors find, for these models, a maximum deviance for the crust-to-core ratio of the moment of inertia, $I_{crust}/I$, of $\sim 3\%$ as compared to GR, concluding that these models are not suitable candidates to explain such glitches. Further aspects of the quadratic $f(R)$ model in slowly rotating concerning orbital and epicyclic frequencies have been addressed by Staykov et al. \cite{Staykov:2015kwa}.

Slowly rotating stars in scalar-tensor theories with the function (\ref{eq:Aansatz}), where we recall that in the static case the parameter $\beta$ is constrained by observations of binary systems to be $\beta \lesssim -4.5$ \cite{J0348+0432,Demorest10}, have been addressed in the literature. In \cite{Pani:2014jra} Pani and Berti consider these theories with $V(\varphi)=0$ and struggle to cast the corresponding field equations in suitable form for numerical integration (see the Appendix A of that paper for details). Using FPS, APR and MS1 EOS, the relevant quantities up to second-order in the angular momentum are obtained for $\beta=-4.5$ and $\varphi_{0}(\infty)=10^{-3}$. The main reported results is an instability of the GR-type solutions, while in scalarized solutions the maximum mass remains roughly the same and the moment of inertia slightly grows; these results may be (very) significantly enhanced if one pushes $\beta$ to much smaller values: for $\beta=-6.0$ one finds a twofold moment of inertia for the three EOS above. Corrections to the scalar charge can also reach a factor two. The authors go forward to suggest improvements of this scenario with massive potentials, which is  indeed the scenario  recently studied by Popchev et al. \cite{Popchev:2018fwu}, where the self-interacting potential $V(\varphi)=2m_{\varphi}^2 \varphi^2+\lambda \varphi^4$, with $m_{\varphi}$ the scalar field mass and $\lambda \geq 0$ a parameter, is assumed. Here the authors choose a range of scalar field masses $10^{-16}$eV$\lesssim m_{\varphi} \lesssim 10^{-9}$eV and a pool of different EOS, finding that compactness can deviate up to $\sim 10\%$ as compared to GR ones. Quantitatively similar results for a piecewise APR4 EOS \cite{FPS3} are reported by Staykov et al. \cite{Staykov:2018hhc}.

Elaborating further upon this topic, the effect of anisotropies (\ref{eq:Tmunufluid0}) in nuclear matter at high densities upon the structure of slowly rotating stars of scalar-tensor theories in the Einstein frame (\ref{eq:steinstein}) has been investigated by Silva et al. \cite{Silva:2014fca}. Within GR, these effects may increase the maximum allowed neutron star mass for a given EOS \cite{Horvat:2010xf}. Assuming again a dependence (\ref{eq:Aansatz}) and focusing on the APR EOS for the radial pressure $P_r$, and the Horvat (\ref{eq:Horvat}), and Bowers-Liang (\ref{eq:BL}) EOS for the anisotropy $\sigma \equiv P_r-P_{\perp}$, with the constraints stressed in those equations, the authors of \cite{Silva:2014fca} set standard boundary conditions at the center upon density, pressure, in addition to $\varphi(0)=\varphi_0$, $\bar{\omega}(0)=\bar{\omega}_{0}$ and $\sigma(0)=0$ to integrate the corresponding TOV equations outwards for each EOS above. The moment of inertia formula (\ref{eq:stinertia}) is generalized to this case as
\begin{equation} \label{eq:stinertiaani}
I=\frac{8\pi}{3} \int_0^{r_S} A^4(\varphi) (\rho+P_r)\left(1-\frac{\sigma}{\rho+P_r} \right) r^4 \frac{1}{\sqrt{A(r) B(r)}}\left(\frac{\bar{\omega}}{\Omega} \right)dr
\end{equation}
reducing to the formula for isotropic stars when $\sigma \rightarrow 0$. In GR, the modelling above allows for a significant increase (decrease) of the maximum mass for negative (positive) values of $\{ \lambda_{H},\lambda_{BL} \}$, approaching $3M_{\odot}$ for $\lambda_{H} = -2.0$, together with an increase (decrease) on the corresponding star's radius, and an even more significant increase ($> 50\%$) of the moment of inertia as compared to a perfect fluid modelling. For scalar-tensor theories with the two anisotropic models above, the range of values for scalarization slightly improves: for $\lambda_H=-2$ one finds $ \beta_c \approx -4.15$ and for $\lambda_{BL}=-2$ we have $\beta_c \approx -4.13$. On the other hand, this introduces further increases on both mass and moment of inertia; for instance, for $\lambda_{\{H,BL\}}=-2$ and $\beta \approx -4.3$, an increase of $\sim 10\%$ is observed, which is comparatively much less important than going from the perfect fluid approximation to the anisotropic one in GR itself.

On the other hand, oscillations of slowly rotating neutron and strange star in the quadratic $f(R)$ model (\ref{eq:fquadratic}) in the scalar-tensor representation have been investigated by Staykov et al. \cite{Staykov:2015cfa}. The relevance of such (non-radial) oscillations lies on the fact that they are a generator of gravitational waves \cite{Andersson:1996pn}. Using SLy, APR4, WFF2 EOS and quark-core matter, the authors find, for $\alpha=10^4$, a maximum deviation of $\sim 10\%$ between the frequencies of the fundamental models and those of this model, which falls probably below the accuracy of current detectors.

Some members of the Horndeski's family \cite{Horndeski:1974wa} have been also analyzed. For instance, Silva et al. \cite{Silva:2016smx} consider the model studied by Cisterna \cite{Cisterna:2015yla}, see Eq.(\ref{eq:GJ}), finding evidence of modifications to mass, radius, and momentum of inertia depending on $\eta$ and $Q_{\infty}$, as compared to GR case. The non-minimal derivative coupling sector of this theory given by (\ref{eq:GJ}) is considered by Cisterna et al. \cite{Cisterna:2016vdx}, extending the results discussed in that section to the slowly-rotating case. Using an ansatz for the scalar field extending the one of (\ref{eq:parsca}) as $\phi(t,r)=Qt+ F(r)+\epsilon \phi_1(t,r)$, the corresponding field equations are obtained and solved for SLy4 and $\{$BSk14, BSk19, BSk20, BSk21$\}$ EOS, where only the two last ones are compatible with the $2M_{\odot}$ threshold. For $\eta>0$ ($<0$) the star's compactness as a function of the mass gets larger (smaller) with the normalized charge $Q_p=Q/b_{\infty}^{1/2}$, with a larger (smaller) internal pressure. Regarding the moment of inertia, for $\eta>0$ it behaves as an increasing function of the mass (until some critical value is reached, depending on $Q_p$); for $\eta<0$ there is another critical value of $Q_p$ above which the moment of inertia also increases. Comparison of these results with current observations of maximum masses and moment of inertia yields the constraints $Q_p^2 \eta \leq 0.027$ for BSk21 and $Q_p^2 \eta \leq 0.011$ for BSk20 EOS.

Maselli et al. \cite{Maselli:2016gxk} use the \emph{``Fab-Four"} subclass of the Horndeski family (see \cite{Charmousis:2011bf} for details), which is composed of four pieces (\emph{George, Ringo, John} and \emph{Paul}). As George is equivalent to GR and Ringo to Einstein-dilaton-Gauss-Bonnet theory, the authors studied slowly rotating  stars in the George and John combination, given by the action (\ref{eq:GJ}), and look for two classes of solutions: those with $\eta = \pm 1$ and those with $\eta=0$\footnote{This choice leads to ``stealth" solutions, defined by the intriguing feature that there is no backreaction of the scalar field on the geometry outside the star in this case.}. Choosing APR, SLy4 and GNH3 \cite{Glendenning:1984jr} EOS, besides a polytropic (\ref{eq:poly}) EOS with $K=123M_{\odot}^2$ and $\Gamma=2$, the ansatz for the scalar field (\ref{eq:parsca}) is taken. Using a shooting method, the space of solutions gets parameterized in terms of the value of $Q_{\infty}=Q/A(\infty)^{1/2}$ following a general trend for a canonical neutron star with mass $M=1.4M_{\odot}$: at fixed negative (positive) $\eta$, larger values of $Q_{\infty}$ lead to less (more) compact stars with larger (smaller) values of their maximum masses as compared to GR predictions.  This improvement is moderate, up to $\sim 30\%$.  Indeed, in the SLy4 and GNH3 EOS a value of $Q_{\infty}=0.032$ allows for masses above the $2M_{\odot}$ threshold, while the polytropic EOS requires a larger value of the charge for achieving this effect.  This yields larger (smaller) moment of inertia, with a $\sim 30\%$ of deviation with respect to GR results for $Q_{\infty}=0.064$. In this sense, precise future measurements of the moment of inertia could allow to constrain the value of $Q_{\infty}$ \cite{KramerI}, which must also satisfy the stability condition for this model, which reads
\begin{equation}
Q_{\infty}^2 \vert \eta \vert=\frac{2}{\kappa^2}\left(\frac{1-2\mathcal{C}}{1+2\mathcal{C}}\right)
\end{equation}
For a typical neutron star, $\mathcal{C} \sim 0.2$, the right-hand-side of this equation is $\sim 0.35$ and, thus, taking $\vert \eta \vert=1$ one could explore further the space of values up to $Q_{\infty} \sim 0.18$ for such stars. Finally, a similar analysis for the Paul class of models with stealth scalar field models seems to indicate the impossibility of finding stable stellar structure configurations in that case.

Another subset of beyond Horndeski theories is a quartic galileon coupled to k-essence and characterized by a single parameter $\Upsilon$, which admits slowly rotating, asymptotically de Sitter solutions as analyzed by Sakstein et al. \cite{Sakstein:2016oel}.  Using an array of 32 EOS, their findings in the static case are an enhance of the maximum mass beyond the $2M_{\odot}$ threshold for $\Upsilon \lesssim  -0.03$, which can be as high as $3M_{\odot}$ for larger choices of $\Upsilon$, as opposed to GR case where this mass lies within the range $\sim 1.5M_{\odot}-2M_{\odot}$. Now, for slowly rotating stars, using an approximately universal $\bar{I}-\mathcal{C}$ relation found in \cite{Breu:2016ufb}, the authors report a significant decrease of the moment of inertia as compared to the GR case, for instance, up to $\sim 15\%$ for $\Upsilon=-0.03$ and up to $\sim 30\%$ for $\Upsilon= -0.05$ depending of compactness; thus this theory can be potentially probed via precise measurements of the $\bar{I}-\mathcal{C}$ relation within this range on assumptions on the EOS. The authors go forward to consider hyperon stars modelled by GM2NPH EOS (GR: $M_{\star} \approx 1.5M_{\odot}$, $r_S \approx 12$km): $\Upsilon=-0.05$ raises the maximum mass to $M_{\star} \approx 2.2M_{\odot}$ and a slightly greater radius, solving the hyperon puzzle if $\Upsilon \lesssim -0.04$. Similarly, quark matter with SQM2 EOS (GR: $M_{\star} \approx 2.0M_{\odot}$) yields $M_{\star} \approx 2.7M_{\odot}$ if $\Upsilon=-0.05$.  This enhance of the total mass is common to different EOS in this model, though the precise value of $\Upsilon $ to overcome the $2M_{\odot}$ threshold is obviously EOS-dependent.

Finally, Sullivan and Yunes study in detail the case of ghost-free massive bigravity in \cite{Sullivan:2017kwo} by considering APR, LS220 \cite{Lattimer:1991nc} and Shen EOS, finding the expected degeneracy between such EOS and the modified gravity parameter in this case, $m$, which is related to the graviton mass. The mass-radius and moment of inertia-to-mass relations are explored in detail, finding significant deviations with respect to GR for values of $m>10^{-7}$cm$^{-1}$. These deviations grow quickly; indeed, for $m=4 \times 10^{-7}$cm$^{-1}$ the dimensionless moment of inertia, $\bar{I} \equiv I/M^3$, increases by a factor $\sim 30\%$ for the APR EOS. However, to test massive bigravity via observations of binary pulsars and X-ray observations one needs to bear in mind that in this theory the mass function of an individual star is \emph{not} constant outside its radius, which implies that one would need to compute a combination of the bigravity coupling parameter and the binary masses that are compatible with the measurement of post-Keplerian parameters.

\subsubsection{Fully rotating models} \label{eq:sec:frm} \vspace{0.2cm}

Rapidly rotating neutron stars have also been addressed in the literature\footnote{For an informative discussion on how fast a neutron star can rotate in GR and the derived constraints on masses, radii and moment of inertia, the interested reader can profit from reading Ref.\cite{Margaritis:2019hfq}}, since it is a more reliable approach to describe binary mergers \cite{FrieSter}. In this case, the axially symmetric line element can be expressed for convenience as
\begin{equation}
ds^2=e^{\gamma+\sigma}dt^2+e^{\gamma-\sigma}r^2(d\phi - \omega dt)^2+e^{2\beta}(dr^2 + r^2 d\theta^2)
\end{equation}
where the three metric functions $\{\gamma,\sigma,\beta\}$ depend now on $r$ and $\theta$, and the fluid's angular velocity is $\Omega \equiv u^{\phi}/u^t$.  Doneva et al. \cite{Doneva:2013qva} consider rapidly rotating stars for Brans-Dicke theories and write the explicit form of the fully rotating equations in this case. They are numerically solved for the choices $A(\varphi)=k_0 \varphi$ and the one of (\ref{eq:Aansatz}), using a polytropic EOS (\ref{eq:poly}) with $\Gamma=1.40$ and $K=1186M_{\odot}^2$. For the first case, with an observationally viable value $k_0=4 \times 10^{-3}$, no changes as compared to GR are observed up to numerical error. For the second case, taking a value $\beta=-4.8$, it is found that deviances of mass, radius and moment of inertia as compared to GR increase with rotation, which, for the moment of inertia, peaks at a twofold value of that of GR.  For more realistic EOS, Yazadjiev et al. \cite{Yazadjiev:2015zia} numerically solve the field equations of the quadratic $f(R)$ model (\ref{eq:fquadratic}) in rapidly rotating stars using APR4 and SLy4 EOS. For the latter, they show that while the maximum mass may be raised, for a maximum allowed value of $\alpha=10^4$, by a $\sim 10\%$ in the static case as compared to GR, this is enhanced to a $\sim 16\%$ in the rapidly rotating case, while the increase in the moment of inertia goes from $\sim 41\%$ to $\sim 65\%$. Yazadjiev et al. \cite{Yazadjiev:2017vpg} compute the oscillation fundamental modes of scalar-tensor theories of gravity, finding significant deviations for these frequencies from GR results  within the range $-6.0 < \beta < -4.5$, particularly for scalar fields with non-vanishing masses. As for Einstein-Gauss-Bonnet-dilaton gravity, Kleihaus et al. \cite{Kleihaus:2016dui} consider this scenario with an analytical fit to FPS EOS and a simple (polytropic) DI-II \cite{DiazIban}  EOS, both below the $2M_{\odot}$ threshold. In the static case, the sequence of stable neutron stars limits the value of the (dimensionless) coupling constant of this theory to $\alpha \leq 3.4$. Taking in this work the values $\alpha=0,1,2$,   smaller values with larger radii than in GR with running $\alpha$ are found. In presence of fixed (fast) angular velocity, one also finds decreases of the maximum mass for both EOS and values of $\alpha$, and it is reported that the universal $I-Q$ relation in this theory is basically the same as in GR.

Differentially rotating stars have been also considered\footnote{For a broad discussion of differentially rotating stars in GR and how it allows to significantly increase the maximum masses see \cite{BaumShaShi}.}. For uniformly rotating stars the angular velocity of the fluid, $\Omega$, is assumed to be constant throughout the star, while in the differentially rotating case one goes to the full regime of $\Omega(r,\theta)$.
For scalar-tensor theories described by a potential $V(\varphi)=0$ and a scalar field function given by the action (\ref{eq:Aansatz}), Doneva et al. \cite{Doneva:2018ouu} construct general axially symmetric configurations assuming that the angular velocity satisfies $u^t u_{\phi}\equiv F(\Omega)=A^2_{diff}(\Omega_0-\Omega)$, where $\Omega_0$ is the value of the angular velocity at the star's center \cite{Komatsu:1989zz}, and $A^2_{diff}$ some parameter. Using APR4 EOS, they report that for observationally viable values of $\beta \lesssim -4.5$, scalar-tensor theories of this type allow for maximum masses greater than GR (with a strong dependence on the value of $\beta$) which may raise above $3M_{\odot}$ for larger allowed values of the angular momentum. These models are too constrained to produce realistic accounts of neutron stars binary mergers, but may be useful as tests of modified gravity.

These examples show once more that, due to the dependence on additional free parameters it is not surprising to find that the properties of the corresponding neutron stars are highly degenerate with GR ones, though the enhancements on the moment of inertia are typically much larger than those of the mass-radius relations. The fact that in some models such enhancements may be larger than the uncertainties of the EOS offers some hopes of constraining these theories beyond the EOS degeneracy in the near future.

\subsubsection{Binary neutron star mergers} \vspace{0.2cm}

Modifications to  stellar structure models are also relevant within the merger of neutron star binaries and their gravitational wave signatures emitted during the late-time and plunge phases, allowing new constraints on the EOS \cite{Forbes:2019xaz} and on the underlying theory of gravity using the electromagnetic counterparts of such mergers \cite{Nakar:2019fza}. An available review on this issue can be found in \cite{Berti:2015itd}. Let us briefly summarize some recent developments. For (Brans-Dicke type) scalar-tensor theories, Barausse et al. \cite{Barausse:2012da} simulate a binary merger of unequal mass neutron stars described by polytropic EOS (\ref{eq:poly}) with $\Gamma=2$ and $K=123 M_{\odot}^2$ (GR: $M_{\star} \approx 1.80$) with separation of $60$km and angular velocity $\Omega=1295$rad/s, and choose $M=\{1.58,1.67\}M_{\odot}$ taking the ansatz $\omega(\phi)=-3/2-\kappa^2/(4\beta \log(\varphi))$ (corresponding to the parameterization of \cite{Damour:1993hw}), and go on to solve numerically the full field equations for this setting.  For values of the parameter $-4.5 \lesssim \beta \lesssim -4.2$ and the asymptotic value of the scalar field $\varphi_0 \leq 10^{-5}$, the authors report a significant decrease in the merging frequency of the binary: indeed, the plunge starts at an early separation of $\sim 52$km of each binaries's center, with a frequency $f \sim 586$Hz, which cannot be   attained in GR even using exotic EOS. Let us also mention that in scalar-tensor theories such as in quadratic $f(R)$ gravity, there may exist new ultra long-lived quasinormal modes \cite{Blazquez-Salcedo:2020ibb}, enriching the GR spectrum thanks to the additional scalar degree of freedom of such theories.

 The case of dynamical Chern-Simons gravity is considered by Yagi et al. \cite{Yagi:2013mbt}. Working at leading order in the theory's coupling constant, and using isolated neutron stars in the slowly rotating approximation, it is found that this theory induces corrections to neutron star's moment of inertia (quadratic order in spin), yielding a modified orbital evolution of a binary neutron star, which allows to place constraints to the theory's parameter via observations of double binary pulsar (here represented by the millisecond pulsar J0737-3039 \cite{Kramer:2006nb}, which requires a mass $\approx 1.93M_{\odot}$ for the modelling of the heaviest neutron star). For such isolated stars, the use of EOS such as APR4, SLy, LS220 \cite{Lattimer:1991nc} and Sen \cite{Shen:1998gq} lowers the maximum mass as compared to GR (but still above that threshold), but increases the angular and quadrupolar momenta.  However, for admissible values of the Chern-Simons parameter, this correction turns out to be highly suppressed $\sim \mathcal{O}(10^{-8})$, being out of current observational capabilities in the analysis of binary pulsars. In Okounkova et al. \cite{Okounkova:2019dfo} a numerical analysis of head-on mergers of
binary black hole mergers in dynamical Chern-Simons finds modifications to the damping time and frequency of quasinormal modes increasing as a power law of the spin of the final black hole, which turn out not to be degenerate with the frequencies of a Kerr black hole of the same spin. Let us also mention the existence of present and future opportunities offered from black hole-neutron star binaries in order to constrain the parameter space of modified theories of gravity such as in scalar-tensor theories with spontaneous scalarization \cite{Carson:2019fxr}, or that can be useful in the construction of new templates of gravitational waveforms as in the case of Einstein-Aether theory \cite{Zhang:2019iim} . Finally, the measurement of gravitational waves out of  triple systems may offer another opportunity to test modified theories of gravity, though as argued by Zhao et al. \cite{Zhao:2019kif} for Einstein-Aether and Brans-Dicke theories this might not be available neither for this nor the next generation of detectors. \\

\emph{Universal relations} \\

Let us conclude this section with some comments about tidal deformability and universal (I-Love-Q) relations. In binary systems, the gravitational field of each star deforms both the shape and the gravitational field of the other, an effect which is encoded within tidal deformability. Using the fact that most such binaries are widely separated, it is possible to solve the Einstein equations by regarding one of the stars as a probe moving in the gravitational field of an isolated star, and study the corresponding deformations upon the first one. The output of this calculation is the quadrupole moment induced by the tidal deformation, from which one can extract the tidal deformability parameter \cite{Hinderer:2007mb} which, heuristically, measures how easily a body is deformed when subjected to an external field. Next, using the universal character (EOS-independent) of some empirical relations, such as the I-Love-Q one, it allows to constrain quantities that are experimentally hard to access, such as the neutron star radius \cite{Kumar:2019xgp}. This method has recently shown its power in constraining the common areal radius of the neutron star merger GW170817 detected by the LIGO/VIRGO Collaboration \cite{TheLIGOScientific:2017qsa} within the range $8.9<r_S<13.2$km \cite{De:2018uhw} (simulating thousands of piecewise polytropic EOS \cite{Lattimer:2015nhk}), further supported by the analysis of \cite{Raithel:2018ncd} yielding the neutron star radius $r_S \lesssim 13$km (90$\%$ confidence level), which in turns allows to rule out sufficiently soft EOS in the context of GR \cite{Coughlin:2018miv}, as well as to obtain the information encoded in the change of the gravitational wave phase caused by the tidal deformability exerted by each neutron star \cite{FlanHind2018}. As different modifications of gravity yield different deformability parameter(s), this represents yet another test of  these theories in the strong field regime. A detailed review on this issue, including dynamical Chern-Simons gravity, Einstein-dilaton Gauss-Bonnet gravity, scalar-tensor theories, $f(R)$ gravity and Eddington-inspired Born-Infeld gravity (in metric-affine formulation, see Sec.\ref{sec:Pal}), and the observational constraints on each theory's parameters, is already available by Yagi and Yunes \cite{Yagi:2016bkt} and therefore we refer the reader to this work for details on this subject. The topic is very active and is expected to yield more constraints on modified theories of gravity in the near future, see e.g. \cite{Gupta:2017vsl} for a recent work on dynamical Chern-Simons gravity, and \cite{Yazadjiev:2018xxk} for $f(R)$ theories. \\

This concludes our analysis of stellar structure models of modified theories of gravity formulated in metric formalism.

\section{Stellar structure models in metric-affine formalism} \label{sec:Pal}

In this section we shall busy ourselves with stellar structure models of modified theories of gravity formulated in the metric-affine approach. This means that we shall keep the  metric and the  affine connection as independent entities. To avoid confusions with quantities defined in the metric approach, all geometric quantities built out of the independent connection will be indicated by using mathcal style. Paralleling the analysis of their metric counterparts, we will begin our considerations with the case of $f(\mathcal{R})$ gravity and grow from there to more general cases.

\subsection{Families of metric-affine theories of gravity } \vspace{0.2cm}

\subsubsection{Field equations and scalar-tensor representation of $f(\mathcal{R})$ gravity  } \vspace{0.2cm}

The metric-affine $f(\mathcal{R})$ gravity action is given as
\begin{equation}
\mathcal{S}=\mathcal{S}_{\text{G}}+\mathcal{S}_{m}=\frac{1}{2\kappa^2}\int d^4 x \sqrt{-g}f(\mathcal{R}) +\mathcal{S}_{m}(g_{\mu\nu},\psi_m),\label{action}
\end{equation}
where only the symmetric part of the Ricci tensor  $\mathcal{R}_{\mu\nu}={\mathcal{R}^\alpha}_{\mu\beta\nu} \delta^\beta_\alpha$ contributes to the Ricci scalar $\mathcal{R}=g^{\mu\nu}\mathcal{R}_{\mu\nu}$ because of the contraction with the metric. Recall that
$ {\mathcal{R}^\alpha}_{ \beta\mu\nu}=\p_\m\G_{\n\b}^\a-\p_\n\G_{\m\b}^\a+\G_{\m\l}^\a\G_{\n\b}^\l-\G_{\n\l}^\a\G_{\m\b}^\l$ is the Riemann tensor of an independent connection $\Gamma \equiv \Gamma_{\mu\nu}^{\lambda}$.  The variation of the action (\ref{action}) with respect to the metric tensor provides the modified Einstein field equations in this case as
\begin{equation}
f_{\mathcal{R}}\mathcal{R}_{\mu\nu}-\frac{1}{2}f(\mathcal{R})g_{\mu\nu}=\kappa^2 T_{\mu\nu},\label{structural}
\end{equation}
where, again, $T_{\mu\nu}=-\frac{2}{\sqrt{-g}} \frac{\delta\mathcal{S}_m}{\delta g_{\mu\nu}}$ is the energy momentum tensor of the matter fields. Taking the trace of the field equations (\ref{structural}) yields a structural equation given by
\begin{equation}
f_{\mathcal{R}}\mathcal{R}-2 f(\mathcal{R})=\kappa^2 T.\label{struc}
\end{equation}
where $T\equiv g^{\mu\nu}T_{\mu\nu}$ is the trace of the energy-momentum tensor. This represents a fundamental difference between the metric-affine and metric approaches: instead of a differential equation, like in (\ref{eq:trace-m}), this is an algebraic equation for the curvature scalar $\mathcal{R}$, whose solution can be expressed as $\mathcal{R}=\mathcal{R}(T)$, i.e., a direct relation between curvature and matter that generalizes the linear relation $\mathcal{R}=-\kappa^2T$ of GR. This has relevant implications for the dynamics of these theories since, as opposed to their metric counterparts, where the trace equation (\ref{eq:trace-m}) promotes the function $f\big(R(g)\big)$ to a dynamical degree of freedom, here $f(\mathcal{R})$ is a function of the matter sources which can always be written in terms of them. This procedure can be made explicit as follows. The variation of the action (\ref{action}) with respect to the independent connection $\Gamma$ can be written in simplified form as
\begin{equation}
\nabla_{\alpha}^{\Gamma}(\sqrt{-g}f_{\mathcal{R}}g^{\mu\nu})=0,\label{con}
\end{equation}
where $\nabla_\alpha^{\Gamma}$ is the covariant derivative computed using $\Gamma$. This expression assumes that the matter fields are not coupled to the connection, which leads to an important simplification regarding the torsion tensor (see \cite{Afonso:2017bxr} for details). In fact, for minimally coupled bosonic fields, torsion boils down to a vector field, which can be set to vanish by the projective invariance of the action. As a result, only the symmetric part of the connection remains, and satisfies Eq.(\ref{con}).  The dependence of $f_{\mathcal{R}}$ in this equation on the matter sources allows to propose an ansatz such that
\begin{equation} \label{eq:connaux}
\nabla_{\alpha}^{\Gamma}(\sqrt{-h}h^{\mu\nu})=0
\end{equation}
which is nothing but (\ref{con}) provided that the new rank-two tensor $h_{\mu\nu}$ is conformally related to the space-time metric $g_{\mu\nu}$ as
\begin{equation}
h_{\mu\nu}=f_{\mathcal{R}}g_{\mu\nu}
\end{equation}
This implies that the independent connection $\Gamma$ can be expressed as the Christoffel symbols of $h_{\mu\nu}$, i.e.,
\begin{equation}
\Gamma_{\mu\nu}^{\lambda}=\frac{h^{\lambda\beta}}{2} \left(\partial_{\mu}h_{\nu\beta}+\partial_{\mu}h_{\nu\beta}-\partial_{\beta}h_{\mu\nu} \right)
\end{equation}
Now, contracting in the metric field equations with $h^{\mu\alpha}$, one finds that they can be re-expressed as
\begin{equation}\label{eq_conf}
{\bar{\mathcal{R}}^\mu}_{\ \nu}(h)=\frac{\kappa^2}{f_{\mathcal{R}}^2} \left[ {T^\mu}_{\nu}+\frac{f(\mathcal{R})}{2\kappa^2} \delta^{\mu}_{\nu} \right]
\end{equation}
where ${\bar{\mathcal{R}}^\mu}_{\ \nu}(h)=h^{\mu\alpha}\mathcal{R}_{\alpha\nu}$ is the Ricci tensor of the auxiliary metric $h_{\mu\nu}$. Eq. (\ref{eq_conf}) is a system of Einstein-like field equations for the metric $h_{\mu\nu}$ with all the terms on the right-hand-side being functions of the matter fields (and possibly of the space-time metric $g_{\mu\nu}$ too).

 It is well known (see e.g. \cite{Olmo:2011uz,DeFelice:2010aj}) that the action (\ref{action}) is dynamically equivalent to the constrained system with first-order metric-affine gravitational Lagrangian:
\begin{equation}\label{action1}
 \mathcal{S}(g_{\mu\nu}, \Gamma^\lambda_{\rho\sigma}, \chi)=\frac{1}{2\kappa^2}\int\mathrm{d}^4x\sqrt{-g}\left( f(\chi)+f^\prime(\chi)(\mathcal{R}-\chi) \right)
 +  \mathcal{S}_m(g_{\mu\nu},\psi_m),
\end{equation}
provided that $f''\equiv f_{\chi\chi}\neq 0 $\footnote{In that case the linear Einstein-Hilbert Lagrangian $f(\mathcal{R})=\mathcal{R} -2\Lambda$ is excluded on that level.}. If one now redefines the scalar field as $\Phi\equiv f_{\chi}(\chi)$ and takes into account the constraint equation $\chi=\mathcal{R}$, such that $\mathcal{R}=\mathcal{R}(\Phi)$ we may rewrite the action in dynamically equivalent way as a metric-affine scalar-tensor theory
\begin{equation}\label{actionP}
\mathcal{S}(g_{\mu\nu}, \Gamma^\lambda_{\rho\sigma},\Phi)=\frac{1}{2\kappa^2}\int\mathrm{d}^4x\sqrt{-g}\left(\Phi \mathcal{R} - U(\Phi) \right) +
\mathcal{S}_m(g_{\mu\nu},\psi_m),
\end{equation}
where the potential $U(\Phi)$ includes the information on the form of the function $f(\mathcal{R})$ via the term
\begin{equation}\label{PotentialP}
U(\Phi)=\chi(\Phi)\Phi-f(\chi(\Phi)),
\end{equation}
where $\chi\equiv \mathcal{R}  = \frac{d U(\Phi)}{d\Phi}$ and $\Phi = \frac{d f(\chi)}{d\chi}$. Furthermore, the metric-affine variation of this action provides the system of equations
	\begin{align}
	\label{EOM_P}
	\Phi\left( \mathcal{R}_{\mu\nu} - \frac{1}{2} g_{\mu\nu} \mathcal{R} \right)   +{1\over 2} g_{\mu\nu} U(\Phi) - \kappa^2 T_{\mu\nu} = 0\\
	\label{EOM_connectP}
	\nabla_\lambda^{\Gamma}(\sqrt{-g}\Phi g^{\mu\nu})=0\\
	\label{EOM_scalar_field_P}
	  \mathcal{R}   -  U^\prime(\Phi) =0
	\end{align}
The last equation of this set is automatically satisfied due to the constraint $\mathcal{R} =\chi= U^\prime(\Phi)$. Regarding Eq.(\ref{EOM_connectP}), as we saw before, it implies that
the independent connection $ \Gamma$ is compatible with the new metric
\begin{equation}\label{confMET}
h_{\mu\nu}=\Phi g_{\mu\nu}.
\end{equation}
The $g$-trace of the first
equation, $-\Phi \mathcal{R}+2 U(\Phi)=\kappa^2 T$, can be recast as
\begin{equation}\label{struc2}
  2U(\Phi)-U'(\Phi)\Phi=\kappa^2 T,
\end{equation}
which provides the counterpart of the structural equation (\ref{struc}) and implies the algebraic relation $\Phi=\Phi(T)$, which means that this scalar quantity is not a dynamical degree of freedom. Moreover, it can be shown that the above equations can be
further rewritten as a dynamical system for the metric $h_{\mu\nu}$ as \cite{BSS,SSB}
	\begin{align}
	\label{EOM_P1}
	 \bar{\mathcal{R}}_{\mu\nu} - \frac{1}{2} h_{\mu\nu} \bar{\mathcal{R}}  &  =\kappa \bar T_{\mu\nu}-{1\over 2} h_{\mu\nu} \bar U(\Phi),\\
	 \label{EOM_scalar_field_P1}
	  \Phi \bar{\mathcal{R}} &  -  \Big(\Phi^2\,\bar U(\Phi)\Big)^\prime =0,
	\end{align}
where we have introduced the new ``potential" $\bar U(\Phi)=U(\Phi)/\Phi^2$ and a new ``energy momentum tensor" $\bar T_{\mu\nu}=\Phi^{-1}T_{\mu\nu}$. Now, noting that $h_{\mu\nu} \bar{\mathcal{R}} =\ g_{\mu\nu}\mathcal{R}$ and using the trace of (\ref{EOM_P1}), the last equation can be replaced by
\begin{equation}\label{EOM_P1c}
 \Phi\,\bar U^\prime(\Phi)  + \kappa^2 \bar T = 0\,.
\end{equation}
Therefore, the system of equations (\ref{EOM_P1})-(\ref{EOM_P1c}) corresponds to a scalar-tensor action for the metric $h_{\mu\nu }$ and a non-dynamical
scalar field $\Phi$ as
\begin{equation}\label{action2}
 \mathcal{S}(h_{\mu\nu},\Phi)=\frac{1}{2\kappa^2}\int\mathrm{d}^4x\sqrt{-h}\bigg(\bar{\mathcal{R}}- \bar U(\Phi) \bigg) + \mathcal{S}_m(\Phi^{-1}h_{\mu\nu},\psi_m),
\end{equation}
where the energy-momentum tensor for a perfect fluid
\begin{equation}\label{em_2}
    \bar T^{\mu\nu} =
-\frac{2}{\sqrt{-h}} \frac{\delta \mathcal{S}_M}{\delta  h_{\mu\nu}}  = (\bar\rho+\bar p)\bar u^{\mu}\bar u^{\nu}+ \bar ph^{\mu\nu}=\Phi^{-3}T^{\mu\nu}~,
\end{equation}
can be expressed in terms of the new variables $\bar u^\mu=\Phi^{-{1\over 2}}u^\mu$, $\bar\rho=\Phi^{-2}\rho,\ \bar p=\Phi^{-2}p$, $\bar T_{\mu\nu}= \Phi^{-1}T_{\mu\nu}, \ \bar T= \Phi^{-2} T$
(see e.g. \cite{dabrawski}).
Furthermore, the trace of (\ref{EOM_P1}), provides
\begin{equation}\label{EOM_metric_2}
    \bar{\mathcal{R}}= 2\bar U(\Phi)-\kappa^2 \bar T \ ,
\end{equation}
where, recall (\ref{EOM_P1c}), $\Phi=\Phi(\bar T)$. 

Before concluding, it is worth noting that a transformation of (\ref{action2}) back to the original Jordan frame restores the coupling of the scalar $\Phi$ to the curvature, bringing the action into a more standard scalar-tensor representation \cite{Olmo:2005hc,Olmo:2005zr}
\begin{equation}
\mathcal{S}(g_{\mu\nu},\Phi)=\frac{1}{2\kappa^2}\int\mathrm{d}^4x\sqrt{-g}\bigg(\Phi\bar{\mathcal{R}}+\frac{3}{2\Phi}\partial_\mu \Phi \partial^\mu \Phi-  U(\Phi) \bigg) + \mathcal{S}_m(g_{\mu\nu},\psi_m) \ .
\end{equation}
In this form the theory can be interpreted as a Brans-Dicke theory with parameter $\omega=-3/2$, which is an exceptional case. In fact, the field equation of the Brans-Dicke scalar takes the general form
\begin{equation}
(3+2\omega)\Box \Phi+2U(\Phi)-\Phi U_\Phi=\kappa^2 T \ .
\end{equation}
Though this scalar field is dynamical in the general case, for the specific choice $\omega=-3/2$ the coefficient in front of the differential operator $\Box \Phi$ vanishes and the field becomes algebraically determined by the matter, $\Phi=\Phi(T)$. This explains why there is no scalar propagating degree of freedom in the Palatini version of $f(\mathcal{R})$. As a consequence, the number of polarizations of gravitational waves in these theories is the same as in GR. 

\subsubsection{Eddington-inspired Born-Infeld gravity } \label{eq:secEiBI} \vspace{0.2cm}

The $f(\mathcal{R})$ family of gravity theories provides a simple and nontrivial framework to explore extensions of GR but  is unable to access an important phenomenological sector associated to the Ricci tensor. In fact, those theories are only sensitive to the trace $\mathcal{R}=g^{\mu\nu}\mathcal{R}_{\mu\nu}$, leaving aside other traces of the object $M^\mu_{\ \nu}\equiv g^{\mu\alpha}\mathcal{R}_{\alpha\nu}$ that could be relevant at high energies, such as $\mathcal{R}^\mu_{\ \alpha}\mathcal{R}^\alpha_{\ \mu}$ and higher orders. In this direction, we now consider  a gravitational theory that has attracted a great deal of attention in the last few years, originally introduced by Vollick \cite{Vollick:2003qp} and popularized by Ba\~{n}ados and Ferreira \cite{banados} under the name  Eddington-inspired Born-Infeld (EiBI) theory. For a recent review of this model and its astrophysical and cosmological applications see Beltr\'an Jim\'{e}nez et al. \cite{BeltranJimenez:2017doy}. The action of EiBI  is given by\footnote{We introduce here parenthesis in the Ricci tensor to stress that we are working with the symmetric part of this object, which will be implicitly understood so from now on (and parenthesis removed). }
\begin{equation} \label{eq:actionEiBI}
 \mathcal{S}_{EiBI}[g,\Gamma,\psi_m]=\frac{1}{\kappa^2 \epsilon}\int d^4x\left[ \sqrt{-|g_{\mu\nu}+\epsilon \mathcal{R}_{(\mu\nu)}(\Gamma)|}
 -\lambda\sqrt{-g} \right]+
 \mathcal{S}_m[g,\Gamma,\psi_m],
\end{equation}
where $\epsilon$ is the typical scale of this theory. An expansion of this action for fields $ \vert \mathcal{R}_{\mu\nu} \vert \ll \epsilon^{-1}$  yields \cite{pani},
\begin{equation} \label{eq:quadgrav}
\mathcal{S}[g,\Gamma,\psi_m]=\frac{1}{\kappa^2}\int d^4x\sqrt{g}\left[\mathcal{R}-2\Lambda+\frac{\epsilon}{4}(\mathcal{R}^2-2\mathcal{R}_{\mu\nu}\mathcal{R}^{\mu\nu})+\mathcal{O}(\epsilon^2)\right]
+\mathcal{S}_m[g,\Gamma,\psi_m]
\end{equation}
which is nothing but GR with  and effective cosmological constant term $\Lambda=\frac{\lambda-1}{\epsilon}$ and supplemented by quadratic curvature corrections. Variation of the action (\ref{eq:actionEiBI}) with respect to metric $g_{\mu\nu}$ and connection $\Gamma$ yields the system of equations
\begin{align}
 \frac{\sqrt{|q|}}{\sqrt{|g|}}q^{\mu\nu}-\lambda g^{\mu\nu}&=-\epsilon\kappa^2 T^{\mu\nu},\\
 g_{\mu\nu}+\epsilon\mathcal{R}_{\mu\nu}&\equiv q_{\mu\nu},\label{metr}
\end{align}
where we have introduced $q_{\mu\nu}$ as the metric compatible with $\Gamma$, i.e. it satisfies Eq.(\ref{eq:connaux}) with $h_{\mu\nu} \to q_{\mu\nu}$. Note also that $q^{\mu\nu}$ denotes the inverse of $q_{\mu\nu}$. Combining the above field equations, we find
\begin{equation}\label{eqI}
 \sqrt{q}q^{\mu\nu}=\lambda\sqrt{g}g^{\mu\nu}-\epsilon\kappa^2\sqrt{g}T^{\mu\nu}
\end{equation}
which together with (\ref{metr}) provides the simplest set of equations in order to analyze the theory. At this stage, it is useful to note that (\ref{eqI}) establishes an algebraic relation between the two metrics, $q^{\mu\nu}$ and $g^{\mu\nu}$, and the matter fields. By introducing the relation
\begin{equation} \label{eq:defmat}
q_{\mu\nu}=g_{\mu\alpha}{\Omega^\alpha}_\nu
\end{equation}
where from now on the object ${\Omega^\alpha}_\nu$ will be termed as the \emph{deformation matrix} (and hats used when needed to stress this matrix character), from (\ref{eqI}) the above relation turns into
\begin{equation}\label{eq:Omega-EiBI}
|\hat\Omega|^{\frac{1}{2}}[\Omega^{-1}]^\mu_{\ \nu}=\lambda \delta^\mu_\nu-\epsilon \kappa^2 {T^\mu}_\nu \ .
\end{equation}
and vertical bars accompanying $\hat\Omega$ denote its determinant. This equation means that the deformation ${\Omega^\alpha}_\nu$ that relates the two metrics is fully determined by ${T^\mu}_\nu$. Using this result in (\ref{metr}), one finds that the field equations for the metric $q_{\mu\nu}$ can be written as
\begin{equation}\label{eom_EiBI}
{\bar{\mathcal{R}}^\mu}_{\ \nu}(q)=\frac{\kappa^2}{|\hat\Omega|^{\frac{1}{2}}} \left[ {T^\mu}_{\nu}+\frac{(|\hat\Omega|^{\frac{1}{2}}-\lambda)}{\epsilon\kappa^2} \delta^{\mu}_{\nu} \right] \ ,
\end{equation}
which puts forward that the right-hand side is determined by the matter sources. This equation should be compared with (\ref{eq_conf}) noticing that the EiBI Lagrangian density can be written
\begin{equation}
\mathcal{L}_{EiBI}=\frac{{(|\hat\Omega|^{\frac{1}{2}}-\lambda)}}{\epsilon\kappa^2}
\end{equation}
and, therefore, is a function of the matter sources as well.

Though Eq.(\ref{eom_EiBI}) allows to write the left-hand side of the equations in Einstein-like form for the metric $q_{\mu\nu}$, in general, the right-hand side still contains the metric $g_{\mu\nu}$, which complicates the algebraic analysis of the equations. Restricting the matter sources to be a perfect fluid,  Delsate et al. \cite{delsate2} found that the correspondence with Einstein equations can be extended also to the right-hand side by defining new matter variables, generalizing in this way Eq.(\ref{em_2}) to  EiBI theory.

\subsubsection{Ricci-based gravity theories}\label{sec:Fyou} \vspace{0.2cm}

The models introduced so far can be further generalized to a family of theories encompassing those defined via scalars built out of the (symmetrized) Ricci tensor\footnote{As recently shown by Beltr\'an  Jim\'{e}nez and Delhom \cite{BeltranJimenez:2019acz}, the symmetrization requirement enforces the projective invariance of the theory, namely, under transformations $\Gamma_{\mu\nu}^{\lambda} \to \Gamma_{\mu\nu}^{\lambda} + \xi_{\mu} \delta_{\nu}^{\lambda}$ , which is nothing but a manifestation of the freedom we have in parametrizing particle's geodesic paths \cite{EisenBook}. This requirement safeguards RBGs from the presence of ghost-like instabilities \cite{Jimenez:2020dpn}. For the consequences of projective invariance for the interpretation of some geometrical aspects of solutions of RBGs see \cite{Bejarano:2019zco}.} and its contractions with the metric. This family of metric-affine theories is known as  Ricci-Based Gravity theories
(RBGs) \cite{Afonso:2017bxr}. Assuming minimal coupling to the matter fields, their action is defined as (again we assume symmetrization in the Ricci tensor)
\begin{equation}\label{eq:f-action}
\mathcal{S}_{RBG}=\int d^4x \sqrt{-g} \mathcal{L}_G[g_{\mu\nu},\mathcal{R}_{\mu\nu}(\Gamma)]+\mathcal{S}_m[g_{\mu\nu},\psi_m]   \ ,
\end{equation}
where $\mathcal{L}_G[g_{\m\n},\mathcal{R}_{\mu\nu}(\G)]$ is the gravitational Lagrangian density. For the function $\mathcal{L}_G$ to be a scalar, its functional dependence must be via traces of powers of the object ${M^\mu}_\nu\equiv g^{\mu\alpha}\mathcal{R}_{\alpha\nu}(\G)$ \cite{Jimenez:2014fla}. An Einstein-frame representation for Eq.(\ref{eq:f-action}) can be obtained by introducing (ten) auxiliary fields $\Sigma_{\mu\nu}$ such that
\begin{equation}
\tilde{\mathcal{S}}\left[g,\Gamma,\Sigma,\psi_m \right]= \int d^4x \sqrt{-g}\left[\mathcal{L}_G(g,\Sigma)+\frac{\partial \mathcal{L}_G}{\partial \Sigma_{\mu\nu}}\big(\mathcal{R}_{\mu\nu}(\Gamma)- \Sigma_{\mu\nu}\big) \right] +  \mathcal{S}_m[g,\psi_m]  \ , \label{actionleg0}
\end{equation}
and defining  $\sqrt{-q}q^{\mu\nu}/2\kappa^2\equiv \sqrt{-g} \frac{\partial \mathcal{L}_G}{\partial \Sigma_{\mu\nu}}$ to express it as
\begin{equation}
\tilde{\mathcal{S}}\left[g,\Gamma,\Sigma,\psi_m \right]= \frac{1}{2\kappa^2}\int d^4x \big[\sqrt{-q} q^{\mu\nu} \big(\mathcal{R}_{\mu\nu}(\Gamma)- \Sigma_{\mu\nu}\big) + \sqrt{-g} \big(\mathcal{L}_G (g,\Sigma) + \mathcal{L}_m\big) \big] \ , \label{actionleg}
\end{equation}
where $q_{\mu\nu}$ is a function of $g_{\mu\nu}$ and $\Sigma_{\mu\nu}$. Note that (\ref{actionleg}) is a convenient representation of (\ref{actionleg0}), being completely equivalent to it. Variation of (\ref{actionleg0}) with respect to $\Sigma_{\mu\nu}$ leads to $\mathcal{R}_{\mu\nu}(\Gamma)=\Sigma_{\mu\nu}$,  while variation of (\ref{actionleg}) with respect to $\Gamma_{\mu\nu}^{\lambda}$ yields (after standard manipulations to deal with the torsion degrees of freedom \cite{Afonso:2017bxr}) $\nabla_{\mu}^{\Gamma}(\sqrt{-q}q^{\alpha\beta})=0$, which implies that $\Gamma$ is Levi-Civita  of $q_{\mu\nu}$ and, hence, $\mathcal{R}_{\mu\nu}(q)=\Sigma_{\mu\nu}$. Variation of (\ref{actionleg}) with respect to $g_{\mu\nu}$ yields $2 g^{\mu\alpha} \frac{\partial \mathcal{L}_G[g,\Sigma]}{\partial g^{\alpha\nu}}= {T^\mu}_{\nu}+\mathcal{L}_G(g,\Sigma) \,{\delta^\mu}_{\nu}$, which establishes an algebraic relation of the form $\Sigma_{\mu\nu}=\Sigma_{\mu\nu}(T_{\alpha\beta},g_{\alpha\beta})$. Now, given that $\mathcal{L}_G$ is a function of ${M^\mu}_\nu\equiv g^{\mu\alpha}\Sigma_{\alpha\nu}$, one finds that $ g^{\mu\alpha} \frac{\partial \mathcal{L}_G[g,\Sigma]}{\partial g^{\alpha\nu}}=\frac{\partial \mathcal{L}_G[g,\Sigma]}{\partial \Sigma^{\mu\alpha}}\Sigma_{\alpha\nu}$, which allows to write $q^{\mu\alpha}\mathcal{R}_{\alpha\nu}(q)$  as
\begin{equation}
{\mathcal{R}^\mu}_{\;\nu} (q) =\kappa^2\frac{\sqrt{-g}}{\sqrt{-q}}\Big[{T^\mu}_\nu+\mathcal{L}_G(g,\Sigma) {\delta^\mu}_{\nu}\Big] \ . \label{eq:Riccifeq}
\end{equation}
Using the fundamental relation (\ref{eq:defmat}) with ${\Omega^\alpha}_\nu$ a model-dependent on-shell function of  $T_{\alpha\beta}$ and $g_{\alpha\beta}$, one arrives at an Einstein-frame representation of the field equations
\begin{equation}\label{eq:GmnGeneral}
{G^\mu}_{\;\nu}(q)=\frac{\kappa^2}{|\hat{\Omega}|^{1/2}}\left[{T^\mu}_\nu-{\delta^\mu}_\nu\left(\mathcal{L}_G+\tfrac{T}{2}\right)\right] \ ,
\end{equation}
where ${G^\mu}_{\;\nu}(q) \equiv  q^{\mu\alpha} \mathcal{R}_{\alpha\nu}(q) - \frac{1}{2} {\delta^\mu}_{\nu}  \mathcal{R}(q)$. Note that both $\mathcal{L}_G$ and $|\hat{\Omega}|$ are functions of the metric $g_{\mu\nu}$ and ${T^\mu}_\nu$\footnote{Due to the fact that via this procedure the affine connection as been fully integrated out of the field equations, sometimes RBGs are interpreted as just GR with non-minimal couplings in the matter fields. We thank Jos\'e Beltr\'an-Jim\'enez for pointing out this to us.}.

From the above equations it is easy to see that in vacuum, ${T^\mu}_\nu=0$, the tensor ${\mathcal{R}^\mu}_{\;\nu}$ is proportional to the identity and that any traces constructed out of it must be constant, with $-{\kappa^2\mathcal{L}_G}/{|\hat{\Omega}|^{1/2}}$ evaluated in vacuum representing the effective cosmological constant of the theory considered.  This result clearly shows that any RBG of the type considered above will yield ghost-free, second-order field equations. Note that Eqs.  (\ref{em_2}) and (\ref{eom_EiBI}), corresponding to the field equations of $f(\mathcal{R})$ and EiBI theories, are just particular cases of  (\ref{eq:Riccifeq}), with the deformation matrix given by ${\Omega^\alpha}_\nu\equiv f_{\mathcal{R}}\delta^\alpha_\nu$ in the $f(\mathcal{R})$ case, and by (\ref{eq:Omega-EiBI}) in the EiBI case\footnote{A perturbative version of this correspondence has been used to place stringent constraints for generic RBGs, and for EiBI gravity in particular, in \cite{Delhom:2019wir}.}.

The generic representation of the RBG field equations given by (\ref{eq:GmnGeneral}) indicates that if the dependence on the metric $g_{\mu\nu}$ on their right-hand side could be explicitly replaced in favor of $q_{\mu\nu}$ and the matter fields, then the system of equations could be interpreted as a genuine set of Einstein field equations. Moreover, given that the left-hand side satisfies the Bianchi identity $\nabla_\mu^{(q)}{G^\mu}_\nu(q)=0$, where $\nabla_\mu^{(q)}$ is the covariant derivative associated to the metric $q_{\alpha\beta}$, the right-hand side must also be conserved. For this to happen, the effective energy-momentum tensor
\begin{equation} \label{eq:effectivetau}
\tau^\mu_{\ \nu}\equiv \frac{1}{|\hat{\Omega}|^{1/2}}\left[{T^\mu}_\nu-{\delta^\mu}_\nu\left(\mathcal{L}_G+\tfrac{T}{2}\right)\right]
\end{equation}
must have precisely the same structure as the energy-momentum tensor of well known matter fields.  This observation has been used recently \cite{Afonso:2018bpv,Afonso:2018mxn,Afonso:2018hyj,Delhom:2019zrb} to define a strategy that allows to construct the effective Einstein-frame matter Lagrangian associated to a given RBG. In particular, it has been shown that the field equations of RBGs in which canonical scalar/electromagnetic fields are minimally coupled to $g_{\mu\nu}$ turn out to be in correspondence with the equations of GR with a minimal coupling of $q_{\mu\nu}$ to nonlinear scalar/electromagnetic fields. In other words, the nonlinearities of the RBG gravitational sector can be effectively transferred to the matter sector of GR. Once the solutions of the corresponding (not equivalent) Einstein problem have been found, the solutions of the RBG problem can be generated by means of the algebraic transformation (\ref{eq:defmat}) inverted as $g_{\mu\nu}=q_{\mu\alpha}[\Omega^{-1}]^\alpha_{\ \nu}$, where the matrix $\hat\Omega$ can be written (on-shell) in terms of the Einstein-frame matter sources.

An important observation is now in order regarding the implications of the relation $g_{\mu\nu}=q_{\mu\alpha}[\Omega^{-1}]^\alpha_{\ \nu}$. In this relation, the metric $q_{\mu\alpha}$ has been obtained by solving a set of genuine Einstein field equations minimally coupled to some matter source. In order for the geometry defined by $g_{\mu\nu}$ to be well behaved, one should demand sufficient continuity and differentiability of the matter sources (on the relevant domain) because otherwise the Riemann tensor of $g_{\mu\nu}$ could contain divergences generated by the derivatives of $[\Omega^{-1}]^\alpha_{\ \nu}$. When one deals with elementary fields, such as scalars or electromagnetic fields, their continuity and differentiability is generally guaranteed everywhere. However, when a fluid description is considered, the matter profiles and boundary conditions must be handled with care in order to avoid undesired effects and properly match solutions across boundaries (see in this regard Section \ref{sec:junction}). Whenever such situations may happen, one should adopt an improved description of the matter sources/boundary conditions or establish the domain of validity of the approximations considered.

\subsubsection{Other metric-affine models } \vspace{0.2cm}

A hybrid version of $f(\mathcal{R})$ theories was proposed by Harko et al. \cite{hg} in the form
\begin{equation} \label{eq:hybrid}
\mathcal{S}=\frac{1}{2\kappa^2}\int d^4x\sqrt{-g}[R+f(\mathcal{R})]+\mathcal{S}_m(g_{\mu\nu},\psi_m)
\end{equation}
where $R$ is the usual curvature constructed with the Christoffel symbols of the space-time metric, while $\mathcal{R}$ is built with the independent connection $\Gamma$. Similarly to pure metric-affine gravity, the variation of the action (\ref{eq:hybrid}) with respect to the independent connection $\Gamma$ yields an equation which
tells us that it is compatible with the conformal metric,
$h_{\mu\nu}=f_{\mathcal{R}}(\mathcal{R})g_{\mu\nu}$. On the other hand, the variation with respect to the spacetime metric $g_{\mu\nu}$ yields the result
\begin{equation} \label{eq:fehybrid}
 G_{\mu\nu}(g)+f_{\mathcal{R}}(\mathcal{R})\mathcal{R}_{\mu\nu}-\frac{1}{2}f(\mathcal{R})g_{\mu\nu}=\kappa^2T_{\mu\nu},
\end{equation}
where $G_{\mu\nu}(g)=R_{\mu\nu}(g)-\frac{1}{2}g_{\mu\nu}R(g)$ is the standard Einstein tensor. The action (\ref{eq:hybrid}) also possesses a scalar-tensor representation; indeed, it can be re-expressed as a (kind of) Brans-Dicke theory with a potential $V(\phi)$ and the BD parameter $\omega=-3/2$ as
\begin{equation}
\mathcal{S}=\frac{1}{2\kappa^2}\int d^4x\sqrt{-g}\left[(1+\phi)R+\frac{3}{2}\partial_\mu\phi\partial^\mu\phi-V(\phi)\right]+\mathcal{S}_m(g_{\mu\nu},\psi_m)
\end{equation}
with the field equations (\ref{eq:fehybrid}) written now as
\begin{eqnarray}
 G_{\mu\nu}&=&\frac{\kappa^2}{1+\phi}\left\{T_{\mu\nu}-\kappa^{-2}\left[
 \frac{1}{2}g_{\mu\nu}(V+2\Box\phi)
+\nabla_\mu\nabla_\nu\phi-\frac{3}{2\phi}\partial_\mu\phi\partial_\nu\phi+\frac{3}{4\phi}g_{\mu\nu}(\partial\phi)^2
 \right]\right\},\nonumber\\
 &-&\Box\phi+\frac{1}{2\phi}\partial_\mu\phi\partial^\mu\phi+\frac{\phi[2V-(1+\phi)V_\phi]}{3}=\frac{\kappa^2\phi}{3}T.
\end{eqnarray}

On the other hand, analogously to the case of $f(R)$ gravity which can be considered either in metric or metric-affine formalism, the $f(R,T)$ gravity originally proposed in \cite{tg} can be also extended to the metric-affine formulation \cite{Barrientos:2018cnx,Wu:2018idg}. The action is thus
\begin{equation}
\mathcal{S}=\frac{1}{2\kappa^2}\int d^4x\sqrt{-g}f(\mathcal{R},T)+\int d^4x\sqrt{-g}\mathcal{L}_m(g_{\mu\nu},\psi_m)
\end{equation}
Variation of this action with respect to metric and connection yields
\begin{align}
 f_{\mathcal{R}}(\mathcal{R},T)\mathcal{R}_{\mu\nu}-\frac{1}{2}g_{\mu\nu}f(\mathcal{R},T)&=\kappa^2 T_{\mu\nu}-f_{\mathcal{T}}(\mathcal{R},T)T_{\mu\nu}-
 f_{\mathcal{T}}(\mathcal{R},T)\Theta_{\mu\nu}, \label{eq:metricfrt}\\
\nabla_\lambda^{\Gamma}(f_{\mathcal{R}}(\mathcal{R},T)\sqrt{-g}g^{\mu\nu})&=0,
\end{align}
where $\Theta_{\mu\nu}:=g^{\alpha\beta}\frac{\delta T_{\alpha\beta}}{\delta g^{\mu\nu}}$. Immediately one notices that the independent connection $\Gamma$ is Levi-Civita of the conformally related metric
\begin{equation}
 h_{\mu\nu}=f_{\mathcal{R}}(\mathcal{R},T)g_{\mu\nu}.
\end{equation}
Using this relation, one can rewrite the modified field equations (\ref{eq:metricfrt}) in terms of space-time metric $g_{\mu\nu}$ as
\begin{align}
 G_{\mu\nu}(g)&=\frac{\kappa^2-f_{\mathcal{T}}(\mathcal{R},T)}{f_{\mathcal{R}}(\mathcal{R},T)}T_{\mu\nu}-\frac{f_{\mathcal{T}}(\mathcal{R},T)}{f_{\mathcal{R}}(\mathcal{R},T)}\Theta_{\mu\nu}-
 \frac{1}{2}g_{\mu\nu}\left(\mathcal{R}-\frac{f(\mathcal{R},T)}{f_{\mathcal{R}}(\mathcal{R},T)}\right)
 -\frac{3}{2f_{\mathcal{R}}^2(\mathcal{R},T)}[\nabla_\mu f_{\mathcal{R}}(\mathcal{R},T)][\nabla_\nu f_{\mathcal{R}}(\mathcal{R},T)]\nonumber\\
 &-
 \frac{1}{2}g_{\mu\nu}(\nabla f_{\mathcal{R}}(\mathcal{R},T))^2+\frac{1}{f_{\mathcal{R}}(\mathcal{R},T)}(\nabla_\mu\nabla_\nu-g_{\mu\nu}\Box)f_{\mathcal{R}}(\mathcal{R},T).
\end{align}

\subsection{Physics of compact stars } \vspace{0.2cm}

Having discussed the families of models whose astrophysical predictions we will be interested in, let us discuss in this section several relevant aspects, subtleties, and difficulties when building stellar structure models in metric-affine formalism.

\subsubsection{Vacuum solutions and TOV equations} \vspace{0.2cm}

Unlike for metric $f(R)$ gravity, for the metric-affine formulation of $f(\mathcal{R})$ gravity,   Birkhoff's theorem holds, which has been seen by some authors as a quick test to confirm that such theories automatically pass\footnote{Microscopic systems might be sensitive to infrared curvature corrections, which can be used to rule out such models  \cite{Olmo:2011uz,Olmo:2008ye,Olmo:2006zu}. \label{fn:subtleties}} Solar System tests \cite{barraco, al1, al2}.  The situation, however, is more subtle, as will be seen in section \ref{sec:PalfR}. The point is that from the trace equation (\ref{struc}) one sees that for vacuum solutions ($T=0$)  the metric-affine scalar curvature $\mathcal{R}$ must be a constant.  As a result, the conformal factor that relates $g_{\mu\nu}$ and $h_{\mu\nu}$ also becomes a constant and the independent connection boils down to  Levi-Civita of the spacetime metric $g_{\mu\nu}$, which reduces the modified field equations (\ref{structural}) to
\begin{equation}
 \mathcal{R}_{\mu\nu}-\Lambda(\mathcal{R}_0)g_{\mu\nu}=0 \ .
\end{equation}
Here the effective cosmological constant, $\Lambda(\mathcal{R}_0)\equiv f(\mathcal{R}_0)/2f_{\mathcal{R}}$, depends on the constant vacuum value of $\mathcal{R}$, that is, $\mathcal{R}(T=0)=\mathcal{R}_0$. Its value depends on the choice of the $f(\mathcal{R})$ function but, in any case, it leads to a  unique exterior solution for each theory, which for spherically symmetric configurations is Schwarzschild if $\Lambda(\mathcal{R}_0)=0$ or Schwarzschild-(anti-)de Sitter when $\Lambda(\mathcal{R}_0)\neq0$. More generally, this property of the vacuum field equations indicates that the orbital motion of compact binary systems must have the same characteristics as in GR,  up to a possible rescaling of the masses as claimed in \cite{hannu}.

 It is worth stressing that the recovery of the vacuum GR equations is a general property of the RBG family of metric-affine theories of gravity including, as particular examples, GR itself, $f(\mathcal{R})$, $f(\mathcal{R},\mathcal{R}_{\mu\nu}\mathcal{R}^{\mu\nu})$, EiBI gravity,  and so on. This can be clearly seen from the system of Einstein-like equations (\ref{eq:GmnGeneral}) which, due to the  extra matter-dependent terms on the right-hand-side, makes them recover the GR vacuum equations and solutions when  ${T^\mu}_{\nu}=0$. Moreover, this also implies that these theories do not propagate extra degrees of freedom beyond the two-polarizations of the gravitational field, which makes them automatically compatible with the propagation of gravitational waves at the speed of light in vacuum (see e.g. \cite{Jana:2017ost} for a recent discussion in the case of EiBI gravity).

Considering now the spherically symmetric line element (\ref{eq:ds2}) and defining the stellar mass function in this case as
\begin{equation}
 m_\text{tot}(r)\equiv \frac{r}{2}(1-B^{-1})
\end{equation}
we may write the generalized TOV equations for Palatini $f(R)$ following \cite{kain} as
\begin{align}\label{tov_palat}
 P'=&-\frac{1}{1+\gamma}\frac{(P+\rho)}{r(r-2m_\text{tot})}
 \left(m_\text{tot}+\frac{4\pi r^3P}{f_{\mathcal{R}}}-\frac{\alpha}{2}(r-2m_\text{tot})\right),\\
 m'_\text{tot}=&\frac{1}{1+\gamma}\left( \frac{4\pi r^2\rho}{f_{\mathcal{R}}}+\frac{\alpha+\beta}{2}-
 \frac{m_\text{tot}}{r}(\alpha+\beta-\gamma) \right), \label{tov_palat2}
\end{align}
where
\begin{align}
 \alpha&\equiv r^2\left(\frac{3}{4}\left(\frac{f_{\mathcal{R}}'}{f_{\mathcal{R}}}\right)^2+
 \frac{2f_{\mathcal{R}}'}{rf_{\mathcal{R}}}+\frac{B}{2}\left(\mathcal{R}-\frac{f}{f_{\mathcal{R}}}\right)\right),\\
 \beta&\equiv r^2\left( \frac{f_{\mathcal{R}}''}{f_{\mathcal{R}}}-\frac{3}{2}\left(\frac{f_{\mathcal{R}}'}{f_{\mathcal{R}}}\right)^2 \right),\;\;\;\;\;
 \gamma\equiv \frac{rf_{\mathcal{R}}'}{2f_{\mathcal{R}}}.
\end{align}
After providing an EOS, the above TOV equations determine density $\rho(r)$, pressure $P(r)$, and the total mass $m_\text{tot}(r)$, where the latter contributes to the forms of the metric components $A(r)$ and $B(r)$  appearing in (\ref{eq:ds2}). Moreover, the exterior mass parameter $M$ is given by \cite{kain}
\begin{equation}
 M=m_\text{tot}(r_S)-\frac{\mathcal{R}_0r_S^3}{6} \ ,
\end{equation}
where $r_S$ represents the surface of the star, interpreted here as the region where the pressure drops below a reasonably small threshold (approximated as $P\to 0$, as in the GR case). An immediate observation follows: the matching of the stellar interior with the exterior vacuum gives rise to a non-trivial relation between the gravitational mass and the internal density and pressure of the star. In fact, given that (\ref{tov_palat2}) involves up to second-order radial derivatives of $f_{\mathcal{R}\mathcal{R}}$ via the function $\beta$, and that $\mathcal{R}$ is a function of the matter density and pressure, given an EOS it follows that $\rho(r)$ and $P(r)$ should have, at least,  smooth derivatives up to second order (see also the discussion at the end of section \ref{sec:Fyou}). Therefore, depending on how these functions behave in their transient from inside the sources to the outside, the resulting exterior solution can be characterized by different parameters. Therefore, although one may deal with different forms of $f(\mathcal{R})$ providing the same vacuum solutions, the interior ones may differ from each other depending on how the transient takes place. This is well known for some popular models, such as the inverse $1/\mathcal{R}$, and the quadratic one, which we will discuss in detail later. Nonetheless, as observed in \cite{kain}, $f_{\mathcal{R}}$ cannot differ significantly from $1$ inside a star like the Sun in order not to be in conflict with Solar system experiments. Since the contribution of $f_{\mathcal{R}}$ to the exterior mass $M$ is negligible (if one considers it as a cosmological constant), the model (\ref{eq:1/Rmodel}), which is ruled out in the metric formalism by Solar System tests, not only seems to pass this requirement but also seems to have no impact on the internal stellar physics (see, however, footnote \ref{fn:subtleties} for some subtleties in this regard).  A different situation can happen if one adds quadratic corrections such as $\alpha \mathcal{R}^2$ because the curvature becomes large inside the star. Therefore, the effect of such a modification should be able to provide constraints on model parameters, which will be discussed later.

\subsubsection{Matching at the stellar surfaces} \vspace{0.2cm}

As first pointed out by Barausse et. al. \cite{barausse1, barausse3}, when matching the interior solution of an $f(R)$ theory with the external vacuum using as matter source a polytropic EOS (\ref{eq:poly}), the terms including first and second derivatives of the pressure appearing in (\ref{tov_palat}) may lead to pathologies at the stellar surface, understood here again as the region where the pressure vanishes. Indeed, for  a range of values of the polytropic index, $\frac{3}{2}<\Gamma<2$, one finds that curvature scalars at the star's surface diverge.

The fact that the interior metric has a nontrivial dependence on the first and second derivatives of the energy density was interpreted in \cite{barausse1, barausse3} as the cause of this anomalous behavior. It is further argued that all modified theories of gravity in which the metric depends on higher-order derivatives of the matter fields will suffer the same problem\footnote{This also includes models with scalar fields algebraically related to matter.}. Whether this divergence of curvature scalars at the star's surface induces problems with tidal forces was discussed in \cite{barausse2}, focusing on the case of physical interest $\Gamma=5/3$, which describes both a monatomic isentropic gas and a degenerate non-relativistic electron gas. Their discussion goes as follows. Defining $C=\frac{df_{\mathcal{R}}}{dp}(p+\rho)$, the following quantities at the star's surface are obtained:
\begin{eqnarray}
 f_{\mathcal{R}}''(r_S)&=&\frac{(\mathcal{R}_0r^3_S-8m_\text{tot})C'}{8r_S(r_S-2m_\text{tot})},\\
 m'_\text{tot}(r_S)&=&\frac{2f_{\mathcal{R}}(r_S)\mathcal{R}_0r^2_S+(\mathcal{R}_0r^3_S-8m_\text{tot})C'}{16f_{\mathcal{R}}(r_S)},
\end{eqnarray}
These expressions diverge for
$\frac{3}{2}<\Gamma<2$ (because $C'\rightarrow\infty$ when the surface is approached) together with the Riemann tensor of
the metric $g_{\mu\nu}$ and its invariants. Moreover, considering the geodesic separation vector $(0,\eta^r\frac{\partial}{\partial r},0,0)$ the geodesic deviation
equation reads explicitly \cite{barausse2}
\begin{equation}
 \frac{D^2\eta^r}{\D\tau^2}=R^r_{\,ttr}(u^t)^2\eta^r\sim(c_0+c_1 f_{\mathcal{R}}'')(u^t)^2\eta^r,
\end{equation}
where $c_0$ and $c_1$ are functions of $f$, $f'$ and $f_{\mathcal{R}}$, both being finite at the surface. Thus one may compare the GR and metric-affine tidal accelerations in the radial direction as
\begin{equation}
 \left|\frac{a_\text{metric-affine}}{a_{GR}}\right|\simeq\frac{|c_1f_{\mathcal{R}}''|(r_S-2m_\text{tot})r^2_S}{8m_\text{tot}}
 \simeq\frac{c_1\mathcal{C}'r_S}{8}
\end{equation}
where the last approximation is done using the cosmological accelerated expansion value for $\mathcal{R}_0$. For the metric-affine model $f(\mathcal{R})=\mathcal{R}-\frac{\mu^4}{\mathcal{R}}+\alpha \mathcal{R}^2$, whose matter part is modeled by a perfect fluid with polytropic EOS (\ref{eq:poly}), the obtained ratio is large even in the layers below the surface. Therefore, the authors claim that the length scale on which tidal forces diverge due to curvature divergences is much larger than the length scale at which the fluid approximation stops working. Furthermore, it is claimed that abandoning the fluid approximation can make the situation even worse and, hence, they conclude that the appearance of the singularities is not caused by the simplified matter description but rather is a feature of metric-affine gravity.

Similar arguments and conclusions were reached by Pani et al. \cite{pani}  considering the EiBI model. Expanding the right-hand side of the field equations in powers of the parameter $\epsilon$, one finds
\begin{align}\label{eqEi}
 \mathcal{R}_{\mu\nu}=\Lambda g_{\mu\nu}+T_{\mu\nu}-\frac{1}{2}Tg_{\mu\nu}+\epsilon\left[S_{\mu\nu}-\frac{1}{4}Sg_{\mu\nu}\right]
 +\frac{\epsilon}{2}[\nabla_\mu\nabla_\nu\tau-2\nabla^\alpha\nabla_{(\mu}\tau_{\alpha\nu)}+\Box \tau_{\mu\nu}]+\mathcal{O}(\epsilon^2),
\end{align}
where covariant derivatives are defined here with respect to the connection associated to the space-time metric $g_{\mu\nu}$, and we have defined  $\tau_{\mu\nu}\equiv T_{\mu\nu}-\frac{1}{2}g_{\mu\nu}T+\Lambda g_{\mu\nu}$, and $S_{\mu\nu}=T^\alpha_\mu T_{\alpha\nu}-\frac{1}{2}TT_{\mu\nu}$. Eq.(\ref{eqEi}) contains second derivatives of $T_{\mu\nu}$, which were interpreted as if one had to  deal with at least third derivatives of the matter fields, while in GR  only their first derivatives appear\footnote{One should note, however, that using the equations of motion of the matter fields, any higher-order derivatives of the matter appearing in this expansion could be written in terms of the fields and their first derivatives. }. The presence of higher-order derivatives of the matter fields is also directly visible in the Newtonian limit of the theory: the modified Poisson equation takes the form
\begin{equation}\label{eq:newEiBI}
 \nabla^2\Phi=4 \pi \rho+\frac{\epsilon}{4}\nabla^2\rho
\end{equation}
whose solution is
\begin{equation}
 \Phi=\Phi_N+\frac{\epsilon}{4}\rho,
\end{equation}
where $\Phi_N$ is the standard Newtonian potential. This equation shows that the gravitational potential $\Phi$ is algebraically related to the energy density, which may produce discontinuities in the metric and divergences in the curvature scalars (if the function $\rho$ is not sufficiently smooth).

Turning back to the complete EiBI theory, and taking a line element for the auxiliary metric in the interior geometry of the form
\begin{equation}
 ds_q^2 \equiv q_{\mu\nu}dx^\mu dx^\nu=-s(r)dt^2+w(r)dr^2+r^2d\Omega^2, \label{eq:qlineBI} \ ,
\end{equation}
assuming an asymptotically flat spacetime ($\Lambda_{eff}=0$), the curvature scalar $R_g$ can be written in terms of the coefficients $s$ and $w$ in Eq.(\ref{eq:qlineBI}) and the matter fields (using an EOS) as:
\begin{equation}
 R_g=R_g(s,s',s'',w,w',P,P',P'').
\end{equation}
Thus, if the function $P(r)$ is continuous but not differentiable at the stellar surface, the derivatives of $P$ would be discontinuous, introducing Dirac-deltas in the curvature. Evaluating explicitly the behavior of $R_g$ at the stellar surface $r=r_S$, using the polytropic EOS (\ref{eq:poly}), which yields, for any $\epsilon\neq0$, the result
\begin{equation}
 R_g(r_S)\sim
  \begin{cases}
   \gamma_\Gamma,        \quad &0<\Gamma\leq3/2\nonumber\\
   \gamma_\Gamma P^{-2+3/\Gamma},   \quad    & 3/2<\Gamma<2\\
    \gamma_\Gamma P^{-1/\Gamma},   \quad &\Gamma\geq2\nonumber.
  \end{cases}
\end{equation}
where for $0<\Gamma\leq3/2$ one has $R_g=\gamma_\Gamma=0$ but for $\Gamma>3/2$ one finds instead
\begin{equation}
 \gamma_\Gamma=
  \begin{cases}
   \frac{\kappa(2-\Gamma)}{2\eta K^{3/\Gamma}\Gamma^2},        \quad &3/2\leq\Gamma<2\nonumber\\
   \frac{-8\kappa^2}{[8+\kappa/K]^3\eta K^{5/2}},  \quad     &\Gamma=2\\
    \frac{8(1-\Gamma)K^{1/\Gamma}}{\kappa\eta},   \quad &\Gamma>2\nonumber
  \end{cases}
\end{equation}
with $\eta=r^3_S(r_S-2M)/M^2$ and $M$ being the total mass defined as usual by $w(r_S)=[1-2M/r_S]^{-1}$.
Thus, for values of the polytropic parameter $\Gamma>3/2$ the curvature scalar diverges at the surface because of the higher derivatives of the matter fields. This was interpreted again as that the degenerate gas of nonrelativistic electrons and a monatomic isentropic gas ($\Gamma=5/3$) are incompatible with the structure of EiBI gravity.

A further contribution to this issue appears in \cite{sham}. Here the authors write $R_g$ as a function of the speed of sound $c_s^2=dp/d\rho$ and its first derivative and find an anomalous behavior of EiBI stars if phase transitions are allowed. Let us write the TOV equation in EiBI gravity as
\begin{align}
 \frac{dP}{dr}=&-\Theta\left[ \frac{2}{\rho+P}+\frac{\epsilon}{2}\left(\frac{3}{b^2}+\frac{1}{a^2c^2_s} \right) \right]^{-1} \left[1-\frac{2m}{r} \right]^{-1},\\
 \frac{dm}{dr}=&\frac{1}{4\epsilon}\left(2-\frac{3}{ab}+\frac{a}{b^3} \right)r^2
\end{align}
with the functions $a$ and $b$ defined as
\begin{equation}\label{ab}
a\equiv\sqrt{1+8\pi\epsilon\rho},\,\,\,\, b\equiv\sqrt{1-8\pi\epsilon P}
\end{equation}
and the function $\Theta$ as
\begin{equation}
 \Theta\equiv\left[\frac{1}{2\epsilon}\left(\frac{1}{ab}+\frac{a}{b^3}-2\right)r+\frac{2m}{r^2} \right].
\end{equation}
Due to the presence of the sound speed in this formula, its behavior can strongly affect the equilibrium configuration of a compact object. In particular, there are EOS with a first-order phase transition for which there exists an interval such that the pressure remains the same while the energy density increases, leading to $c_s^2=0$. Thus, expanding the TOV equations around $c_s^2\approx0$ up to order $c^2_s$ one finds
\begin{equation}
 \frac{dP}{dr}\approx-\frac{2c_s^2a^2\Theta}{\epsilon}\left[1-\frac{2m}{r}\right]^{-1}
\end{equation}
From this formula it is readily seen that $d\rho/dr=c^{-2}_sdP/dr$ is non-vanishing when $c_s^2\>0$. Therefore, for positive $\epsilon$ one deals with a region whose thickness is proportional to $\epsilon$ in which the energy density is a continuous function instead of discontinuous, as it happens in GR ($\epsilon=0$) at the transition point. This anomalous ``mixed phase" region in which one deals with the constant pressure confirms the results of \cite{vitor}: the existence of a constant pressure shell in compact stars in EiBI gravity allows to consider the formation of pressureless stars. However, in the case of the negative parameter $\epsilon$, both $dP/dr$ and $d\rho/dr$ become positive while $c^2_s\approx0$. Thus, the energy density decreases initially up to the region when a first-order phase transition occurs at $c^2_s=0$ and then rises again, which leads to the fact that the surface of the star cannot be defined (as the energy density will not drop to zero). Consequently, the equilibrium of compact stars which is characterized by EOS with first-order phase transitions does not exist for $\epsilon<0$. The same result is found for soft EOS.

Curvature divergences at boundary layers are also found for the case of $f(\mathcal{R},T)$ gravity by Barrientos and Rubilar \cite{barri} using polytropic EOS (\ref{eq:poly}). This result was obtained for the simplified class of models $f({\mathcal{R}},T)=f_1({\mathcal{R}})+f_2(T)$. The equations differ from those of the $f(\mathcal{R})$ gravity case considered above by an additional term proportional to the product $H(d\rho/dp) \equiv \frac{\partial f(\mathcal{R},T)}{\partial T}(d\rho/dp)$ which could eventually improve the behavior at the surface. Two cases are considered: $H(T=0)=H_0=\text{const}$ and $H(T)\sim T^n,\,\,n\geq1$. In the first one it is possible to find solutions in which the metric and its first derivative are continuous across the surface since the crucial functions which enter the curvature scalar converge at the surface in the range of the polytropic parameter $3/2<\Gamma<2$ which is troublesome in the case of $f(\mathcal{R})$ gravity. The power-law case, however, turns out to be problematic in the same way as before because the term proportional to $H$ vanishes at the surface leaving the equations in the same form as in $f(\mathcal{R})$ gravity for $3/2<\Gamma<2$.

\subsubsection{Shortcomings with stellar surfaces or limitations of the models?} \vspace{0.2cm}

The shortcomings discussed above and the conclusions about the non-viability of $f(\mathcal{R})$ theories and the EiBI model have been contested in the literature by different authors and on different grounds. In \cite{olmo1} the author faces the criticisms of \cite{barausse1, barausse2, barausse3} by arguing that, for the metric-affine quadratic model
\begin{equation} \label{eq:maquadratic}
f(\mathcal{R})=\mathcal{R}+\lambda \mathcal{R}^2
 \end{equation}
and assuming Planck scale corrections in the quadratic contribution, the value of the surface density at which   curvature scalars begin to grow is many orders of magnitude below any physically accessible density scale. In fact, it is shown in \cite{olmo1} that a single electron in the universe would suffice to prevent the development of such anomalies.

This example crudely illustrates the importance of considering more realistic descriptions of stellar atmospheres and scenarios in which the exterior solution is not completely empty in order to avoid  undesired behaviors in metric-affine theories.  In fact, though polytropic models can be excellent descriptions of the matter inside many stars and may provide sufficiently good gross estimates for their mass and radii, stellar atmospheres are certainly not as simple as polytropes \cite{Hubeny2014,Potekhin:2004jr}. Moreover, in order to account for {\it all} the observable features of stars, such as spectral signatures, polarization, or anisotropic temperature distributions, one must take into account the existence of radiation fields, finite temperature, the chemical composition of the ionized gas, electric repulsion, magnetic fields, vacuum polarization, \ldots which requires considerably much more work than dealing with a simple polytrope. In particular, a typical neutron star polytropic model yields an estimated radius of a few kilometers, while its electromagnetic observable features may be entirely determined by the outermost few centimeters. In the Sun, the thickness of the atmosphere is of several hundred kilometers, which is negligible as compared to the solar radius but turns out to be essential to extract indirect information about the processes that take place inside. One should thus not blame a new theory just because a helpful simplification that works in GR and is useful to extract certain information, such as the mass-radius relation, does not fit with the differentiability requirements of the new equations. In $f(\mathcal{R})$, EiBI, and any other RBG, the matter distributions must comply with the differentiability requirements discussed in section \ref{sec:Fyou}. Thus, though the issues with polytropic equations of state might be inconvenient, they should certainly not be regarded as fundamental enough to rule out a whole family of gravity theories. Similar arguments can be used to address the issue of potential phase transitions in the interior of neutron stars \cite{sham}.

Regarding the role of higher-order derivatives of the matter fields in the field equations of metric-affine theories \cite{barausse1, barausse3}, one should bear in mind that, despite appearances, the field equations of RBGs can be put into correspondence with those of GR coupled to nonlinear  (but second-order) matter sources \cite{Afonso:2018bpv,Afonso:2018mxn,Afonso:2018hyj,Delhom:2019zrb}. This shows that there exists a choice of variables in which the relevant equations have the same order as in GR. In addition, there are known solutions of scalar and electrically charged compact objects which do not exhibit any pathologies as they transit from high-density to low-density regions (see the review \cite{BeltranJimenez:2017doy}, section 4.2, and also \cite{Afonso:2019fzv}). This further supports the view that the coupling of metric-affine theories to elementary matter fields is well defined and free of the pathologies associated to specific polytropic models (which are just statistical descriptions and not fundamental ones) in the low-density limit.

To further clarify and detail the above discussion, let us now consider the field equations of $f_{\mathcal{R}}$ in terms of the Einstein frame metric  $h_{\mu\nu}$ as in (\ref{eq_conf}), that is
\begin{equation}
 G_{\mu\nu}(h)=\frac{\kappa^2}{f_{\mathcal{R}}}T_{\mu\nu}-\frac{|\mathcal{R} f_{\mathcal{R}}-f|}{2(f_{\mathcal{R}})^2}h_{\mu\nu}.
\end{equation}
This way, the apparently troublesome higher-derivative terms in (\ref{eq_pal1}) are avoided in the field equations but still impose constraints on the matter sources because  $f_{\mathcal{R}}$ has to be continuous and differentiable up to the second
order, $\partial^2g_{\mu\nu}\sim(\partial^2f_{\mathcal{R}})h_{\mu\nu}\sim(\partial f_{\mathcal{R}})^2h_{\mu\nu}$ to guarantee a smooth transit from one geometry to the other. This also imposes that the energy-momentum trace $T$ must be
differentiable up to this order, too. In practical terms, this implies that a  rough matter profile such as
\begin{equation}
 \rho= \left\{
  \begin{array}{lr}
    \rho_c & \;\text{if}\; r < r_S\\
    0 & \;\text{if}\; r \ge r_S
  \end{array}
\right. \ ,
\end{equation}
which is frequently used in GR for spherically symmetric objects, is not valid on its metric-affine extension because it leads to divergent derivatives at the boundary. For this reason, it needs to be modified in order to smoothly interpolate between the interior $\rho_c$ and the exterior vacuum. It should be stressed (again) that this modelling of matching the internal metric to an external Schwarzschild one at the stellar surface is useful from the point of view of analyzing global properties of the corresponding star, but must be taken with care when extracting conclusions about the inconsistence of the gravitational model. Indeed a realistic modelling of stellar atmospheres requires to take into account many aspects not captured by the polytropic simplification, such as gas pressure and turbulent pressure, ionization equilibrium of atoms, dissociation equilibrium of molecules, overall metallicity, and so on \cite{Hubeny2014,Potekhin:2004jr,Gustafsson:2008jx}.

For convenience, let us rewrite the TOV equation of $f(\mathcal{R}$) following  \cite{olmo1} as
\begin{equation}
 \frac{dP}{dr}=-\frac{P^{(0)}_r}{(1-\alpha)}\frac{2}{\left(1+\sqrt{1+\beta P^{(0)}_r}\right)}
\end{equation}
with the definitions
\begin{align}
P^{(0)}_r&\equiv\frac{(\rho+P)}{r[r-2m(r)]}\left[ m(r)+\left(\kappa^2P-\frac{V}{2}\right)\frac{r^3}{2f_{\mathcal{R}}} \right],\\
\alpha&\equiv\frac{(\rho+P)}{2}\frac{f_{\mathcal{R}}^P}{f_{\mathcal{R}}},\\
\beta&\equiv\frac{3r^2[r-2m(r)]}{2(1-\alpha)^2}\left(\frac{f_{\mathcal{R}}^P}{f_{\mathcal{R}}}\right)^2(\rho+P),\\
 f_{\mathcal{R}}^P&\equiv\frac{\kappa^2f_{\mathcal{R}} \mathcal{R}}{\mathcal{R}f_{\mathcal{R} \mathcal{R}}-f_{\mathcal{R}}}\left(3-\frac{d\rho}{dP}\right) \ .
\end{align}
For the quadratic model (\ref{eq:starom}), assuming its parameter $\lambda$ to be of order of the Planck scale, $l^2_P$,  introduces a density scale $\rho_\lambda=(\kappa^2\lambda)^{-1}\sim2\times10^{92}$\,g/cm$^3$. Near the star's surface, using a polytrope (\ref{eq:poly})
one gets
\begin{align}
 P^{(0)}_r&\simeq\frac{m(r)}{r(r-2m(r))}\left(\frac{P}{K}\right)^\frac{1}{\Gamma},\\
 \alpha&\simeq\mp\frac{1}{\rho_\lambda}\frac{1}{\Gamma K}\left(\frac{P}{K}\right)^\frac{(2-\Gamma)}{\Gamma},\\
 \beta&\simeq\frac{6r^2[r-2m(r)]}{(1-\alpha)^2}\frac{1}{(\Gamma K\rho_\lambda)^2}\left(\frac{P}{K}\right)^\frac{(3-2\Gamma)}{\Gamma}
\end{align}
which is well behaved for $\Gamma<2$ when $P\rightarrow0$. However, the term $f_{\mathcal{R}}^{PP}P^2_r$ which appears in the modified field equations
gives rise to the following contribution
\begin{equation}
 f_{\mathcal{R}}^{PP}P^2_r\simeq\mp\frac{2}{\rho_\lambda}\frac{(1-\Gamma)}{(\Gamma K)^2}\frac{m(r)^2}{r^2(r-2m(r))^2}\left(\frac{P}{K}\right)^\frac{(3-2\Gamma)}{\Gamma}
\end{equation}
which indeed diverges for $\Gamma>3/2$ when $P\rightarrow0$ irrespective of the form of the function $f(\mathcal{R})$, as was already mentioned in \cite{barausse1}. In the analysis of \cite{olmo1} this term is strongly suppressed everywhere by the factor $1/\rho_\lambda\sim10^{-92}$,  while in \cite{barausse1} they consider it to be of order $1/\rho_\lambda\sim10^{-18}$. In the Planck scale case, the problematic term becomes of order unity when the energy density is  of order $\rho_\lambda\sim10^{-210}$g/cm$^3$ which is far below any physically attainable energy density. However, in the case considered in \cite{barausse1}, where the quadratic gravity parameter $\lambda$ in Eq.(\ref{eq:maquadratic}) was chosen to be $\lambda=(0.15\,km)^2$, the density scale at which the undesired terms begin to grow was found to be of order $10^{12}$g/cm$^3$, which is already within the domain of validity of  the polytropic EOS. This observation makes it evident that, instead of ruling out all metric-affine models,  one can establish explicit constraints on the allowed values of the parameter $\lambda$ \cite{olmo1}. For instance, one can assume that the polytropic EOS cannot be trusted below some critical value of energy density $\rho_s\sim10^{-n}$gr/cm$^3$ (as it happens in the region close to the surface, where the density is much lower than in the center of the star). In such a case it can be shown that the following constraint holds
\begin{equation}
 \lambda \ll 10^{4-n/3} \; \text{cm}^2
\end{equation}
giving us the allowed values for the quadratic gravity parameter $\lambda$.

Within EiBI gravity, a different way out of the surface singularities issue has been considered by Kim in \cite{kim}, arguing that the gravitational backreaction on the fluid particles due to curvature divergences is such that the pressure changes in a way that can cure the divergence. This is shown in the case of non-relativistic degenerate Fermi gas which can be approximated by a polytropic EOS with $\Gamma=5/3$, which falls within the troublesome range. The first observation concerns the validity of this approximation, which can only be used when the temperature is smaller than the Fermi energy, while the surface singularity arises in the low density regime. A second criticism is that this EOS is obtained from the motion of the Fermi particles in a flat spacetime and therefore although the energy density can be treated as
a local quantity based on the principle of general covariance, the pressure might turn out not to be local in the high curvature regime. This means that, if any high curvature effects modify the microscopic structure of the fluid, then the covariant assumption for the pressure might not hold. Note, in this sense, that it has been found that RBGs can indeed modify elementary particle interactions by means of new terms generated by the dependence of the metric on the local energy-momentum distributions \cite{Latorre:2017uve}. In regions of divergent curvature, such effects would become very important, possibly leading to modified relations between pressure and density. Accordingly, near the star's surface,
the geodesic deviation equation for $\Gamma=5/3$ and a spherically symmetric line element takes the form
\begin{equation}
 a^1\approx \frac{\epsilon\rho''\frac{F_0}{B_0}}{4}X^1,\;\;\;a^0=a^3=0,
\end{equation}
where $F_0$ and $B_0$ are the values of the
functions at the star's surface $r_S$. The obtained expression takes the form of Hooke's law with frequency:
\begin{equation}
 f=\frac{1}{2\pi}\sqrt{\frac{F(r_S)R_g}{2}}.
\end{equation}
In the case of $3/2<\Gamma<2$ the Ricci scalar $R_g$ is positive and the frequency can be written as
\begin{equation}
 f\approx\frac{\epsilon_0}{4\pi r_S}\left(\frac{M}{r_S}\right)
 \left(\frac{\sqrt{\epsilon}\rho^{3/2-\Gamma}}{K}\right)
\end{equation}
 and hence any two geodesics (e.g. a wall and a particle) along $X^\mu$ will cross each other $f$ times \cite{kim}. Here, $\epsilon_0$ is a dimensionless constant  depending on the species of particle constituting the fluid.
Furthermore, according to the oscillations appearing in the geodesic deviation, the pressure  can be estimated as
\begin{equation}
 P_{gd}\approx fnl P_F,
\end{equation}
where $l$ is the characteristic scale of the oscillations (which will be set to be the star's radius $r_S$) $n$ is the particle
number
density, and $P_F=\hbar(3\pi^2n)^{1/3}$ is the Fermi momentum. Thus, one gets $P_{gd}\propto \rho^{7/6}$ and hence the final expression for the pressure is the sum $P_F+P_{gd}$. When one deals with low densities, the term $P_{gd}$ dominates: thus for the case of the polytropic fluid with $\Gamma=7/6$ the surface singularity issue disappears although it is still present when the curvature becomes very small because of the dominance of the
degenerate Fermi gas EOS. This can be corrected by reconsidering the nature of the scale $l$, which motivates a new EOS of the form:
\begin{equation}
 P=K\rho^\Gamma+\mu\rho^{3/2},\;\;\text{for}\;\;\frac{3}{2}<\Gamma<2,
\end{equation}
where $\mu=\frac{\epsilon_0}{4\pi}\frac{M}{r_S}\sqrt{\epsilon}$.
It should be noticed that $\mu$ decreases with
increasing radius and disappears in the GR limit, $\mu\rightarrow0$. With this new EOS, the curvature scalar at the star's surface is now
\begin{equation}
 R_g\approx\frac{16\pi^2}{9\epsilon^2_0(1-2M/r_S)}\frac{1}{r_S^2},
\end{equation}
being larger than in GR, where $ R_g\sim M/r^3_S$. As the effect is geometric, other polytropic types of matter described by values of $\Gamma$ within this range may be treated in this way.  The situation differs significantly in the case of polytropic EOS within the range $2<\Gamma<3$, because the curvature $R_g$ at the surface becomes negative. Due to this, Hooke's law yields infinite repulsion acting on the particles, which causes that a surface cannot be formed. But it should be noted that this kind of matter can exist in the core of the star while a different one, described by another EOS, may describe the surface.

The issue with polytropic EOS was also considered by Pannia et al. in \cite{pannia}. As already discussed, the generalized TOV equation includes the first and second derivatives of the energy density which, however, are not captured by most tables of EOS used in the literature. While these EOS are enough to study stellar structure in GR, in the context of metric-affine gravity extra information is needed in order to prevent artificial discontinuities and divergences associated to these first and second-order derivatives. The healing proposal of \cite{pannia} is to use some analytic approximation of the EOS (illustrated here with the Sly and FPS)
to obtain their corresponding derivatives. In order to control them during the numerical integration, a new parametrization of the EOS was introduced (PLYT), accounting for the core and the crust regions of the neutron star. This way polytropic relations are connected in a continuous and analytic way, somewhat mimicking the phase-transition between both regions. For instance, using the analytical representation of Ref.\cite{Sly4}:
\begin{align}
 \tilde P&=\frac{a_1+a_2\tilde\rho+a_3\tilde\rho^3}{1+a_4\tilde\rho}f(a_5(\tilde\rho-a_6))
 +(a_7+a_8\tilde\rho)f(a_9(a_{10}-\tilde\rho))\nonumber\\
 &+(a_{11}+a_{12}\tilde\rho)f(a_{13}(a_{14}-\tilde\rho))
 +(a_{15}+a_{16}\tilde\rho)f(a_{17}(a_{18}-\tilde\rho)),
\end{align}
where $\tilde P=\text{log}(P/\text{dyn cm}^{-2})$, $\tilde\rho=\text{log}(\rho\text{g cm}^{-3})$, and $f(x)=1/(e^x+1)$, while the constants $a_i$ are fitted to the tabulated EOS, one may examine the stability in the context of EiBI gravity as done by Sham et al. in \cite{sham1}. In order to check the stability of the star, one needs to obtain the frequencies of radial oscillation modes from the linearized conservation
equation $\nabla^\mu T_{\mu\nu}=0$ as (see the details of the procedure
in Sec. \ref{sec:stGR})
\begin{equation}
 \frac{F}{B}(\rho+P)\omega^2\Xi\chi=\delta P'+\frac{1}{2}(\rho+P)\delta F+\frac{1}{2}(a+c_s^2)F'\delta P,
\end{equation}
where now the metric coefficients depend on both $t$ and $r$ while $\chi\equiv r^2(\rho+P)Q_1\xi$ with $\Xi$ the Lagrangian displacement, and $Q_1$ an expression depending on the metric coefficients and the parameter $\epsilon$ (see the Appendix of \cite{sham1} for details). It can be shown that it takes the form of the eigenvalue equation
\begin{equation}
 \chi''=-W_1\chi-W_2\chi',
\end{equation}
where the functions $W_i$ depend only on the background quantities and the frequency squared $\omega^2$. Since the Lagrangian displacement $\Xi$ must vanish at the center of the star (because of spherical symmetry), one gets
\begin{equation}
 \chi(0)=0
\end{equation}
while at the stellar's surface $\delta P=0$, which provides the boundary condition as $\chi(r_S)=0$. In order to find $\omega^2$, a shooting method can be applied. The results, which are $M-\rho_c$ and $\omega^2-\rho_c$ relations, show that the maximum mass configurations are shifted with respect to GR: negative values of $\epsilon$ decrease the mass while positive ones increase it, which is connected to the strengthening/weakening of the gravitational interactions due to EiBI gravity corrections. Moreover, the behavior of $\omega^2$ is qualitatively similar as in the GR case: the stellar mass increases with the central density until it reaches a maximum while the mode frequency passes through zero at the value of $\rho_c$ which corresponds to the maximum mass configuration. This point is a critical density which informs about dynamical instability, that is, beyond that point the stellar models are unstable with respect to radial perturbations. Because of the fact that for the considered EOS one finds stellar models possessing central densities smaller that the critical values,  compact stars in EiBI gravity are stable in these ranges of $\rho_c$ if the mode frequency is $\omega^2>0$.

It is also worth mentioning another similarity with GR. As shown in \cite{sham1}, for $\epsilon=0.1$ and the APR EOS, the central density is extended to such a point that the condition $P_c \epsilon <1$ (for $\epsilon<0$) does not hold. One observes that there exists a second critical point at higher densities: in the region between it and the point where the condition above is violated one has $dM/d\rho_c>0$ but stellar models there are unstable. Therefore, the criterion $dM/d\rho_c>0$ neither guarantee the star's stability in EiBI gravity nor in GR. Moreover, it turns out that the EiBI parameter $\epsilon$ does not have any influence on the fundamental mode frequency, but the frequencies of higher-order modes are strongly sensitive to it. This could provide another constraint on the value of this parameter via possible detections of higher-order radial oscillation modes. In  \cite{sotani} it was clearly demonstrated how the $f$ and $p_1$ mode frequencies depend on the $\epsilon$ parameter with the values $8\pi \rho_s\epsilon=\pm0.03$. Indeed, the strong dependence of the frequencies on $\epsilon$ offers an avenue to distinguish the theory from GR independently on the EOS used (here Shen and FPS) as the EOS dependence is very weak. Thus, for $\epsilon<0$ ($>0$) one deals with frequencies higher (lower) than in GR. In particular, the $f$ mode oscillations, which are acoustic waves, are the most promising tool to distinguish between both theories since we deal with clearly different expectations on their values in GR and EiBI theories. It was also shown the difficulties in distinguishing the theories using mass-radius relations unless the EOS is better constrained from the other astronomical observations.

Another approach to deal with the shortcomings of metric-affine gravity was discussed in \cite{fatibene}, retaking the argument that polytropic EOS are just an approximation rather than a fundamental description of the behaviour of matter, which encourages the search for a more realistic representation of such surfaces. In this sense, bearing in mind that some metric-affine theories are the simplest representation (besides GR itself) of the Ehlers-Pirani-Schild (EPS) interpretation \cite{eps}, one could match the inner and exterior solutions with respect to the conformally related metric (\ref{con}); in such a case the latter becomes the one responsible for the free fall description. Using Darmois junction conditions \cite{dyer} rather than just matching the metric coefficients allows to avoid the assumption on a coordinate system around the matching surface. Therefore, it is shown that the matching is possible for the conformal metric in a similar way as it is in GR, providing instead a singularity appearing in the conformal transformation making it not differentiable (it is just continuous). They showed that the conformal transformation preserves the polytropic EOS at small pressure, as well as the spherical symmetric ansatz and the form of the energy-momentum tensor. Since one deals with the Einstein-like form of the field equations for the conformal metric, and the matching is done with respect to it, the above mentioned problems on the surface of the stars do not appear in the case of polytropic models.

The same approach was used by Wojnar in \cite{aneta2} to show that in metric-affine gravity one may deal with stable star configurations in a similar manner as it happens in GR. It was indeed proved that for any arbitrary function $f(\mathcal{R})$ in a scalar-tensor representation, where $\phi=f_{\mathcal{R}}$, a particular stellar configuration in equilibrium is described by the generalized TOV equations of the form
 \begin{align}
M(r)&=\int^r_0 4\pi \tilde{r}^2\frac{Q(\tilde{r})}{\phi(\tilde{r})^2}d\tilde{r},\label{masaP}\\
   \frac{d}{dr}\left( \frac{\Pi(r)}{\phi^2(r)}\right)&=-\frac{M}{\tilde{B}r^2}\left(\frac{\Pi+Q}{\phi({r})^2}\right)
   \left(1+4\pi r^3\frac{\Pi}{\phi({r})^2M}\right),\label{pressP}
\end{align}
where $\tilde{B}(r)=\phi^{-1}B(r)$ while
\begin{align}\label{genQUAN}
 Q=\rho+\frac{1}{2\kappa^2}U,\;\;\;\;\Pi=P-\frac{1}{2\kappa^2}U.
\end{align}
It should also be noticed that the radial coordinate $r$ is actually a conformal one (the one of the Einstein frame), so in order to examine stars in metric-affine gravity with the above generalized TOV equations, one should take into account the conformal transformation $r^2\rightarrow\phi(r) r^2$. Moreover, the equilibrium is stable with respect to radial oscillation if the quantity $M$ is a minimum with respect to all variations of $\bar{Q}(r) \equiv Q/\phi^2(r)$. Thus, the constrained extremum must be a minimum and hence one deals with $\delta^2 M>0$. Since the mass function $M$ is expressed as (\ref{masaP}) or, alternatively, as
\begin{equation}\label{nowem2}
M(r)=\int^r_0\left( 4\pi \tilde{r}^2\frac{\rho}{\phi(\tilde{r})^2}(\tilde{r})+\frac{\tilde{r}^2U(\tilde{r})}{4\phi(\tilde{r})^2} \right)d\tilde{r},
\end{equation}
we immediately see the analogue of GR where in the case of stability one deals with positive-defined (for all perturbations) second order terms of $\delta\rho$. Here, we will consider the positive-defined terms which are second order in $\delta\bar{Q}$. This must be so since, should any of the squared frequencies $\omega_n^2$ from all perturbation modes take negative values, then the frequency would become purely imaginary and the time variation of this mode would grow exponentially leading to an unstable system. Therefore we notice again that the stability problem is similar to the one in GR but, apart from the standard examination of the stability conditions with respect to an EOS, in metric-affine gravity one has to first choose a $f(\mathcal{R})$ model since the generalized density $Q$, potential $U$ and the coupling
$\phi$ depend on it, and later find the relation $\mathcal{R}=\mathcal{R}(T)$ which introduces dependencies on all the mentioned quantities on the chosen EOS.

\subsubsection{Junction conditions}\label{sec:junction} \vspace{0.2cm}

Before closing the discussion of this section, we will consider a fundamental aspect related to the matching of interior and exterior solutions across a given surface which, to our knowledge, has not yet been properly addressed in the literature and is essential for this discussion. For this purpose, let us now write the modified field equations (\ref{structural}) and (\ref{con}) with respect to the metric $g_{\mu\nu}$ only (see e.g. \cite{Olmo:2011uz,DeFelice:2010aj, dabrawski} for details of this procedure) as
\begin{align}\label{eq_pal1}
 G_{\mu\nu}(g)=\frac{\kappa^2}{f_{\mathcal{R}}}T_{\mu\nu}-\frac{1}{2}g_{\mu\nu}\left(R-\frac{f}{f_{\mathcal{R}}}\right)+
 \frac{1}{f_{\mathcal{R}}}\left(\nabla_\mu\nabla_\nu-g_{\mu\nu}\Box\right)f_{\mathcal{R}}-
 \frac{3}{2f_{\mathcal{R}}^2}\left( (\nabla_\mu f_{\mathcal{R}})(\nabla_\nu f_{\mathcal{R}}) -
 \frac{1}{2}g_{\mu\nu}(\nabla f_{\mathcal{R}})^2\right),
\end{align}
where covariant derivatives are taken in this case with respect to the Levi-Civita connection of $g_{\mu\nu}$. Given that $\mathcal{R}$ is a function of $T$, the above equations involve terms of the form $\nabla_\mu T \nabla_\nu T\sim \nabla_\mu T \nabla^\mu T$ multiplied by $(f_{\mathcal{R}\mathcal{R}}\partial_T \mathcal{R})^2$, $f_{\mathcal{R}\mathcal{R}\mathcal{R}}(\partial_T \mathcal{R})^2$, and $f_{\mathcal{R}\mathcal{R}}\partial_{T}^2\mathcal{R}$, and also $\nabla_\mu \nabla_\nu T\sim \Box T$ multiplied by $f_{\mathcal{R}\mathcal{R}}\partial_T \mathcal{R}$. Paralleling the analysis of Senovilla \cite{Senovilla:2013vra}, whose notation we borrow, we now explore the implications of matching an interior solution with an exterior solution at a timelike hypersurface $\Sigma$. The Einstein tensor and the energy-momentum tensor distributions can be written as
\begin{eqnarray}
\underline{G}_{\ \mu\nu}&=&G_{\mu\nu}^+\underline{\theta}+G_{\mu\nu}^-(1-\underline{\theta})+\mathcal{G}_{\mu\nu}\underline{\delta}^\Sigma \\
\underline{T}_{\ \mu\nu}&=&T_{\mu\nu}^+\underline{\theta}+T_{\mu\nu}^-(1-\underline{\theta})+\tau_{\mu\nu}\underline{\delta}^\Sigma
\end{eqnarray}
where $\underline{\theta}$ represents a step distribution, $\underline{\delta}^\Sigma$ a Dirac delta distribution, $\tau_{\mu\nu}$ describes the matter sources on $\Sigma$, and $\mathcal{G}_{\mu\nu}=-[K_{\mu\nu}]+\sigma_{\mu\nu}[K^\rho_\rho]$, with $\sigma_{\mu\nu}=g_{\mu\nu}-n_\mu n_\nu$ being the first fundamental form on $\Sigma$, $K_{\mu\nu}={\sigma_\mu}^\alpha {\sigma_\nu}^\beta\nabla_\alpha n_\beta$ the second fundamental form, $[K_{\mu\nu}]=K_{\mu\nu}^+-K_{\mu\nu}^-$, and $n^\mu$ the normal to $\Sigma$. From the above it follows that $\underline{T}=T^+\underline{\theta}+T^-(1-\underline{\theta})+\tau \underline{\delta}^\Sigma$. From this one finds that
\begin{equation}
\underline{\nabla_\mu T}=\nabla_\mu T^+\underline{\theta}+\nabla_\mu T^-(1-\underline{\theta})+\underline{\delta}^\Sigma n_\mu [T]+\nabla_\mu(\tau \underline{\delta}^\Sigma) \ ,
\end{equation}
where $[T]=T^+-T^-$, $\tau$ represents the trace of $\tau_{\mu\nu}$ (the energy-momentum tensor on $\Sigma$), and we have used that $\nabla_\mu \underline{\theta}=n_\mu \underline{\delta}^\Sigma$. Given that products of two $\underline{\delta}^\Sigma$ are not well defined distributions, in order to make sense of the equations (\ref{eq_pal1}) as distributions, one must demand that $[T]=0$ and $\tau=0$.  From this it then follows that
\begin{equation}
\underline{\nabla_\mu \nabla_\nu T}=\nabla_\mu \nabla_\nu T^+\underline{\theta}+\nabla_\mu \nabla_\nu T^-(1-\underline{\theta})+\underline{\delta}^\Sigma n_\mu n_{\nu}n^{\alpha} [\nabla_\alpha T] \ ,
\end{equation}
and after elementary algebra, one finds that the singular part of the field equations can be written as
\begin{equation}\label{eq:GmnSigma}
f_{\mathcal{R}}\mathcal{G}_{\mu\nu}+f_{\mathcal{R}\mathcal{R}}\mathcal{R}_T n^\alpha[\nabla_\alpha T]\sigma_{\mu\nu}=\kappa^2\tau_{\mu\nu} \ ,
\end{equation}
where $f_{\mathcal{R}}, f_{\mathcal{R}\mathcal{R}}$, and $\mathcal{R}_T\equiv \partial_T \mathcal{R}$, must be computed on $\Sigma$ subject to the conditions $[T]=0$ and $\tau=0$.  Note that a discontinuity in $\nabla_\alpha T$ across $\Sigma$ is now allowed and does not spoil the consistency of the distributional equations.
Given that the trace of the above equation must vanish in order to comply with the requirement $\tau=0$, one finds that
\begin{equation}\label{eq:Kconstraint}
[K^\rho_{\rho}]=-\frac{3}{2}\left.\frac{f_{\mathcal{R}\mathcal{R}}\mathcal{R}_T}{f_{\mathcal{R}}}\right|_{\Sigma} n^\alpha[\nabla_\alpha T] \ ,
\end{equation}
which allows to write the equations (\ref{eq:GmnSigma}) as
\begin{equation}\label{eq:GmnSigma2}
-[K_{\mu\nu}]+\frac{1}{3}[K^{\rho}_{\rho}]\sigma_{\mu\nu}=\frac{\kappa^2}{f_{\mathcal{R}}}\tau_{\mu\nu} \ ,
\end{equation}
The constraint (\ref{eq:Kconstraint}) on $[K^\rho_{\rho}]$ does not appear in the case of pure GR and is clearly different from the condition $[K^\rho_{\rho}]=0$ that arises in the metric version of $f(R)$ theories.  Thus, the junction conditions in Palatini $f(R)$ are different from those in GR even though their vacuum equations agree in the bulk.

Our analysis above puts forward that a consistent matching between the interior solution of a star and the vacuum exterior solution in Palatini $f(\mathcal{R})$ requires junction conditions which are non-standard. In particular, consistency requires that the matter at the matching surface has vanishing trace, $\tau=0$, that the trace of the full energy-momentum tensor be continuous, $[T]=0$, while a certain discontinuity in $\nabla_\mu T$ is allowed.  For the analysis of stellar objects, therefore, one should bear in mind these requirements in order to consistently define the matching surface between the interior and the empty exterior of a star in idealized situations. In order to construct specific models, one still needs to look at the conservation equations at the matching surface, but this is an aspect beyond the scope of this review \cite{OlRu}. The relevant point here is that if the above conditions are not satisfied, then the attempted matching would not be consistent with the field equations. In particular, the matching attempted in \cite{barausse1,barausse2,barausse3} involved divergent derivatives of the pressure below the matching surface and zero outside (a divergent $n^\alpha[\nabla_\alpha T]$), thus leading to ill defined equations for the discontinuity in the second fundamental form  $[K_{\mu\nu}]$ on $\Sigma$ (see (\ref{eq:GmnSigma2}) above). This is an unambiguous result that casts serious doubts on the conclusions of those works.

To conclude, it is worth noting that unlike in the metric formulation \cite{Senovilla:2013vra}, the quadratic $f(\mathcal{R})$ model is not an exception to the general rule due to the nonvanishing term $(f_{\mathcal{R}\mathcal{R}}\partial_T \mathcal{R})^2$. As a result, all Palatini $f(\mathcal{R})$ theories without exception must satisfy the above junction conditions. For other Palatini theories, the peculiarities of the junction conditions and the consistent construction of viable models is a challenge that should be explored on a case-by-case basis.

\subsection{Relativistic stars}   \vspace{0.2cm}

In this section we shall describe specific stellar structure models and their predictions for relativistic and non-relativistic stars. We will also compare them to those  obtained in the metric formalism, in GR, and with observational constraints.

\subsubsection{$f(R)$ gravity}\label{sec:PalfR} \vspace{0.2cm}

To begin with, one may look at the $1/R$ model of Carroll et al. \cite{Carroll:2003wy}, whose metric formulation is known to be in conflict with solar system tests \cite{olmo2,al3, rug, barr, all}. There was the hope that the metric-affine version of this model could pass local tests because of the generic recovery of the Einstein field equations when $f(R)$ theories are considered in vacuum \cite{Vollick:2003aw}.  However, the model turns out to have subtle effects in microscopic systems due to strong fluctuations in the function $f_{\mathcal{R}}$ when going from regions of high energy density to regions of low or vanishing density \cite{Olmo:2008ye} (see also \cite{Olmo:2006zu,Flanagan:2003rb,Vollick:2003ic}). Such large variations in  $f_{\mathcal{R}}$ are difficult to perceive if one considers continuum fluid models and does not  look carefully at the transient when the matter distributions transit to vacuum configurations. A detailed analysis, however, shows that the dependence of $f_{\mathcal{R}}$ on $T$ is given by
\begin{equation}\label{eq:1/RPal}
f_{\mathcal{R}}=1+\frac{1}{36\mu^4}\left(\kappa^2T-\sqrt{(\kappa^2T)^2+12\mu^4}\right)^2 \ ,
\end{equation}
where $\mu$ is the model parameter [see (\ref{eq:1/Rmodel})].
From this it follows that for large matter densities, $(\kappa^2T)^2\gg 12\mu^4$, we have $f_{\mathcal{R}}\approx 1$, while in the limit $(\kappa^2T)^2\ll 12\mu^4$ we get $f_{\mathcal{R}}\approx 4/3$, which confirms the existence of order unity variations in the effective Newton's constant $G_{eff}=G/f_{\mathcal{R}}$ when going from inside to outside bodies \cite{Flanagan:2003rb,Olmo:2006zu} (and also around the zeroes of atomic wavefunctions \cite{Olmo:2008ye}, which triggers undesired instabilities).

The stellar structure of this $f({\mathcal{R}})$ theory was studied in  \cite{kain} assuming a pressureless configuration ($T=-\rho$) with constant energy density $\rho$. From Eq.(\ref{struc}), it follows that this scenario yields  the standard Einstein equations with a rescaled gravitational constant $G/f_{\mathcal{R}}(\rho)$ depending only on $\rho$, i.e.:
 \begin{equation}
  G_{\mu\nu}+\Lambda(\rho) g_{\mu\nu}=\frac{\kappa^2}{f_{\mathcal{R}}(\rho)}T_{\mu\nu}
 \end{equation}
and an effective cosmological constant of the form
 \begin{equation}
 \Lambda(\rho)\equiv\frac{1}{2}\left(R_\rho-\frac{f(\rho)}{f_{\mathcal{R}}(\rho)}\right).\end{equation}
In \cite{kain}, the exterior SdS metric with cosmological constant
$\Lambda_0=\sqrt{3}\mu^2/4$, was matched to the above interior solution. In particular, an expression for the exterior mass at the surface $r=r_S$ was found in the form
\begin{equation}\label{mass_palat}
 M=m(r_S)+\frac{\Lambda_\rho-\Lambda_0}{6}r_S^3.
\end{equation}
where the interior mass function $m(r)$ is supposed to be a solution of
\begin{equation}
 G^{\;\nu}_\mu+\delta^{\;\nu}_\mu\Lambda_\rho=\frac{\kappa^2}{f_{\mathcal{R}}(\rho)}\rho u_\mu u^\nu,\;\;\;\rho=\text{const}
\end{equation}
defined as
\begin{equation}\label{mass_palat2}
 m(r)=\int^r_0 dr' \frac{4\pi r'^2\rho}{f_{\mathcal{R}}(\rho)}.
\end{equation}
It was then claimed that (\ref{mass_palat}) with (\ref{mass_palat2}) for $r=r_S$ is not a trivial relation between the exterior mass parameter $M$ and the interior density $\rho$, unlike in GR. However, given that the analysis did not take care of the continuity and differentiability requirements of the matter profiles at the matching surface, the validity of this expression is dubious.

On the other hand, the authors observed that since the local pressure and other thermodynamical properties describing the star are determined by the local energy density, $f_{\mathcal{R}}(\rho)$ should not differ much from one inside a star in order not to change the Solar physics, and this is in fact supported by (\ref{eq:1/RPal}) in the large density region. Assuming that the effective cosmological constant  $\Lambda({\rho})$ is compatible with the estimates for dark energy (which sets the scale $\mu$), the mass shift is then negligible. These results justify the  conclusion of \cite{kain} that this model does not change the stellar physics. A similar conclusion was obtained in the case of the Fermi gas EOS studied by Reijonen \cite{reij}, which in both the non-relativistic $P\sim\rho^\frac{5}{3}$ and
ultra-relativistic $P\sim\rho^\frac{4}{3}$ limits can be approximated by polytropic EOS.

The problem with the above conclusions, however, comes from the fact that the continuum fluid approximation is not valid in scenarios with infrared curvature corrections. In fact, since radiation fields do not contribute to the trace $T$, the variations of $f_{\mathcal{R}}(T)$ will be mainly due to changes in the matter density and, given that the average distance between nuclei (at least in the outer layers of stars) is much larger than their own nuclear size, the function $f_{\mathcal{R}}(\rho)$ will be transitioning between $1$ and $4/3$ from atomic nucleus to atomic nucleus. This represents fluctuations of order unity in $f_{\mathcal{R}}(\rho)$ at microscopic scales which invalidates the picture provided by a continuous fluid approximation in which $f_{\mathcal{R}}(\rho)$ is fixed to unity within the high density region \cite{Avelino:2012qe}. In other words, $\langle f_{\mathcal{R}}(\rho)\rangle\neq f_{\mathcal{R}}(\langle\rho\rangle)$, which prevents the use of the field equations with the macroscopic averaging of the sources.  In particular, for the $1/R$ model, since most of the space is empty except at the locations of ions and electrons, one expects $\langle f_{\mathcal{R}}(\rho)\rangle\approx 4/3$, while for a fluid with average density higher than $\sim \mu^2$, we have  $f_{\mathcal{R}}(\langle\rho\rangle)\approx 1$. Similar conclusions are applicable to any model with infrared corrections, in which the function $f_{\mathcal{R}}(\rho)$ undergoes large fluctuations around a low (cosmic) density scale. For models with ultraviolet corrections, however, the averaging procedure is expected to be valid in general.

The quadratic model (\ref{eq:starom}) was numerically investigated in \cite{pannia} for different values of the parameter $\lambda>0$ with the Sly, FPS and PLY EOS. The focus was on the mass-radius relation assuming a certain small cutoff in the pressure as a way to define the surface of the solutions. Though tabulated EOS usually only provide values for the density and pressure, in order to deal with the differentiability requirements of $f(\mathcal{R})$ theories, interpolating techniques were used to generate smooth functions.  Interestingly, a strong correlation between the mass profiles and the second derivatives of the EOS was found, which becomes even more prominent in the crust-core transition region. Here, the mass parameter decreases with the energy density ($dm/d\rho<0$); the effect is more noticeable with increasing $\lambda$ and does not disappear with the EOS parametrization introduced in \cite{pannia}. Although one obtains lower maximum masses than in GR, the differences are not large enough in order to constrain $\lambda$ from the observations of neutron stars at the $2M_{\odot}$ threshold. It could be possible to constrain $\lambda$ via realistic EOS requiring $dm/d\rho>0$ in the whole interior, but there is no EOS for matter in the high density region so far which is constrained for modified theories of gravity taking into account the relations
between the derivatives of $\rho$ and $P$.

The opposite situation with the mass parameter is found in both models when one applies the EOS corresponding to a relativistic gas of deconfined quarks (\ref{quark}) describing a strange star \cite{panoto}. Using it in the TOV equations and the matching conditions, one deals with the standard GR equations but with the energy density, pressure and bag constant rescaled by a factor $1/f_{\mathcal{R}}(\rho)$. In both models one observes the same shift of the mass and the radius depending on quadratic model parameter as in the $1/\mathcal{R}$ model. However, similarly to the previous result, potentially observable effects would correspond to values of models' parameters which are ruled out by cosmological investigations. Indeed, those scales which are appropriate for cosmology introduce negligible effects for astrophysics of compact stars within this parametrization.

Let us also mention that neutron stars in metric-affine $f(R,T)$ models were recently considered by Bhatti et al. \cite{YousaffRT}, finding the associated TOV equations and some constraints for stability.

\subsubsection{Radiative spherical collapse in $f(\mathcal{R})$ } \vspace{0.2cm}

For the polynomial metric-affine models
\begin{equation} \label{eq:starom}
f(\mathcal{R})=\mathcal{R}+\lambda\mathcal{R}^n \ ,
\end{equation}
the radiating spherical collapse was examined by Sharif and Yousaf \cite{sharif}. The interior of the star is described by the non-static spherical metric \cite{sharif2}
\begin{equation}
 ds^2_-=-A^2(t,r)dt^2+B^2(t,r)dr^2+C^2(t,r)d\Omega^2
\end{equation}
which was matched with the exterior one \cite{olmo2005}
\begin{equation}
 ds_+^2=-\left(1-\frac{2M}{r}-\frac{\mathcal{R}_0r^2}{12}\right)d\nu^2-2d\nu dr+r^2d\theta^2+r^2\text{sin}^2\theta d\phi^2
\end{equation}
where $\nu$ is the retarded time and $M$ is the total fluid's mass. The authors then considered the standard Darmois junction conditions \cite{darmois}
\begin{eqnarray}
 M(t,r)&=&M+\frac{r^3 \mathcal{R}_0}{24},\\
 P_r&=&q+\frac{\mathcal{R}_0}{2\kappa}\left(f_{\mathcal{R}}-\frac{f}{\mathcal{R}_0}\right)-\frac{\mathcal{R}_0r}{12},
\end{eqnarray}
where $P_r$ denotes in this context the radial pressure. Consider now a sphere made of locally anisotropic fluid  including a dissipation effect through a diffusion (heat) approximation, which is described by
the fluid energy-momentum tensor \cite{sharif3}
\begin{align}
 T_{\mu\nu}=(\rho+P_\perp)u_\mu u_\nu+q_\alpha u_\beta+ P_\perp g_{\mu\nu}
 +(P_r-P_\perp)\chi_\mu\chi_\nu+q_\nu u_\mu,
\end{align}
where $q_\mu$ is the heat conducting vector, the radial unit four-vector is normalized as $\chi^\mu=B^{-1}\delta^\mu_1$, and the fluid velocity as $u^\mu=A^{-1}\delta^\mu_0$. One then applies the perturbation technique (see \cite{sharif} for details) in order to examine the dynamical instability of a radiating anisotropic star and hence all matter and metric coefficients are in hydrostatic equilibrium at $t=0$. Accordingly, all metric coefficients and matter quantities are perturbed as $A(t,r)=A_0(r)+\eta T(t)a(r)$ and so on, while the heat vector transforms as $q(t,r)=\eta\bar q(t,r)$  and the structure functions as
\begin{align}
 f(t,r)&=\mathcal{R}_0(r)+\lambda \mathcal{R}^n_0+\eta T(t)e(r)(1+\eta n \mathcal{R}_0^{n-1}),\\
 f_{\mathcal{R}}(t,r)&=1+\eta n\mathcal{R}^{n-1}_0+\eta T(t)e(r)n\lambda(n-1)\mathcal{R}^{n-2}_0,
\end{align}
where $\mathcal{R}_0(r)$ has a quite complex form, presented in \cite{sharif} together with the dynamical equations after applying the perturbed quantities. The solution indicating  unstable collapsing stages has an exponential form with the frequency that depends on the quantities introduced during the perturbation procedure. To study collapse, one chooses an EOS of the form \cite{har}
\begin{equation}
 \bar p_i=\Gamma_1\frac{p_{i0}}{\rho_0+p_{i0}}\bar\rho,
\end{equation}
where $\Gamma_1$ is an adiabiatic index that accounts for fluid stiffness (which tells us how pressure changes with variations in the matter density). The collapse equations are then reduced to describe a collapsing stellar body in both the Newtonian and post-Newtonian approximations. Therefore, for the Newtonian limit one gets an upper bound  on the adiabatic index for an unstable system: if during the collapse the system's stiffness increases making the adiabatic index equal or bigger than the bound, the process will cease collapse allowing the system to move into hydrostatic equilibrium. A similar situation occurs in the case of post-Newtonian limit, however the condition on $\Gamma_1$ is much more complex. In both approximations the matter stiffness parameter depends on the radial profile of pressure, heat flux, energy density, and the $f(\mathcal{R})$ model. Analogously to GR, in metric-affine $f(\mathcal{R})$  gravity one also deals with the contribution of heat dissipation, which decreases the stability of the spherical isotropic and anisotropic stellar object making the collapse to speed up.

\subsubsection{Eddington-inspired Born-Infeld gravity} \vspace{0.2cm}

The astrophysical phenomenology of EiBI gravity was extensively discussed by Beltr\'an et al. in \cite{BeltranJimenez:2017doy}. Here we will incorporate some new elements and further stress some others present there, whose main focus will be again  the masses, radius, and compactness of relativistic stars following the spirit of the present work.

The first interesting difference of EiBI gravity as compared to GR is the fact that in this theory there exist pressureless stars \cite{vitor}, that is, with $P=0$. The solutions have a positive binding
energy and compactness $\mathcal{C} \sim 0.3$ for the dimensionless central density $\epsilon\rho_c\sim200$. Since dark matter particles are described by such an EOS, one deals with self-gravitating objects made of dark matter only, and reaching the typical compactness of neutron stars \cite{vitor, delsate}. In the case of polytropic stars one obtains larger and more massive stars than in GR for $\epsilon>0$, and the binding energy also increases (the situation is reversed for $\epsilon<0$). Moreover, one also distinguishes maxima of the curves $M=M(\rho_b)$, where $\rho_b$ is the central baryonic density. In GR they correspond to marginally stable equilibrium configurations and all solutions after the first maximum happen to be unstable against radial perturbations. Because in the non-relativistic case, namely, when $\epsilon\rho_c \ll 1$, the solutions
are stable (see Sec.\ref{nRel}) for $\epsilon>0$, they are also likely to be in the relativistic case before they reach the maximum: thus, the stability problem seems to be similar to the one posed in GR. Polytropic stars were also considered by Harko et al. \cite{harko} for the case $n=3$, $K=1.23\times10^{15}$cm$^5$/(s$^2$gr$^{\frac{1}{3}}$) and $\rho_c=8\times10^{14}\text{gr/cm}^3$. For those values, in order for the function $b$ in Eq.(\ref{ab}) to be a real one, one immediately gets an upper limit for the dimensionless parameter
$\epsilon_0<7.692$ defined as
\begin{equation}
 \epsilon=\frac{c^2}{8\pi \rho_c}\epsilon_0.
\end{equation}
Within this constraint, a significant difference as compared to GR is found: polytropic stars are more compact in EiBI gravity; indeed their maximum masses may be up to $2.5$ times those of GR. Besides polytropic EOS, stiff $P\sim\rho$, radiation $P\sim\rho/3$ and quark EOS are also considered in \cite{harko}. For the stiff one, the (rescaled) EiBI parameter is constrained as $\epsilon_0<1$, but no remarkable differences in the mass and radius of the star are found; for instance, for $\epsilon_0\leq0.2$ the predictions of GR and EiBI basically coincide, while for the range $\epsilon_0\in(0.3,0.9999)$ the differences are still very small. In the case of  radiation, one finds the constraint $\epsilon_0<3$, and while the radius is almost the same as in GR, the maximum mass increases up to $3.84M_\odot$, which is $\sim 22\%$ larger than the GR value. Finally, for the quark star EOS (\ref{quark}) with $a=1/3$ and $B=100$MeV/fm$^3$, the EiBI mass is $\sim 25\%$ larger than the GR value with almost no change in the radius. The constraint on the EiBI parameter depends on the bag constant $B$ and has the upper limit $\epsilon_0<3.33$.

A neutron star with hyperons in its core was discussed within EiBI gravity by Qauli et al. \cite{qauli} using an extension of the standard realistic mean field model. In order to obtain a neutron star overcoming the $2 M_\odot$ threshold with hyperons, the parameter $\epsilon$ must satisfy $\epsilon\gtrsim4\times10^6\text{m}^2$. Together with the lower bound obtained by Avelino in \cite{avelino} one has the allowed range $4\times10^6\text{m}^2\lesssim\epsilon\lesssim6\times10^6\text{m}^2$ to consistently avoid the hyperon puzzle \cite{qauli}. In addition, it is found that since $\epsilon$ varies the pressure and energy density in the neutron star's core, its compactness also changes without violating the causality constraint. For positive values of the parameter, one finds higher masses and larger radii of the stars and also larger compactness (an effect reversed for $\epsilon<0$). Moreover, the compactness, maximum masses and radii of canonical neutron and quark stars were used in \cite{qauli2} in order to find upper and lower limits of the parameter $\epsilon$ using different EOS. Using the data on compactness \cite{ozel} and the recent maximum masses from gravitational wave constraint analysis \cite{rezz} for neutron stars with hyperons, the available range to be compatible with all these observations becomes  $16 \times 10^6\text{m}^2\leq\epsilon\leq 47 \times 10^6\text{m}^2$. In the case of a quark star with additional scalar Coulomb term a similar analysis sets this range as $9.5  \times 10^6\text{m}^2\leq\epsilon\leq 13 \times 10^6\text{m}^2$ for consistency of this scenario. These ranges differ largely depending on the EOS used: for both types of stars, in the case of stiffer EOS the upper limit of the parameter from the compactness constraint becomes smaller while the lower limit from maximum masses constraint becomes larger. It was also shown that, for larger values of $\epsilon$, neither a neutron nor a quark star cross causality constraints
because of the saturation of their compactness after passing a certain large value of the parameter $\epsilon$.

The model was also studied in the context of higher-dimensional braneworlds by Prasetyo et al. \cite{prasetyo}, assuming that the matter fields live on the brane while the bulk is empty. Using the Einstein-frame representation of the field equations, they use EOS which include hyperons under the form of the BSR23 parameter set of the ERMF model, and playing with the two parameters of the theory, $\epsilon$ and $\lambda$ (the latter associated to the brane tension), they find neutron stars compatible with observational constraints (that is, $M \sim 2.1M_\odot$, $r_S \sim 10$km) for the (cosmologically and astrophysically acceptable) range
\begin{equation}
 0<\epsilon<6\times 10^6\text{m}^2,\;\;\lambda>>1\text{ MeV}^4.
\end{equation}
The TOV equations, as opposed to other higher dimensional descriptions of neutron star's structure, turn out to be dependent on the parameter $\omega$ coming from the anisotropic stress seen by observers living on the brane. This parameter also influences the radius of the star \cite{castro}.
Moreover, it was shown that the braneworld theory introduces additional corrections in apparent pressure and apparent energy density.

It was found in \cite{harko} that it is possible to obtain an exact solution for the specific case $a^2=3b^2$ (see Eq.(\ref{ab})). Making this relation explicit, one immediately gets an EOS of the form
\begin{equation}\label{eosEX}
P=-\frac{1}{3}\rho+\frac{1}{12\pi\epsilon}.
\end{equation}
Since the field equations are greatly simplified for the above case, using in this case the hydrostatic equilibrium equation and the boundary condition
$P(0)=P_c$, one finds the expression
\begin{equation}
P(r)=\frac{1}{\kappa^2\epsilon}\left[\left(\kappa^2\epsilon P_c+\sqrt{3}-1\right)\sqrt{1-\frac{r^2}{3\epsilon}}-\sqrt{3}+1 \right].
\end{equation}
It is now straightforward to express the radius of the star $P(r_S)=0$ as
\begin{equation}
 r_S^2=\frac{6\kappa^2\epsilon^2 P_c(4\pi\epsilon P_c+\sqrt{3}-1)}{(\kappa^2\epsilon P_c+\sqrt{3}-1)^2},
\end{equation}
which is a monotonically decreasing function from $P_c$ to $0$ within the interval $r\in[0,r_S]$. It should be noticed that the size of the star described by this exact solution cannot exceed the maximal size $r_0=\sqrt{3\epsilon}$. Moreover, inverting the relation (\ref{eosEX}) one writes $\rho(r)=-3P(r)+\frac{1}{4\pi\epsilon}$ and because of the requirement on the positivity of the energy density $\rho>0$, the following condition for the central pressure $P_c$ is found:
\begin{equation}
P_c<\frac{1}{12\pi\epsilon}
\end{equation}
which yields a bound on the pressure at the center of the star related to $\epsilon$. In that particular example, one also finds the exact form for the space-time metric $g_{\mu\nu}$, which is given by
\begin{equation}
 ds^2=-\frac{\sqrt{3}s_cb^2(r)}{b_c^4}dt^2+\frac{1}{\sqrt{3}b^2(r)}\left[\frac{dr^2}{\left(1-\frac{r^2}{3\epsilon}\right)}+r^2d\Omega^2\right],
\end{equation}
where $s_c=s(0)$, $b_c=b(0)$, and
\begin{equation}
b^2(r)=\sqrt{3}-(\kappa^2\epsilon P_c+\sqrt{3}-1)\sqrt{1-\frac{r^2}{3\epsilon}}.
\end{equation}

It has been recently noted by Danarianto and Sulaksono \cite{Danarianto:2019mxf}  that overturning/cracking instabilities may occur in an isotropic EiBI stellar structure, using the correspondence between metric-affine theories and GR developed in \cite{Afonso:2018bpv,Afonso:2018mxn,Afonso:2018hyj}. To discuss this point, let us first rewrite the EiBI equations in the auxiliary metric $q_{\mu\nu}$
as \cite{delsate2}
\begin{align}
 \mathcal{G}^\mu_{\;\nu}(q) \equiv \mathcal{R}^\mu_{\;\nu}-\frac{1}{2}\delta^\mu_{\;\nu}\mathcal{R}
 =\kappa^2(\tau T^\mu_{\;\nu}+\mathcal{P}\delta^\mu_{\;\nu})\equiv\kappa^2\mathcal{T}^\mu_{\;\nu},
\end{align}
where $\mathcal{P}=(\tau-1)/(\kappa^2\kappa)-\tau T/2$ is the isotropic pressure and $\mathcal{T}^\mu_{\;\mu}$  is introduced as an effective energy momentum tensor. Assuming a perfect fluid description of a self-gravitating sphere, the effective density $\tilde{\rho}$ and pressure $\tilde P$ are defined as
 \begin{align}
  \mathcal{T}^0_{\;0}&=\frac{-a^2+3b^2-2ab^3}{2\kappa^2\epsilon ab^3}\equiv-\tilde\rho,\\
  \mathcal{T}^i_{\;i}&=\frac{a^2+b^2-2ab^3}{2\kappa^2\epsilon ab^3}\equiv\tilde P
 \end{align}
Assuming the ansatz on the auxiliary metric
(\ref{eq:qlineBI}) of the form $w^2(r)\equiv\big[1-2m_q(r)/r\big]^{-1}$ with $m_q(r)$ being the auxiliary mass, one has that the TOV equations
for auxiliary quantities take the form
\begin{eqnarray}
\frac{d\tilde P}{dr}&=&-\frac{m_q +4\pi\tilde P r^3}{r(r-2m_q)}(\tilde\rho+\tilde P) \label{eq:TOVEiBI1},\\
\frac{d m_q}{dr}&=&\frac{\kappa^2}{2} \tilde\rho r^2 \label{eq:TOVEiBI2},
\end{eqnarray}
where the tilde variables refer to the Einstein frame, while for the mass in the auxiliary frame reads $m_q=\frac{m-\frac{r}{2}(1-ab)}{ab}$. Taylor-expanding both sides of the TOV equations around $\kappa^2 \epsilon \rho \to 0$ and/or $\kappa^2 \epsilon P \to 0$, the authors find the well known form of the TOV equation with the apparent anisotropic pressure factor $\Delta_E$ (see \cite{Danarianto:2019mxf} for the explicit form of this factor)
\begin{equation}
\frac{dP}{dr}=-\frac{ m +4\pi P r^3}{r(r-2 m)}(\rho+ P)-\Delta_E
\end{equation}
while the mass  reads
\begin{equation}
\frac{d m}{dr}=\frac{\gamma}{2}\left[ \rho+\frac{k\gamma}{8}(6P\rho+3P^2-5\rho^2) \right]r^2.
\end{equation}
The numerical analysis of the equations of the isotropic model with the barotropic EOS introduced by Carriere et al. \cite{Carriere:2002bx} shows that for the parameter $ \vert \epsilon \vert \gtrsim 5 \times 10^6$ m$^2$ one needs to go to the 3rd order of $\Delta_E$ in order to consider the convergence rate of the expansion on mass and compactness, finding that for a slight dependence on the EOS, the series is convergent. Having a well behaved expansion allows to examine if cracking/overturning instabilities appear when a self-gravitating body splits or compresses when its equilibrium state is perturbed. Thus, an instability will occur when the perturbed term changes its sign, yielding cracking (overturning) instability when the sign changes from positive (negative) to negative (positive). It was shown that in the case of neutron stars the overturning instability occurs near its center ($r\sim 1-3$)km for $\epsilon>0$, with similar results for different EOS, while for $\epsilon <0$ no instability is present. The bottom line of this analysis is that instabilities appear due to the behaviour of the apparent radial and tangential speed of sound in the region very close to the star's center.

Let us conclude this part of the section by mentioning that the case of slowly rotating stars in EiBI gravity was briefly discussed in \cite{delsate, vitor} using the FPS EOS. Their main finding is that the
moment of inertia strongly depends on the EiBI parameter $\epsilon$ in such a way that for $\epsilon > 0$ the larger  its value the larger the moment of inertia (and the other way around for $\epsilon>0$). \\

\textit{Constraints on compactness and Buchdahl's limit} \\

Interestingly, although EiBI gravity possesses a lot of similarities with GR in the case of the stellar structure properties, as for example in the stability problem discussed already in Sec. \ref{sec:stGR}, there are also important features which differ greatly between both theories. This is the case of Buchdahl's stability bound \cite{Buchdahlpaper} which in GR is an absolute limit which does not depend on the EOS, while it was shown in \cite{feng} to be EOS-dependent in EiBI theory.
In order to show this point, let us write explicitly Eq.(\ref{eq:TOVEiBI1}) as
\begin{equation}\label{etov}
 -\kappa^2\epsilon\left[\frac{(a^2-b^2)(b^2/c^2_s+3a^2)+4a^2b^2}{4a^2b^2(a^2-b^2)}\right]\frac{dP}{dr}=\frac{m_q+4\pi r^3\tilde P}{r(r-2m_q)}
\end{equation}
for the physical pressure $P$. Though this equation does not constrain the sign of $\epsilon$ as far as stellar structure equilibrium is concerned, the authors only explored the case $\epsilon>0$ to avoid potential difficulties in the event of having first-order phase transitions in the stellar interior.  Following \cite{feng}, taking $g_{rr}=(1-2m(r)/r)^{-1}$ and $q_{rr}=(1-2m_q(r)/r)^{-1}$, the relation between frames leads to the identification
\begin{equation}\label{phys}
m(r)=m_q(r)+\frac{1}{2}(1-ab)(r-2m_q).
\end{equation}
An effective density is then associated with the above mass $m$ as $\rho_\text{eff}\equiv m'(r)/4\pi r^2$ or, explicitly:
\begin{equation}
 \rho_\text{eff}=ab\tilde\rho+\frac{1-ab}{\kappa^2 r^2}+\frac{\epsilon}{2r}\left(\frac{a^2c^2_s-b^2}{ab}\right)\frac{d\rho}{dr} \ .
\end{equation}
At this point the authors require the condition $ab=1$ at the star's center in order to keep $\rho_\text{eff}$ finite, though the actual physical meaning of this quantity is far from clear because the quantity that is directly measurable is the $\rho(r)$ that appears in ${T^\mu}_{\;\nu}$.  Moreover, the key point on Buchdahl's stability bound in GR is the monotonically non-increasing character of the energy density $\rho$, but that might not be enough to have the same behavior
for $\tilde\rho(\rho,p)$ and $\rho_\text{eff}(\rho,p)$. Due to this, the following relation was examined:
\begin{equation}
 \frac{d\tilde\rho}{dr}=\left[\frac{3a^2(a^2-b^2)c_s^2+(3b^2+a^2)b^2}{4a^3b^5}\right]\frac{d\rho}{dr}
\end{equation}
whose numerator has to be positive in order to have the non-increasing monotonicity of $\tilde\rho$ inherited from that of $\rho$. For $\epsilon>0$, this is satisfied if  the null energy condition $\rho+P \geq0$ holds, because $a^2-b^2=8\pi\epsilon(\rho+P)$ [see (\ref{ab}) for definitions]. It turns out that the monotonically non-increasing character of $\rho$ does not lead to the same monotonicity of $\rho_{eff}$, but depends on the term appearing in the
derivative $d \rho_{eff}/dr$ as  \cite{feng}
\begin{equation}
 \frac{1}{2r}\left(1-\frac{2m_q}{r}\right)\rho''+\left(\frac{2m_q}{r^3}-\kappa^2\tilde\rho\right)\rho'>0.
\end{equation}
Interpreting $\rho_{eff}$ as an energy density contributing to the effective physical mass $m(r)$ and assuming its non-increasing monotonicity constraints the possible types of EOS in EiBI gravity because of the $\rho$-dependence on the EOS and the modified TOV equation (\ref{etov}). It should be noticed that those three densities become the same in the GR limit $\epsilon\rightarrow0$ but have a completely different nature and status for any finite $\epsilon$. One can show that
\begin{equation}
m(r)\geq\frac{r^3}{r^3_S}m(r_S),
\end{equation}
where we recall that the star surface is defined as the point where $P(r_S)=0$ supplemented with the conditions $P_r=0$ and $P_{rr}=0$ in order to guarantee a smooth matching with the exterior. Together with (\ref{phys}), this gives
\begin{equation}
 m_q(r)+\frac{1}{2}(1-ab)[r-2m_q(r)]\geq\frac{r^3}{r^3_S}m(r_S).
\end{equation}
Moreover, if $\tilde\rho$ is a monotonically non-increasing function, one has two more inequalities
\begin{eqnarray}
m(r)&+&\frac{1}{2ab}(ab-1)[r-2m(r)]\geq\frac{r^3}{r_0^3}m(r_S),\\
 \left(\frac{m_q}{r^3}\right)' &\leq & 0.
\end{eqnarray}
Using all these inequalities it is possible to show that the Buchdahl's stability in EiBI gravity for $ab>1$, after imposing the requirement for $\tilde\rho$ and $\rho_\text{eff}$ to be finite and monotonically non-increasing, together with the functions $s^2(r)$ and $w^2(r)$ of the auxiliary line element (\ref{eq:qlineBI}) being positive
definite, reads
\begin{equation}
 r_S\left(1-\frac{1}{2}g-\frac{1}{2}g^2\right)\geq\frac{9}{4} m(r_S),
\end{equation}
where the function $g$ was defined as
\begin{equation}
 g\equiv\frac{m(r_S)}{r^3_S}\int^{r_S}_0\frac{\sqrt{ab}-1}{\sqrt{1-\frac{2r^2}{r_S^3}m(r_S)}}rdr.
\end{equation}
This function is positive definite if in the interior of the star one has $ab>1$, which means that the lower bound of a stable radius is larger than the $9/4$ bound of GR due to the repulsive effect of EiBI gravity. Computing $g$ by the mean value theorem, by which one has
\begin{equation}
\rho_{eff}(\bar r)=\frac{\int_0^2 4\pi\xi^2\rho_{eff}(0)d\xi}{\int_0^r 4\pi\xi^2d\xi}\equiv\bar\rho_{eff}(r)
\end{equation}
in the interval $\bar r\in(0,r)$, and expanding it in series of $\epsilon$ one finds
\begin{equation}
 \frac{r_S}{2m(r_S)}\geq\frac{9}{8}+\frac{3\pi}{8}\epsilon (\bar\rho-\bar p)+\mathcal{O}(\epsilon^2),
\end{equation}
which can be used to constrain $\epsilon$. In the limit $\epsilon \to 0$ this equation provides the standard Buchdahl limit, $\mathcal{C} \leq 4/9$, as expected. Now,  for a typical density $\bar{\rho}\sim10^{15}\text{gr/cm}^3$, one finds that in order to obtain a  mass $M \sim2M_\odot$ and a   radius $r_S \sim12$km  the parameter is constrained as $\epsilon\lesssim10^9\text{m}^2$. As we have seen, explicit stellar structure models provide more stringent constraints on $\epsilon$. As it happens in GR, considering an anisotropic fluid as the matter source would introduce further changes to the bound presented here, though we will leave this issue here.

\subsubsection{Hybrid gravity} \vspace{0.2cm}

To conclude our section on relativistic stars, let us briefly address hybrid gravity. Stellar structure models are described in this case by the modified TOV equations obtained recently by Danila et al. \cite{hybr} as (here we make explicit Newton's constant)
\begin{align}
 \frac{dm_\text{eff}}{dr}&=-\frac{rf(\Phi)+3\Phi'2}{1+\Phi'r/2}m_\text{eff}+\frac{4\pi r^2}{\kappa^2[1+\Phi'r/2]}
 \left[ 2\frac{\Phi'}{r}+\frac{U}{2}+f(\Phi)+\kappa^2_\text{eff}\rho \right],\\
 \frac{dp}{dr}&=-\frac{(\rho +p)[(\kappa^2 p e^{-\Phi}-U/2)r}{r(1-2G_0m_\text{eff}r)(2+\Phi'r)}
 + \frac{[(\rho +p)(1+r(2+r[2+rh(\Phi)]\Phi'))+1]}{r(2+\Phi'r)}
\end{align}
where a rescaled scalar field is defined by $\phi=e^\Phi-1$, the effective gravitational constant $G_\text{eff}=G_0e^{-\Phi}$ and the potential $V(\phi)=e^\Phi U(\Phi)$. The form of the modified Klein-Gordon equation in the new variables can be found in \cite{hybr}. Four types of EOS are examined: stiff, radiation, quark matter and the Bose-Einstein condensate superfluid, $P\sim\rho^2$ while the potential is of Higgs type: $U(\Phi)=-\mu^2/2\Phi^2+\xi/4\Phi^4$, with $\mu^2$ and $\xi$ being constants. The authors consider the simple hybrid model $f(\mathcal{R})=c_1\mathcal{R}+c_2(c_1,\mu,\xi)$, where the constant $c_1$ must be determined via initial/boundary conditions. In both the stiff and radiation EOS one obtains more massive stars with similar profiles as their GR counterparts with shifted maximum masses: for   stiff EOS one finds $M_\star \approx (3.278 -3.968) M_\odot$, and for  radiation one has $M_\star \approx (2.0278-2.442) M_\odot$. As for the mass profiles of quark stars one finds $M_\star \approx ( 2.025-2.706)M_\odot$, though in this case the scalar field does not vanish at the surface as opposed to the previous two EOS. The same feature is found for Bose-Einstein condensate, where the scalar field shows an involved behaviour inside the star, depending strongly on the Higgs parameters. For this case one finds maximum masses $M_\star \approx (2.003-2.231)M_\odot$. In the last two cases the potential has negative values at the interior of the star and a complex evolution pattern. Moreover, it is suggested that black holes from $3.8 \text{ to }6$ solar masses could actually be hybrid metric-Palatini stars since the model provides much more massive compact stars than those of GR. Distinguishing between them could be possible via the study of thin accretion disks around rapidly rotating stars and Kerr-type black holes  in hybrid gravity because their radiation properties may be substantially different (see the discussion
in Chavanis et al. \cite{chavanis}).

\subsection{Non-relativistic stars}\label{nRel} \vspace{0.2cm}

\subsubsection{Lane-Emden equation in $f(R)$ gravity}  \vspace{0.2cm}

Similarly as it was presented in the section \ref{sec:LEsection} for the metric formalism, one may examine non-relativistic stars using the (generalized) Lane-Emden equation for polytropic EOS. In order to do it, let us bring again our favourite quadratic model (\ref{eq:maquadratic}). In the scalar-tensor formalism the transformation function reads $\phi=1+2\kappa^2 \lambda\rho=1+2\alpha\theta^n$ after applying the dimensionless variables (\ref{eq:redefpoly}), where $\rho_c$ is the star's central density.
As shown in \cite{fatibene}, the generalized energy density and pressure (\ref{genQUAN}) for small values of the pressure $P$ are
\begin{align}
 \bar{Q}&=\bar{\rho}+\frac{\bar{U}}{2\kappa^2}=\frac{4\rho+\lambda\kappa^2\rho^2}{4(1+\lambda\kappa^2\rho)^2}\sim \rho,\\
 \bar{\Pi}&=\bar{P}-\frac{\bar{U}}{2\kappa^2}=\frac{4K\rho^\gamma-\lambda\kappa^2\rho^2}{4(1+\lambda\kappa^2\rho)^2}\sim K\rho^\Gamma
 \end{align}
 and thus the polytropic EOS structure (\ref{eq:poly}) is preserved by the conformal transformation (\ref{confMET}). Applying again the Newtonian regime to the  TOV equations (\ref{masaP}) and (\ref{pressP}) as presented in \cite{aneta3}, the modified Lane-Emden equation in the Einstein frame reads
\begin{equation}
\frac{1}{\bar{\xi}}\frac{d^2}{d\bar{\xi}^2}\left(\left[1+\frac{2\alpha}{n+1}\theta^n\right]\bar{\xi}\theta\right)=-\theta^n
\end{equation}
and hence after applying the conformal transformation $\bar{\xi}^2=\Phi\xi^2$, the equation in Jordan frame becomes
\begin{equation}\label{almost}
\xi^2\theta^n\Phi+\displaystyle\frac{\Phi^{-1/2}}{1+\frac12 \xi\Phi_{\xi}/\Phi}\displaystyle\frac{d}{d\xi}\left(\displaystyle\frac{\xi^2\Phi^{3/2}}{1+\frac12 \xi\Phi_{\xi}/\Phi}\displaystyle\frac{d\theta}{d\xi} \right)=0.
\end{equation}
Up to linear terms in $\alpha$, the equation above takes the form
\begin{equation}\label{em}
 \frac{1}{\xi^2}\frac{d}{d\xi}\left[(1+2\alpha\theta^n)\xi^2\frac{d\theta}{d\xi}+\alpha\xi^3\theta^{2n}\right]=-\theta^n+3\alpha\theta^{2n}
\end{equation}
which for $\alpha=0$ reduces to the GR expression  (\ref{le}). It possesses an exact solution for the case $n=0$ (incompressible stars) extending the GR case (\ref{eq:eta0}) as:
\begin{equation}
 \theta(\xi)=1-\frac{\xi^2}{6(1+2\alpha)}
\end{equation}
This equation implies that, for  positive values of the quadratic model parameter, $\alpha$, one gets smaller radius for the corresponding star, until $\alpha=-\frac{1}{2}$ is reached, beyond which no physical solution is found. For $n=\{1;\frac{3}{2};3\}$ numerical analysis indicates that for negative $\alpha$ one finds larger stars with higher masses. However, the behaviour for positive $\alpha$ depends on the type of star: $n=1$ stars are
significantly smaller and less massive, but $n=3$ stars yield increases of radius but lowering of masses. This means that, for instance, dwarfs stars are less dense in this case. Furthermore, in Sergyeyev et al \cite{artur} the exact solutions of (\ref{almost}) for $n=\{0,1\}$ were found to be
\begin{equation}
 \theta_{n=0}=1-\frac{\xi^2}{6},\;\;\;\;\;\theta_{n=1}=\frac{15-\xi^2}{2\alpha(10+\xi^2)},
\end{equation}
where the last expression was used to compare the masses, radii and central densities of this model with those of GR. It was also shown that the quantities $\omega_n$ and $\delta_n$ are modified according to the conformal transformation, differing from their GR forms. For the quadratic metric-affine $f(\mathcal{R})$ model (\ref{eq:starom}) they are explicitly obtained as
\begin{equation}
 \omega_n=-\frac{\xi^2\Phi^\frac{3}{2}}{1+\frac{1}{2}\xi\frac{\Phi_\xi}{\Phi}}\frac{d\theta}{d\xi}\mid_{\xi=\xi_R},\label{omega} \hspace{0.1cm}\\
 \delta_n=-\frac{\xi_R}{3\frac{\Phi^{-\frac{1}{2}}}{1+\frac{1}{2}\xi\frac{\Phi_\xi}{\Phi}}\frac{d\theta}{d\xi}\mid_{\xi=\xi_R}},
\end{equation}
and were used to show that the mass and radius are $\approx 1.23$ times bigger than in the GR case, while the central densities remain almost the same.

The set of equations above were recently used by Olmo et al. \cite{Olmo:2019qsj} to find the MMSM for the  quadratic model (\ref{eq:maquadratic}) following pretty much the same trail of thought and techniques discussed in Sec. \ref{sec:RIM}. It was indeed found in  \cite{Olmo:2019qsj} that this mass reads
\begin{equation} \label{result}
M_{-1}^{MMSM}=0.290 \frac{\gamma_{3/2}^{1.32} \omega_{3/2}^{0.09}}{\delta_{3/2}^{0.51}} I(\eta,\alpha) \ ,
\end{equation}
where we have introduced the function
\begin{equation} \label{eq:Ifunc}
I(\eta,\alpha)=\frac{(\alpha_d + \eta)^{1.509}}{\eta^{1.325}} \left(1-1.31\alpha\frac{\left(\frac{\alpha_d+\eta}{\eta}\right)^4}{\delta_{3/2}\kappa_{-2} }\right)^{0.111} \ .
\end{equation}
and the constant $\alpha \equiv \kappa^2 \lambda \rho_c$. This introduces a fundamental difference as compared to the discussion of the MMSM in the metric formalism, in that here the star's central density plays a role in the determination of that mass. This is a consequence of the dependence on the local energy density of the matter sources typically ascribed to metric-affine theories of gravity. It was also found in that the work that, should one use the GR values for the  quantities $\delta_{3/2}$, $\omega_{3/2}$ and $\gamma_{3/2}$, the resulting expression would be misleading from a comparison case-to-case (that is, by fixing $\alpha$ first and then computing the value of the MMSM after finding $\delta_{3/2},\omega_{3/2},\gamma_{3/2}$ for that $\alpha$). For $\alpha=0$ one finds the GR value $M_{MMSM}^{GR} \approx0.0922M_\odot$, which is consistent with the mass reported in \cite{Crisostomi:2019yfo}. Departing from GR, it is found that positive (negative) values of $\alpha$ provide larger (smaller) values of the MMSM; for $\alpha=0.010$ one finds $M\approx0.0933M_\odot$ which is near the bound $(0.093\pm0.0008)M_\odot$ corresponding to the lowest mass star ever observed - that of the M-dwarf star G1 866C \cite{Segransan:2000jq} and thus significantly higher values than this one would rule out this quadratic gravity model. On the other hand, the branch $\alpha<0$ is safe since it lowers the MMSM.

\subsubsection{Eddington-inspired Born-Infeld gravity}  \vspace{0.2cm}

Examining the non-relativistic limit of EiBI gravity one finds the modified hydrostatic equilibrium \cite{vitor, delsate}
\begin{equation}\label{newt_mod}
 \frac{dP}{dr}=-\frac{m(r)\rho}{r^2}-\epsilon\rho\frac{\rho'}{4},
\end{equation}
which implies that in the case of constant density profiles there is no correction to the standard Newtonian equations. As already mentioned, the theory supports pressureless stars, which is the case of non-interacting particles. This fact allows to consider stars consisting of self-gravitating exotic dark matter. Together with the standard mass conservation $dm/dr=4\pi r^2\rho(r)$, one finds that Eq.(\ref{newt_mod}) can be solved as
\begin{equation}\label{sol_ei}
 \rho(r)=\rho_c\frac{\sin{\bar{\omega}r}}{\bar{\omega}r},\;\;\text{where}\;\;\bar{\omega}=4\sqrt{\pi/\epsilon} \ .
\end{equation}
Since in the star's interior the Newtonian potential is constant, this allows to match it continuously to the vacuum potential $M/r$ at the radius $r=\pi/\bar{\omega}$. The mass of the star is then $M=4\pi^2\rho_c/\bar{\omega}^3$. Note, however, that though the Newtonian potential is continuous, the matter density profile and its derivatives are abruptly set to vanish at $r\geq \pi/\bar{\omega}$, which could be in conflict with the differentiability requirements of RBGs. Nonetheless, note that the energy density would be continuous and the discontinuity in the derivative of $\rho$ would be finite, which is compatible with the junction conditions of $f(R)$ theories derived in Section \ref{sec:junction}. It seems plausible that similar conditions apply to the EiBI case, making these solutions still worth of interest.

One also can note that the modified Poisson equation (\ref{eq:newEiBI})
also modifies the radial acceleration
\begin{equation}
 a(r)=-\frac{m(r)}{r^2}-\frac{\epsilon}{4}\frac{d\rho}{dr}.
\end{equation}
This correction to the GR formula acts as repulsive force if $\epsilon>0$ because inside the star the term $d\rho/dr$ is negative. Because of that, for $\epsilon>0$ the theory may support more massive stars. Moreover, it can be shown that it corresponds to an effective polytropic fluid with an EOS given by \cite{delsate}.
\begin{equation}
 P_{\text{eff}}=K\rho^2\;\;\text{with}\;\;K=\frac{\epsilon}{8}.
\end{equation}
Using this result, white dwarfs are employed by Banerjee et al. \cite{Baner} to put constraints on the theory's parameter. In agreement with the results of \cite{delsate}, they confirm that for $\epsilon>0$ there is no Chandrasekhar limit  and, moreover, there exists a critical radius $\sim\sqrt{\epsilon}$ below which a white dwarf cannot exist. For  $\epsilon<0$, the mass is lower than the Chandrasekhar limit of Newtonian
gravity and, therefore, there  is a bound on $\epsilon$ depending on the central density above which one does not have a stable white dwarf  supported by a polytrope, namely
\begin{equation}
 \epsilon>-4K\left(1+\frac{1}{n}\right)\rho_c^{\frac{1}{n}-1}.
\end{equation}
The discovery of highly over-luminous supernovae (SNe Ia), raises the suggestion of the existence of super Chandrasekhar white dwarfs as their progenitors, with a mass in the range $2.1-2.8M_\odot$ \cite{wd1,wd2,wd3,wd4,wd5}, offering further avenues to test EiBI gravity (and other modified theories of gravity). Taking as a reference value the maximum of this interval, namely  corresponding to an assumed white dwarf progenitor of SN 2009dc \cite{wd5}, one finds the constraint
\begin{equation}
 \epsilon<0.35\times10^2\text{m}^5\text{kg}^{-1}\text{s}^{-2}
\end{equation}
which is consistent with the most stringent bounds \cite{delsate} coming from neutron stars $\epsilon<10^{-2}\text{m}^5\text{kg}^{-1}\text{s}^{-2}$. Further constraints from the observational data of twelve white dwarfs
\cite{obse} using $\chi^2$ tests results in the bounds (in $\text{m}^5\text{kg}^{-1}\text{s}^{-2}$ units)
\begin{align}
 -0.7&\times10^3<\epsilon<1.66\times10^3\;\;\;\text{up to}\;\;1\sigma\;\text{confidence level};\\
 -1.598&\times10^3<\epsilon<4.858\times10^3\;\;\;\text{up to}\;\;5\sigma \;\text{confidence level}
\end{align}
However, should one also want to explain the suggested existence of super-Chandrasekhar white dwarfs with masses up to $2.8M_{\odot}$ (for a progenitor carbon-oxygen dwarf), then a further constraint can be obtained in the positive $\epsilon$ branch (since it is the one in which the maximum mass of white dwarfs can be larger than in GR) then one finds the constraint $\epsilon<0.35 \times 10^2$ (at $5\sigma$). In addition, another $\chi^2$ analysis of the data of a compilation of brown dwarf masses and radii obtained in \cite{Bayliss2016} yields the constraints \cite{Rosyadi:2019hdb}
\begin{align}
 -1.51&\times10^2<\epsilon<0.81\times10^2\;\;\;\text{up to}\;\;1\sigma\;\text{confidence level};\\
 -1.59&\times10^2<\epsilon<1.16\times10^2\;\;\;\text{up to}\;\;5\sigma \;\text{confidence level}
\end{align}

On the other hand, the Lane-Emden equation for EiBI gravity was obtained in Wibisono et al. \cite{wibi} as
\begin{equation}
 \frac{1}{\xi^2}\frac{d}{d\xi}\left[\xi^2\theta^{\frac{n-1}{1-2n}}\frac{d\theta}{d\xi}\left(
 \theta^{\frac{1-n}{1-2n}}+\frac{\epsilon n}{16\pi r^2_c}\right)\right]=-\theta^n
\end{equation}
which also depends on $\rho_c$ as is the case of metric-affine quadratic $f(\mathcal{R})$ gravity (\ref{almost}). A stellar solution exists if the following constraint is
satisfied \cite{delsate}:
\begin{equation}
 \epsilon >-|\epsilon_c|=-4K\Gamma\rho_c^{\Gamma-2},
\end{equation}
For instance, taking an energy density at the center $\rho_c=8\times10^{14}$g/cm$^{-3}$ one deals with $\epsilon_c=4.53\times10^8\text{m}^2$. For $\Gamma=\frac{4}{3}$
negative (positive) values
of the theory parameter make the dimensionless radius $\xi$ larger (smaller) than the one of  GR. Moreover, it is also found that for fixed mass and radius of the star the central density depends on the EiBI parameter  $\epsilon$ as
\begin{equation}
 \rho_c(\epsilon)=-\frac{\xi_{r_S}}{3\omega_n(\epsilon)}\frac{3M}{4r^3_S}
\end{equation}
which decreases when $\epsilon$ grows. Under the assumption that the real fundamental mode of a radial perturbation is equivalent to $dM/d\rho_c=0$,
one finds
\begin{equation}\label{stab_bound}
 \frac{dM}{d\rho_c}\sim \left[\frac{3}{2}\Gamma_c-2+\frac{\rho_c}{\omega_\gamma(\epsilon)}\frac{d\omega_\gamma(\epsilon)}{d\rho_c}\right]=0
\end{equation}
which leads to the critical value of the polytropic index $\Gamma_c=4/3$ in the GR case, that is, when the last term in the above equation is zero. In the EiBI case, the stability bound of polytropic stars is shifted according to the free parameter governing the last term in (\ref{stab_bound}). Together with the analysis of the binding energy it was shown that for $\epsilon>0$ there is a wider range of the $\Gamma$ parameter for which the stars are stable. Moreover, obtaining the configuration entropy profile they find that the polytropic stars are stable for all positive values of $\epsilon\geq|\epsilon_c|$, and that there is no Chandrasekhar limit in such a case. In the case of $\epsilon\leq-0.1$ polytropes turn out to be unstable.

Unlike the pressureless model, which is stable (radial oscillatory frequencies $\omega^2>0$) in both the non-relativistic and relativistic cases,  the stability of  polytropes depends on the sign of $\epsilon$. This can be seen by studying the modified eigenvalue equation derived from (\ref{newt_mod}) after the assumption that the fields are time-dependent $\sim e^{i\omega t}$ \cite{vitor}:
\begin{equation}
 \frac{4\Xi P'}{r}+\frac{\kappa\rho}{4}\left[\frac{2}{r}\Xi\rho'-\Xi'\rho'-\left(\frac{\rho}{r^2}(r^2\Xi)'\right)'\right]-
 \left[\frac{\gamma P}{r^2}(r^2\Xi)'\right]'=\rho\Xi\omega^2,
\end{equation}
where $\gamma$ is the adiabatic index of the perturbation while $\Xi$ the Lagrangian displacement. The marginal case in Newtonian gravity for any $n$, that is, the polytropic models with $\Gamma=4/3$, turn out to be stable for $\epsilon>0$ and unstable otherwise. Thus, independently of $\Gamma$, positive values of the EiBI parameter $\epsilon$ stabilize the model solutions.

There exists yet another interesting feature in the Newtonian limit of EiBI gravity for positive parameter $\epsilon$  \cite{vitor}:
the only
modification to the standard Eulerian equations, which govern the fluid dynamics, appears in the one for the fluid velocity $u(\mathbf{x})$, where $\mathbf{x}=(t,r)$, as
\begin{equation}
 \partial_tu(\mathbf{x})+u(\mathbf{x})\partial_ru(\mathbf{x})=-\frac{M(\mathbf{x})}{r^2}-\epsilon\frac{\partial_r\rho(\mathbf{x})}{4}.
\end{equation}
Close to the center the approximated fields can be written as
\begin{align}
 \rho(\mathbf{x})=&\rho_0(t)+\rho_1(t)r+\rho_2(t)r^2+\mathcal{O}(r^3),\\
 u(\mathbf{x})=&u_0(t)+u_1(t)r+u_2(t)r^2+\mathcal{O}(r^3)
\end{align}
indicating the gravitational collapse when they diverge at some $t$. Thus, with $\rho_1(t)=u_0(t)=u_2(t)=0$ and $\rho_2(t)=\eta\rho_0^{5/3}(t)$ from the EiBI field equations, one
deals with ($\eta<0$ is a constant)
\begin{equation}
 t(\rho_o)-t(\rho_i)=\int^{\rho_0}_{\rho_i}\frac{dxx^{-\frac{4}{3}}}{\sqrt{24\pi}}\sqrt{x^\frac{1}{3}-\rho_i^\frac{1}{3}+\frac{\epsilon\eta}{8\pi}(x-\rho_i)}.
\end{equation}
As the collapse may take place when $t(\rho_0\rightarrow\infty)\ge t(\rho_i)$, this is only possible if $\epsilon\le0$; otherwise the collapse does not occur.

This concludes our analysis of stellar structure models in the metric-affine formalism.

\section{Conclusion and perspectives}

In this review we have provided a broad and systematic analysis of stellar structure models within the context of modified theories of gravity. The main goal behind this work was to identify the most relevant modifications in the astrophysical predictions of such stellar models as compared to those of GR, which may be used as observational discriminators on the viability of any such theories, in particular, when combined with other astrophysical and/or cosmological tests.

We started by splitting our analysis into models formulated in the metric approach, where the affine connection is the Levi-Civita one of the metric, and the metric-affine formulation, where the affine connection is an independent field, and devoted independent sections to each of them. In the metric approach we began our considerations with the case of $f(R)$ theories, which are one of the most thoroughly studied models in the literature, and discussed several of their features, including the scalar-field representation of their field equations, the troubles with identifying the correct stellar mass due to the contribution of the scalar field outside of the matter sources, and the observational viability of these theories in the Post-Newtonian regime regarding solar system constraints, which have been the subject of many investigations. Next, we focused our attention upon non-relativistic stars, which are governed by the Newtonian limit of the Tolman-Oppenheimer-Volkoff equation, namely, the (generalized) Lane-Emden equation. At this step we introduced other families of theories that are of great interest in the literature, such as the Horndeski family and its beyond Horndeski and DHOST extensions, and discussed their phenomenology for white and brown dwarfs, with particular emphasis on Chandrasekhar's mass for white dwarfs, and the minimum mass required for stable hydrogen burning in high-mass brown dwarfs. Subsequently, the predictions of such families of theories and many others (with particular interest on those with a strong theoretical support and consistency), were discussed in detail within the context of relativistic stars, where the degeneracy of the astrophysical predictions with the equation of state for nuclear matter at supranuclear densities within neutron stars was the pivotal point of many of our considerations. Indeed, for each such model we described their associated phenomenology derived from their generalized TOV equations on each case, whose main astrophysical outcome is the mass-radius relation and the maximum allowed mass for a given EOS in the case of static stars, and the moment of inertia for slowly and fast rotating stars, both of which are observationally accessible quantities. Other aspects such as stability, oscillations, etc, were also addressed. Finally, we also included a short discussion on binary neutron star mergers and universal relations, with references to more detailed literature on the subject.

Next, we considered stellar structure models in modified theories of gravity in the metric-affine formulation, where the research is comparatively more scarce. First we highlighted the fact that the independent connection (when matter is minimally coupled to the action),
 for $f(\mathcal{R})$ theories, can always be solved in favour of the matter sources (and possibly the metric itself),  and extended this result to any generalization constructed with scalars out of contractions of the (symmetric part of) Ricci tensor with the metric (Ricci-Based Gravities). Moreover, we showed that the corresponding field equations can always be cast in Einstein-like form with all the matter-dependent contributions on the right-hand-side. This property guarantees that the  vacuum solutions of these theories reduce to those of GR, which makes them consistent with gravitational wave observations, but it  does not automatically ensure their consistency with local tests of gravity. The reason lies on the fact that in models with infrared corrections the averaged equations of motion are not equivalent to the equations of motion considering averaged matter distributions. Such models are generically inconsistent with microscopic systems while models with higher curvature corrections are compatible with weak-field tests.  Then we discussed some of the main results regarding compact stars within these theories. The difficulty of defining a consistent stellar surface was highlighted, together with different viewpoints in this respect. The idea that polytropic models can be used to rule out {\it all} these theories because they lead to curvature divergences at the matching surface with the Schwarzschild solution has been contested by several authors and on different grounds. A direct analysis of the junction conditions in the particular case of $f(\mathcal{R})$ theories introduces an unexpected turn in this problem. In fact, though the vacuum solutions for these theories coincide with those of GR, the junction conditions at the stellar surface are different. Consistent stellar models with an exterior Schwarzschild geometry, therefore, must satisfy non-standard junction conditions, which opens new research avenues in this field. Finally, paralleling the analysis of the metric formalism, for relativistic stars we discussed the main results regarding mass-radius relations and the associated maximum masses for the cases of $f(\mathcal{R})$, EiBI gravity and some hybrid models, while for non-relativistic stars our main focus was the description of some of the results found in the literature for white dwarfs and their modifications on Chandrasekhar's mass, as well as some stability issues of the corresponding stars.

Resulting from the analysis above, there are several main results for stellar structure models of modified theories of gravity that can be highlighted. For relativistic stars, typically these theories shift the mass-radius curves of GR to either higher masses and larger radii or the opposite depending on the sign of their additional parameters, with some theories of gravity in combination with specific EOS having both kinds of effects depending on the range of values chosen for the model parameters, and with a few exceptions to this general rule. To discuss the viability of the corresponding predictions of these theories we took as a reference value the $2M_{\odot}$ threshold resulting from the most massive neutron stars observed so far, which became a cornerstone in guiding us on the compatibility of modified theories of gravity with observational data. Unfortunately, as the well known uncertainty of the EOS of strongly-interacting matter at supranuclear densities already present in GR is intertwined in this context with the new gravitational parameter(s) coming from the modified gravity side, most of the reported analysis here can only hope to determine the space of parameters of each gravitational theory which, at a given EOS, can be made compatible with such a threshold. In general, most increases on such a maximum mass are meager for many such theories, up to current available theoretical analysis and within current observational bounds for the allowed strength of the parameters of each modified theory of gravity. Indeed, in many such cases, the current observational probes introduce experimental errors in the determination of neutron star masses and radii that are frequently of the same order or larger than the increases  produced by the modified gravity side, therefore questioning the usefulness of the latter. There are some exceptions of theories and theoretical/numerical modelings to this rule, where quite significant increases of the maximum mass above the $2M_{\odot}$ threshold are found using realistic EOS. This, in turn, may bring back to life EOS that were already ruled out in the context of GR due to their inability to reach to this mass threshold. These increases may have an impact on our interpretation of astrophysical observations; indeed, as the available energy for release out of gravitational wave in binary mergers can be (much) bigger than in GR, this can alter the conclusions by the LIGO/VIRGO Collaboration on the nature of the progenitor compact objects (either black holes or neutron stars), where GR is assumed. Further hopes of observational discriminators for these theories can be placed on the modifications to the moment of inertia, where the deviations with respect to GR predictions tend to be much more significant than those of the masses/radii. Indeed, we reported several works with slowly rotating stars where significant increases $\sim 40 \%$ in the moment of inertia are found, while the mass is only increased $\sim 10 \%$, and the consideration of fast rotating models may further enhace this result. This opens a window to more relevant observational predictions of these theories by going to the fully rotating scenario, and the comparison of their predictions with future probes of moment of inertia and universal  relations.

Though relativistic stars have been widely explored in the literature as they represent the astrophysical scenarios where the strongest curvatures and largest densities can be reached at the neutron star's center, non-relativistic stars offer a much less explored scenario, which nonetheless has a great potential as a complementary test of modified gravity. This is so because the astrophysical structure of such stars can be  well modelled analytically (up to some limitations) using polytropic EOS, and are much more easily constrained using known nuclear physics phenomenology in combination with astrophysical data. Indeed, their weaker dependence on unknown non-gravitational physics makes them particularly suitable to test the effects of additional gravitational parameters without the contaminations from strong uncertainties on the EOS. In this sense, there are two robust predictions of GR that are modified by many such gravitational extensions of GR (in both metric and metric-affine formalisms), namely, the  Chandrasekhar's $1.4 M_{\odot}$ limit for white dwarfs, and the $\sim 0.08-0.09 M_{\odot}$ limit for minimum stable hydrogen burning mass for high-mass brown dwarfs. Indeed, when the gravitational force inside matter sources is weakened due to the new gravitational contributions, such masses typically increase, thus allowing the corresponding gravitational parameters to be constrained by present and future observations of the masses of a larger pool of such two types of stars. Combining for the same theory its predictions for non-relativistic and relativistic  stars seems a suitable scenario to place more stringent constraints upon the parameter(s) of the theory at hand. This is well illustrated here by the current results on some classes of Horndeski theories of gravity with a single parameter $\Upsilon$: the combination of several such tests has narrowed its range down to $-0.22<\Upsilon<0.027$\footnote{This can be combined with other kinds of tests, for instance, those of galaxy clusters profiles \cite{Sakstein:2016ggl}.}. This is a promising path that deserves to be explored in other well motivated gravitational candidates.

To conclude,  despite being comparatively much less explored than the cosmological settings \cite{Koyama:2015vza,Frusciante}, and perhaps not currently attracting as much attention as the fast-growing field of gravitational wave astronomy out of rotating black hole mergers \cite{Barack:2018yly}, models of stellar structure offer a suitable, yet to be fully exploited, window to test the dynamics of modified theories of gravity and their consistency with astrophysical observations of both relativistic and non-relativistic stars. As the corresponding probes in this context grow more numerous and their experimental precision improves in the future, these stellar models may help us to enlighten the observational viability of modified theories of gravity, particularly when combined with other tests. We hope to have provided the reader with a quick taste on the lessons and challenges offered by these scenarios, and that he/she has profited from the discussions presented here.

\section*{Acknowledgments}
\addcontentsline{toc}{section}{Acknowledgments}

We are indebted to Victor I. Afonso, Jos\'e Beltr\'an-Jim\'enez, Francisco Cabral, Joaqu\'in D\'iaz-Alonso, Francisco S. N. Lobo, Sergei Odintsov, Emanuele Orazi, Diego S\'aez-G\'omez, Jeremy Sakstein, and Hermano Velten for providing us with feedback and clarifications related to the contents of this work at different levels. In particular, we are deeply thankful to Adri\'a Delhom for his effort in proofreading this entire work and still keeping friends with us.
GJO is funded by the Ramon y Cajal contract RYC-2013-13019 (Spain).
DRG is funded by the \emph{Atracci\'on de Talento Investigador} programme of the Comunidad de Madrid (Spain) No. 2018-T1/TIC-10431, and acknowledges further support from the Ministerio de Ciencia, Innovaci\'on y Universidades (Spain) project No. PID2019-108485GB-I00/AEI/10.13039/501100011033, and by the Funda\c{c}\~ao para a Ci\^encia e
a Tecnologia (FCT, Portugal) research projects Nos. PTDC/FIS-OUT/29048/2017 and PTDC/FIS-PAR/31938/2017. 
AW is supported by the European Union through the ERDF CoE grant TK133 and by FAPES (Brazil).
DRG and AW thank the Department of Physics and IFIC of the University of Valencia for their hospitality during different stages of the elaboration of this work.
This work is supported by the Spanish project  FIS2017-84440-C2-1-P (MINECO/FEDER, EU), the project H2020-MSCA-RISE-2017 Grant FunFiCO-777740, the project SEJI/2017/042 (Generalitat Valenciana), the Consolider Program CPANPHY-1205388, the Severo Ochoa grant SEV-2014-0398 (Spain)  and the Edital 006/2018 PRONEX (FAPESQ-PB/CNPQ, Brazil) Grant No. 0015/2019.
This article is based upon work from COST Actions CA15117 (\emph{Cosmology and Astrophysics Network for Theoretical Advances and Training Actions}) and CA18108 (\emph{Quantum gravity phenomenology in the multi-messenger approach}), supported by COST (European Cooperation in Science and Technology).

\addcontentsline{toc}{section}{References}

\bibliographystyle{elsarticle-num}

\begin{thebibliography}{99}



  \bibitem{Will:2014kxa}
  C.~M.~Will,
  The Confrontation between General Relativity and Experiment,
  Living Rev.\ Rel.\  {\bf 17} (2014) 4
  \href{https://arxiv.org/pdf/1403.7377.pdf}{[arXiv:1403.7377 [gr-qc]]}.

  \bibitem{Bull:2015stt}
  P.~Bull {\it et al.},
  Beyond $\Lambda$CDM: Problems, solutions, and the road ahead,
  Phys.\ Dark Univ.\  {\bf 12} (2016) 56
  \href{https://arxiv.org/pdf/1512.05356.pdf}{[arXiv:1512.05356 [astro-ph.CO]]}.

  \bibitem{Barack:2018yly}
  L.~Barack {\it et al.},
  Black holes, gravitational waves and fundamental physics: a roadmap, Class. Quant. Grav. \textbf{36} (2019) 143001
  \href{https://arxiv.org/pdf/1806.05195.pdf}{[arXiv:1806.05195 [gr-qc]]}.

  \bibitem{Akiyama:2019eap}
  K.~Akiyama {\it et al.} [Event Horizon Telescope Collaboration],
  First M87 Event Horizon Telescope Results. VI. The Shadow and Mass of the Central Black Hole,
  Astrophys.\ J.\  {\bf 875} (2019) L6
  \href{https://arxiv.org/ftp/arxiv/papers/1906/1906.11243.pdf}{[arXiv:1906.11243 [astro-ph.GA]]}.

  \bibitem{Burgess:2003jk}
  C.~P.~Burgess,
  Quantum gravity in everyday life: General relativity as an effective field theory,
  Living Rev.\ Rel.\  {\bf 7} (2004) 5
  \href{https://arxiv.org/pdf/gr-qc/0311082.pdf}{[gr-qc/0311082]}.

  \bibitem{Senovilla:2014gza}
  J.~M.~M.~Senovilla and D.~Garfinkle,
  The 1965 Penrose singularity theorem,
  Class.\ Quant.\ Grav.\  {\bf 32} (2015)  124008
  \href{https://arxiv.org/pdf/1410.5226.pdf}{[arXiv:1410.5226 [gr-qc]]}.


\bibitem{TheLIGOScientific:2017qsa}
  B.~P.~Abbott {\it et al.} [LIGO Scientific and Virgo Collaborations],
  GW170817: Observation of Gravitational Waves from a Binary Neutron Star Inspiral,
  Phys.\ Rev.\ Lett.\  {\bf 119} (2017)  161101
  \href{https://arxiv.org/ftp/arxiv/papers/1710/1710.05832.pdf}{[arXiv:1710.05832 [gr-qc]]}.

  \bibitem{Monitor:2017mdv}
  B.~P.~Abbott {\it et al.} [LIGO Scientific and Virgo and Fermi-GBM and INTEGRAL Collaborations],
  Gravitational Waves and Gamma-rays from a Binary Neutron Star Merger: GW170817 and GRB 170817A,
  Astrophys.\ J.\  {\bf 848} (2017) L13
  \href{https://arxiv.org/ftp/arxiv/papers/1710/1710.05834.pdf}{[arXiv:1710.05834 [astro-ph.HE]]}.

  \bibitem{Ezquiaga:2017ekz}
  J.~M.~Ezquiaga and M.~Zumalacárregui,
  Dark Energy After GW170817: Dead Ends and the Road Ahead,
  Phys.\ Rev.\ Lett.\  {\bf 119} (2017) 251304
  \href{https://arxiv.org/pdf/1710.05901.pdf}{[arXiv:1710.05901 [astro-ph.CO]]}.

     \bibitem{Cottam:2007cd}
  J.~Cottam, F.~Paerels, M.~Mendez, L.~Boirin, W.~H.~G.~Lewin, E.~Kuulkers and J.~M.~Miller,
  The Burst Spectra of EXO 0748-676 during a Long 2003 XMM-Newton Observation,
  Astrophys.\ J.\  {\bf 672} (2008) 504
  \href{https://arxiv.org/pdf/0709.4062.pdf}{[arXiv:0709.4062 [astro-ph]]}.

  \bibitem{Schaffner:1995th}
  J.~Schaffner and I.~N.~Mishustin,
  Hyperon rich matter in neutron stars,
  Phys.\ Rev.\ C {\bf 53} (1996) 1416
  \href{https://arxiv.org/pdf/nucl-th/9506011.pdf}{[nucl-th/9506011]}.

\bibitem{Alford:2004pf}
  M.~Alford, M.~Braby, M.~W.~Paris and S.~Reddy,
  Hybrid stars that masquerade as neutron stars,
  Astrophys.\ J.\  {\bf 629} (2005) 969
  \href{https://arxiv.org/pdf/nucl-th/0411016.pdf}{[nucl-th/0411016]}.

     \bibitem{Baym:2017whm}
  G.~Baym, T.~Hatsuda, T.~Kojo, P.~D.~Powell, Y.~Song and T.~Takatsuka,
  From hadrons to quarks in neutron stars: a review,
  Rept.\ Prog.\ Phys.\  {\bf 81} (2018) 056902
  \href{https://arxiv.org/pdf/1707.04966.pdf}{[arXiv:1707.04966 [astro-ph.HE]]}.

\bibitem{Alcock:1986hz}
  C.~Alcock, E.~Farhi and A.~Olinto,
  Strange stars,
  Astrophys.\ J.\  {\bf 310} (1986) 261.

  \bibitem{Jetzer:1991jr}
  P.~Jetzer,
  Boson stars,
  Phys.\ Rept.\  {\bf 220} (1992) 163.

  \bibitem{STBook}
S. L. Shapiro and S. A. Teukolsky, \textit{Black Holes, White Dwarfs and Neutron Stars:
The Physics of Compact Objects} (John Wiley, New York 1983).

\bibitem{Lombardo:2000ec}
  U.~Lombardo and H.~J.~Schulze,
  Superfluidity in neutron star matter,
  Lect.\ Notes Phys.\  {\bf 578} (2001) 30
  \href{https://arxiv.org/pdf/astro-ph/0012209.pdf}{[astro-ph/0012209]}.


  \bibitem{Lattimer:2004pg}
  J.~M.~Lattimer and M.~Prakash,
  The physics of neutron stars,
  Science {\bf 304} (2004) 536
  \href{https://arxiv.org/pdf/astro-ph/0405262.pdf}{[astro-ph/0405262]}.


\bibitem{Chatterjee:2015pua}
  D.~Chatterjee and I.~Vidaña,
  Do hyperons exist in the interior of neutron stars?,
  Eur.\ Phys.\ J.\ A {\bf 52} (2016)  29
 \href{https://arxiv.org/pdf/1510.06306.pdf}{[arXiv:1510.06306 [nucl-th]]}.

\bibitem{Glendenning:2001pe}
  N.~K.~Glendenning,
  Phase transitions and crystalline structures in neutron star cores,'
  Phys.\ Rept.\  {\bf 342} (2001) 393.

\bibitem{Chatziioannou:2019yko}
  K.~Chatziioannou and S.~Han,
  Studying strong phase transitions in neutron stars with gravitational waves, Phys. Rev. D \textbf{101} (2020)  044019
  \href{https://arxiv.org/pdf/1911.07091.pdf}{arXiv:1911.07091 [gr-qc]}.
  
  \bibitem{Bauswein:2018bma}
A.~Bauswein, N.~U.~F.~Bastian, D.~B.~Blaschke, K.~Chatziioannou, J.~A.~Clark, T.~Fischer and M.~Oertel,
Identifying a first-order phase transition in neutron star mergers through gravitational waves,
Phys. Rev. Lett. \textbf{122} (2019)061102
\href{https://arxiv.org/pdf/1809.01116.pdf}{[arXiv:1809.01116 [astro-ph.HE]]}.

\bibitem{massNS}
F. Ozel and P. Freire,
  Masses, Radii, and the Equation of State of Neutron Stars,
  Ann.\ Rev.\ Astron.\ Astrophys.\  {\bf 54} (2016) 401
    \href{https://arxiv.org/pdf/1603.02698.pdf}{[arXiv:1603.02698 [astro-ph.HE]]}.


\bibitem{J0348+0432}
J.~Antoniadis {\it et al.},
  A Massive Pulsar in a Compact Relativistic Binary,
  Science {\bf 340} (2013) 6131
  \href{https://arxiv.org/pdf/1304.6875.pdf}{[arXiv:1304.6875 [astro-ph.HE]]}.

\bibitem{Crawford:2006xb}
  F.~Crawford, M.~S.~E.~Roberts, J.~W.~T.~Hessels, S.~M.~Ransom, M.~Livingstone, C.~R.~Tam and V.~M.~Kaspi,
  A Survey of 56 Mid-latitude EGRET Error Boxes for Radio Pulsars,
  Astrophys.\ J.\  {\bf 652} (2006) 1499
  \href{https://arxiv.org/pdf/astro-ph/0608225.pdf}{[astro-ph/0608225]}.

 \bibitem{LinShaCas18}
M.~Linares, T.~Shahbaz and J.~Casares,
  Peering into the dark side: Magnesium lines establish a massive neutron star in PSR J2215+5135,
  Astrophys.\ J.\  {\bf 859} (2018) 54
  \href{https://arxiv.org/pdf/1805.08799.pdf}{[arXiv:1805.08799 [astro-ph.HE]]}.

\bibitem{Cromartie:2019kug}
  H.~T.~Cromartie {\it et al.},
  A very massive neutron star: relativistic Shapiro delay measurements of PSR J0740+6620, Nature Astron. \textbf{4} (2019) 72
  \href{https://arxiv.org/pdf/1904.06759.pdf}{[arXiv:1904.06759 [astro-ph.HE]]}.


\bibitem{Gamba:2019kwu}
  R.~Gamba, J.~S.~Read and L.~E.~Wade,
  The impact of the crust equation of state on the analysis of GW170817, Class. Quant. Grav. \textbf{37} (2020) 025008
  \href{https://arxiv.org/pdf/1902.04616.pdf}{[arXiv:1902.04616 [gr-qc]]}.


   \bibitem{Lattimer:2004nj}
  J.~M.~Lattimer and B.~F.~Schutz,
  Constraining the equation of state with moment of inertia measurements,
  Astrophys.\ J.\  {\bf 629} (2005) 979
  \href{https://arxiv.org/pdf/astro-ph/0411470.pdf}{[astro-ph/0411470]}.


\bibitem{Ozel:2016oaf}
  F.~Ozel and P.~Freire,
  Masses, Radii, and the Equation of State of Neutron Stars,
  Ann.\ Rev.\ Astron.\ Astrophys.\  {\bf 54} (2016) 401
  \href{https://arxiv.org/pdf/1603.02698.pdf}{[arXiv:1603.02698 [astro-ph.HE]]}.

  \bibitem{Eksi:2014wia}
  K.~Y.~Ekşi, C.~Güngör and M.~M.~Türkoğlu,
  What does a measurement of mass and/or radius of a neutron star constrain: Equation of state or gravity?,
  Phys.\ Rev.\ D {\bf 89} (2014)  063003
 \href{https://arxiv.org/pdf/1402.0488.pdf}{[arXiv:1402.0488 [astro-ph.HE]]}.

  \bibitem{Miller:2014aaa}
  M.~C.~Miller and J.~M.~Miller,
  The Masses and Spins of Neutron Stars and Stellar-Mass Black Holes,
  Phys.\ Rept.\  {\bf 548} (2014) 1
  \href{https://arxiv.org/pdf/1408.4145.pdf}{[arXiv:1408.4145 [astro-ph.HE]]}.

  \bibitem{Psaltis:2013fha}
  D.~Psaltis, F.~Özel and D.~Chakrabarty,
  Prospects for Measuring Neutron-Star Masses and Radii with X-Ray Pulse Profile Modeling,
  Astrophys.\ J.\  {\bf 787} (2014) 136
  \href{https://arxiv.org/pdf/1311.1571.pdf}{[arXiv:1311.1571 [astro-ph.HE]]}.

\bibitem{Raithel:2016vtt}
  C.~A.~Raithel, F.~Ozel and D.~Psaltis,
  Model-Independent Inference of Neutron Star Radii from Moment of Inertia Measurements,
  Phys.\ Rev.\ C {\bf 93} (2016) 032801
   Addendum: [Phys.\ Rev.\ C {\bf 93} (2016)  049905]
  \href{https://arxiv.org/pdf/1603.06594.pdf}{[arXiv:1603.06594 [astro-ph.HE]]}.

  \bibitem{Abbott:2018exr}
  B.~P.~Abbott {\it et al.} [LIGO Scientific and Virgo Collaborations],
  GW170817: Measurements of neutron star radii and equation of state, Phys.\ Rev.\ Lett.\  {\bf 121} (2018)   161101
  \href{https://arxiv.org/pdf/1805.11581.pdf}{[arXiv:1805.11581 [gr-qc]]}.

   \bibitem{Bauswein:2017vtn}
  A.~Bauswein, O.~Just, H.~T.~Janka and N.~Stergioulas,
  Neutron-star radius constraints from GW170817 and future detections,
  Astrophys.\ J.\  {\bf 850} (2017)  L34 \href{https://arxiv.org/pdf/1710.06843.pdf}{[arXiv:1710.06843 [astro-ph.HE]]}.

  \bibitem{Ruiz:2017due}
  M.~Ruiz, S.~L.~Shapiro and A.~Tsokaros,
  GW170817, General Relativistic Magnetohydrodynamic Simulations, and the Neutron Star Maximum Mass,
  Phys.\ Rev.\ D {\bf 97} (2018) 021501
  \href{https://arxiv.org/pdf/1711.00473.pdf}{[arXiv:1711.00473 [astro-ph.HE]]}.

  \bibitem{Shibata:2019ctb}
  M.~Shibata, E.~Zhou, K.~Kiuchi and S.~Fujibayashi,
  Constraint on the maximum mass of neutron stars using GW170817 event, Phys.\ Rev.\ D {\bf 100} (2019)  023015
  \href{https://arxiv.org/pdf/1905.03656.pdf}{[arXiv:1905.03656 [astro-ph.HE]]}.

  \bibitem{Abbott:2018wiz}
  B.~P.~Abbott {\it et al.} [LIGO Scientific and Virgo Collaborations],
  Properties of the binary neutron star merger GW170817, Phys.\ Rev.\ X {\bf 9} (2019)  011001
  \href{https://arxiv.org/pdf/1805.11579.pdf}{[arXiv:1805.11579 [gr-qc]]}.

  \bibitem{Landry:2018jyg}
  P.~Landry and B.~Kumar,
  Constraints on the moment of inertia of PSR J0737-3039A from GW170817,
  Astrophys.\ J.\  {\bf 868} (2018) L22
  \href{https://arxiv.org/pdf/1807.04727.pdf}{[arXiv:1807.04727 [gr-qc]]}.

\bibitem{Wang:2018nye}
  Y.~Z.~Wang {\it et al.},
 GW170817: The energy extraction process of the off-axis relativistic outflow and the constraint on the equation of state of neutron stars, Astrophys.\ J.\  {\bf 877} (2019)  2
\href{https://arxiv.org/pdf/1811.02558.pdf}{[arXiv:1811.02558 [astro-ph.HE]]}.

\bibitem{Carney:2018sdv}
  M.~F.~Carney, L.~E.~Wade and B.~S.~Irwin,
  Comparing two models for measuring the neutron star equation of state from gravitational-wave signals,
  Phys.\ Rev.\ D {\bf 98} (2018)  063004
  \href{https://arxiv.org/pdf/1805.11217.pdf}{[arXiv:1805.11217 [gr-qc]]}.

  \bibitem{Biswas:2019gkw}
  B.~Biswas and S.~Bose,
  Tidal deformability of an anisotropic compact star: Implications of GW170817,
  Phys.\ Rev.\ D {\bf 99} (2019) 104002
  \href{https://arxiv.org/pdf/1903.04956.pdf}{[arXiv:1903.04956 [gr-qc]]}.

  \bibitem{Vivanco:2019qnt}
  F.~H.~Vivanco, R.~Smith, E.~Thrane, P.~D.~Lasky, C.~Talbot and V.~Raymond,
  Measuring the neutron star equation of state with gravitational waves: the first forty binary neutron star mergers, Phys.\ Rev.\ D {\bf 100} (2019)  103009
  \href{https://arxiv.org/pdf/1909.02698.pdf}{[arXiv:1909.02698 [gr-qc]]}.

  \bibitem{Essick:2019ldf}
  R.~Essick, P.~Landry and D.~E.~Holz,
  Nonparametric Inference of Neutron Star Composition, Equation of State, and Maximum Mass with GW170817,  Phys. Rev. D \textbf{101} (2020) 063007 \href{https://arxiv.org/pdf/1910.09740.pdf}{[arXiv:1910.09740 [astro-ph.HE]]}.

  \bibitem{Llanes-Estrada:2019wmz}
  F.~J.~Llanes-Estrada and E.~Lope-Oter,
  Hadron matter in neutron stars in view of gravitational wave observations, Prog.\ Part.\ Nucl.\ Phys.\  {\bf 109} (2019) 103715
  \href{https://arxiv.org/pdf/1907.12760.pdf}{[arXiv:1907.12760 [nucl-th]]}.

  \bibitem{Kumar:2019xgp}
  B.~Kumar and P.~Landry,
  Inferring neutron star properties from GW170817 with universal relations, Phys.\ Rev.\ D {\bf 99} (2019)  123026
  \href{https://arxiv.org/pdf/1902.04557.pdf}{[arXiv:1902.04557 [gr-qc]]}.

  \bibitem{Fasano:2019zwm}
  M.~Fasano, T.~Abdelsalhin, A.~Maselli and V.~Ferrari,
  Constraining the neutron star equation of state using multi-band independent measurements of radii and tidal deformabilities, Phys.\ Rev.\ Lett.\  {\bf 123} (2019) 141101
  \href{https://arxiv.org/pdf/1902.05078.pdf}{[arXiv:1902.05078 [astro-ph.HE]]}.



\bibitem{Cottam:2002cu}
  J.~Cottam, F.~Paerels and M.~Mendez,
  Gravitationally redshifted absorption lines in the x-ray burst spectra of a neutron star,
  Nature {\bf 420} (2002) 51
  \href{https://arxiv.org/ftp/astro-ph/papers/0211/0211126.pdf}{[astro-ph/0211126]}.


 \bibitem{Hinderer:2007mb}
  T.~Hinderer,
  Tidal Love numbers of neutron stars,
  Astrophys.\ J.\  {\bf 677} (2008) 1216
  \href{https://arxiv.org/pdf/0711.2420.pdf}{[arXiv:0711.2420 [astro-ph]]}.


     \bibitem{Paschalidis:2017qmb}
V.~Paschalidis, K.~Yagi, D.~Alvarez-Castillo, D.~B.~Blaschke and A.~Sedrakian,
  Implications from GW170817 and I-Love-Q relations for relativistic hybrid stars,
  Phys.\ Rev.\ D {\bf 97} (2018)  084038
  \href{https://arxiv.org/pdf/1712.00451.pdf}{[arXiv:1712.00451 [astro-ph.HE]]}.

 \bibitem{Lattimer:2006xb}
  J.~M.~Lattimer and M.~Prakash,
  Neutron Star Observations: Prognosis for Equation of State Constraints,
  Phys.\ Rept.\  {\bf 442} (2007) 109
 \href{https://arxiv.org/pdf/astro-ph/0612440.pdf}{[astro-ph/0612440]}.


 \bibitem{Rezzollabook}
 L. Rezzolla, P. Pizzochero, D. I. Jones, N. Rea and I. Vidana, (Eds.), \emph{The Physics and Astrophysics of Neutron Stars} (Springer, 2018).

  \bibitem{Berti:2015itd}
  E.~Berti {\it et al.},
  Testing General Relativity with Present and Future Astrophysical Observations,
  Class.\ Quant.\ Grav.\  {\bf 32} (2015) 243001
  \href{https://arxiv.org/pdf/1501.07274.pdf}{[arXiv:1501.07274 [gr-qc]]}.



 \bibitem{Shao:2019gjj}
  L.~Shao,
  Degeneracy in Studying the Supranuclear Equation of State and Modified Gravity with Neutron Stars, AIP Conf.\ Proc.\  {\bf 2127} (2019)  020016
  \href{https://arxiv.org/pdf/1901.07546.pdf}{[arXiv:1901.07546 [gr-qc]]}.


    \bibitem{Chandra}
S. Chandrasekhar,
The highly collapsed configurations of a stellar mass,
Mon. Not. R. Astron. Soc. \textbf{95} (1935) 207.

 \bibitem{Burrows:1992fg}
  A.~Burrows and J.~Liebert,
  The Science of brown dwarfs,
  Rev.\ Mod.\ Phys.\  {\bf 65} (1993) 301.

  \bibitem{Burrows2}
 A.~Burrows, W.~B.~Hubbard, J.~I.~Lunine and J.~Liebert,
  The theory of brown dwarfs and extrasolar giant planets,
  Rev.\ Mod.\ Phys.\  {\bf 73} (2001) 719
  \href{https://arxiv.org/pdf/astro-ph/0103383.pdf}{[astro-ph/0103383]}.

\bibitem{Afonso:2017bxr}
  V.~I.~Afonso, C.~Bejarano, J.~Beltran Jimenez, G.~J.~Olmo and E.~Orazi,
  The trivial role of torsion in projective invariant theories of gravity with non-minimally coupled matter fields,
  Class.\ Quant.\ Grav.\  {\bf 34}, 235003 (2017)
  \href{https://arxiv.org/pdf/1705.03806.pdf}{[arXiv:1705.03806 [gr-qc]]}.

\bibitem{Olmo:2011uz}
  G.~J.~Olmo,
  Palatini Approach to Modified Gravity: f(R) Theories and Beyond,
  Int.\ J.\ Mod.\ Phys.\ D {\bf 20}, 413 (2011)
  \href{https://arxiv.org/pdf/1101.3864.pdf}{[arXiv:1101.3864 [gr-qc]]}.


\bibitem{DeFelice:2010aj}
  A.~De Felice and S.~Tsujikawa,
  f(R) theories,
  Living Rev.\ Rel.\  {\bf 13} (2010) 3
  \href{https://arxiv.org/pdf/1002.4928.pdf}{[arXiv:1002.4928 [gr-qc]]}.

  \bibitem{Sakstein:2018fwz}
J.~Sakstein,
Astrophysical tests of screened modified gravity,
Int. J. Mod. Phys. D \textbf{27} (2018) 1848008
\href{https://arxiv.org/pdf/2002.04194.pdf}{[arXiv:2002.04194 [astro-ph.CO]]}.

 \bibitem{CLreview}
S.~Capozziello and M.~De Laurentis, Extended Theories of Gravity, Phys.\ Rept.\  {\bf 509} (2011) 167
\href{https://arxiv.org/pdf/1108.6266.pdf}{[arXiv:1108.6266 [gr-qc]].}

\bibitem{Clifton:2011jh}
  T.~Clifton, P.~G.~Ferreira, A.~Padilla and C.~Skordis,
  Modified Gravity and Cosmology,
  Phys.\ Rept.\  {\bf 513} (2012) 1
  \href{https://arxiv.org/pdf/1106.2476.pdf}{[arXiv:1106.2476 [astro-ph.CO]]}.

\bibitem{Nojiri:2017ncd}
  S.~Nojiri, S.~D.~Odintsov and V.~K.~Oikonomou,
  Modified Gravity Theories on a Nutshell: Inflation, Bounce and Late-time Evolution,
  Phys.\ Rept.\  {\bf 692} (2017) 1
  \href{https://arxiv.org/pdf/1705.11098.pdf}{[arXiv:1705.11098 [gr-qc]]}.

  \bibitem{Joyce:2014kja}
A.~Joyce, B.~Jain, J.~Khoury and M.~Trodden,
Beyond the Cosmological Standard Model,
Phys. Rept. \textbf{568} (2015) 1
\href{https://arxiv.org/pdf/1407.0059.pdf}{[arXiv:1407.0059 [astro-ph.CO]]}.

  \bibitem{BeltranJimenez:2017doy}
  J.~Beltran Jimenez, L.~Heisenberg, G.~J.~Olmo and D.~Rubiera-Garcia,
  Born–Infeld inspired modifications of gravity,
  Phys.\ Rept.\  {\bf 727} (2018) 1
  \href{https://arxiv.org/pdf/1704.03351.pdf}{[arXiv:1704.03351 [gr-qc]]}.

  \bibitem{Heisenberg:2018vsk}
  L.~Heisenberg,
  A systematic approach to generalisations of General Relativity and their cosmological implications,
  Phys.\ Rept.\  {\bf 796} (2019) 1
  \href{https://arxiv.org/pdf/1807.01725.pdf}{arXiv:1807.01725 [gr-qc]}.

\bibitem{Ezquiaga:2018btd}
  J.~M.~Ezquiaga and M.~Zumalacárregui,
  Dark Energy in light of Multi-Messenger Gravitational-Wave astronomy,
  Front.\ Astron.\ Space Sci.\  {\bf 5} (2018) 44
 \href{https://arxiv.org/pdf/1807.09241.pdf}{[arXiv:1807.09241 [astro-ph.CO]]}.

\bibitem{Bejarano:2019zco}
  C.~Bejarano, A.~Delhom, A.~Jiménez-Cano, G.~J.~Olmo and D.~Rubiera-Garcia,
  Geometric inequivalence of metric and Palatini formulations of General Relativity, Phys. Lett. B \textbf{802} (2020) 135275
  \href{https://arxiv.org/pdf/1907.04137.pdf}{[arXiv:1907.04137 [gr-qc]]}.


  \bibitem{GlenBook}
N. K. Glendenning, \emph{Compact Stars: Nuclear Physics, Particle Physics, and General Relativity}, 2nd Edition (Astronomy and Astrophysics Library, 2000).


  \bibitem{Liu:2018jhd}
  H.~L.~Liu and G.~L.~Lü,
  Properties of white dwarfs in Einstein-$\Lambda$ gravity, JCAP \textbf{02} (2019) 040
  \href{https://arxiv.org/pdf/1805.00333.pdf}{[arXiv:1805.00333 [gr-qc]]}.

\bibitem{T}
R. C. Tolman, Static Solutions of Einstein's Field Equations for Spheres of Fluid, Phys. Rev. \textbf{55} (1939) 364.

\bibitem{OV}
J. R. Oppenheimer and G. M. Volkoff, On Massive Neutron Cores, Phys. Rev. \textbf{55} (1939) 374.

\bibitem{ZhLuWa17}
L. Zhu, J.-L. Lu, and L. Wang, Gen. Rel. Grav. \textbf{50} (2018) 11.

\bibitem{Carter:1998rn}
  B.~Carter and D.~Langlois,
  Relativistic models for superconducting superfluid mixtures,
  Nucl.\ Phys.\ B {\bf 531} (1998) 478
  \href{https://arxiv.org/pdf/gr-qc/9806024.pdf}{[gr-qc/9806024]}

\bibitem{KippWeiBook}
R. Kippenhahn, A. Weigert, and A. Weiss, \emph{Stellar Structure and Evolution} (Springer, 2012).

 \bibitem{Herrera:1997plx}
  L.~Herrera and N.~O.~Santos,
  Local anisotropy in self-gravitating systems,  Phys.\ Rept.\  {\bf 286} (1997) 53.


\bibitem{Raposo:2018rjn}
  G.~Raposo, P.~Pani, M.~Bezares, C.~Palenzuela, and V.~Cardoso,
  Anisotropic stars as ultracompact objects in General Relativity, Phys. Rev. D \textbf{99} (2019) 104072
  \href{https://arxiv.org/pdf/1811.07917.pdf}{[arXiv:1811.07917 [gr-qc]]}.


\bibitem{stifflimit}
A. Dobado, F. J. Llanes-Estrada, and J. A. Oller, Existence of two-solar-mass neutron star constrains gravitational constant
$G_N$ at strong field, Phys.
Rev. C \textbf{85} (2012) 012801.

\bibitem{Bombaci:2016xzl}
  I.~Bombaci,
  The Hyperon Puzzle in Neutron Stars,
  JPS Conf.\ Proc.\  {\bf 17} (2017) 101002
 \href{https://arxiv.org/pdf/1601.05339.pdf}{[arXiv:1601.05339 [nucl-th]]}.


\bibitem{Demorest10}
  P.~Demorest, T.~Pennucci, S.~Ransom, M.~Roberts and J.~Hessels,
  Shapiro Delay Measurement of A Two Solar Mass Neutron Star,
  Nature {\bf 467} (2010) 1081 \href{https://arxiv.org/pdf/1010.5788.pdf}{[arXiv:1010.5788 [astro-ph.HE]]}.

\bibitem{RhoRuf}
C. E. Rhoades and R. Ruffini, Maximum Mass of a Neutron Star, Phys. Rev. Lett. \textbf{32} (1974) 324.


  \bibitem{Buchdahlpaper}
  H. A. Buchdahl,
  General Relativistic Fluid Spheres,
  Phys. Rev. \textbf{116} (1959) 1027.


\bibitem{WeinbergBook}
S. Weinberg, \emph{Gravitation and Cosmology: Principles and Applications of the General Theory of Relativity} (John Wiley $\&$ Sons, 1972).

  \bibitem{Mak:2001gg}
  M.~K.~Mak, P.~N.~Dobson, Jr. and T.~Harko,
  Maximum mass radius ratio for compact general relativistic objects in Schwarzschild-de Sitter geometry,
  Mod.\ Phys.\ Lett.\ A {\bf 15} (2000) 2153
  \href{https://arxiv.org/pdf/gr-qc/0104031v1.pdf}{[gr-qc/0104031]}.

\bibitem{Andreasson:2012dj}
  H.~Andreasson, C.~G.~Boehmer and A.~Mussa,
  Bounds on M/R for Charged Objects with positive Cosmological constant,
  Class.\ Quant.\ Grav.\  {\bf 29} (2012) 095012
 \href{https://arxiv.org/pdf/1201.5725.pdf}{[arXiv:1201.5725 [gr-qc]]}.


  \bibitem{Urbano:2018nrs}
  A.~Urbano and H.~Veermäe,
  On gravitational echoes from ultracompact exotic stars, JCAP \textbf{04} (2019) 011
  \href{https://arxiv.org/pdf/1810.07137.pdf}{[arXiv:1810.07137 [gr-qc]]}.

  \bibitem{Tsuchida:1998jw}
  T.~Tsuchida, G.~Kawamura and K.~Watanabe,
  A Maximum mass-to-size ratio in scalar tensor theories of gravity,
  Prog.\ Theor.\ Phys.\  {\bf 100} (1998) 291
  \href{https://arxiv.org/pdf/gr-qc/9802049.pdf}{[gr-qc/9802049]}.

      \bibitem{Dadhich:2016fku}
  N.~Dadhich and S.~Chakraborty,
  Buchdahl compactness limit for a pure Lovelock static fluid star,
  Phys.\ Rev.\ D {\bf 95} (2017) 064059
  \href{https://arxiv.org/pdf/1606.01330.pdf}{[arXiv:1606.01330 [gr-qc]]}.

  \bibitem{chandras} S. Chandrasekhar, Dynamical Instability of Gaseous Masses Approaching the Schwarzschild Limit in General Relativity, Phys. Rev. Lett. \textbf{12} (1964) 437.

\bibitem{sagert} I.~Sagert, M.~Hempel, C.~Greiner and J.~Schaffner-Bielich,
  Compact stars for undergraduates,
  Eur.\ J.\ Phys.\  {\bf 27} (2006) 577
  \href{https://arxiv.org/pdf/astro-ph/0506417.pdf}{[astro-ph/0506417]}.

\bibitem{Olmo:2006eh}
  G.~J.~Olmo,
  Limit to general relativity in f(R) theories of gravity,
  Phys.\ Rev.\ D {\bf 75} (2007) 023511
  \href{https://arxiv.org/pdf/gr-qc/0612047.pdf}{[gr-qc/0612047]}.

  \bibitem{stt}
  Y. Fujii and K.-i. Maeda, \emph{The Scalar-Tensor Theory of Gravitation}, Cambridge Monographs on Mathematical
Physics (Cambridge University Press, 2003).



   \bibitem{Brax:2017wcj}
  P.~Brax, A.~C.~Davis and R.~Jha,
  Neutron Stars in Screened Modified Gravity: Chameleon vs Dilaton,
  Phys.\ Rev.\ D {\bf 95} (2017) 083514
  \href{https://arxiv.org/pdf/1702.02983.pdf}{[arXiv:1702.02983 [gr-qc]]}.

  \bibitem{Zhang:2017srh}
  X.~Zhang, T.~Liu and W.~Zhao,
  Gravitational radiation from compact binary systems in screened modified gravity,
  Phys.\ Rev.\ D {\bf 95} (2017)  104027
  \href{https://arxiv.org/pdf/1702.08752.pdf}{[arXiv:1702.08752 [gr-qc]]}.


\bibitem{Vainshtein:1972sx}
  A.~I.~Vainshtein,
  To the problem of nonvanishing gravitation mass,
  Phys.\ Lett.\  {\bf 39B} (1972) 393.


\bibitem{Flanagan:2004bz}
  E.~E.~Flanagan,
  The Conformal frame freedom in theories of gravitation,
  Class.\ Quant.\ Grav.\  {\bf 21} (2004) 3817
  \href{https://arxiv.org/pdf/gr-qc/0403063.pdf}{[gr-qc/0403063]}.

  \bibitem{Damour:1992we}
  T.~Damour and G.~Esposito-Farese,
  Tensor multiscalar theories of gravitation,
  Class.\ Quant.\ Grav.\  {\bf 9} (1992) 2093.


   \bibitem{Resco:2016upv}
  M.~Aparicio Resco, A.~de la Cruz-Dombriz, F.~J.~Llanes Estrada and V.~Zapatero Castrillo,
  On neutron stars in $f(R)$ theories: Small radii, large masses and large energy emitted in a merger,
  Phys.\ Dark Univ.\  {\bf 13} (2016) 147
  \href{https://arxiv.org/pdf/1602.03880.pdf}{[arXiv:1602.03880 [gr-qc]]}.


\bibitem{Silva:2018yxz}
  H.~O.~Silva and N.~Yunes,
  Neutron star pulse profiles in scalar-tensor theories of gravity, Phys.\ Rev.\ D {\bf 99} (2019)  044034
  \href{https://arxiv.org/pdf/1808.04391.pdf}{[arXiv:1808.04391 [gr-qc]]}.

\bibitem{Horbatsch:2011nh}
  M.~W.~Horbatsch and C.~P.~Burgess,
  Model-Independent Comparisons of Pulsar Timings to Scalar-Tensor Gravity,
  Class.\ Quant.\ Grav.\  {\bf 29} (2012) 245004
  \href{https://arxiv.org/pdf/1107.3585.pdf}{[arXiv:1107.3585 [gr-qc]]}.

  \bibitem{WillsBook}
  C.~M.~Will, \emph{Theory and Experiment in Gravitational Physic"} (Cambridge University Press, 2018)


  \bibitem{Will:1972zz}
  C.~M.~Will and K.~Nordtvedt, Jr.,
  Conservation Laws and Preferred Frames in Relativistic Gravity. I. Preferred-Frame Theories and an Extended PPN Formalism,
  Astrophys.\ J.\  {\bf 177} (1972) 757.

\bibitem{Sotiriou:2008rp}
  T.~P.~Sotiriou and V.~Faraoni,
  f(R) Theories Of Gravity,
  Rev.\ Mod.\ Phys.\  {\bf 82} (2010) 451
  \href{https://arxiv.org/pdf/0805.1726.pdf}{[arXiv:0805.1726 [gr-qc]]}.

  \bibitem{Capozziello:2005bu}
  S.~Capozziello and A.~Troisi,
  PPN-limit of fourth order gravity inspired by scalar-tensor gravity,
  Phys.\ Rev.\ D {\bf 72} (2005) 044022
 \href{https://arxiv.org/pdf/astro-ph/0507545.pdf}{[astro-ph/0507545]}.


\bibitem{chiba}
T. Chiba, $1/R$ Gravity and Scalar-Tensor Gravity,  Phys. Lett. B \textbf{575} (2003) 1 \href{https://arxiv.org/pdf/astro-ph/0307338.pdf}{[arXiv:astro-ph/0307338]}.

\bibitem{Carroll:2003wy}
  S.~M.~Carroll, V.~Duvvuri, M.~Trodden and M.~S.~Turner,
  Is cosmic speed - up due to new gravitational physics?,
  Phys.\ Rev.\ D {\bf 70} (2004) 043528
  \href{https://arxiv.org/pdf/astro-ph/0306438.pdf}{[arXiv:astro-ph/0306438]}.

\bibitem{Erickcek}
A. L. Erickcek, T. L. Smith, M. Kamionkowski, Solar system tests do rule out 1/R gravity.
Phys. Rev. D \textbf{74} (2006) 121501 \href{https://arxiv.org/pdf/astro-ph/0610483v1.pdf}{[arXiv:astro-ph/0610483]}.



\bibitem{Bertotti:2003rm}
  B.~Bertotti, L.~Iess and P.~Tortora,
  A test of general relativity using radio links with the Cassini spacecraft,
  Nature {\bf 425} (2003) 374.

\bibitem{Suvorov:2018frf}
  A.~G.~Suvorov,
  Monopolar and quadrupolar gravitational radiation from magnetically deformed neutron stars in modified gravity,
  Phys.\ Rev.\ D {\bf 98} (2018) 084026
  \href{https://arxiv.org/pdf/1810.02975.pdf}{[arXiv:1810.02975 [astro-ph.HE]]}.

 \bibitem{Abbott:2017ylp}
  B.~P.~Abbott {\it et al.} [LIGO Scientific and Virgo Collaborations],
  First search for gravitational waves from known pulsars with Advanced LIGO,
  Astrophys.\ J.\  {\bf 839} (2017)  12
   Erratum: [Astrophys.\ J.\  {\bf 851} (2017)  71]
  \href{https://arxiv.org/pdf/1701.07709.pdf}{[arXiv:1701.07709 [astro-ph.HE]]}.

\bibitem{Kainulainen:2007bt}
  K.~Kainulainen, J.~Piilonen, V.~Reijonen and D.~Sunhede,
  Spherically symmetric spacetimes in f(R) gravity theories,
  Phys.\ Rev.\ D {\bf 76} (2007) 024020
  \href{https://arxiv.org/pdf/0704.2729.pdf}{[arXiv:0704.2729 [gr-qc]]}.

  \bibitem{Zhang:2007ne}
  P.~J.~Zhang,
  The behavior of f(R) gravity in the solar system, galaxies and clusters,
  Phys.\ Rev.\ D {\bf 76} (2007) 024007 \href{https://arxiv.org/pdf/astro-ph/0701662.pdf}{[astro-ph/0701662]}.

  \bibitem{Chiba:2006jp}
T.~Chiba, T.~L.~Smith and A.~L.~Erickcek,
Solar System constraints to general f(R) gravity,
Phys. Rev. D \textbf{75} (2007) 124014
\href{https://arxiv.org/pdf/astro-ph/0611867.pdf}{[arXiv:astro-ph/0611867 [astro-ph]]}.

\bibitem{Nojiri:2003ft}
S.~Nojiri and S.~D.~Odintsov,
Modified gravity with negative and positive powers of the curvature: Unification of the inflation and of the cosmic acceleration,
Phys. Rev. D \textbf{68} (2003) 123512
\href{https://arxiv.org/pdf/hep-th/0307288.pdf}{[arXiv:hep-th/0307288 [hep-th]]}.

  \bibitem{Glampedakis:2015sua}
  K.~Glampedakis, G.~Pappas, H.~O.~Silva and E.~Berti,
  Post-Tolman-Oppenheimer-Volkoff formalism for relativistic stars,
  Phys.\ Rev.\ D {\bf 92} (2015) 024056
  \href{https://arxiv.org/pdf/1504.02455.pdf}{[arXiv:1504.02455 [gr-qc]]}.


 \bibitem{Faulkner:2006ub}
  T.~Faulkner, M.~Tegmark, E.~F.~Bunn and Y.~Mao,
  Constraining f(R) Gravity as a Scalar Tensor Theory,
  Phys.\ Rev.\ D {\bf 76} (2007) 063505
  \href{https://arxiv.org/pdf/astro-ph/0612569.pdf}{[astro-ph/0612569]}.


\bibitem{mult} T. Multamaki, I. Vilja, Phys. Rev. D \textbf{76} (2007) 064021 \href{https://arxiv.org/pdf/astro-ph/0612775v1.pdf}{[arXiv:astro-ph/0612775]}.


\bibitem{hen}
K. Henttunen, T. Multamaki, I. Vilja,
Stellar configurations in f (R) theories of gravity,
Phys. Rev. D \textbf{77} (2008) 024040  \href{https://arxiv.org/pdf/0705.2683v2.pdf}{[arXiv:0705.2683 [astro-ph]]}.

\bibitem{PolyBook}
 G. P. Horedt, \emph{Polytropes: Applications in Astrophysics and Related Fields} (Springer, 2004).


  \bibitem{KippenWeig}
  R. Kippenhahn and A. Weigert, \emph{Stellar structure and Evolution}, Astronomy and Astrophysics Library (Springer, New York, 1990).

\bibitem{Cikintoglu:2017jfh}
  S.~Cikintoglu,
  Vacuum solutions around spherically symmetric and static objects in the Starobinsky model,
  Phys.\ Rev.\ D {\bf 97} (2018) 044040 \href{https://arxiv.org/pdf/1708.00345.pdf}{[arXiv:1708.00345 [gr-qc]]}.

  \bibitem{Astashenok:2018bol}
  A.~V.~Astashenok, K.~Mosani, S.~D.~Odintsov and G.~C.~Samanta,
  Gravitational collapse in General Relativity and in $R^2$-gravity: A comparative study
  Int.\ J.\ Geom.\ Meth.\ Mod.\ Phys.\  {\bf 16} (2019)  1950035
 \href{https://arxiv.org/pdf/1812.10441.pdf}{[arXiv:1812.10441 [gr-qc]]}.

\bibitem{sunil} R. Goswami, A. M. Nzioki, S. D. Maharaj, S. G. Ghosh, Collapsing spherical stars in f(R) gravity, Phys. Rev. D \textbf{90} (2014) 084011  \href{https://arxiv.org/pdf/1409.2371.pdf}{[arXiv:1409.2371 [gr-qc]]}.

\bibitem{OS39}
J. R. Oppenheimer and H. Snyder, On Continued Gravitational Contraction, Phys. Rev. \textbf{56} (1939) 455.

\bibitem{Datt38}
B. Datt, Z. Phys. \textbf{108} (1938) 314 [Reprinted as  Gen. Rel. Grav. \textbf{31}(1999) 1615].

\bibitem{HansenBook}
C. J. Hansen, S. D. Kawaler, V. Trimble, \emph{Stellar Interiors -
Physical Principles, Structure, and Evolution}, A\&A
Library (1994) 2nd Edition (Spinger-Verlag, New York).

\bibitem{capo} S. Capozziello, M. De Laurentis, S. D. Odintsov, A. Stabile, Hydrostatic equilibrium and stellar structure in $f(R)$-gravity, Phys. Rev. D, \textbf{83} (2011) 064004  \href{https://arxiv.org/pdf/1101.0219.pdf}{[arXiv:1101.0219 [gr-qc]]}.


\bibitem{Farinelli:2013pza}
  R.~Farinelli, M.~De Laurentis, S.~Capozziello and S.~D.~Odintsov,
  Numerical solutions of the modified Lane–Emden equation in $f(R)$-gravity,
  Mon.\ Not.\ Roy.\ Astron.\ Soc.\  {\bf 440} (2014) 2909
  \href{https://arxiv.org/pdf/1311.2744.pdf}{[arXiv:1311.2744 [astro-ph.SR]]}.

\bibitem{andre}
R.~Andr\'e and G.~M.~Kremer,
  Stellar structure model in hydrostatic equilibrium in the context of $f(\mathcal{R})$-gravity,
  Res.\ Astron.\ Astrophys.\  {\bf 17} (2017) 122 \href{https://arxiv.org/pdf/1707.07675.pdf}{[arXiv:1707.07675 [gr-qc]]}.


    \bibitem{Naf:2010zy}
  J.~Naf and P.~Jetzer,
  On the 1/c Expansion of f(R) Gravity,
  Phys.\ Rev.\ D {\bf 81} (2010) 104003
  \href{https://arxiv.org/pdf/1004.2014.pdf}{[arXiv:1004.2014 [gr-qc]]}.

\bibitem{Teide1}
  R.~Rebolo, E.~L.~Martin, G.~Basri, G.~W.~Marcy and M.~R.~Z.~Osorio,
  Brown dwarfs in the Pleiades cluster confirmed by the lithium test,
  Astrophys.\ J.\  {\bf 469} (1996) L53
\href{https://arxiv.org/pdf/astro-ph/9607002.pdf}{[arXiv:astro-ph/9607002]}.

\bibitem{RG1}
K.~Ohnaka,
  Spatially resolved, high-spectral resolution observation of the K giant Aldebaran in the CO first overtone lines with VLTI/AMBER,
  Astron.\ Astrophys.\  {\bf 553} (2013) A3
  \href{https://arxiv.org/pdf/1303.4763.pdf}{[arXiv:1303.4763 [astro-ph.SR]]}.

\bibitem{RG2}
 T.~Tsuji,
  Cool luminous stars: the hybrid nature of their infrared spectra,
  Astron.\ Astrophys.\  {\bf 489} (2008) 1271
  \href{https://arxiv.org/pdf/0807.4387.pdf}{[arXiv:0807.4387 [astro-ph]]}.

\bibitem{RG3}
A.~Richichi and V.~Roccatagliata,
  Aldebaran's angular diameter: How well do we know it?,
  Astron.\ Astrophys.\
   [Astron.\ Astrophys.\  {\bf 433} (2005) 305]
  \href{https://arxiv.org/ftp/astro-ph/papers/0502/0502181.pdf}{[astro-ph/0502181]}.

\bibitem{SiriusB}
J. B. Holberg, M. A. Barstow, F. C. Bruhweiler, A. M. Cruise, and A. J. Penny, Sirius B: A New, More Accurate View, The Astrophysical Journal \textbf{497}, 935 (1998).

\bibitem{zhao}
  X.~F.~Zhao,
  The properties of the massive neutron star PSR J0348+0432,
  Int.\ J.\ Mod.\ Phys.\ D {\bf 24} (2015) 1550058  \href{https://arxiv.org/pdf/1712.08860.pdf}{[arXiv:1712.08860 [nucl-th]]}.



   \bibitem{Horndeski:1974wa}
  G.~W.~Horndeski,
  Second-order scalar-tensor field equations in a four-dimensional space,
  Int.\ J.\ Theor.\ Phys.\  {\bf 10} (1974) 363.

    \bibitem{Langlois:2015cwa}
  D.~Langlois and K.~Noui,
  Degenerate higher derivative theories beyond Horndeski: evading the Ostrogradski instability,
  JCAP {\bf 1602} (2016) 034
  \href{https://arxiv.org/pdf/1510.06930.pdf}{[arXiv:1510.06930 [gr-qc]]}.

  \bibitem{Zumalacarregui:2013pma}
  M.~Zumalac\'arregui and J.~García-Bellido,
  Transforming gravity: from derivative couplings to matter to second-order scalar-tensor theories beyond the Horndeski Lagrangian,
  Phys.\ Rev.\ D {\bf 89} (2014) 064046
  \href{https://arxiv.org/pdf/1308.4685.pdf}{[arXiv:1308.4685 [gr-qc]]}.

  \bibitem{Gleyzes:2014dya}
  J.~Gleyzes, D.~Langlois, F.~Piazza and F.~Vernizzi,
  Healthy theories beyond Horndeski,
  Phys.\ Rev.\ Lett.\  {\bf 114} (2015) 211101
  \href{https://arxiv.org/pdf/1404.6495.pdf}{[arXiv:1404.6495 [hep-th]]}.

  \bibitem{Gleyzes:2014qga}
  J.~Gleyzes, D.~Langlois, F.~Piazza and F.~Vernizzi,
  Exploring gravitational theories beyond Horndeski,
  JCAP {\bf 1502} (2015) 018
  \href{https://arxiv.org/pdf/1408.1952.pdf}{[arXiv:1408.1952 [astro-ph.CO]]}.


    \bibitem{Koyama:2013paa}
  K.~Koyama, G.~Niz and G.~Tasinato,
  Effective theory for the Vainshtein mechanism from the Horndeski action,
  Phys.\ Rev.\ D {\bf 88} (2013) 021502
 \href{https://arxiv.org/pdf/1305.0279.pdf}{[arXiv:1305.0279 [hep-th]]}.


\bibitem{DeFelice:2011th}
  A.~De Felice, R.~Kase and S.~Tsujikawa,
  Vainshtein mechanism in second-order scalar-tensor theories,
  Phys.\ Rev.\ D {\bf 85} (2012) 044059
  \href{https://arxiv.org/pdf/1111.5090.pdf}{[arXiv:1111.5090 [gr-qc]]}.

    \bibitem{Deffayet:2009wt}
  C.~Deffayet, G.~Esposito-Farese and A.~Vikman,
  Covariant Galileon,
  Phys.\ Rev.\ D {\bf 79} (2009) 084003
 \href{https://arxiv.org/pdf/0901.1314.pdf}{[arXiv:0901.1314 [hep-th]]}.

  \bibitem{Koyama:2015oma}
  K.~Koyama and J.~Sakstein,
  Astrophysical Probes of the Vainshtein Mechanism: Stars and Galaxies,
  Phys.\ Rev.\ D {\bf 91} (2015) 124066
  \href{https://arxiv.org/pdf/1502.06872.pdf}{[arXiv:1502.06872 [astro-ph.CO]]}.

   \bibitem{Kobayashi:2014ida}
  T.~Kobayashi, Y.~Watanabe and D.~Yamauchi,
  Breaking of Vainshtein screening in scalar-tensor theories beyond Horndeski,
  Phys.\ Rev.\ D {\bf 91} (2015) 064013
  \href{https://arxiv.org/pdf/1411.4130.pdf}{[arXiv:1411.4130 [gr-qc]]}.


\bibitem{Saito:2015fza}
  R.~Saito, D.~Yamauchi, S.~Mizuno, J.~Gleyzes and D.~Langlois,
  Modified gravity inside astrophysical bodies,
  JCAP {\bf 1506} (2015) 008
 \href{https://arxiv.org/pdf/1503.01448.pdf}{[arXiv:1503.01448 [gr-qc]]}.

 \bibitem{Wibisono:2017dkt}
  C.~Wibisono and A.~Sulaksono,
  Information-entropic method for studying the stability bound of nonrelativistic polytropic stars within modified gravity theories,
  Int.\ J.\ Mod.\ Phys.\ D {\bf 27} (2018) 1850051
  \href{https://arxiv.org/pdf/1712.07587.pdf}{[arXiv:1712.07587 [gr-qc]]}.

  \bibitem{Sakstein:2015zoa}
  J.~Sakstein,
  Hydrogen Burning in Low Mass Stars Constrains Scalar-Tensor Theories of Gravity,
  Phys.\ Rev.\ Lett.\  {\bf 115} (2015) 201101
  \href{https://arxiv.org/pdf/1510.05964.pdf}{[arXiv:1510.05964 [astro-ph.CO]]}.

  \bibitem{Paxton:2010ji}
  B.~Paxton, L.~Bildsten, A.~Dotter, F.~Herwig, P.~Lesaffre and F.~Timmes,
  Modules for Experiments in Stellar Astrophysics (MESA),
  Astrophys.\ J.\ Suppl.\  {\bf 192} (2011) 3
  \href{https://arxiv.org/pdf/1009.1622.pdf}{[arXiv:1009.1622 [astro-ph.SR]]}.


  \bibitem{Jain:2015edg}
  R.~K.~Jain, C.~Kouvaris and N.~G.~Nielsen,
  White Dwarf Critical Tests for Modified Gravity,
  Phys.\ Rev.\ Lett.\  {\bf 116} (2016) 151103
  \href{https://arxiv.org/pdf/1512.05946.pdf}{[arXiv:1512.05946 [astro-ph.CO]]}.


  \bibitem{Saltas:2018mxc}
  I.~D.~Saltas, I.~Sawicki and I.~Lopes,
  White dwarfs and revelations,
  JCAP {\bf 1805} (2018) 028
  \href{https://arxiv.org/pdf/1803.00541.pdf}{[arXiv:1803.00541 [astro-ph.CO]]}.

  \bibitem{KoesterChan}
  D. Koester and G. Chanmugam,
  REVIEW: Physics of white dwarf stars,
  Reports on Progress in Physics \textbf{53} (1990) 837.

\bibitem{Salpeter}
E.~E.~Salpeter,
  Energy and Pressure of a Zero-Temperature Plasma,
  Astrophys.\ J.\  {\bf 134} (1961) 669.

  \bibitem{Parsons et al}
    S. G. Parsons et al.
    Testing the white dwarf mass-radius relationship with eclipsing binaries,
    Mon. Not. Roy. Astron. Soc. \textbf{470} (2017) 4473 \href{https://arxiv.org/pdf/1706.05016.pdf}{[arXiv:1706.05016 [astro-ph.SR]]}.

  \bibitem{Babichev:2016jom}
  E.~Babichev, K.~Koyama, D.~Langlois, R.~Saito and J.~Sakstein,
  Relativistic Stars in Beyond Horndeski Theories,
  Class.\ Quant.\ Grav.\  {\bf 33} (2016)  235014
  \href{https://arxiv.org/pdf/1606.06627.pdf}{[arXiv:1606.06627 [gr-qc]]}.


  \bibitem{Chowdhury:2018qrf}
  S.~Chowdhury and T.~Sarkar,
  Small Anisotropy in Stellar Objects in Modified Theories of Gravity,
  \href{https://arxiv.org/pdf/1811.07685.pdf}{[arXiv:1811.07685 [astro-ph.SR]]}.


  \bibitem{HeinHil}
  H. Heintzmann, W. Hillebrandt,
  Neutron stars with an anisotropic equation of state - Mass, redshift and stability
	Astronomy and Astrophysics, \textbf{38} (1975) 55.
	
	\bibitem{Cermeno:2018qed}
  M.~Cermeno, J.~Carro, A.~L.~Maroto and M.~A.~Perez-Garcia,
  Modified Gravity at Astrophysical Scales, Astrophys.\ J.\  {\bf 872} (2019) 130
  \href{https://arxiv.org/pdf/1811.11171.pdf}{[arXiv:1811.11171 [astro-ph.SR]]}.

  \bibitem{Hachisu}
I.~Hachisu and M.~Kato,
  A theoretical light-curve model for the recurrent nova v394 coronae austrinae,
  Astrophys.\ J.\  {\bf 540} (2000) 447
  \href{https://arxiv.org/pdf/astro-ph/0003471.pdf}{[astro-ph/0003471]}.

    \bibitem{Banerjee:2017uwz}
  S.~Banerjee, S.~Shankar and T.~P.~Singh,
  Constraints on Modified Gravity Models from White Dwarfs,
  JCAP {\bf 1710} (2017)  004
  \href{https://arxiv.org/pdf/1705.01048.pdf}{[arXiv:1705.01048 [gr-qc]]}.

  \bibitem{Stelle:1977ry}
  K.~S.~Stelle,
  Classical Gravity with Higher Derivatives,
  Gen.\ Rel.\ Grav.\  {\bf 9} (1978) 353.


  \bibitem{Moffat:2005si}
  J.~W.~Moffat,
  Scalar-tensor-vector gravity theory,
  JCAP {\bf 0603} (2006) 004
  \href{https://arxiv.org/pdf/gr-qc/0506021.pdf}{[gr-qc/0506021]}.

  \bibitem{Sun:2019niy}
  X.~Sun and S.~Y.~Zhou,
  Relativistic stars in mass-varying massive gravity, Phys. Rev. D \textbf{101} (2020) 044060
  \href{https://arxiv.org/pdf/1912.00685.pdf}{[arXiv:1912.00685 [gr-qc]]}.

\bibitem{Kalita:2019yaj}
  S.~Kalita, B.~Mukhopadhyay and T.~R.~Govindarajan,
  Violation of Chandrasekhar mass-limit in noncommutative geometry: A strong possible explanation for the super-Chandrasekhar limiting mass white dwarfs,
  \href{https://arxiv.org/pdf/1912.00900.pdf}{[arXiv:1912.00900 [gr-qc]]}.

  \bibitem{Howell:2006vn}
  D.~A.~Howell {\it et al.} [SNLS Collaboration],
  The type Ia supernova SNLS-03D3bb from a super-Chandrasekhar-mass white dwarf star,
  Nature {\bf 443} (2006) 308
  \href{https://arxiv.org/pdf/astro-ph/0609616.pdf}{[astro-ph/0609616]}.

  \bibitem{Holberg:2012pu}
  J.~B.~Holberg, T.~D.~Oswalt and M.~A.~Barstow,
  Observational Constraints on the Degenerate Mass-Radius Relation,
  Astron.\ J.\  {\bf 143} (2012) 68
  \href{https://arxiv.org/ftp/arxiv/papers/1201/1201.3822.pdf}{[arXiv:1201.3822 [astro-ph.SR]]}.

  \bibitem{Carvalho:2017pgk}
  G.~A.~Carvalho, R.~V.~Lobato, P.~H.~R.~S.~Moraes, J.~D.~V.~Arbañil, R.~M.~Marinho, E.~Otoniel and M.~Malheiro,
  Stellar equilibrium configurations of white dwarfs in the f(R, T) gravity,
  Eur.\ Phys.\ J.\ C {\bf 77} (2017)  871
  \href{https://arxiv.org/pdf/1706.03596.pdf}{[arXiv:1706.03596 [gr-qc]]}.

  \bibitem{Vennes}
  S. Vennes, P.A. Thejll, R.G. Galvan, J. Dupuis, Hot White Dwarfs in the Extreme Ultraviolet Explorer Survey. II. Mass Distribution, Space Density, and Population Age, Astrophys.
J. \textbf{480} (1997) 714.


  \bibitem{Crisostomi:2019yfo}
  M.~Crisostomi, M.~Lewandowski and F.~Vernizzi,
 Vainshtein regime in Scalar-Tensor gravity: constraints on DHOST theories,
  Phys.\ Rev.\ D {\bf 100} (2019) 024025
  \href{https://arxiv.org/pdf/1903.11591.pdf}{[arXiv:1903.11591 [gr-qc]]}.


  \bibitem{Garay:1994en}
  L.~J.~Garay,
  Quantum gravity and minimum length,
  Int.\ J.\ Mod.\ Phys.\ A {\bf 10} (1995) 145
  \href{https://arxiv.org/pdf/gr-qc/9403008.pdf}{[gr-qc/9403008]}.

  \bibitem{Rashidi:2015rro}
  R.~Rashidi,
  Generalized uncertainty principle and the maximum mass of ideal white dwarfs,
  Annals Phys.\  {\bf 374} (2016) 434
  \href{https://arxiv.org/pdf/1512.06356.pdf}{[arXiv:1512.06356 [gr-qc]]}.

  \bibitem{Ong:2018nzk}
  Y.~C.~Ong and Y.~Yao,
  Generalized Uncertainty Principle and White Dwarfs Redux: How Cosmological Constant Protects Chandrasekhar Limit,
  Phys.\ Rev.\ D {\bf 98} (2018) 126018
  \href{https://arxiv.org/pdf/1809.06348.pdf}{[arXiv:1809.06348 [gr-qc]]}.

  \bibitem{Mathew:2017drw}
  A.~Mathew and M.~K.~Nandy,
  Effect of minimal length uncertainty on the mass–radius relation of white dwarfs,
  Annals Phys.\  {\bf 393} (2018) 184
  \href{https://arxiv.org/pdf/1712.03953}{[arXiv:1712.03953 [gr-qc]]}.
  
  \bibitem{Mathew:2020wnx}
A.~Mathew and M.~K.~Nandy,
Existence of Chandrasekhar's limit in GUP white dwarfs,
\href{https://arxiv.org/pdf/2002.08360.pdf}{[arXiv:2002.08360 [gr-qc]]}.


  \bibitem{MacielBook}
  W. J. Maciel, \emph{Energy Production. In: Introduction to
Stellar Structure} (Springer, 2016).

\bibitem{Sakstein:2015aac}
  J.~Sakstein,
  Testing Gravity Using Dwarf Stars,
  Phys.\ Rev.\ D {\bf 92} (2015) 124045
  \href{https://arxiv.org/pdf/1511.01685.pdf}{[arXiv:1511.01685 [astro-ph.CO]]}.
  
  \bibitem{Spiegel:2010ju}
D.~S.~Spiegel, A.~Burrows and J.~A.~Milsom,
The Deuterium-Burning Mass Limit for Brown Dwarfs and Giant Planets,
Astrophys. J. \textbf{727} (2011) 57
\href{https://arxiv.org/pdf/1008.5150.pdf}{[arXiv:1008.5150 [astro-ph.EP]]}.

\bibitem{Rosyadi:2019hdb}
A.~Rosyadi, A.~Sulaksono, H.~Kassim and N.~Yusof,
Brown dwarfs in Eddington-inspired Born-Infeld and beyond Horndeski theories,
Eur. Phys. J. C \textbf{79} (2019) 1030.

\bibitem{Bayliss2016}
D. Bayliss et al., Astrophys. J \textbf{153} (2016) 1.


\bibitem{Kumarnum}
S. S. Kumar, The Structure of Stars of Very Low Mass, Astrophysical Journal \textbf{137} (1963) 1121.

\bibitem{Auddy}

S. Auddy, S. Basu, and S. R. Valluri, Analytic Models of Brown Dwarfs and the Substellar Mass Limit, Adv. Astron. 2016, 5743272 (2016).


    \bibitem{Segransan:2000jq}
  D.~Segransan, X.~Delfosse, T.~Forveille, J.-L.~Beuzit, S.~Udry, C.~Perrier and M.~Mayor,
  Accurate masses of very low mass stars. 3. 16 New or improved masses,
  Astron.\ Astrophys.\  {\bf 364} (2000) 665
  \href{https://arxiv.org/pdf/astro-ph/0010585.pdf}{[arXiv:astro-ph/0010585]}.

  \bibitem{Saumon:1995bn}
  D.~Saumon, W.~B.~Hubbard, A.~Burrows, T.~Guillot, J.~I.~Lunine and G.~Chabrier,
  A Theory of extrasolar giant planets,
  Astrophys.\ J.\  {\bf 460} (1996) 993
  \href{https://arxiv.org/pdf/astro-ph/9510046.pdf}{[arXiv:astro-ph/9510046]}.


\bibitem{Brown:2018dum}
  A.~G.~A.~Brown {\it et al.} [Gaia Collaboration],
  Gaia Data Release 2 : Summary of the contents and survey properties,
  Astron.\ Astrophys.\  {\bf 616} (2018) A1
  \href{https://arxiv.org/pdf/1804.09365.pdf}{[arXiv:1804.09365 [astro-ph.GA]]}.



   \bibitem{Sakstein:2016lyj}
  J.~Sakstein, M.~Kenna-Allison and K.~Koyama,
  Stellar Pulsations in Beyond Horndeski Gravity Theories,
  JCAP {\bf 1703} (2017) 007
 \href{https://arxiv.org/pdf/1611.01062.pdf}{[arXiv:1611.01062 [gr-qc]]}.

 \bibitem{Sakstein:2013pda}
  J.~Sakstein,
  Stellar Oscillations in Modified Gravity,
  Phys.\ Rev.\ D {\bf 88} (2013)  124013
  \href{https://arxiv.org/pdf/1309.0495.pdf}{[arXiv:1309.0495 [astro-ph.CO]]}.



  \bibitem{Sakstein:2014nfa}
  J.~Sakstein, B.~Jain and V.~Vikram,
  Detecting modified gravity in the stars,
  Int.\ J.\ Mod.\ Phys.\ D {\bf 23} (2014)  1442002
  \href{https://arxiv.org/pdf/1409.3708.pdf}{[arXiv:1409.3708 [astro-ph.CO]]}.

  \bibitem{Chang:2010xh}
  P.~Chang and L.~Hui,
  Stellar Structure and Tests of Modified Gravity,
  Astrophys.\ J.\  {\bf 732} (2011) 25
  \href{https://arxiv.org/pdf/1011.4107.pdf}{[arXiv:1011.4107 [astro-ph.CO]]}.


\bibitem{Bauswein:2013jpa}
  A.~Bauswein, T.~W.~Baumgarte and H.-T.~Janka,
  Prompt merger collapse and the maximum mass of neutron stars,
  Phys.\ Rev.\ Lett.\  {\bf 111} (2013) 131101 \href{https://arxiv.org/pdf/1307.5191.pdf}{[arXiv:1307.5191 [astro-ph.SR]]}.

 \bibitem{Annala:2017llu}
  E.~Annala, T.~Gorda, A.~Kurkela and A.~Vuorinen,
  Gravitational-wave constraints on the neutron-star-matter Equation of State,
  Phys.\ Rev.\ Lett.\  {\bf 120} (2018) 172703
	\href{https://arxiv.org/pdf/1711.02644.pdf}{[arXiv:1711.02644 [astro-ph.HE]]}.
	
	\bibitem{Radice:2017lry}
  D.~Radice, A.~Perego, F.~Zappa and S.~Bernuzzi,
  GW170817: Joint Constraint on the Neutron Star Equation of State from Multimessenger Observations,
  Astrophys.\ J.\  {\bf 852} (2018)  L29
  \href{https://arxiv.org/pdf/1711.03647.pdf}{[arXiv:1711.03647 [astro-ph.HE]]}.

  \bibitem{GM1}
N. K. Glendenning and S. A. Moszkowski, Reconciliation of neutron-star masses and binding of the $\Lambda$ in hypernuclei, Phys. Rev. Lett. \textbf{67} (1991) 2414.

\bibitem{FPS1}
V. R. Pandharipande, D. G. Ravenhall D.G. \emph{Hot nuclear matter in Nuclear Matter and Heavy Ion Collisions}, Eds. M.
Soyeur, H. Flocard, B. Tamain, and M. Porneuf,  (Dordrecht: Reidel), 103132 (1989). Dordrecht: Reidel), 103132 (1989).

\bibitem{FPS2}
B. Friedman and V. R. Pandharipande, Hot and cold, nuclear and neutron matter, Nucl. Phys. A \textbf{361} (1981) 502.

\bibitem{Lorenz:1992zz}
  C.~P.~Lorenz, D.~G.~Ravenhall and C.~J.~Pethick, Neutron star crusts,
  Phys.\ Rev.\ Lett.\  {\bf 70} (1993) 379.


  \bibitem{FPS3}
 J.~S.~Read, B.~D.~Lackey, B.~J.~Owen and J.~L.~Friedman,
  Constraints on a phenomenologically parameterized neutron-star equation of state,
  Phys.\ Rev.\ D {\bf 79} (2009) 124032
  \href{https://arxiv.org/pdf/0812.2163.pdf}{[arXiv:0812.2163 [astro-ph]]}.



\bibitem{SLy1}
E.~Chabanat, P.~Bonche, P.~Haensel, J.~Meyer and R.~Schaeffer,
  A Skyrme parametrization from subnuclear to neutron star densities. 2. Nuclei far from stablities,
  Nucl.\ Phys.\ A {\bf 635} (1998) 231
   Erratum: [Nucl.\ Phys.\ A {\bf 643} (1998) 441].

\bibitem{Sly2}
F.~Douchin and P.~Haensel,
  Inner edge of neutron star crust with SLY effective nucleon-nucleon interactions,
  Phys.\ Lett.\ B {\bf 485} (2000) 107
  \href{https://arxiv.org/pdf/astro-ph/0006135.pdf}{[arXiv:astro-ph/0006135]}.

\bibitem{Sly3}
F.~Douchin and P.~Haensel,
  A unified equation of state of dense matter and neutron star structure,
  Astron.\ Astrophys.\  {\bf 380} (2001) 151
  \href{https://arxiv.org/pdf/astro-ph/0111092.pdf}{[arXiv:astro-ph/0111092]}.

\bibitem{Sly4}
P.~Haensel and A.~Y.~Potekhin,
  Analytical representations of unified equations of state of neutron-star matter,
  Astron.\ Astrophys.\  {\bf 428} (2004) 191
  \href{https://arxiv.org/pdf/astro-ph/0408324.pdf}{[arXiv:astro-ph/0408324]}.

\bibitem{CamenzindBook}
M. Camenzind, \textit{Compact Objects in Astrophysics} (Springer, 2007).

  \bibitem{Lattimer12}
J. Lattimer, The Nuclear Equation of State and Neutron Star Masses, Annu. Rev. Nucl. Part. Sci. \textbf{62} (2012) 485.


  \bibitem{Akmal:1998cf}
  A.~Akmal, V.~R.~Pandharipande and D.~G.~Ravenhall,
  The Equation of state of nucleon matter and neutron star structure,
  Phys.\ Rev.\ C {\bf 58} (1998) 1804
  \href{https://arxiv.org/pdf/nucl-th/9804027.pdf}{[arXiv:nucl-th/9804027]}.



\bibitem{Wiringa:1988tp}
  R.~B.~Wiringa, V.~Fiks and A.~Fabrocini,
  Equation of state for dense nucleon matter,
  Phys.\ Rev.\ C {\bf 38} (1988) 1010.

  \bibitem{Goriely:2010bm}
  S.~Goriely, N.~Chamel and J.~M.~Pearson,
  Further explorations of Skyrme-Hartree-Fock-Bogoliubov mass formulas. XII: Stiffness and stability of neutron-star matter,
  Phys.\ Rev.\ C {\bf 82} (2010) 035804
  \href{https://arxiv.org/pdf/1009.3840.pdf}{[arXiv:1009.3840 [nucl-th]]}.

  \bibitem{Chamel:2011aa}
  N.~Chamel, A.~F.~Fantina, J.~M.~Pearson and S.~Goriely,
  Masses of neutron stars and nuclei,
  Phys.\ Rev.\ C {\bf 84} (2011) 062802
  \href{https://arxiv.org/pdf/1112.2878.pdf}{[arXiv:1112.2878 [nucl-th]]}.

\bibitem{Potekhin:2013qqa}
  A.~Y.~Potekhin, A.~F.~Fantina, N.~Chamel, J.~M.~Pearson and S.~Goriely,
  Analytical representations of unified equations of state for neutron-star matter,
  Astron.\ Astrophys.\  {\bf 560} (2013) A48 \href{https://arxiv.org/pdf/1310.0049.pdf}{[arXiv:1310.0049 [astro-ph.SR]]}.



\bibitem{GM1nph}
M.~Oertel, C.~Providência, F.~Gulminelli and A.~R.~Raduta,
  Hyperons in neutron star matter within relativistic mean-field models,
  J.\ Phys.\ G {\bf 42} (2015) 075202
 \href{https://arxiv.org/pdf/1412.4545.pdf}{[arXiv:1412.4545 [nucl-th]]}.

\bibitem{MPA}
H. Muther, M. Prakash, T. L. Ainsworth,  The Equation of State of Nuclear Matter and Neutron Stars Properties, Phys.
Lett. B \textbf{199}, 469 (1987).


  \bibitem{Gungor:2011vq}
  C.~Gungor and K.~Y.~Eksi,
  Analytical Representation for Equations of State of Dense Matter,
  \href{https://arxiv.org/pdf/1108.2166.pdf}{[arXiv:1108.2166 [astro-ph.SR]]}.


  \bibitem{Mueller:1996pm}
  H.~Mueller and B.~D.~Serot,
  Relativistic mean field theory and the high density nuclear equation of state,
  Nucl.\ Phys.\ A {\bf 606} (1996) 508
  \href{https://arxiv.org/pdf/nucl-th/9603037.pdf}{[arXiv:nucl-th/9603037]}.

\bibitem{JL79}
R. L. Jaffe, F. E. Low, Phys. Rev. D. \textbf{19} (1979) 2105.

\bibitem{Bonanno:2011ch}
  L.~Bonanno and A.~Sedrakian,
  Composition and stability of hybrid stars with hyperons and quark color-superconductivity,
  Astron.\ Astrophys.\  {\bf 539} (2012) A16
  \href{https://arxiv.org/pdf/1108.0559.pdf}{[arXiv:1108.0559 [astro-ph.SR]]}.

\bibitem{Lattimer:2012nd}
  J.~M.~Lattimer,
  The nuclear equation of state and neutron star masses,
  Ann.\ Rev.\ Nucl.\ Part.\ Sci.\  {\bf 62} (2012) 485
  \href{https://arxiv.org/pdf/1305.3510.pdf}{[arXiv:1305.3510 [nucl-th]]}.

    \bibitem{Read:2008iy}
  J.~S.~Read, B.~D.~Lackey, B.~J.~Owen and J.~L.~Friedman,
  Constraints on a phenomenologically parameterized neutron-star equation of state,
  Phys.\ Rev.\ D {\bf 79} (2009) 124032
  \href{https://arxiv.org/pdf/0812.2163.pdf}{[arXiv:0812.2163 [astro-ph]]}.

  \bibitem{Yagi:2016bkt}
  K.~Yagi and N.~Yunes,
  Approximate Universal Relations for Neutron Stars and Quark Stars,
  Phys.\ Rept.\  {\bf 681} (2017) 1
  \href{https://arxiv.org/pdf/1608.02582.pdf}{[arXiv:1608.02582 [gr-qc]]}.

    \bibitem{Breu:2016ufb}
  C.~Breu and L.~Rezzolla,
  Maximum mass, moment of inertia and compactness of relativistic stars,
  Mon.\ Not.\ Roy.\ Astron.\ Soc.\  {\bf 459} (2016) 646  \href{https://arxiv.org/pdf/1601.06083.pdf}{[arXiv:1601.06083 [gr-qc]]}.

\bibitem{Oter:2019kig}
  E.~L.~Oter, A.~Windisch, F.~J.~Llanes-Estrada and M.~Alford,
  nEoS: Neutron Star Equation of State from hadron physics alone, J.\ Phys.\ G {\bf 46} (2019) 084001
 \href{https://arxiv.org/pdf/1901.05271.pdf}{[arXiv:1901.05271 [gr-qc]]}.


  \bibitem{Damour:1993hw}
  T.~Damour and G.~Esposito-Farese,
  Nonperturbative strong field effects in tensor - scalar theories of gravitation,
  Phys.\ Rev.\ Lett.\  {\bf 70} (1993) 2220.

   \bibitem{Podkowka:2018gib}
  D.~M.~Podkowka, R.~F.~P.~Mendes and E.~Poisson,
  Trace of the energy-momentum tensor and macroscopic properties of neutron stars,
  Phys.\ Rev.\ D {\bf 98} (2018)  064057
  \href{https://arxiv.org/pdf/1807.01565.pdf}{[arXiv:1807.01565 [gr-qc]]}.


 \bibitem{Cooney:2009rr}
  A.~Cooney, S.~DeDeo and D.~Psaltis,
  Neutron Stars in f(R) Gravity with Perturbative Constraints,
  Phys.\ Rev.\ D {\bf 82} (2010) 064033 \href{https://arxiv.org/pdf/0910.5480.pdf}{[arXiv:0910.5480 [astro-ph.HE]]}.

\bibitem{Orellana:2013gn}
  M.~Orellana, F.~Garcia, F.~A.~Teppa Pannia and G.~E.~Romero,
  Structure of neutron stars in $R$-squared gravity,
  Gen.\ Rel.\ Grav.\  {\bf 45} (2013) 771 \href{https://arxiv.org/pdf/1301.5189v1.pdf}{[arXiv:1301.5189 [astro-ph.CO]]}.

  \bibitem{Chatterjee:2018prm}
  D.~Chatterjee, J.~Novak and M.~Oertel,
  Magnetic field distribution in magnetars, Phys.\ Rev.\ C {\bf 99} (2019) 055811
  \href{https://arxiv.org/pdf/1808.01778.pdf}{[arXiv:1808.01778 [nucl-th]]}.



  \bibitem{Ryu:2010zzb}
  C.~Y.~Ryu, K.~S.~Kim and M.~K.~Cheoun,
  Medium effects of magnetic moments of baryons on neutron stars under strong magnetic fields,
  Phys.\ Rev.\ C {\bf 82} (2010) 025804.

  \bibitem{Cheoun:2013tsa}
  M.~K.~Cheoun, C.~Deliduman, C.~Güngör, V.~Keleş, C.~Y.~Ryu, T.~Kajino and G.~J.~Mathews,
  Neutron stars in a perturbative $f(R)$ gravity model with strong magnetic fields,
  JCAP {\bf 1310} (2013) 021
  \href{https://arxiv.org/pdf/1304.1871.pdf}{[arXiv:1304.1871 [astro-ph.HE]]}.

\bibitem{Yazadjiev:2014cza}
  S.~S.~Yazadjiev, D.~D.~Doneva, K.~D.~Kokkotas and K.~V.~Staykov,
  Non-perturbative and self-consistent models of neutron stars in R-squared gravity,
  JCAP {\bf 1406} (2014) 003 [\href{https://arxiv.org/pdf/1402.4469v2.pdf}{arXiv:1402.4469 [gr-qc]}].


  \bibitem{Arapoglu:2010rz}
  A.~S.~Arapoglu, C.~Deliduman and K.~Y.~Eksi,
  Constraints on Perturbative f(R) Gravity via Neutron Stars,
  JCAP {\bf 1107} (2011) 020 \href{https://arxiv.org/pdf/1003.3179.pdf}{[arXiv:1003.3179 [gr-qc]]}.

  \bibitem{Hebeler:2013nza}
  K.~Hebeler, J.~M.~Lattimer, C.~J.~Pethick and A.~Schwenk,
  Equation of state and neutron star properties constrained by nuclear physics and observation,
  Astrophys.\ J.\  {\bf 773} (2013) 11
  \href{https://arxiv.org/pdf/1303.4662.pdf}{[arXiv:1303.4662 [astro-ph.SR]]}.

    \bibitem{Astashenok:2017dpo}
   A.~V.~Astashenok, S.~D.~Odintsov and A.~de la Cruz-Dombriz,
  The realistic models of relativistic stars in $f(R) = R + \alpha R^2$ gravity,
  Class.\ Quant.\ Grav.\  {\bf 34} (2017)  205008 \href{https://arxiv.org/pdf/1704.08311.pdf}{[arXiv:1704.08311 [gr-qc]]}.



  \bibitem{Sbisa:2019mae}
  F.~Sbisà, P.~O.~Baqui, T.~Miranda, S.~E.~Jor\'as and O.~F.~Piattella,
  Neutron star masses in $R^{2}$-gravity, Phys.\ Dark Univ.\ C {\bf 27} (2020) 100411
  \href{https://arxiv.org/pdf/1907.08714.pdf}{[arXiv:1907.08714 [gr-qc]]}.

  \bibitem{Astashenok:2018iav}
  A.~V.~Astashenok, A.~S.~Baigashov and S.~A.~Lapin,
  Neutron Stars in frames of $R^{2}$-gravity and Gravitational Waves,
  Int.\ J.\ Geom.\ Meth.\ Mod.\ Phys.\  {\bf 16} (2018) 1950004
  \href{https://arxiv.org/pdf/1812.10439.pdf}{[arXiv:1812.10439 [gr-qc]]}.

  \bibitem{Miyatsu:2013hea}
  T.~Miyatsu, S.~Yamamuro and K.~Nakazato,
  A new equation of state for neutron star matter with nuclei in the crust and hyperons in the core,
  Astrophys.\ J.\  {\bf 777} (2013) 4
  \href{https://arxiv.org/pdf/1308.6121.pdf}{[arXiv:1308.6121 [astro-ph.HE]]}.

  \bibitem{Mansour:2017ovd}
  H.~Mansour, B.~S.~Lakhal and A.~Yanallah,
  Weakly Charged Compact Stars in $f \left( R \right)$ gravity,
  JCAP {\bf 1806} (2018) 006
  \href{https://arxiv.org/pdf/1710.09294.pdf}{[arXiv:1710.09294 [gr-qc]]}.


  \bibitem{Arbanil:2013pua}
  J.~D.~V.~Arbañil, J.~P.~S.~Lemos and V.~T.~Zanchin,
  Polytropic spheres with electric charge: compact stars, the Oppenheimer-Volkoff and Buchdahl limits, and quasiblack holes,
  Phys.\ Rev.\ D {\bf 88} (2013) 084023
  \href{https://arxiv.org/pdf/1309.4470.pdf}{[arXiv:1309.4470 [gr-qc]]}.

  \bibitem{Folomeev:2018ioy}
  V.~Folomeev,
  Anisotropic neutron stars in $R^2$ gravity,
  Phys.\ Rev.\ D {\bf 97} (2018) 124009
  \href{https://arxiv.org/pdf/1802.01801.pdf}{[arXiv:1802.01801 [gr-qc]]}.

  \bibitem{Horvat:2010xf}
   D.~Horvat, S.~Ilijic and A.~Marunovic,
  Radial pulsations and stability of anisotropic stars with quasi-local equation of state,
  Class.\ Quant.\ Grav.\  {\bf 28} (2011) 025009
 \href{https://arxiv.org/pdf/1010.0878.pdf}{[arXiv:1010.0878 [gr-qc]]}.

  \bibitem{Bowers:1974tgi}
  R.~L.~Bowers and E.~P.~T.~Liang,
  Anisotropic Spheres in General Relativity,
  Astrophys.\ J.\  {\bf 188} (1974) 657.

  \bibitem{Feola:2019zqg}
  P.~Feola, X.~J.~Forteza, S.~Capozziello, R.~Cianci and S.~Vignolo,
  The mass-radius relation for neutron stars in $f(R)=R+\alpha R^2$ gravity: a comparison between purely metric and torsion formulations, Phys. Rev. D \textbf{101} (2020) 044037
  \href{https://arxiv.org/pdf/1909.08847.pdf}{[arXiv:1909.08847 [astro-ph.HE]]}.


  \bibitem{Astashenok:2013vza}
  A.~V.~Astashenok, S.~Capozziello and S.~D.~Odintsov,
  Further stable neutron star models from f(R) gravity,
  JCAP {\bf 1312} (2013) 040 \href{https://arxiv.org/pdf/1309.1978v2.pdf}{[arXiv:1309.1978 [gr-qc]]}.

\bibitem{Cognola08}
G.~Cognola, E.~Elizalde, S.~Nojiri, S.~D.~Odintsov, L.~Sebastiani and S.~Zerbini,
  A Class of viable modified f(R) gravities describing inflation and the onset of accelerated expansion,
  Phys.\ Rev.\ D {\bf 77} (2008) 046009 \href{https://arxiv.org/pdf/0712.4017.pdf}{[arXiv:0712.4017 [hep-th]]}.

\bibitem{Camezind}
M. Camenzind, \emph{Compact Objects in Astrophysics} (Springer, 2007).

    \bibitem{Alavirad:2013paa}
  H.~Alavirad and J.~M.~Weller,
  Modified gravity with logarithmic curvature corrections and the structure of relativistic stars,
  Phys.\ Rev.\ D {\bf 88} (2013) 124034
  \href{https://arxiv.org/pdf/1307.7977.pdf}{[arXiv:1307.7977 [gr-qc]]}.


  \bibitem{Capozziello:2015yza}
  S.~Capozziello, M.~De Laurentis, R.~Farinelli and S.~D.~Odintsov,
  Mass-radius relation for neutron stars in f(R) gravity,
  Phys.\ Rev.\ D {\bf 93} (2016) 023501 \href{https://arxiv.org/pdf/1509.04163v3.pdf}{[arXiv:1509.04163 [gr-qc]]}.



  \bibitem{Astashenok:2014pua}
  A.~V.~Astashenok, S.~Capozziello and S.~D.~Odintsov,
  Maximal neutron star mass and the resolution of the hyperon puzzle in modified gravity,
  Phys.\ Rev.\ D {\bf 89} (2014) 103509
  \href{https://arxiv.org/pdf/1401.4546.pdf}{[arXiv:1401.4546 [gr-qc]]}.



  \bibitem{Astashenok:2014gda}
  A.~V.~Astashenok, S.~Capozziello and S.~D.~Odintsov,
  Magnetic Neutron Stars in f(R) gravity,
  Astrophys.\ Space Sci.\  {\bf 355} (2015)  333
  \href{https://arxiv.org/pdf/1405.6663.pdf}{[arXiv:1405.6663 [gr-qc]]}.


\bibitem{Astashenok:2014nua}
  A.~V.~Astashenok, S.~Capozziello and S.~D.~Odintsov,
  Extreme neutron stars from Extended Theories of Gravity,
  JCAP {\bf 1501} (2015)  001
  \href{https://arxiv.org/pdf/1408.3856.pdf}{[arXiv:1408.3856 [gr-qc]]}.

  \bibitem{Clifton05}
T.~Clifton and J.~D.~Barrow,
  The Power of General Relativity,
  Phys.\ Rev.\ D {\bf 72} (2005),  103005 [Erratum: [Phys.\ Rev.\ D {\bf 90} (2014) 029902] \href{https://arxiv.org/pdf/gr-qc/0509059.pdf}{[arXiv:gr-qc/0509059]}.

     \bibitem{DeLaurentis:2018odx}
  M.~De Laurentis,
  Noether's stars in $f(\cal {R})$ gravity,
  Phys.\ Lett.\ B {\bf 780} (2018) 205
  \href{https://arxiv.org/pdf/1802.09073.pdf}{[arXiv:1802.09073 [gr-qc]]}.


  \bibitem{HS07}
 W.~Hu and I.~Sawicki,
  Models of f(R) Cosmic Acceleration that Evade Solar-System Tests,
  Phys.\ Rev.\ D {\bf 76} (2007) 064004
  \href{https://arxiv.org/pdf/0705.1158.pdf}{[arXiv:0705.1158 [astro-ph]]}.

  \bibitem{Tsujikawa:2007xu}
  S.~Tsujikawa,
  Observational signatures of f(R) dark energy models that satisfy cosmological and local gravity constraints,
  Phys.\ Rev.\ D {\bf 77} (2008) 023507
  \href{https://arxiv.org/pdf/0709.1391.pdf}{[arXiv:0709.1391 [astro-ph]]}.


  \bibitem{Kase:2019dqc}
  R.~Kase and S.~Tsujikawa,
  Neutron stars in $f(R)$ gravity and scalar-tensor theories, JCAP {\bf 1909} (2019)  054
  \href{https://arxiv.org/pdf/1906.08954.pdf}{[arXiv:1906.08954 [gr-qc]]}.




    \bibitem{Horbatsch:2010hj}
  M.~W.~Horbatsch and C.~P.~Burgess,
  Semi-Analytic Stellar Structure in Scalar-Tensor Gravity,
  JCAP {\bf 1108} (2011) 027
  \href{https://arxiv.org/pdf/1006.4411.pdf}{[arXiv:1006.4411 [gr-qc]]}.


  \bibitem{Cisterna:2015yla}
  A.~Cisterna, T.~Delsate and M.~Rinaldi,
  Neutron stars in general second order scalar-tensor theory: The case of nonminimal derivative coupling,
  Phys.\ Rev.\ D {\bf 92} (2015)  044050
  \href{https://arxiv.org/pdf/1504.05189.pdf}{[arXiv:1504.05189 [gr-qc]]}.


  \bibitem{Charmousis:2011bf}
  C.~Charmousis, E.~J.~Copeland, A.~Padilla and P.~M.~Saffin,
  General second order scalar-tensor theory, self tuning, and the Fab Four,
  Phys.\ Rev.\ Lett.\  {\bf 108} (2012) 051101
  \href{https://arxiv.org/pdf/1106.2000.pdf}{[arXiv:1106.2000 [hep-th]]}.

\bibitem{Lattimer:2000nx}
  J.~M.~Lattimer and M.~Prakash,
  Neutron star structure and the equation of state,
  Astrophys.\ J.\  {\bf 550} (2001) 426
  \href{https://arxiv.org/pdf/astro-ph/0002232.pdf}{[arXiv:astro-ph/0002232]}.

   \bibitem{Cheong:2018gzn}
  P.~C.~K.~Cheong and T.~G.~F.~Li,
  Numerical Studies on Core Collapse Supernova in Self-interacting Massive Scalar-Tensor Gravity, Phys.\ Rev.\ D {\bf 100} (2019) 024027
  \href{https://arxiv.org/pdf/1812.04835.pdf}{[arXiv:1812.04835 [gr-qc]]}.



    \bibitem{Silva:2014ora}
  H.~O.~Silva, H.~Sotani, E.~Berti and M.~Horbatsch,
  Torsional oscillations of neutron stars in scalar-tensor theory of gravity,
  Phys.\ Rev.\ D {\bf 90} (2014)  124044
  \href{https://arxiv.org/pdf/1410.2511.pdf}{[arXiv:1410.2511 [gr-qc]]}.

   \bibitem{Harada:1998ge}
  T.~Harada,
  Neutron stars in scalar tensor theories of gravity and catastrophe theory,
  Phys.\ Rev.\ D {\bf 57} (1998) 4802
  \href{https://arxiv.org/pdf/gr-qc/9801049.pdf}{[arXiv:gr-qc/9801049]}.

   \bibitem{Doneva:2018ouu}
  D.~D.~Doneva, S.~S.~Yazadjiev, N.~Stergioulas and K.~D.~Kokkotas,
  Differentially rotating neutron stars in scalar-tensor theories of gravity, Phys.\ Rev.\ D {\bf 98} (2018) 104039
  \href{https://arxiv.org/pdf/1807.05449.pdf}{[arXiv:1807.05449 [gr-qc]]}.


      \bibitem{Palenzuela:2015ima}
  C.~Palenzuela and S.~L.~Liebling,
  Constraining scalar-tensor theories of gravity from the most massive neutron stars,
  Phys.\ Rev.\ D {\bf 93} (2016)  044009
  \href{https://arxiv.org/pdf/1510.03471.pdf}{[arXiv:1510.03471 [gr-qc]]}.


  \bibitem{Mendes:2016fby}
  R.~F.~P.~Mendes and N.~Ortiz,
  Highly compact neutron stars in scalar-tensor theories of gravity: Spontaneous scalarization versus gravitational collapse,
  Phys.\ Rev.\ D {\bf 93} (2016) 124035
  \href{https://arxiv.org/pdf/1604.04175.pdf}{[arXiv:1604.04175 [gr-qc]]}.




  \bibitem{Kobyakov:2013eta}
  D.~Kobyakov and C.~J.~Pethick,
  Dynamics of the inner crust of neutron stars: hydrodynamics, elasticity and collective modes,
  Phys.\ Rev.\ C {\bf 87} (2013) 055803
   Erratum: [Phys.\ Rev.\ C {\bf 94} (2016) 059902]
  \href{https://arxiv.org/pdf/1303.1315.pdf}{[arXiv:1303.1315 [nucl-th]]}.


\bibitem{Freire:2012mg}
  P.~C.~C.~Freire {\it et al.},
  The relativistic pulsar-white dwarf binary PSR J1738+0333 II. The most stringent test of scalar-tensor gravity,
  Mon.\ Not.\ Roy.\ Astron.\ Soc.\  {\bf 423} (2012) 3328
 \href{https://arxiv.org/pdf/1205.1450.pdf}{[arXiv:1205.1450 [astro-ph.GA]]}.

  \bibitem{Sotani:2017pfj}
  H.~Sotani and K.~D.~Kokkotas,
  Maximum mass limit of neutron stars in scalar-tensor gravity,
  Phys.\ Rev.\ D {\bf 95} (2017) 044032
  \href{https://arxiv.org/pdf/1702.00874.pdf}{[arXiv:1702.00874 [gr-qc]]}.

  \bibitem{Oyamatsu:2006vd}
  K.~Oyamatsu and K.~Iida,
  The Symmetry energy at subnuclear densities and nuclei in neutron star crusts,
  Phys.\ Rev.\ C {\bf 75} (2007) 015801
  \href{https://arxiv.org/pdf/nucl-th/0609040.pdf}{[arXiv:nucl-th/0609040]}.


  \bibitem{Mendes:2018qwo}
  R.~F.~P.~Mendes and N.~Ortiz,
  New class of quasinormal modes of neutron stars in scalar-tensor gravity,
  Phys.\ Rev.\ Lett.\  {\bf 120} (2018) 201104
  \href{https://arxiv.org/pdf/1802.07847.pdf}{[arXiv:1802.07847 [gr-qc]]}.

   \bibitem{Watts:2006mr}
  A.~L.~Watts and T.~E.~Strohmayer,
  Neutron star oscillations and QPOs during magnetar flares,
  Adv.\ Space Res.\  {\bf 40} (2007) 1446
  \href{https://arxiv.org/pdf/astro-ph/0612252.pdf}{[arXiv:astro-ph/0612252]}.

  \bibitem{BaumShaShi}
  T. W. Baumgarte, S. L. Shapiro, and M. Shibata, The
Astrophysical Journal Letters \textbf{528} (2000) L29.


\bibitem{Hendi:2015pua}
  S.~H.~Hendi, G.~H.~Bordbar, B.~Eslam Panah and M.~Najafi,
  Dilatonic Equation of Hydrostatic Equilibrium and Neutron Star Structure,
  Astrophys.\ Space Sci.\  {\bf 358} (2015)  30
  \href{https://arxiv.org/pdf/1503.01011.pdf}{[arXiv:1503.01011 [gr-qc]]}.

\bibitem{Horbatsch:2015bua}
  M.~Horbatsch, H.~O.~Silva, D.~Gerosa, P.~Pani, E.~Berti, L.~Gualtieri and U.~Sperhake,
  Tensor-multi-scalar theories: relativistic stars and 3 + 1 decomposition,
  Class.\ Quant.\ Grav.\  {\bf 32} (2015)  204001
  \href{https://arxiv.org/pdf/1505.07462.pdf}{[arXiv:1505.07462 [gr-qc]]}.

  \bibitem{Novak:1997hw}
  J.~Novak,
  Spherical neutron star collapse in tensor - scalar theory of gravity,
  Phys.\ Rev.\ D {\bf 57} (1998) 4789
  \href{https://arxiv.org/pdf/gr-qc/9707041.pdf}{[gr-qc/9707041]}.

  \bibitem{Yazadjiev:2019oul}
  S.~S.~Yazadjiev and D.~D.~Doneva,
  Dark compact objects in massive tensor-multi-scalar theories of gravity,
  Phys.\ Rev.\ D {\bf 99} (2019)  084011
  \href{https://arxiv.org/pdf/1901.06379.pdf}{[arXiv:1901.06379 [gr-qc]]}.
  
  \bibitem{Doneva:2020afj}
D.~D.~Doneva and S.~S.~Yazadjiev,
Nontopological spontaneously scalarized neutron stars in tensor-multiscalar theories of gravity,
Phys. Rev. D \textbf{101} (2020)  104010
\href{https://arxiv.org/pdf/2004.03956.pdf}{[arXiv:2004.03956 [gr-qc]]}.

  \bibitem{Doneva:2019krb}
  D.~D.~Doneva and S.~S.~Yazadjiev,
  Mixed configurations of tensor-multi-scalar solitons and neutron stars, Phys. Rev. D \textbf{101} (2020)  024009
  \href{https://arxiv.org/pdf/1909.00473.pdf}{arXiv:1909.00473 [gr-qc]}.

 \bibitem{Doneva:2019ltb}
  D.~D.~Doneva and S.~S.~Yazadjiev,
  Topological neutron stars in tensor-multi-scalar theories of gravity, Phys. Rev. D \textbf{101} (2020)  064072
  \href{https://arxiv.org/pdf/1911.06908.pdf}{arXiv:1911.06908 [gr-qc]}.

 \bibitem{aneta1}
 A.~Wojnar, H.~Velten, Equilibrium and stability of relativistic stars in extended theories of gravity,
 Eur.\ Phys.\ J.\ C {\bf 76} (2016) 697
 \href{https://arxiv.org/pdf/1604.04257.pdf}{[arXiv:astro-ph/1604.04257]}.

\bibitem{mim}S. Capozziello, S., F. S. Lobo and J. P. Mimoso,  Energy conditions in modified gravity, Phys.\ Lett.\ B {\bf 730} (2014) 280
\href{https://arxiv.org/abs/1312.0784}{[arXiv:gr-qc/1312.0784]}

\bibitem{mim2} S. Capozziello, F. S. N. Lobo, J. P. Mimoso, Generalized energy conditions in Extended Theories of Gravity, Phys. Rev. D \textbf{91}  (2015) 124019
\href{https://arxiv.org/abs/1407.7293}{[arXiv:gr-qc/1407.7293]}.


\bibitem{mim3} J. P. Mimoso, F. S. Lobo and S. Capozziello, Extended Theories of Gravity with Generalized Energy Conditions, Journal of Physics: Conference Series Vol. \textbf{600} 012047 (2015), IOP Publishing,
\href{https://arxiv.org/abs/1412.6670}{[arXiv:gr-qc/1412.6670]}



    \bibitem{Kobayashi:2018xvr}
  T.~Kobayashi and T.~Hiramatsu, Relativistic stars in degenerate higher-order scalar-tensor theories after GW170817,
  Phys.\ Rev.\ D {\bf 97} (2018)  104012
\href{https://arxiv.org/pdf/1803.10510.pdf}{[arXiv:1803.10510 [gr-qc]]}.

\bibitem{Langlois:2017dyl}
  D.~Langlois, R.~Saito, D.~Yamauchi and K.~Noui,
  Scalar-tensor theories and modified gravity in the wake of GW170817,
  Phys.\ Rev.\ D {\bf 97} (2018) 061501
  \href{https://arxiv.org/pdf/1711.07403.pdf}{[arXiv:1711.07403 [gr-qc]]}.
  
  \bibitem{Saltas:2019ius}
I.~D.~Saltas and I.~Lopes,
Obtaining Precision Constraints on Modified Gravity with Helioseismology,
Phys. Rev. Lett. \textbf{123} (2019)  091103
\href{https://arxiv.org/pdf/1909.02552.pdf}{[arXiv:1909.02552 [astro-ph.CO]]}.

  \bibitem{deRham:2016wji}
  C.~de Rham and A.~Matas,
  Ostrogradsky in Theories with Multiple Fields,
  JCAP {\bf 1606} (2016)  041
  \href{https://arxiv.org/pdf/1604.08638.pdf}{[arXiv:1604.08638 [hep-th]]}.

  \bibitem{Crisostomi:2017pjs}
  M.~Crisostomi and K.~Koyama,
  Self-accelerating universe in scalar-tensor theories after GW170817,
  Phys.\ Rev.\ D {\bf 97} (2018)  084004
  \href{https://arxiv.org/pdf/1712.06556.pdf}{[arXiv:1712.06556 [astro-ph.CO]]}.

  \bibitem{Babichev:2013usa}
  E.~Babichev and C.~Deffayet,
  An introduction to the Vainshtein mechanism,
  Class.\ Quant.\ Grav.\  {\bf 30} (2013) 184001
  \href{https://arxiv.org/pdf/1304.7240.pdf}{[arXiv:1304.7240 [gr-qc]]}.



\bibitem{Sakstein:2016ggl}
  J.~Sakstein, H.~Wilcox, D.~Bacon, K.~Koyama and R.~C.~Nichol,
  Testing Gravity Using Galaxy Clusters: New Constraints on Beyond Horndeski Theories,
  JCAP {\bf 1607} (2016)  019
  \href{https://arxiv.org/pdf/1603.06368.pdf}{[arXiv:1603.06368 [astro-ph.CO]]}.

  \bibitem{Bettoni:2016mij}
  D.~Bettoni, J.~M.~Ezquiaga, K.~Hinterbichler and M.~Zumalacárregui,
  Speed of Gravitational Waves and the Fate of Scalar-Tensor Gravity,
  Phys.\ Rev.\ D {\bf 95} (2017) 084029
  \href{https://arxiv.org/pdf/1608.01982.pdf}{[arXiv:1608.01982 [gr-qc]]}.

  \bibitem{Chagoya:2018lmv}
  J.~Chagoya and G.~Tasinato,
  Compact objects in scalar-tensor theories after GW170817,
  JCAP {\bf 1808} (2018) 006
  \href{https://arxiv.org/pdf/1803.07476.pdf}{[arXiv:1803.07476 [gr-qc]]}.

  \bibitem{Chagoya:2017fyl}
  J.~Chagoya, G.~Niz and G.~Tasinato,
  Black Holes and Neutron Stars in Vector Galileons,
  Class.\ Quant.\ Grav.\  {\bf 34} (2017) 165002
  \href{https://arxiv.org/pdf/1703.09555.pdf}{[arXiv:1703.09555 [gr-qc]]}.

  \bibitem{Momeni:2016srq}
  D.~Momeni, M.~Faizal, K.~Myrzakulov and R.~Myrzakulov,
  Compact stars in vector–tensor-Horndeski theory of gravity,
  Eur.\ Phys.\ J.\ C {\bf 77} (2017)  37
  \href{https://arxiv.org/pdf/1606.05526.pdf}{[arXiv:1606.05526 [physics.gen-ph]]}.



  \bibitem{Pani:2011xm}
  P.~Pani, E.~Berti, V.~Cardoso and J.~Read,
  Compact stars in alternative theories of gravity. Einstein-Dilaton-Gauss-Bonnet gravity,
  Phys.\ Rev.\ D {\bf 84} (2011) 104035 \href{https://arxiv.org/pdf/1109.0928.pdf}{[arXiv:1109.0928 [gr-qc]]}.

  \bibitem{Panotopoulos:2019zxv}
  G.~Panotopoulos and Á.~Rincón,
  Relativistic strange quark stars in Lovelock gravity, Eur. Phys. J. Plus \textbf{134} (2019)  472
  \href{https://arxiv.org/pdf/1907.03545.pdf}{[arXiv:1907.03545 [gr-qc]]}.


  \bibitem{Gross:1986mw}
  D.~J.~Gross and J.~H.~Sloan,
  The Quartic Effective Action for the Heterotic String,
  Nucl.\ Phys.\ B {\bf 291} (1987) 41.

  \bibitem{Pani:2009wy}
  P.~Pani and V.~Cardoso,
  Are black holes in alternative theories serious astrophysical candidates? The Case for Einstein-Dilaton-Gauss-Bonnet black holes,
  Phys.\ Rev.\ D {\bf 79} (2009) 084031
 \href{https://arxiv.org/pdf/0902.1569.pdf}{[arXiv:0902.1569 [gr-qc]]}.



\bibitem{Silva:2017uqg}
  H.~O.~Silva, J.~Sakstein, L.~Gualtieri, T.~P.~Sotiriou and E.~Berti,
  Spontaneous scalarization of black holes and compact stars from a Gauss-Bonnet coupling,
  Phys.\ Rev.\ Lett.\  {\bf 120} (2018) 131104
  \href{https://arxiv.org/pdf/1711.02080.pdf}{[arXiv:1711.02080 [gr-qc]]}.

\bibitem{Doneva:2017duq}
  D.~D.~Doneva and S.~S.~Yazadjiev,
  Neutron star solutions with curvature induced scalarization in the extended Gauss-Bonnet scalar-tensor theories,
  JCAP {\bf 1804} (2018)  011
  \href{https://arxiv.org/pdf/1712.03715.pdf}{[arXiv:1712.03715 [gr-qc]]}.

    \bibitem{Blazquez-Salcedo:2015ets}
  J.~L.~Blázquez-Salcedo, L.~M.~González-Romero, J.~Kunz, S.~Mojica and F.~Navarro-Lérida,
  Axial quasinormal modes of Einstein-Gauss-Bonnet-dilaton neutron stars,
  Phys.\ Rev.\ D {\bf 93} (2016)  024052
  \href{https://arxiv.org/pdf/1511.03960.pdf}{[arXiv:1511.03960 [gr-qc]]}.

  \bibitem{Hinterbichler:2011tt}
K.~Hinterbichler,
Theoretical Aspects of Massive Gravity,
Rev. Mod. Phys. \textbf{84} (2012) 671
\href{https://arxiv.org/pdf/1105.3735.pdf}{[arXiv:1105.3735 [hep-th]]}.


  \bibitem{deRham:2014zqa}
  C.~de Rham,
  Massive Gravity,
  Living Rev.\ Rel.\  {\bf 17} (2014) 7
  \href{https://arxiv.org/pdf/1401.4173.pdf}{[arXiv:1401.4173 [hep-th]]}.

  \bibitem{Katsuragawa:2015lbl}
  T.~Katsuragawa, S.~Nojiri, S.~D.~Odintsov and M.~Yamazaki,
  Relativistic stars in de Rham-Gabadadze-Tolley massive gravity,
  Phys.\ Rev.\ D {\bf 93} (2016) 124013
  \href{https://arxiv.org/pdf/1512.00660.pdf}{[arXiv:1512.00660 [gr-qc]]}.

  \bibitem{Kareeso:2018xum}
  P.~Kareeso, P.~Burikham and T.~Harko,
  Mass - radius ratio bounds for compact objects in Massive Gravity theory, Eur.\ Phys.\ J.\ C {\bf 78} (2018)  941
  \href{https://arxiv.org/pdf/1802.01017.pdf}{[arXiv:1802.01017 [gr-qc]]}.

   \bibitem{Wiringa:1994wb}
  R.~B.~Wiringa, V.~G.~J.~Stoks and R.~Schiavilla,
  An Accurate nucleon-nucleon potential with charge independence breaking,
  Phys.\ Rev.\ C {\bf 51} (1995) 38
\href{https://arxiv.org/pdf/nucl-th/9408016.pdf}{[arXiv:nucl-th/9408016]}.

\bibitem{Stoks:1994wp}
  V.~G.~J.~Stoks, R.~A.~M.~Klomp, C.~P.~F.~Terheggen and J.~J.~de Swart,
  Construction of high quality N N potential models,
  Phys.\ Rev.\ C {\bf 49} (1994) 2950
  \href{https://arxiv.org/pdf/nucl-th/9406039.pdf}{[arXiv:nucl-th/9406039]}.

   \bibitem{Kase:2017egk}
  R.~Kase, M.~Minamitsuji and S.~Tsujikawa,
  Relativistic stars in vector-tensor theories,
  Phys.\ Rev.\ D {\bf 97} (2018) 084009
  \href{https://arxiv.org/pdf/1711.08713.pdf}{[arXiv:1711.08713 [gr-qc]]}.

  \bibitem{DeFelice:2016cri}
  A.~De Felice, L.~Heisenberg, R.~Kase, S.~Tsujikawa, Y.~l.~Zhang and G.~B.~Zhao,
  Screening fifth forces in generalized Proca theories,
  Phys.\ Rev.\ D {\bf 93} (2016) 104016
  \href{https://arxiv.org/pdf/1602.00371.pdf}{[arXiv:1602.00371 [gr-qc]]}.


  \bibitem{Heisenberg:2014rta}
  L.~Heisenberg,
  Generalization of the Proca Action,
  JCAP {\bf 1405} (2014) 015
  \href{https://arxiv.org/pdf/1402.7026.pdf}{[arXiv:1402.7026 [hep-th]]}.

  \bibitem{Jimenez:2016isa}
  J.~Beltran Jimenez and L.~Heisenberg,
  Derivative self-interactions for a massive vector field,
  Phys.\ Lett.\ B {\bf 757} (2016) 405
  \href{https://arxiv.org/pdf/1602.03410.pdf}{[arXiv:1602.03410 [hep-th]]}.

  \bibitem{Damour:1996ke}
  T.~Damour and G.~Esposito-Farese,
  Tensor - scalar gravity and binary pulsar experiments,
  Phys.\ Rev.\ D {\bf 54} (1996) 1474
  \href{https://arxiv.org/pdf/gr-qc/9602056.pdf}{[arXiv:gr-qc/9602056]}.

  \bibitem{Nicolis:2008in}
  A.~Nicolis, R.~Rattazzi and E.~Trincherini,
  The Galileon as a local modification of gravity,
  Phys.\ Rev.\ D {\bf 79} (2009) 064036
  \href{https://arxiv.org/pdf/0811.2197.pdf}{[arXiv:0811.2197 [hep-th]]}.

  \bibitem{Magueijo:2002xx}
  J.~Magueijo and L.~Smolin,
  Gravity's rainbow,
  Class.\ Quant.\ Grav.\  {\bf 21} (2004) 1725
  \href{https://arxiv.org/pdf/gr-qc/0305055.pdf}{[arXiv:gr-qc/0305055]}.

    \bibitem{Hendi:2015vta}
  S.~H.~Hendi, G.~H.~Bordbar, B.~E.~Panah and S.~Panahiyan,
  Modified TOV in gravity's rainbow: properties of neutron stars and dynamical stability conditions,
  JCAP {\bf 1609} (2016)  013
  \href{https://arxiv.org/pdf/1509.05145.pdf}{[arXiv:1509.05145 [hep-th]]}.



 \bibitem{Hendi:2017ibm}
  S.~H.~Hendi, G.~H.~Bordbar, B.~Eslam Panah and S.~Panahiyan,
  Neutron stars structure in the context of massive gravity,
  JCAP {\bf 1707} (2017) 004
  \href{https://arxiv.org/pdf/1701.01039.pdf}{[arXiv:1701.01039 [gr-qc]]}.


  \bibitem{EslamPanah:2018evk}
  B.~Eslam Panah and H.~L.~Liu,
  White dwarfs in massive gravity, Phys.\ Rev.\ D {\bf 99} (2019) 104074
  \href{https://arxiv.org/pdf/1805.10650.pdf}{[arXiv:1805.10650 [gr-qc]]}.


  \bibitem{Hassan:2011vm}
  S.~F.~Hassan and R.~A.~Rosen,
  On Non-Linear Actions for Massive Gravity,
  JHEP {\bf 1107} (2011) 009
  \href{https://arxiv.org/pdf/1103.6055.pdf}{[arXiv:1103.6055 [hep-th]]}.

  \bibitem{Enander:2015kda}
  J.~Enander and E.~Mortsell,
  On stars, galaxies and black holes in massive bigravity,
  JCAP {\bf 1511} (2015) 023
  \href{https://arxiv.org/pdf/1507.00912.pdf}{[arXiv:1507.00912 [astro-ph.CO]]}.

 \bibitem{Aoki:2016eov}
  K.~Aoki, K.~i.~Maeda and M.~Tanabe,
  Relativistic stars in bigravity theory,
    Phys.\ Rev.\ D {\bf 93} (2016) 064054
  \href{https://arxiv.org/pdf/1602.02227.pdf}{[arXiv:1602.02227 [gr-qc]]}.

  \bibitem{Huang:2012pe}
  Q.~G.~Huang, Y.~S.~Piao and S.~Y.~Zhou,
  Mass-Varying Massive Gravity,
  Phys.\ Rev.\ D {\bf 86} (2012) 124014
  \href{https://arxiv.org/pdf/1206.5678.pdf}{[arXiv:1206.5678 [hep-th]]}.



\bibitem{Maluf:2013gaa}
  J.~W.~Maluf,
  The teleparallel equivalent of general relativity,
  Annalen Phys.\  {\bf 525} (2013) 339
  \href{https://arxiv.org/pdf/1303.3897.pdf}{[arXiv:1303.3897 [gr-qc]]}.

\bibitem{Ilijic:2018ulf}
  S.~Ilijic and M.~Sossich,
  Compact stars in $f(T)$ extended theory of gravity,
  Phys.\ Rev.\ D {\bf 98} (2018) 064047
  \href{https://arxiv.org/pdf/1807.03068.pdf}{[arXiv:1807.03068 [gr-qc]]}.

  \bibitem{DeBenedictis:2016aze}
  A.~DeBenedictis and S.~Ilijic,
  Spherically symmetric vacuum in covariant $F(T) = T + \frac{\alpha}{2}T^{2} + \mathcal{O}(T^{\gamma})$ gravity theory,
  Phys.\ Rev.\ D {\bf 94} (2016)  124025
  \href{https://arxiv.org/pdf/1609.07465.pdf}{[arXiv:1609.07465 [gr-qc]]}.

  \bibitem{Deb:2018voz}
  D.~Deb, S.~Ghosh, S.~K.~Maurya, M.~Khlopov and S.~Ray,
  Anisotropic compact stars in $f(T)$ gravity under Karmarkar condition,
  \href{https://arxiv.org/pdf/1811.11797.pdf}{[arXiv:1811.11797 [gr-qc]]}.


  \bibitem{Sebastiani:2016ras}
  L.~Sebastiani, S.~Vagnozzi and R.~Myrzakulov,
  Mimetic gravity: a review of recent developments and applications to cosmology and astrophysics,
  Adv.\ High Energy Phys.\  {\bf 2017} (2017) 3156915
  \href{https://arxiv.org/pdf/1612.08661.pdf}{[arXiv:1612.08661 [gr-qc]]}.

   \bibitem{Astashenok:2015qzw}
  A.~V.~Astashenok and S.~D.~Odintsov,
  From neutron stars to quark stars in mimetic gravity,
  Phys.\ Rev.\ D {\bf 94} (2016) 063008
  \href{https://arxiv.org/pdf/1512.07279.pdf}{[arXiv:1512.07279 [gr-qc]]}.

\bibitem{Xu:2019gua}
  R.~Xu, J.~Zhao and L.~Shao,
  Neutron Star Structure in the Minimal Gravitational Standard-Model Extension and the Implication to Continuous Gravitational Waves, Phys. Lett. B \textbf{803} (2020) 135283
  \href{https://arxiv.org/pdf/1909.10372.pdf}{[arXiv:1909.10372 [gr-qc]]}.

  \bibitem{Horava:2009uw}
  P.~Horava,
  Quantum Gravity at a Lifshitz Point,
  Phys.\ Rev.\ D {\bf 79} (2009) 084008
  \href{https://arxiv.org/pdf/0901.3775.pdf}{[arXiv:0901.3775 [hep-th]]}.

  \bibitem{Liu:2010hzb}
  M.~Liu, J.~Lu, B.~Yu and J.~Lu,
  Solar system constraints on asymptotically flat IR modified Horava gravity through light deflection,
  Gen.\ Rel.\ Grav.\  {\bf 43} (2011) 1401
  \href{https://arxiv.org/pdf/1010.6149.pdf}{[arXiv:1010.6149 [gr-qc]]}.

  \bibitem{Kim:2018dbs}
  K.~Kim, J.~J.~Oh, C.~Park and E.~J.~Son,
  Neutron Star Structure in Ho\v{r}ava-Lifshitz Gravity,
  \href{https://arxiv.org/pdf/1810.07497.pdf}{[arXiv:1810.07497 [gr-qc]]}.

  \bibitem{Eling:2004dk}
  C.~Eling, T.~Jacobson and D.~Mattingly,
  Einstein-Aether theory,
  \href{https://arxiv.org/pdf/gr-qc/0410001.pdf}{[arXiv:gr-qc/0410001]}.

  \bibitem{Eling:2007xh}
  C.~Eling, T.~Jacobson and M.~Coleman Miller,
  Neutron stars in Einstein-aether theory,
  Phys.\ Rev.\ D {\bf 76} (2007) 042003
   [Erratum: Phys.\ Rev.\ D {\bf 80} (2009) 129906]
  \href{https://arxiv.org/pdf/0705.1565.pdf}{[arXiv:0705.1565 [gr-qc]]}.

  \bibitem{Farhi}
  E. Farhi and R. L. Jaffe, Phys. Rev. D \textbf{30} (1984) 2379.

  \bibitem{Barausse:2019yuk}
  E.~Barausse,
  Neutron star sensitivities in Ho\v{r}ava gravity after GW170817, Phys.\ Rev.\ D {\bf 100} (2019) 084053
  \href{https://arxiv.org/pdf/1907.05958.pdf}{[arXiv:1907.05958 [gr-qc]]}.
  
  

\bibitem{Randall:1999ee}
L.~Randall and R.~Sundrum,
A Large mass hierarchy from a small extra dimension,
Phys. Rev. Lett. \textbf{83} (1999) 3370
\href{https://arxiv.org/pdf/hep-ph/9905221.pdf}{[arXiv:hep-ph/9905221 [hep-ph]]}.

\bibitem{Randall:1999vf}
L.~Randall and R.~Sundrum,
An Alternative to compactification,
Phys. Rev. Lett. \textbf{83} (1999) 4690
\href{https://arxiv.org/pdf/hep-th/9906064.pdf}{[arXiv:hep-th/9906064 [hep-th]]}.

\bibitem{Wiseman:2001xt}
T.~Wiseman,
Relativistic stars in Randall-Sundrum gravity,
Phys. Rev. D \textbf{65} (2002) 124007
\href{https://arxiv.org/pdf/hep-th/0111057.pdf}{[arXiv:hep-th/0111057 [hep-th]]}.

\bibitem{Creek:2006je}
S.~Creek, R.~Gregory, P.~Kanti and B.~Mistry,
Braneworld stars and black holes,
Class. Quant. Grav. \textbf{23} (2006) 6633
\href{https://arxiv.org/pdf/hep-th/0606006.pdf}{[arXiv:hep-th/0606006 [hep-th]]}.

\bibitem{Ovalle:2007bn}
J.~Ovalle,
Searching exact solutions for compact stars in braneworld: A Conjecture,
Mod. Phys. Lett. A \textbf{23} (2008), 3247
\href{https://arxiv.org/pdf/gr-qc/0703095.pdf}{[arXiv:gr-qc/0703095 [gr-qc]]}.

\bibitem{Ovalle:2013xla}
J.~Ovalle and F.~Linares,
Tolman IV solution in the Randall-Sundrum Braneworld,
Phys. Rev. D \textbf{88} (2013)104026
\href{https://arxiv.org/pdf/1311.1844.pdf}{[arXiv:1311.1844 [gr-qc]]}.

\bibitem{Garcia-Aspeitia:2014pna}
M.~A.~García-Aspeitia and L.~Ureña-López,
Stellar stability in brane-worlds revisited,
Class. Quant. Grav. \textbf{32} (2015) 025014
\href{https://arxiv.org/pdf/1405.3932.pdf}{[arXiv:1405.3932 [gr-qc]]}.

  \bibitem{Bertolami:2008ab}
  O.~Bertolami, F.~S.~N.~Lobo and J.~Paramos,
  Non-minimum coupling of perfect fluids to curvature,
  Phys.\ Rev.\ D {\bf 78} (2008) 064036
  \href{https://arxiv.org/pdf/0806.4434.pdf}{[arXiv:0806.4434 [gr-qc]]}.

    \bibitem{Harko:2011kv}
  T.~Harko, F.~S.~N.~Lobo, S.~Nojiri and S.~D.~Odintsov,
  $f(R,T)$ gravity,
  Phys.\ Rev.\ D {\bf 84} (2011) 024020
  \href{https://arxiv.org/pdf/1104.2669.pdf}{[arXiv:1104.2669 [gr-qc]]}.

  \bibitem{Faraoni:2009rk}
  V.~Faraoni,
  The Lagrangian description of perfect fluids and modified gravity with an extra force,
  Phys.\ Rev.\ D {\bf 80} (2009) 124040
  \href{https://arxiv.org/pdf/0912.1249.pdf}{[arXiv:0912.1249 [astro-ph.GA]]}.

\bibitem{Avelino:2018rsb}
  P.~P.~Avelino and R.~P.~L.~Azevedo,
  Perfect fluid Lagrangian and its cosmological implications in theories of gravity with nonminimally coupled matter fields,
  Phys.\ Rev.\ D {\bf 97} (2018)  064018
  \href{https://arxiv.org/pdf/1802.04760.pdf}{[arXiv:1802.04760 [gr-qc]]}.

  \bibitem{Moraes:2015uxq}
  P.~H.~R.~S.~Moraes, J.~D.~V.~Arba\~nil and M.~Malheiro,
  Stellar equilibrium configurations of compact stars in $f(R,T)$ gravity,
  JCAP {\bf 1606} (2016) 005
 \href{https://arxiv.org/pdf/1511.06282.pdf}{[arXiv:1511.06282 [gr-qc]]}.

  \bibitem{Das:2016mxq}
  A.~Das, F.~Rahaman, B.~K.~Guha and S.~Ray,
  Compact stars in $f(R,\mathcal {T})$ gravity,
  Eur.\ Phys.\ J.\ C {\bf 76} (2016) 654
  \href{https://arxiv.org/pdf/1608.00566.pdf}{[arXiv:1608.00566 [gr-qc]]}.


  \bibitem{Deb:2018gzt}
  D.~Deb, S.~V.~Ketov, M.~Khlopov and S.~Ray,
  Study on charged strange stars in $f\left(R,\mathcal{T}\right)$ gravity, JCAP {\bf 1910} (2019)  070
  \href{https://arxiv.org/pdf/1812.11736.pdf}{[arXiv:1812.11736 [gr-qc]]}.

  \bibitem{Usov:2004iz}
  V.~V.~Usov,
  Electric fields at the quark surface of strange stars in the color-flavor locked phase,
  Phys.\ Rev.\ D {\bf 70} (2004) 067301
  \href{https://arxiv.org/pdf/astro-ph/0408217.pdf}{[arXiv:astro-ph/0408217]}.



   \bibitem{Zubair:2015gsb}
  M.~Zubair, G.~Abbas and I.~Noureen,
  Possible formation of compact stars in $f(R,T)$ gravity,
  Astrophys.\ Space Sci.\  {\bf 361} (2016)  8
  \href{https://arxiv.org/pdf/1512.05202.pdf}{[arXiv:1512.05202 [physics.gen-ph]]}.


  \bibitem{Maurya:2019hds}
  S.~K.~Maurya and F.~Tello-Ortiz,
  Charged anisotropic compact star in $f(R,\mathcal{T})$ gravity: A minimal geometric deformation gravitational decoupling approach, Phys. Dark Univ. \textbf{27} (2020) 100442
  \href{https://arxiv.org/pdf/1905.13519.pdf}{[arXiv:1905.13519 [gr-qc]]}.

   \bibitem{Maurya:2019sfm}
  S.~K.~Maurya, A.~Errehymy, D.~Deb, F.~Tello-Ortiz and M.~Daoud,
  Study on anisotropic strange stars in $f(R,T)$ gravity: An embedding approach under simplest linear functional of matter-geometry coupling, Phys.\ Rev.\ D {\bf 100} (2019)  044014
  \href{https://arxiv.org/pdf/1907.10149.pdf}{[arXiv:1907.10149 [gr-qc]]}.

  \bibitem{Carvalho:2019gzs}
  G.~A.~Carvalho, S.~I.~d.~Santos, P.~H.~R.~S.~Moraes and M.~Malheiro,
  Strange stars in energy-momentum-conserved $f(R,T)$ gravity,
 \href{https://arxiv.org/pdf/1911.02484.pdf}{arXiv:1911.02484 [gr-qc]}.
 
 \bibitem{Mathew:2020zic}
A.~Mathew, M.~Shafeeque and M.~K.~Nandy,
Stellar structure of quark stars in a modified Starobinsky gravity,
\href{https://arxiv.org/pdf/2006.06421.pdf}{[arXiv:2006.06421 [gr-qc]]}.

  \bibitem{Rastall1}
P. Rastall, Generalization of the Einstein Theory, Phys. Rev. D \textbf{6} (1972) 3357.

\bibitem{Rastall2}
P. Rastall, Can. J. Phys. \textbf{54} 66 (1976) 66.

\bibitem{Oliveira:2015lka}
  A.~M.~Oliveira, H.~E.~S.~Velten, J.~C.~Fabris and L.~Casarini,
  Neutron Stars in Rastall Gravity,
  Phys.\ Rev.\ D {\bf 92} (2015) 044020
  \href{https://arxiv.org/pdf/1506.00567.pdf}{[arXiv:1506.00567 [gr-qc]]}.

  \bibitem{Hansraj:2018reh}
  S.~Hansraj and A.~Banerjee,
  Equilibrium stellar configurations in Rastall theory and linear equation of state,
  \href{https://arxiv.org/pdf/1807.00812.pdf}{[arXiv:1807.00812 [gr-qc]]}.

  \bibitem{lazo}  M. J. Lazo, J. Paiva, J. T. S. Amaral, and G. S. F. Frederico, From an Action Principle for Action-dependent Lagrangians toward non-conservative Gravity: accelerating Universe without dark energy, Phys. Rev. D \textbf{95}, 101501 (2017)
    \href{https://arxiv.org/pdf/1705.04604.pdf}{[arXiv:1705.04604 [gr-qc]]}.

  \bibitem{julio} J.C. Fabris, H. Velten, A. Wojnar,
  On the existence of static spherically-symmetric objects in action-dependent Lagrangian theories, Phys.\ Rev.\ D {\bf 99} (2019)  124031
  \href{https://arxiv.org/pdf/1903.12193.pdf}{[arXiv:1903.12193 [gr-qc]]}

    \bibitem{Santos:2011ye}
  E.~Santos,
  Neutron stars in generalized f(R) gravity,
  Astrophys.\ Space Sci.\  {\bf 341} (2012) 411 \href{https://arxiv.org/pdf/1104.2140.pdf}{[arXiv:1104.2140 [gr-qc]]}.

    \bibitem{Deliduman:2011nw}
  C.~Deliduman, K.~Y.~Eksi and V.~Keles,
  Neutron star solutions in perturbative quadratic gravity,
  JCAP {\bf 1205} (2012) 036
  \href{https://arxiv.org/pdf/1112.4154.pdf}{[arXiv:1112.4154 [gr-qc]]}.

  \bibitem{Glendenning:1997ak}
  N.~K.~Glendenning and J.~Schaffner-Bielich,
  First order kaon condensate,
  Phys.\ Rev.\ C {\bf 60} (1999) 025803
  \href{https://arxiv.org/pdf/astro-ph/9810290.pdf}{[arXiv:astro-ph/9810290]}.


\bibitem{Shamir:2017yza}
  M.~F.~Shamir and S.~Zia,
  Physical attributes of anisotropic compact stars in $f(R,G)$ gravity,
  Eur.\ Phys.\ J.\ C {\bf 77} (2017) 448
  \href{https://arxiv.org/pdf/1705.06582.pdf}{[arXiv:1705.06582 [physics.gen-ph]]}.

  \bibitem{Shamir:2017rjz}
  M.~F.~Shamir and M.~Ahmad,
  Emerging anisotropic compact stars in $f(\mathcal {G},T)$ gravity,
  Eur.\ Phys.\ J.\ C {\bf 77} (2017) 674
  \href{https://arxiv.org/pdf/1705.06910.pdf}{[arXiv:1705.06910 [gr-qc]]}.


  \bibitem{Shamir:2018mjs}
  M.~F.~Shamir and S.~Zia,
  Analysis of Charged Compact Stars in Modified Gravity,
  Int.\ J.\ Mod.\ Phys.\ D {\bf 27} (2018)  1850082
  \href{https://arxiv.org/pdf/1804.09251.pdf}{[arXiv:1804.09251 [gr-qc]]}.

   \bibitem{Sert:2018rjv}
  O.~Sert, F.~Çeliktas and M.~Adak,
  Anisotropic Stars in the Non-minimal $Y(R)F^2$ Gravity,
  Eur.\ Phys.\ J.\ C {\bf 78} (2018) 824
  \href{https://arxiv.org/pdf/1805.07730.pdf}{[arXiv:1805.07730 [gr-qc]]}.

\bibitem{Sharif:2019txi}
  M.~Sharif and A.~Waseem,
  Stellar Evolution of Compact Stars in Curvature-Matter Coupling Gravity, PTEP {\bf 2019} (2019)  053E02
  \href{https://arxiv.org/pdf/1904.05885.pdf}{[arXiv:1904.05885 [gr-qc]]}.

   \bibitem{UG}
  Y. Jack Ng and H. van Dam,
  Possible solution to the cosmological-constant problem,
  Phys. Rev. Lett. \textbf{65} (1990) 1972.

  \bibitem{Astorga-Moreno:2019uin}
  J.~A.~Astorga-Moreno, J.~Chagoya, J.~C.~Flores-Urbina and M.~A.~Garcia-Aspeitia,
  Stellar dynamics in unimodular gravity, JCAP \textbf{09} (2019) 005
  \href{https://arxiv.org/pdf/1905.11253.pdf}{[arXiv:1905.11253 [gr-qc]]}.

\bibitem{Arapoglu:2019mun}
  A.~S.~Arapoğlu, K.~Y.~ Ek{\c s}i and A.~Emrah Yükselci,
  Neutron Star Structure in the Presence of Nonminimally Coupled Scalar Fields, Phys.\ Rev.\ D {\bf 99} (2019) 064055
  \href{https://arxiv.org/pdf/1903.00391.pdf}{[arXiv:1903.00391 [gr-qc]]}.

\bibitem{Nari:2018aqs}
  N.~Nari and M.~Roshan,
  Compact stars in Energy-Momentum Squared Gravity,
  Phys.\ Rev.\ D {\bf 98} (2018)  024031
  \href{https://arxiv.org/pdf/1802.02399.pdf}{[arXiv:1802.02399 [gr-qc]]}.
  
  \bibitem{Akarsu:2018zxl}
Ö.~Akarsu, J.~D.~Barrow, S.~Çıkıntoğlu, K.~Y.~Ekşi and N.~Katırcı,
Constraint on energy-momentum squared gravity from neutron stars and its cosmological implications,
Phys. Rev. D \textbf{97} (2018) 124017
\href{https://arxiv.org/pdf/1802.02093.pdf}{[arXiv:1802.02093 [gr-qc]]}.

    \bibitem{Armengol:2016mzu}
  F.~G.~Lopez Armengol and G.~E.~Romero,
  Neutron stars in Scalar-Tensor-Vector Gravity,
  Gen.\ Rel.\ Grav.\  {\bf 49} (2017)  27
  \href{https://arxiv.org/pdf/1611.05721.pdf}{[arXiv:1611.05721 [gr-qc]]}.



\bibitem{Parker:1993dk}
  L.~Parker and J.~Z.~Simon,
  Einstein equation with quantum corrections reduced to second order,
  Phys.\ Rev.\ D {\bf 47} (1993) 1339
  \href{https://arxiv.org/pdf/gr-qc/9211002.pdf}{[arXiv:gr-qc/9211002]}.

 \bibitem{BirrelDavies}
 N. D. Birrell and P. C. W. Davies, \emph{Quantum Fields in Curved Space}, Cambridge Monographs on Mathematical Physics (Cambridge Univ. Press, Cambridge, UK, 1984).

  \bibitem{Carballo-Rubio:2017tlh}
  R.~Carballo-Rubio,
  Stellar equilibrium in semiclassical gravity,
  Phys.\ Rev.\ Lett.\  {\bf 120} (2018) 061102
  \href{https://arxiv.org/pdf/1706.05379.pdf}{[arXiv:1706.05379 [gr-qc]]}.

\bibitem{Barcelo:2007yk}
  C.~Barcelo, S.~Liberati, S.~Sonego and M.~Visser,
  Fate of gravitational collapse in semiclassical gravity,
  Phys.\ Rev.\ D {\bf 77} (2008) 044032
  \href{https://arxiv.org/pdf/0712.1130.pdf}{[arXiv:0712.1130 [gr-qc]]}.

  \bibitem{Stergioulas:2003yp}
  N.~Stergioulas,
  Rotating Stars in Relativity,
  Living Rev.\ Rel.\  {\bf 6} (2003) 3
  \href{https://arxiv.org/pdf/gr-qc/0302034.pdf}{[arXiv:gr-qc/0302034]}.

   \bibitem{Hartle:1967he}
  J.~B.~Hartle,
  Slowly rotating relativistic stars. 1. Equations of structure,
  Astrophys.\ J.\  {\bf 150} (1967) 1005.


  \bibitem{Kehl:2016mgp}
  M.~S.~Kehl, N.~Wex, M.~Kramer and K.~Liu,
  Future measurements of the Lense-Thirring effect in the Double Pulsar,
  \href{https://arxiv.org/pdf/1605.00408.pdf}{[arXiv:1605.00408 [astro-ph.HE]]}.

    \bibitem{Friedman:1997uh}
  J.~L.~Friedman and S.~M.~Morsink,
  Axial instability of rotating relativistic stars,
  Astrophys.\ J.\  {\bf 502} (1998) 714
  \href{https://arxiv.org/pdf/gr-qc/9706073.pdf}{[arXiv:gr-qc/9706073]}.


  \bibitem{Staykov:2014mwa}
  K.~V.~Staykov, D.~D.~Doneva, S.~S.~Yazadjiev and K.~D.~Kokkotas,
  Slowly rotating neutron and strange stars in $R^2$ gravity,
  JCAP {\bf 1410} (2014) 006
  \href{https://arxiv.org/pdf/1407.2180.pdf}{[arXiv:1407.2180 [gr-qc]]}.


  \bibitem{GondekRosinska:2008nf}
  D.~Gondek-Rosinska and F.~Limousin,
  The final phase of inspiral of strange quark star binaries,
  \href{https://arxiv.org/pdf/0801.4829.pdf}{[arXiv:0801.4829 [gr-qc]]}.


\bibitem{Staykov:2015mma}
  K.~Staykov, K.~Y.~ Ek{\c s}i, S.~S.~Yazadjiev, M.~M.~T{\"
u}rko{\u g}lu and A.~S.~Arapoğlu,
  Moment of inertia of neutron star crust in alternative and modified theories of gravity,
  Phys.\ Rev.\ D {\bf 94} (2016) 024056
  \href{https://arxiv.org/pdf/1507.05878.pdf}{[arXiv:1507.05878 [gr-qc]]}.



  \bibitem{Staykov:2015kwa}
  K.~V.~Staykov, D.~D.~Doneva and S.~S.~Yazadjiev,
  Orbital and epicyclic frequencies around neutron and strange stars in $R^2$ gravity,
  Eur.\ Phys.\ J.\ C {\bf 75}  (2015) 607
  \href{https://arxiv.org/pdf/1508.07790.pdf}{[arXiv:1508.07790 [gr-qc]]}.


\bibitem{Pani:2014jra}
  P.~Pani and E.~Berti,
  Slowly rotating neutron stars in scalar-tensor theories,
  Phys.\ Rev.\ D {\bf 90} (2014)  024025
 \href{https://arxiv.org/pdf/1405.4547.pdf}{[arXiv:1405.4547 [gr-qc]]}.

\bibitem{Popchev:2018fwu}
  D.~Popchev, K.~V.~Staykov, D.~D.~Doneva and S.~S.~Yazadjiev,
  Moment of inertia - mass universal relations for neutron stars in scalar-tensor theory with self-interacting massive scalar field, Eur. Phys. J. C \textbf{79} (2019) 178
  \href{https://arxiv.org/pdf/1812.00347.pdf}{[arXiv:1812.00347 [gr-qc]]}.

\bibitem{Staykov:2018hhc}
  K.~V.~Staykov, D.~Popchev, D.~D.~Doneva and S.~S.~Yazadjiev,
  Static and slowly rotating neutron stars in scalar–tensor theory with self-interacting massive scalar field,
  Eur.\ Phys.\ J.\ C {\bf 78} (2018) 586
  \href{https://arxiv.org/pdf/1805.07818.pdf}{[arXiv:1805.07818 [gr-qc]]}.

  \bibitem{Silva:2014fca}
  H.~O.~Silva, C.~F.~B.~Macedo, E.~Berti and L.~C.~B.~Crispino,
  Slowly rotating anisotropic neutron stars in general relativity and scalar-tensor theory,
  Class.\ Quant.\ Grav.\  {\bf 32} (2015) 145008
  \href{https://arxiv.org/pdf/1411.6286.pdf}{[arXiv:1411.6286 [gr-qc]]}.

\bibitem{Staykov:2015cfa}
  K.~V.~Staykov, D.~D.~Doneva, S.~S.~Yazadjiev and K.~D.~Kokkotas,
  Gravitational wave asteroseismology of neutron and strange stars in R$^2$ gravity,
  Phys.\ Rev.\ D {\bf 92} (2015) 043009
  \href{https://arxiv.org/pdf/1503.04711.pdf}{[arXiv:1503.04711 [gr-qc]]}.

\bibitem{Andersson:1996pn}
  N.~Andersson and K.~D.~Kokkotas,
  Gravitational waves and pulsating stars: What can we learn from future observations?,
  Phys.\ Rev.\ Lett.\  {\bf 77} (1996) 4134
 \href{https://arxiv.org/pdf/gr-qc/9610035.pdf}{[arXiv:gr-qc/9610035]}.



\bibitem{Silva:2016smx}
  H.~O.~Silva, A.~Maselli, M.~Minamitsuji and E.~Berti,
  Compact objects in Horndeski gravity,
  Int.\ J.\ Mod.\ Phys.\ D {\bf 25} (2016)  1641006
  \href{https://arxiv.org/pdf/1602.05997.pdf}{[arXiv:1602.05997 [gr-qc]]}.

   \bibitem{Cisterna:2016vdx}
  A.~Cisterna, T.~Delsate, L.~Ducobu and M.~Rinaldi,
  Slowly rotating neutron stars in the nonminimal derivative coupling sector of Horndeski gravity,
  Phys.\ Rev.\ D {\bf 93} (2016) 084046
  \href{https://arxiv.org/pdf/1602.06939.pdf}{[arXiv:1602.06939 [gr-qc]]}.

\bibitem{Maselli:2016gxk}
  A.~Maselli, H.~O.~Silva, M.~Minamitsuji and E.~Berti,
  Neutron stars in Horndeski gravity,
  Phys.\ Rev.\ D {\bf 93} (2016)  124056
  \href{https://arxiv.org/pdf/1603.04876.pdf}{[arXiv:1603.04876 [gr-qc]]}.

    \bibitem{Glendenning:1984jr}
  N.~K.~Glendenning,
  Neutron Stars Are Giant Hypernuclei?,
  Astrophys.\ J.\  {\bf 293} (1985) 470.

  \bibitem{KramerI}
  M. Kramer and N. Wex, The double pulsar system: A unique laboratory for gravity, Class. Quant. Grav. \textbf{26} (2009) 073001.

  \bibitem{Sakstein:2016oel}
  J.~Sakstein, E.~Babichev, K.~Koyama, D.~Langlois and R.~Saito,
  Towards Strong Field Tests of Beyond Horndeski Gravity Theories,
  Phys.\ Rev.\ D {\bf 95} (2017) 064013
  \href{https://arxiv.org/pdf/1612.04263.pdf}{[arXiv:1612.04263 [gr-qc]]}.

\bibitem{Sullivan:2017kwo}
  A.~Sullivan and N.~Yunes,
  Slowly-Rotating Neutron Stars in Massive Bigravity,
  Class.\ Quant.\ Grav.\  {\bf 35} (2018) 045003
  \href{https://arxiv.org/pdf/1709.03311.pdf}{[arXiv:1709.03311 [gr-qc]]}.


    \bibitem{Lattimer:1991nc}
  J.~M.~Lattimer and F.~D.~Swesty,
  A Generalized equation of state for hot, dense matter,
  Nucl.\ Phys.\ A {\bf 535} (1991) 331.

  \bibitem{Margaritis:2019hfq}
  C.~Margaritis, P.~S.~Koliogiannis and C.~C.~Moustakidis,
  Speed of sound constraints on maximally-rotating neutron stars, Phys. Rev. D \textbf{101} (2020)  043023
  \href{https://arxiv.org/pdf/1910.05767.pdf}{[arXiv:1910.05767 [nucl-th]]}.

   \bibitem{FrieSter}
  J. L. Friedman and N. Stergioulas, \emph{Rotating Relativistic Stars} (Cambridge Monographs on Mathematical Physics) (Cambridge
University Press, 2013).


  \bibitem{Doneva:2013qva}
  D.~D.~Doneva, S.~S.~Yazadjiev, N.~Stergioulas and K.~D.~Kokkotas,
  Rapidly rotating neutron stars in scalar-tensor theories of gravity,
  Phys.\ Rev.\ D {\bf 88} (2013)  084060
  \href{https://arxiv.org/pdf/1309.0605.pdf}{[arXiv:1309.0605 [gr-qc]]}.

\bibitem{Yazadjiev:2015zia}
  S.~S.~Yazadjiev, D.~D.~Doneva and K.~D.~Kokkotas,
  Rapidly rotating neutron stars in R-squared gravity,
  Phys.\ Rev.\ D {\bf 91} (2015) 084018
  \href{https://arxiv.org/pdf/1501.04591.pdf}{[arXiv:1501.04591 [gr-qc]]}.

\bibitem{Yazadjiev:2017vpg}
  S.~S.~Yazadjiev, D.~D.~Doneva and K.~D.~Kokkotas,
  Oscillation modes of rapidly rotating neutron stars in scalar-tensor theories of gravity,
  Phys.\ Rev.\ D {\bf 96} (2017)  064002
  \href{https://arxiv.org/pdf/1705.06984.pdf}{[arXiv:1705.06984 [gr-qc]]}.

  \bibitem{Kleihaus:2016dui}
  B.~Kleihaus, J.~Kunz, S.~Mojica and M.~Zagermann,
  Rapidly Rotating Neutron Stars in Dilatonic Einstein-Gauss-Bonnet Theory,
  Phys.\ Rev.\ D {\bf 93} (2016) 064077
 \href{https://arxiv.org/pdf/1601.05583.pdf}{[arXiv:1601.05583 [gr-qc]]}.

 \bibitem{DiazIban}
 J. Diaz-Alonso and J.M. Ibanez-Cabanell, Astrophys. J. \textbf{291} (1985) 308.





  \bibitem{Komatsu:1989zz}
  H.~Komatsu, Y.~Eriguchi and I.~Hachisu,
  Rapidly rotating general relativistic stars. I - Numerical method and its application to uniformly rotating polytropes,
  Mon.\ Not.\ Roy.\ Astron.\ Soc.\  {\bf 237} (1989) 355.


\bibitem{Forbes:2019xaz}
  M.~M.~Forbes, S.~Bose, S.~Reddy, D.~Zhou, A.~Mukherjee and S.~De,
  Constraining the neutron-matter equation of state with gravitational waves, Phys.\ Rev.\ D {\bf 100} (2019) 083010
  \href{https://arxiv.org/pdf/1901.05271.pdf}{[arXiv:1904.04233 [astro-ph.HE]]}.

    \bibitem{Nakar:2019fza}
E.~Nakar,
The electromagnetic counterparts of compact binary mergers,
\href{https://arxiv.org/pdf/1912.05659.pdf}{[arXiv:1912.05659 [astro-ph.HE]]}.


   \bibitem{Barausse:2012da}
  E.~Barausse, C.~Palenzuela, M.~Ponce and L.~Lehner,
  Neutron-star mergers in scalar-tensor theories of gravity,
  Phys.\ Rev.\ D {\bf 87} (2013) 081506
  \href{https://arxiv.org/pdf/1212.5053.pdf}{[arXiv:1212.5053 [gr-qc]]}.
  
  \bibitem{Blazquez-Salcedo:2020ibb}
J.~L.~Bl\'azquez-Salcedo, F.~S.~Khoo and J.~Kunz,
Ultra long lived quasinormal modes of neutron stars in $R^2$ gravity,
\href{https://arxiv.org/pdf/2001.09117.pdf}{[arXiv:2001.09117 [gr-qc]]}.

  \bibitem{Yagi:2013mbt}
  K.~Yagi, L.~C.~Stein, N.~Yunes and T.~Tanaka,
  Isolated and Binary Neutron Stars in Dynamical Chern-Simons Gravity,
  Phys.\ Rev.\ D {\bf 87} (2013) 084058
   Erratum: [Phys.\ Rev.\ D {\bf 93} (2016) 089909]
  \href{https://arxiv.org/pdf/1302.1918.pdf}{[arXiv:1302.1918 [gr-qc]]}.

\bibitem{Kramer:2006nb}
  M.~Kramer {\it et al.},
  Tests of general relativity from timing the double pulsar,
  Science {\bf 314} (2006) 97
  \href{https://arxiv.org/pdf/astro-ph/0609417.pdf}{[arXiv:astro-ph/0609417]}.



  \bibitem{Shen:1998gq}
  H.~Shen, H.~Toki, K.~Oyamatsu and K.~Sumiyoshi,
  Relativistic equation of state of nuclear matter for supernova and neutron star,
  Nucl.\ Phys.\ A {\bf 637} (1998) 435
  \href{https://arxiv.org/pdf/nucl-th/9805035.pdf}{[arXiv:nucl-th/9805035]}.

  \bibitem{Okounkova:2019dfo}
  M.~Okounkova, L.~C.~Stein, M.~A.~Scheel and S.~A.~Teukolsky,
 Numerical binary black hole collisions in dynamical Chern-Simons gravity,  Phys.\ Rev.\ D {\bf 100} (2019) 104026
  \href{https://arxiv.org/pdf/1906.08789.pdf}{[arXiv:1906.08789 [gr-qc]]}.

  \bibitem{Carson:2019fxr}
  Z.~Carson, B.~C.~Seymour and K.~Yagi,
  Future Prospects for Probing Scalar-Tensor Theories with Gravitational Waves from Mixed Binaries, Class. Quant. Grav. \textbf{37} (2020) 065008
  \href{https://arxiv.org/pdf/1907.03897.pdf}{[arXiv:1907.03897 [gr-qc]]}.

  \bibitem{Zhang:2019iim}
  C.~Zhang, X.~Zhao, A.~Wang, B.~Wang, K.~Yagi, N.~Yunes, W.~Zhao and T.~Zhu,
  Gravitational waves from the quasi-circular inspiral of compact binaries in Einstein-aether theory, Phys. Rev. D \textbf{101} (2020) 044002
  \href{https://arxiv.org/pdf/1911.10278.pdf}{[arXiv:1911.10278 [gr-qc]]}.

  \bibitem{Zhao:2019kif}
  X.~Zhao {\it et al.},
  Gravitational waveforms and radiation powers of the triple system PSR J0337+1715 in modified theories of gravity,
  Phys.\ Rev.\ D {\bf 100} (2019)  083012
  \href{https://arxiv.org/pdf/1903.09865.pdf}{[arXiv:1903.09865 [astro-ph.HE]]}.

  \bibitem{De:2018uhw}
  S.~De, D.~Finstad, J.~M.~Lattimer, D.~A.~Brown, E.~Berger and C.~M.~Biwer,
  Tidal Deformabilities and Radii of Neutron Stars from the Observation of GW170817,
  Phys.\ Rev.\ Lett.\  {\bf 121} (2018) 091102
  \href{https://arxiv.org/pdf/1804.08583.pdf}{[arXiv:1804.08583 [astro-ph.HE]]}.

  \bibitem{Lattimer:2015nhk}
  J.~M.~Lattimer and M.~Prakash,
  The Equation of State of Hot, Dense Matter and Neutron Stars,
  Phys.\ Rept.\  {\bf 621} (2016) 127
  \href{https://arxiv.org/pdf/1512.07820.pdf}{[arXiv:1512.07820 [astro-ph.SR]]}.


\bibitem{Raithel:2018ncd}
  C.~Raithel, F.~Özel and D.~Psaltis,
  Tidal deformability from GW170817 as a direct probe of the neutron star radius,
  Astrophys.\ J.\  {\bf 857} (2018) L23
  \href{https://arxiv.org/pdf/1803.07687.pdf}{[arXiv:1803.07687 [astro-ph.HE]]}.

  \bibitem{Coughlin:2018miv}
  M.~W.~Coughlin {\it et al.},
  Constraints on the neutron star equation of state from AT2017gfo using radiative transfer simulations,  Mon.\ Not.\ Roy.\ Astron.\ Soc.\  {\bf 480} (2018)  3871
 \href{https://arxiv.org/pdf/1805.09371.pdf}{[arXiv:1805.09371 [astro-ph.HE]]}.


  \bibitem{FlanHind2018}
   E.~E.~Flanagan and T.~Hinderer,
  Constraining neutron star tidal Love numbers with gravitational wave detectors,
  Phys.\ Rev.\ D {\bf 77} (2008) 021502
  \href{https://arxiv.org/pdf/0709.1915.pdf}{[arXiv:0709.1915 [astro-ph]]}.

  \bibitem{Gupta:2017vsl}
  T.~Gupta, B.~Majumder, K.~Yagi and N.~Yunes,
  I-Love-Q Relations for Neutron Stars in dynamical Chern Simons Gravity,
  Class.\ Quant.\ Grav.\  {\bf 35} (2018) 025009
  \href{https://arxiv.org/pdf/1710.07862.pdf}{[arXiv:1710.07862 [gr-qc]]}.

  \bibitem{Yazadjiev:2018xxk}
  S.~S.~Yazadjiev, D.~D.~Doneva and K.~D.~Kokkotas,
  Tidal Love numbers of neutron stars in $f(R)$ gravity,
  Eur.\ Phys.\ J.\ C {\bf 78} (2018) 818
  \href{https://arxiv.org/pdf/1803.09534.pdf}{[arXiv:1803.09534 [gr-qc]]}.

 \bibitem{BSS} A. Stachowski, M. Szydlowski, A. Borowiec, Starobinsky cosmological model in Palatini formalism, Eur. Phys. J. C  \textbf{77} (2017)  406
 \href{https://arxiv.org/pdf/1608.03196.pdf}{[arXiv:1608.03196 [gr-qc]]}.

\bibitem{SSB} M. Szydlowski, A. Stachowski, A. Borowiec, Emergence of dynamical dark energy from polynomial $f(R)$ theory in Palatini formalism, Eur. Phys. J.  \textbf{77} (2017) 603
 \href{https://arxiv.org/pdf/1707.01948.pdf}{[arXiv:1707.01948 [gr-qc]]}.

  \bibitem{dabrawski}
  M.~P.~Dabrowski, J.~Garecki and D.~B.~Blaschke,
  Conformal transformations and conformal invariance in gravitation,
  Annalen Phys.\  {\bf 18} (2009) 13
  \href{https://arxiv.org/pdf/0806.2683.pdf}{[arXiv:0806.2683 [gr-qc]]}.
  
  
\bibitem{Olmo:2005hc}
G.~J.~Olmo,
Post-Newtonian constraints on f(R) cosmologies in metric and Palatini formalism,
Phys. Rev. D \textbf{72} (2005) 083505
\href{https://arxiv.org/pdf/gr-qc/0505135.pdf}{[arXiv:gr-qc/0505135 [gr-qc]]}.

\bibitem{Olmo:2005zr}
G.~J.~Olmo,
The Gravity Lagrangian according to solar system experiments,
Phys. Rev. Lett. \textbf{95} (2005) 261102
\href{https://arxiv.org/pdf/gr-qc/0505101.pdf}{[arXiv:gr-qc/0505101 [gr-qc]]}.

\bibitem{Vollick:2003qp}
  D.~N.~Vollick,
  Palatini approach to Born-Infeld-Einstein theory and a geometric description of electrodynamics,
  Phys.\ Rev.\ D {\bf 69}, 064030 (2004)
 \href{https://arxiv.org/pdf/gr-qc/0309101.pdf}{[arXiv:gr-qc/0309101]}.

  \bibitem{banados} M. Banados and P. G. Ferreira, 	
Eddington's theory of gravity and its progeny, Phys. Rev. Lett. \textbf{105} (2010) 011101
 \href{https://arxiv.org/pdf/1006.1769.pdf}{[arXiv:1006.1769 [astro-ph]]}.





\bibitem{pani} P.~Pani and T.~P.~Sotiriou,
  Surface singularities in Eddington-inspired Born-Infeld gravity,
  Phys.\ Rev.\ Lett.\  {\bf 109} (2012) 251102
 \href{https://arxiv.org/pdf/1209.2972.pdf}{[arXiv:1209.2972 [gr-qc]]}.

    \bibitem{delsate2} T. Delsate and J. Steinhoff, New insights on the matter-gravity coupling paradigm, Phys. Rev. Lett. \textbf{109}, 021101 (2012)
     \href{https://arxiv.org/pdf/1201.4989.pdf}{[arXiv:1201.4989 [gr-qc]]}.


    \bibitem{BeltranJimenez:2019acz}
  J.~Beltr\'an Jim\'enez and A.~Delhom,
  Ghosts in metric-affine higher order curvature gravity, Eur.\ Phys.\ J.\ C {\bf 79} (2019) 656
  \href{https://arxiv.org/pdf/1901.08988.pdf}{[arXiv:1901.08988 [gr-qc]]}.


   \bibitem{EisenBook}
   L. P. Eisenhart, {\it Non-Riemannian geometry} (American Mathematical Society, New York, 1927).

   \bibitem{Jimenez:2020dpn}
J.~B.~Jim\'enez and A.~Delhom,
Instabilities in Metric-Affine Theories of Gravity,
 \href{https://arxiv.org/pdf/2004.11357.pdf}{[arXiv:2004.11357 [gr-qc]]}.


  \bibitem{Jimenez:2014fla}
J.~Beltr\'an Jim\'enez, L.~Heisenberg and G.~J.~Olmo, Infrared lessons for ultraviolet gravity: the case of massive gravity and Born-Infeld, JCAP \textbf{1411} (2014) 004
\href{https://arxiv.org/pdf/1409.0233.pdf}{[arXiv:1409.0233 [hep-th]]}.

\bibitem{Delhom:2019wir}
A.~Delhom, V.~Miralles and A.~Peñuelas,
Effective interactions in Ricci-Based Gravity models below the non-metricity scale,
Eur. Phys. J. C \textbf{80} (2020)  340
\href{https://arxiv.org/pdf/1907.05615.pdf}{[arXiv:1907.05615 [hep-th]]}.

\bibitem{Afonso:2018bpv}
  V.~I.~Afonso, G.~J.~Olmo and D.~Rubiera-Garcia,
  Mapping Ricci-based theories of gravity into general relativity,
  Phys.\ Rev.\ D {\bf 97},  021503 (2018)
  \href{https://arxiv.org/pdf/1801.10406.pdf}{[arXiv:1801.10406 [gr-qc]]}.

\bibitem{Afonso:2018mxn}
  V.~I.~Afonso, G.~J.~Olmo, E.~Orazi and D.~Rubiera-Garcia,
  Mapping nonlinear gravity into General Relativity with nonlinear electrodynamics,
  Eur.\ Phys.\ J.\ C {\bf 78},  866 (2018)
  \href{https://arxiv.org/pdf/1807.06385.pdf}{[arXiv:1807.06385 [gr-qc]]}.

\bibitem{Afonso:2018hyj}
  V.~I.~Afonso, G.~J.~Olmo, E.~Orazi and D.~Rubiera-Garcia,
  Correspondence between modified gravity and general relativity with scalar fields,
  Phys.\ Rev.\ D {\bf 99}, 044040 (2019)
  \href{https://arxiv.org/pdf/1810.04239.pdf}{[arXiv:1810.04239 [gr-qc]]}.



\bibitem{Delhom:2019zrb}
  A.~Delhom, G.~J.~Olmo and E.~Orazi,
  Ricci-Based Gravity theories and their impact on Maxwell and nonlinear electromagnetic models,
  \href{https://arxiv.org/pdf/1907.04183.pdf}{[arXiv:1907.04183 [gr-qc]]}.


\bibitem{Afonso:2019fzv}
  V.~I.~Afonso, G.~J.~Olmo, E.~Orazi and D.~Rubiera-Garcia,
  New scalar compact objects in Ricci-based gravity theories, JCAP \textbf{12} (2019) 044
  \href{https://arxiv.org/pdf/1906.04623.pdf}{[arXiv:1906.04623 [hep-th]]}.

 \bibitem{hg} T. Harko, T. S. Koivisto, F. S. N. Lobo, G. J. Olmo, Metric-Palatini gravity unifying local constraints and late-time cosmic acceleration, Phys. Rev. D \textbf{85} (2012) 084016
 \href{https://arxiv.org/pdf/1110.1049.pdf}{[arXiv:1110.1049 [gr-qc]]}.

 \bibitem{tg} T. Harko, F. S. N. Lobo, S. Nojiri, S. D. Odintsov, $f(R,T)$ gravity, Phys. Rev. D \textbf{84}  (2011) 024020
 \href{https://arxiv.org/pdf/1104.2669.pdf}{[arXiv:1104.2669 [gr-qc]]}.

 \bibitem{Barrientos:2018cnx}
  E.~Barrientos, F.~S.~N.~Lobo, S.~Mendoza, G.~J.~Olmo and D.~Rubiera-Garcia,
  Metric-affine $f(R,T)$ theories of gravity and their applications,
  Phys.\ Rev.\ D {\bf 97} (2018)104041
  \href{https://arxiv.org/pdf/1803.05525.pdf}{[arXiv:1803.05525 [gr-qc]]}.

  \bibitem{Wu:2018idg}
  J.~Wu, G.~Li, T.~Harko and S.~D.~Liang,
  metric-affine formulation of $f(R,T)$ gravity theory, and its cosmological implications,
  Eur.\ Phys.\ J.\ C {\bf 78} (2018)   430
  \href{https://arxiv.org/pdf/1805.07419.pdf}{[arXiv:1805.07419 [gr-qc]]}.


\bibitem{Olmo:2008ye}
  G.~J.~Olmo,
  Hydrogen atom in Palatini theories of gravity,
  Phys.\ Rev.\ D {\bf 77}, 084021 (2008)
  \href{https://arxiv.org/pdf/0802.4038.pdf}{[arXiv:0802.4038 [gr-qc]]}.


\bibitem{Olmo:2006zu}
  G.~J.~Olmo,
  Violation of the Equivalence Principle in Modified Theories of Gravity,
  Phys.\ Rev.\ Lett.\  {\bf 98}, 061101 (2007) \href{https://arxiv.org/pdf/gr-qc/0612002.pdf}{[arXiv:gr-qc/0612002]}.

\bibitem{barraco} D. Barraco, V. H. Hamity and H. Vucetich,  $f(R)$ cosmology in the first order formalism, Gen. Rel. Grav. \textbf{34} (2002) 533.

\bibitem{al1} G. Allemandi, A. Borowiec and M. Francaviglia, First-Order Non-Linear Gravity, Phys. Rev. D \textbf{70} (2004) 043524
\href{https://arxiv.org/pdf/hep-th/0403264.pdf}{[arXiv:0403264 [hep-th]]}.

\bibitem{al2}  G. Allemandi, A. Borowiec and M. Francaviglia,  Ricci squared Gravity, Phys. Rev. D \textbf{70} (2004) 103503
\href{https://arxiv.org/pdf/hep-th/0407090.pdf}{[arXiv:0407090 [hep-th]]}.

\bibitem{hannu} K. Enqvist, H.J. Nyrhinen, T. Koivisto, Binary systems in Palatini-$f(R)$ gravity, Phys. Rev. D \textbf{88} (2013) 104008
\href{https://arxiv.org/pdf/1308.0988.pdf}{[arXiv:1308.0988 [gr-qc]]}.

\bibitem{Jana:2017ost}
  S.~Jana, G.~K.~Chakravarty and S.~Mohanty,
  Constraints on Born-Infeld gravity from the speed of gravitational waves after GW170817 and GRB 170817A,
  Phys.\ Rev.\ D {\bf 97} (2018)  084011
  \href{https://arxiv.org/pdf/1711.04137.pdf}{[arXiv:1711.04137 [gr-qc]]}.

\bibitem{Senovilla:2013vra}
  J.~M.~M.~Senovilla,
  Junction conditions for F(R)-gravity and their consequences,
  Phys.\ Rev.\ D {\bf 88} (2013)  064015  \href{https://arxiv.org/pdf/1303.1408.pdf}{[arXiv:1303.1408 [gr-qc]]}.

\bibitem{kain}
K. Kainulainen, V. Reijonen, D. Sunhede, The interior spacetimes of stars in Palatini $f(R)$ gravity, Phys. Rev. D. \textbf{76}  (2007)  043503.
\href{https://arxiv.org/pdf/gr-qc/0611132.pdf}{[arXiv:0611132 [gr-qc]]}

\bibitem{Vollick:2003aw}
  D.~N.~Vollick,
  $1/R$ Curvature corrections as the source of the cosmological acceleration,
  Phys.\ Rev.\ D {\bf 68} (2003) 063510 \href{https://arxiv.org/pdf/astro-ph/0306630.pdf}{[arXiv:astro-ph/0306630]]}.


\bibitem{Flanagan:2003rb}
  E.~E.~Flanagan,
  Palatini form of 1/R gravity,
  Phys.\ Rev.\ Lett.\  {\bf 92} (2004) 071101 \href{https://arxiv.org/pdf/astro-ph/0308111.pdf}{[arXiv:gr-qc/0409068]}.

\bibitem{Vollick:2003ic}
  D.~N.~Vollick,
  On the viability of the Palatini form of 1/R gravity,
  Class.\ Quant.\ Grav.\  {\bf 21} (2004) 3813 \href{https://arxiv.org/pdf/gr-qc/0312041.pdf}{[arXiv:gr-qc/0312041]]}.

\bibitem{Avelino:2012qe}
  P.~P.~Avelino,
  Eddington-inspired Born-Infeld gravity: nuclear physics constraints and the validity of the continuous fluid approximation,
  JCAP {\bf 1211} (2012) 022 \href{https://arxiv.org/pdf/1207.4730.pdf}{[arXiv:1207.4730 [astro-ph.CO]]}.


\bibitem{sharif} M. Sharif, Z. Yousaf, Astrophysics and Space Science, \textbf{354} (2014) 481.


  \bibitem{sharif2} M. Sharif, Z. Yousaf, Dynamical instability of the charged expansion-free spherical collapse in
$f(R)$ gravity, Phys. Rev. D \textbf{88} (2013) 024020.


   \bibitem{olmo2005} G. J. Olmo, Post-Newtonian constraints on $f(R)$ cosmologies in metric formalism, Phys. Rev. D \textbf{72} (2005) 083505
    \href{https://arxiv.org/pdf/gr-qc/0505135.pdf}{[arXiv:0505135 [gr-qc]]}.

       \bibitem{darmois} G. Darmois, \emph{Memorial des Sciences Mathematiques}, Gauthier-Villars, Paris (1927).


       \bibitem{sharif3} M. Sharif, Z. Yousaf, Energy Density Inhomogeneities with Polynomial $f(R)$ Cosmology, Astrophysics and Space Science \textbf{352}  (2014) 321
               \href{https://arxiv.org/pdf/1501.03478.pdf}{[arXiv:1501.03478 [gr-qc]]}.

  \bibitem{har} B. K. Harrison, K. S. Thorne, M. Wakano, J. A.  Wheeler, \emph{Gravitation Theory  and  Gravitational  Collapse},
  University  of  Chicago  Press (Chicago, 1965).

\bibitem{vitor} P. Pani, V. Cardoso, T. Delsate, Compact stars in Eddington inspired gravity, Phys. Rev. Lett. \textbf{107} (2011) 031101
\href{https://arxiv.org/pdf/1106.3569.pdf}{[arXiv:1106.3569 [gr-qc]]}.

\bibitem{OlRu}
G. J. Olmo and D. Rubiera-Garcia, Junction conditions in Palatini $f(R)$ gravity and their applications to stellar surfaces in polytropic models, to appear.

\bibitem{barausse1} E. Barausse, T. P. Sotiriou, J. C. Miller, A no-go theorem for polytropic spheres in Palatini $f(R)$ gravity, Class. Quant. Grav. \textbf{25} (2008) 062001
\href{https://arxiv.org/pdf/gr-qc/0703132.pdf}{[arXiv:0703132 [gr-qc]]}.

\bibitem{barausse2} E. Barausse, T. P. Sotiriou, J C. Miller, Curvature singularities, tidal forces and the viability of Palatini $f(R)$ gravity, Class. Quant. Grav. \textbf{25} (2008) 105008
\href{https://arxiv.org/pdf/0712.1141.pdf}{[arXiv:0712.1141 [gr-qc]]}.

\bibitem{barausse3} E. Barausse, T. P. Sotiriou, J C. Miller, Polytropic spheres in Palatini $f(R)$ gravity, EAS Publications Series \textbf{30} (2008) 189
\href{https://arxiv.org/pdf/0801.4852.pdf}{[arXiv:0801.4852 [gr-qc]]}.


\bibitem{sham} Y-H. Sham, P. T. Leung, L-M. Lin, Compact stars in Eddington-inspired Born-Infeld gravity: Anomalies associated with phase transitions, Phys. Rev. D \textbf{87} (2013) 061503
\href{https://arxiv.org/pdf/1304.0550.pdf}{[arXiv:1304.0550 [gr-qc]]}.

\bibitem{barri}
J.~Barrientos O. and G.~F.~Rubilar,
  Surface curvature singularities of polytropic spheres in Palatini $f(R,T)$ gravity,
  Phys.\ Rev.\ D {\bf 93} (2016)   024021.

\bibitem{olmo1} G. J. Olmo, Re-examination of Polytropic Spheres in Palatini $f(R)$ Gravity, Phys. Rev. D \textbf{78} (2008) 104026
\href{https://arxiv.org/pdf/0810.3593.pdf}{[arXiv:0810.3593 [gr-qc]]}.

\bibitem{Hubeny2014}
I. Hubeny and D. Mihalas, {\it Theory of stellar atmospheres}, Princeton University Press (2014).

\bibitem{Potekhin:2004jr}
  A.~Y.~Potekhin, D.~Lai, G.~Chabrier and W.~C.~G.~Ho,
  Electromagnetic polarization in partially ionized plasmas with strong magnetic fields and neutron star atmosphere models,
  Astrophys.\ J.\  {\bf 612}, 1034 (2004)
  \href{https://arxiv.org/pdf/astro-ph/0405383.pdf}{[arXiv:astro-ph/0405383]}.

\bibitem{Gustafsson:2008jx}
B.~Gustafsson, B.~Edvardsson, K.~Eriksson, U.~G.~Jorgensen, A.~Nordlund and B.~Plez,
A grid of MARCS model atmospheres for late-type stars I. Methods and general properties,
Astron.\ Astrophys.\  {\bf 486} (2008) 951
\href{https://arxiv.org/pdf/0805.0554.pdf}{[arXiv:0805.0554 [astro-ph]]}.

\bibitem{kim} H-Ch. Kim, Physics at the surface of a star in Eddington-inspired Born-Infeld gravity, Phys. Rev. D \textbf{89} (2014) 064001
\href{https://arxiv.org/pdf/1312.0705.pdf}{[arXiv:1312.0705 [gr-qc]]}.

\bibitem{Latorre:2017uve}
  A.~D.~I.~Latorre, G.~J.~Olmo and M.~Ronco,
  Observable traces of non-metricity: new constraints on metric-affine gravity,
  Phys.\ Lett.\ B {\bf 780}, 294 (2018) \href{https://arxiv.org/pdf/1709.04249.pdf}{[arXiv:1709.04249 [hep-th]]}.


\bibitem{pannia} F. A. T. Pannia, F. Garcia, S. E. P. Bergliaffa, M. Orellana, M., G. E. Romero, Structure of Compact Stars in $R$-squared Palatini Gravity, Gen. Rel. Grav. \textbf{49} (2017) 25
\href{https://arxiv.org/pdf/1607.03508.pdf}{[arXiv:1312.0705 [gr-qc]]}.


\bibitem{sham1} T-H. Sham, L-M. Lin, P. T. Leung, Radial oscillations and stability of compact stars in Eddington-inspired Born-Infeld gravity, Phys. Rev. D \textbf{86} (2012) 064015
\href{https://arxiv.org/pdf/1208.1314.pdf}{[arXiv:1208.1314 [gr-qc]]}

\bibitem{Chirenti:2015dda}
  C.~Chirenti, G.~H.~de Souza and W.~Kastaun,
  Fundamental oscillation modes of neutron stars: validity of universal relations,
  Phys.\ Rev.\ D {\bf 91} (2015)  044034
  \href{https://arxiv.org/pdf/1501.02970.pdf}{[arXiv:1501.02970 [gr-qc]]}.

\bibitem{sotani} H. Sotani, Stellar oscillations in Eddington-inspired Born-Infeld gravity, Phys. Rev. D \textbf{89} (2014) 124037
\href{https://arxiv.org/pdf/1406.3097.pdf}{[arXiv:1406.3097 [astro-ph]]}.

\bibitem{fatibene} A. Mana, L. Fatibene, M. Ferraris, A further study on Palatini $f(\mathcal{R})$-theories for polytropic stars, JCAP \textbf{2015} (2015) 040
\href{https://arxiv.org/pdf/1505.06575.pdf}{[arXiv:1505.06575 [gr-qc]]}

\bibitem{eps} J. Ehlers, F. A. E. Pirani, A. Schild, \emph{General
Relativity}, ed. L.O. Raifeartaigh (Clarendon, Oxford, 1972).

\bibitem{dyer} C. C. Dyer, C. Oliwa, The ``Swiss cheese" cosmological model has no extrinsic curvature discontinuity: A comment on the paper by G.A. Baker, Jr.,
\href{https://arxiv.org/abs/astro-ph/0004090v1}{[arXiv: astro-ph/0004090]}

\bibitem{aneta2} A. Wojnar, On stability of a neutron star system in Palatini gravity,  Eur. Phys. J. C \textbf{78} (2018) 421
 \href{https://arxiv.org/pdf/1712.01943.pdf}{[arXiv:1712.01943 [gr-qc]]}.

\bibitem{al3}  G. Allemandi, M. Francaviglia, M. L. Ruggiero, A. Tartaglia, Post-Newtonian Parameters from Alternative Theories of Gravity, Gen. Rel. Grav. \textbf{37} (2005) 1891
\href{https://arxiv.org/pdf/gr-qc/0506123.pdf}{[arXiv:0506123 [gr-qc]]}.

\bibitem{rug}  M. L. Ruggiero, L. Iorio, Solar System planetary orbital motions and $f(R)$ Theories of Gravity, JCAP \textbf{0701} (2007) 010
\href{https://arxiv.org/pdf/gr-qc/0607093.pdf}{[arXiv:0607093 [gr-qc]]}.

\bibitem{barr} D. E. Barraco, V. H. Hamity, Spherically symmetric solutions in f(R) theories of gravity obtained using the first order formalism, Phys. Rev. D \textbf{62} (2000) 044027.

\bibitem{olmo2}  G. J. Olmo, The gravity lagrangian according to solar system experiments, Phys. Rev. Lett. \textbf{95} (2005) 261102
\href{https://arxiv.org/pdf/gr-qc/0505101.pdf}{[arXiv:0505101 [gr-qc]]}.

\bibitem{faraoni} V. Faraoni, Solar System experiments do not yet veto modified gravity models,  Phys. Rev. D \textbf{74} (2006) 023529
\href{https://arxiv.org/pdf/gr-qc/0607016.pdf}{[arXiv:0607016 [gr-qc]]}.

\bibitem{sotir} T. P. Sotiriou, The nearly Newtonian regime in Non-Linear Theories of Gravity, Gen. Rel. Grav. \textbf{38} (2006) 1407
\href{https://arxiv.org/pdf/gr-qc/0507027.pdf}{[arXiv:0507027 [gr-qc]]}.

\bibitem{all} G. Allemandi, M. L. Ruggiero, Constraining Extended Theories of Gravity using Solar System Tests, Gen. Rel. Grav. \textbf{39} (2007) 1381
\href{https://arxiv.org/pdf/astro-ph/0610661.pdf}{[arXiv:0507027 [astro-ph]]}.

\bibitem{reij} V. Reijonen, On white dwarfs and neutron stars in Palatini $f(R)$ gravity, \href{https://arxiv.org/pdf/0912.0825.pdf}{[arXiv:0912.0825 [gr-qc]]}.

\bibitem{panoto} G. Panotopoulos, Strange stars in $f(R)$ theories of gravity in the Palatini formalism,  Gen. Rel. Grav. \textbf{49} (2017) 69
\href{https://arxiv.org/pdf/1704.04961.pdf}{[arXiv:1704.04961 [gr-qc]]}.

\bibitem{YousaffRT}
M. Z. Bhatti, Z. Yousaf, Zarnoor, Stability analysis of neutron stars in Palatini $f(R,T)$ gravity, Gen. Rel. Grav. \textbf{51} (2019) 144.

\bibitem{delsate}  P. Pani, T. Delsate, and V. Cardoso, Eddington-inspired Born-Infeld gravity. Phenomenology of non-linear gravity-matter coupling, Phys. Rev. D \textbf{85} (2012) 084020
\href{https://arxiv.org/pdf/1201.2814.pdf}{[arXiv:1201.2814 [gr-qc]]}.

\bibitem{harko} T. Harko, FSN Lobo, MK Mak, SV Sushkov, Structure of neutron, quark and exotic stars in Eddington-inspired Born-Infeld gravity, Phys. Rev. D \textbf{88} (2013) 044032
\href{https://arxiv.org/pdf/1305.6770.pdf}{[arXiv:1305.6770 [gr-qc]]}.


\bibitem{qauli} A. I. Qauli, M. Iqbal, A. Sulaksono, H.S. Ramadhan, Hyperons in neutron stars within Eddington-inspired Born-Infeld theory of gravity, Phys. Rev. D \textbf{93} (2016) 104056
\href{https://arxiv.org/pdf/1605.01152.pdf}{[arXiv:1605.01152 [astro-ph]]}.

\bibitem{avelino} P. P. Avelino, Eddington-inspired Born-Infeld gravity: astrophysical and cosmological constraints, Phys. Rev. D \textbf{85} (2012) 104053
\href{https://arxiv.org/pdf/1201.2544.pdf}{[arXiv:1201.2544 [astro-ph]]}.

\bibitem{qauli2} A. I. Qauli, A. Sulaksono, H. S. Ramadhan, I. Husin, Compactness, masses and radii of compact stars within the
Eddington-inspired Born-Infeld theory,
\href{https://arxiv.org/pdf/1710.03988.pdf}{[arXiv:1710.03988 [gr-qc]]}.


\bibitem{ozel} F. Ozel and P. Freire, Masses, Radii, and Equation of State of Neutron Stars, Ann. Rev. Astron. Astrophys. \textbf{54} (2016) 401
\href{https://arxiv.org/pdf/1603.02698.pdf}{[arXiv:1603.02698 [astro-ph]]}.

\bibitem{rezz} L. Rezzolla, E. R. Most and L. R. Weih, Using gravitational-wave observations and quasi-universal relations to constrain the maximum mass of neutron stars, Astrophys. J. \textbf{852} (2018) L25
\href{https://arxiv.org/pdf/1711.00314.pdf}{[arXiv:1711.00314 [astro-ph]]}.

\bibitem{prasetyo} I. Prasetyo, Neutron stars in the braneworld within the Eddington-inspired Born-Infeld gravity, JCAP \textbf{2018} (2018) 027
\href{https://arxiv.org/pdf/1708.04837.pdf}{[arXiv:1708.04837 [astro-ph]]}.

\bibitem{castro}  L. B. Castro, M. D. Alloy, D. P. Menezes, Mass radius relation of compact stars in the braneworld,
JCAP 1408 (2014) 047 \href{https://arxiv.org/pdf/1403.1099.pdf}{[arXiv:1403.1099 [nucl-th]]}.


  \bibitem{Carriere:2002bx}
  J.~Carriere, C.~J.~Horowitz and J.~Piekarewicz,
  Low mass neutron stars and the equation of state of dense matter,
  Astrophys.\ J.\  {\bf 593} (2003) 463
  \href{https://arxiv.org/pdf/nucl-th/0211015.pdf}{ [arXiv:nucl-th/0211015]}.

  \bibitem{Danarianto:2019mxf}
  M.~D.~Danarianto and A.~Sulaksono,
  Overturning and apparent anisotropic pressure in Eddington-inspired Born-Infeld theory on compact stars,
  Phys.\ Rev.\ D {\bf 100} (2019) 064042.

\bibitem{feng} W-X. Feng, C-Q. Geng, L-W Luo, The Buchdahl Stability Bound in Eddington-inspired Born-Infeld
Gravity, Chin.\ Phys.\ C {\bf 43} (2019)  083107 \href{https://arxiv.org/pdf/1810.06753.pdf}{[arXiv:1810.06753 [gr-qc]]}.

\bibitem{hybr} B. Danila, T. Harko, F. S. N. Lobo, M. K. Mak, Hybrid metric-Palatini stars, Phys. Rev. D \textbf{95} (2017) 044031
\href{https://arxiv.org/pdf/1608.02783.pdf}{[arXiv:1608.02783 [gr-qc]]}.

\bibitem{chavanis} P. H. Chavanis and T. Harko, Bose-Einstein Condensate general relativistic stars, Phys. Rev. D \textbf{86} (2012) 064011
\href{https://arxiv.org/pdf/1108.3986.pdf}{arXiv:1108.3986 [astro-ph]}

\bibitem{aneta3} A. Wojnar, Polytropic stars in Palatini gravity, Eur. Phys. J. C \textbf{79} (2019) 51
\href{https://arxiv.org/pdf/1808.04188.pdf}{[arXiv:1808.04188 [gr-qc]]}.

\bibitem{artur} A. Sergyeyev, A. Wojnar, The Palatini star: exact solutions of the modified Lane-Emden equation, Eur. Phys. J. C \textbf{80} (2020) 313 \href{https://arxiv.org/pdf/1901.10448.pdf}{[arXiv:1901.10448 [gr-qc]]}.

\bibitem{Olmo:2019qsj}
  G.~J.~Olmo, D.~Rubiera-Garcia and A.~Wojnar,
  Minimum main sequence mass in quadratic Palatini $f(R)$ gravity,
  Phys.\ Rev.\ D {\bf 100} (2019) 044020
  \href{https://arxiv.org/pdf/1906.04629.pdf}{[arXiv:1906.04629 [gr-qc]]}.

\bibitem{Baner} S. Banerjee, S. Shankar, T.P. Singh, Constraints on Modified Gravity models from White Dwarfs, JCAP \textbf{2017} (2017) 004
\href{https://arxiv.org/pdf/1705.01048.pdf}{[arXiv:1705.01048 [gr-qc]]}.

\bibitem{wd1} M. Hicken, P. M. Garnavich, J. L. Prieto, S. Blondin, D. L. DePoy, R. P. Kurshner, J. Parrent, The Luminous and carbon-rich supernova 2006GZ: a double degenerate merger?, Astrophys. J. \textbf{669} (2007) L17
\href{https://arxiv.org/pdf/0709.1501.pdf}{[arXiv:0709.1501 [astro-ph]]}.

\bibitem{wd2} D. A. Howell et al., The type Ia supernova SNLS-03D3bb from a super-Chandrasekhar-mass white dwarf star,  Nature \textbf{443} (2006) 308
\href{https://arxiv.org/pdf/astro-ph/0609616.pdf}{[arXiv:0609616 [astro-ph]]}.

\bibitem{wd3} R. A. Scalzo et al., Nearby supernova factory observations of SN 2007IF: first total mass measurement of a super-Chandrasekhar-mass progenitor,  Astrophys. J. \textbf{713} (2010) 1073
\href{https://arxiv.org/pdf/1003.2217.pdf}{[arXiv:1003.2217 [astro-ph]]}.

\bibitem{wd4} J. M. Silverman, M. Ganeshalingam, W. Li, A. V. Filippenko, A. A. Miller, D. Poznanski, Fourteen Months of Observations of the Possible Super-Chandrasekhar Mass Type Ia Supernova 2009dc,  Mon. Not. Roy. Astron. Soc. \textbf{410} (2011) 585
\href{https://arxiv.org/pdf/1003.2417.pdf}{[arXiv:1003.2417 [astro-ph]]}.

\bibitem{wd5} S. Taubenberger \emph{et al.}, High luminosity, slow ejecta and persistent carbon lines: SN 2009dc challenges thermonuclear explosion scenarios, Mon. Not. Roy. Astron. Soc. \textbf{412} (2011) 2735
\href{https://arxiv.org/pdf/1011.5665.pdf}{[arXiv:1011.5665 [astro-ph]]}.

\bibitem{obse} J. B. Holberg, T. D. Oswalt and M. A. Barstow, Observational Constraints on the   Degenerate Mass-Radius Relation, Astrophys. J. \textbf{143} (2012) 68
\href{https://arxiv.org/pdf/1201.3822.pdf}{[arXiv:1201.3822 [astro-ph]]}.


\bibitem{wibi} C. Wibisono, A. Sulaksono, Information-Entropic Method: Stability of Stars and Modified Gravity Theories, Int. J. Mod. Phys. D, \textbf{27} (2018) 1850051
\href{https://arxiv.org/abs/1712.07587v1}{[arXiv:1712.07587 [gr-qc]]}.


\bibitem{Koyama:2015vza}
  K.~Koyama,
  Cosmological Tests of Modified Gravity,
  Rept.\ Prog.\ Phys.\  \textbf{79} (2016)  046902
  \href{https://arxiv.org/pdf/1504.04623.pdf}{[arXiv:1504.04623 [astro-ph.CO]]}.

 \bibitem{Frusciante}
 N. Frusciante, L. Perenon, Effective Field Theory of Dark Energy: a Review, Phys. Rept. \textbf{857} (2020) 1 \href{https://arxiv.org/pdf/1907.03150.pdf}{[arXiv:1907.03150 [astro-ph.CO]]}.






\end{thebibliography}

\end{document}